\RequirePackage{fix-cm}
\documentclass[natbib,smallextended]{svjour3}   
\smartqed  
\usepackage{xcolor}

\usepackage{graphicx}
\usepackage{amssymb}
\usepackage{amsmath}
\usepackage{appendix}
\usepackage{upgreek}
\usepackage[normalem]{ulem}
\usepackage{rotating}
\usepackage{footnote}
\usepackage[caption=false]{subfig}
\usepackage{wrapfig}
\journalname{The Astronomy and Astrophysics Review}

\def\iaucirc{IAU Circ.}       
\def\aj{AJ}                   
\def\araa{ARA\&A}             
\def\apj{ApJ}                 
\def\apjl{ApJ}                
\def\apjs{ApJS}               
\def\apss{Ap\&SS}             
\def\aap{A\&A}                
\def\aaps{A\&AS}              
\def\mnras{MNRAS}             
\def\nat{Nature}              
\def\pasp{PASP}               
\def\memsai{Mem. Soc. Astr. It.}
\def\jcap{JCAP}
\def\ssr{Sp. Sci. Rev.}
\def\pasj{PASJ}
\def\actaa{Acta Astr.}
\def\zap{ZAP}
\def\sovast{Sov. Astr.}
\def\physrep{Phys. Rept.}
\def\baas{Bull. Am. Astr. Soc.}
\def\jrasc{J. Royal. Astr. Soc. Canada}

\begin{document}

\title{Observations of galactic and extragalactic novae}


\author{Massimo Della Valle and Luca Izzo
}

\authorrunning{M. Della Valle \& L. Izzo} 

\institute{M.~Della Valle \at INAF-Capodimonte, Naples, Italy \& ESO, Garching, Germany \\\email{massimo.dellavalle@inaf.it}
              \and
              L.~Izzo \at DARK/NBI, Copenhagen, Denmark \& IAA/CSIC, Granada, Spain 
}
              

\date{Received: date / Accepted: date}

\maketitle

\begin{abstract}
The recent GAIA DR2 measurements of distances to galactic novae have allowed to re-analyse some properties of nova populations in the Milky Way and in external galaxies on new and more solid empirical bases. In some cases we have been able to confirm results previously obtained, such as the concept of nova populations into two classes of objects, that is \emph{disk} and \emph{bulge} novae and their link with the Tololo spectroscopic classification in \emph{Fe II} and \emph{He/N} novae. The recent and robust estimates of nova rates in the Magellanic Clouds galaxies provided by the OGLE team have confirmed the dependence of the normalized nova rate (i.e., the nova rate per unit of luminosity of the host galaxy) with the colors and/or class of luminosity of the parent galaxies. The nova rates in the Milky Way and in external galaxies have been collected from literature and critically discussed. They are the necessary ingredient to asses the contribution of novae to the nucleosynthesis of the respective host galaxies, particularly to explain the origin of the overabundance of lithium observed in young stellar populations. A direct comparison between distances obtained via GAIA DR2 and Maximum Magnitude vs.\ Rate of Decline (MMRD) relationship points out that the MMRD can provide distances with an uncertainty better than 30\%.
Multiwavelength observations of novae along the whole electromagnetic spectrum,  from radio to gamma-rays, have revealed that novae undergo a complex evolution characterized by several emission phases and a non-spherical geometry for the nova ejecta.


 \end{abstract}

\setcounter{tocdepth}{3}
\tableofcontents

\section{Observations of guest stars}

In the ancient chronicles, particularly in the almanacs of the Chinese astronomers, all ``temporary'' or ``transient'' phenomena that suddenly appeared in the sky, to then fade and disappear after days or months, were generally indicated with the name of ``guest stars'' or ``visitor stars''. This definition was very broad and it was used to classify a large number of astronomical events such as comets, planets in opposition or conjunction, and new stars. In those days there were no clues about the different origin of these phenomena and the common idea was that they were somehow related to each others.  In Europe these events were also reported as ``\emph{Stella Nova}'',  the Latin name for ``New Star'' (Stellae Novae in his plural form). Unlike the locution \emph{Supernova} that is a recent term, coined in the 1930s,  Stella Nova in astronomy was used for the first time about 2000 years ago  in the Natural History of Pliny the Elder. In his second book\footnote{Out of 37 that constitute his entire work} dedicated to Astronomy and Meteorology, Pliny reports:

\begin{quote}
 ``Hipparchus himself \dots detected a ``\emph{new star}'' that led him to wonder whether this was a frequent occurrence''  {\normalfont and} ``he dared to prepare a list of stars for posterity to indicate their positions and magnitudes, in order that from that time onward it might be possible to discern not only whether stars perish and are born, transit and in motion, and also whether they increase and decrease in magnitude''
 \end{quote}

To preserve the Aristotle's dogma on the perfection and immutability of the heavens, in which Stellae Novae obviously did not fit, the ancient astronomers considered nova stars sub-lunar phenomena or even Earth's atmosphere occurrences. Therefore, in this view, these objects did not belong to the ``octave sphere'' the so called ``sphere of fixed stars'': the skies beyond Saturn were indeed immutable by definition. Despite these ideological prejudices, the observation of new stars was never abandoned, and therefore today we know, at least approximately, the positions in the sky of many remnants of those stellar explosions and some of them have been targeted with modern instrumentation (see, e.g., \citealt{Shara2017a}). The reason for which the ancient observers keep alive the interest in observing these phenomena was simple. Until  few centuries ago, very little or even nothing was known about the real nature of the stars and planets. According to the Italian poet, Giacomo Leopardi (1798--1837):
\begin{quote}
``the knowledge of the effects and the ignorance of the causes produced astrology'',
\end{quote}
therefore, we should not be surprised to learn that in those times astrology and astronomy were indistinguishable. They were both part of a culture that considered the earthly events announced in the sky through the manifestation of strange and unusual phenomena, such the sudden appearance of comets or ``new stars''. We can only imagine what was the apprehension or excitement in the naive and unprepared observers of those strange phenomena. Due to the way in which nova stars appeared in the sky, suddenly and without any warning signs, these events were often considered omens of misfortune, so their observational follow-up was extremely important: it could help in predicting the ``unknown'' that a little later  would have happened on Earth. This climate is well represented by the chronicles of observations made in the Benedictine Abbey of St. Gall, today in Switzerland, about the ``nova star'' of 1006 that we know today to have been a Supernova explosion indeed. The monks reported an immense light that \emph{blinded sight} and aroused \emph{a certain fear}. 

Novae and Supernovae were not considered two distinct families of objects until last century.  The rapidly increasing number of nova discovery in Spiral Nebulae due to systematic searches \citep{Ritchey1917b,Ritchey1917a,Ritchey1918,Curtis1917b,Curtis1917a,Pease1917,Duncan1918,Sanford1918,Sanford1919} started to cast the first doubts about a common origin of these phenomena.  In those years, novae played a pivotal role in the debate about the galactic or extra-galactic nature of the nebulae.  For example, in his paper on ``Novae in Spiral Nebulae and the Island Universe Theory'' Heber Curtis at the end of his short note \citep{Curtis1917c} correctly inferred: 

\begin{quote}
``If we assume equality of absolute magnitude for galactic and spiral novae, then the latter, being apparently 10 magnitudes the fainter, are of the order of 100 times as far away as the former.''
\end{quote}

However, this simple conclusion was questioned by the strange case of S Andromedae (And). This \emph{nova} occurred in M31 in 1885 and today is classified as sub-luminous SN-Ia. The 6th magnitude achieved by S And at peak luminosity was much brighter than the apparent magnitudes of all other nova stars in Andromeda Galaxy and in other spiral Nebulae. As noted by \citet{Shapley1917}: 

\begin{quote}
``\dots the average [magnitude] for the eleven [novae] given in the table is at least magnitude $-7$, and the maximum for S Andromedae is at least $-15$, a luminosity  nearly a hundred million times that of our Sun, the equivalent of the light emission of a million stars of zero absolute magnitude, and probably, therefore, much greater than the total light of all the stars seen with naked eye \dots'' 
\end{quote}
an argument that led him to conclude:

\begin{quote}
 `` \dots hence stellar luminosities of these orders seem out of the question and accordingly the close comparability of spirals containing such Novae to our Galaxy appears inadmissible \dots ''.
 \end{quote}
 
\begin{figure}
    \centering
    \includegraphics[width=9cm]{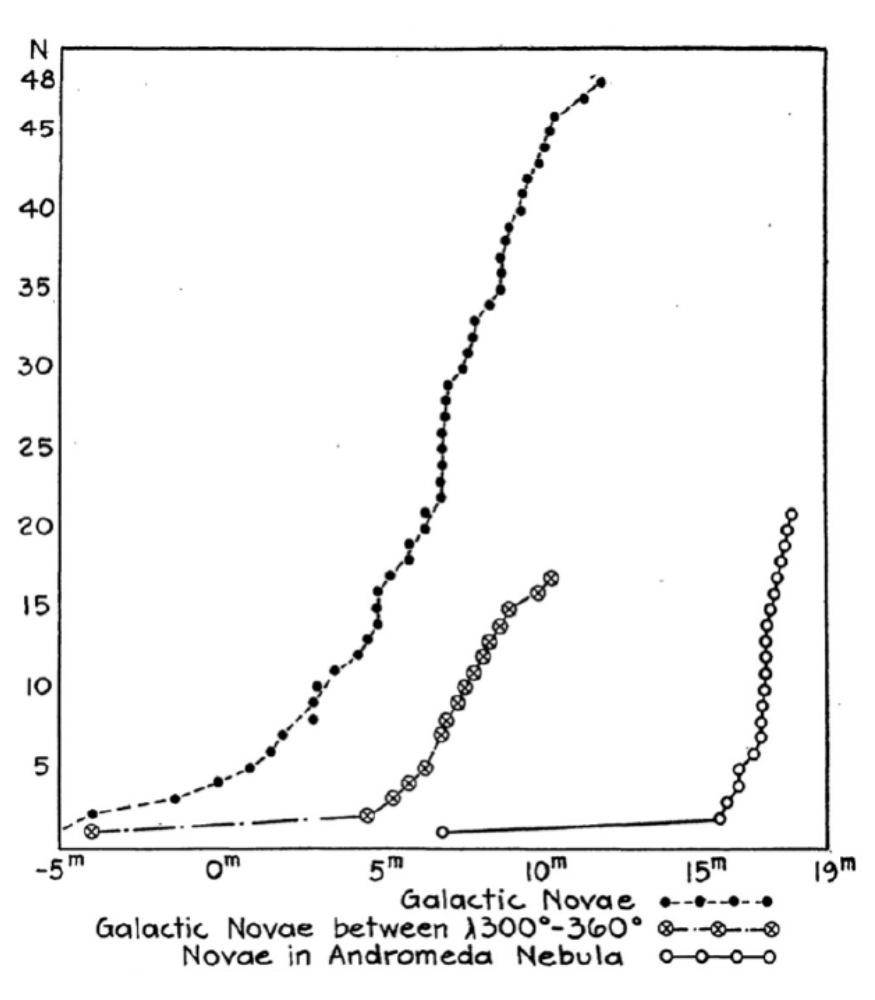}
    \caption{Number of novae brighter than magnitude $m$. Image reproduced with permission from \citet{Lundmark1923}.}
    \label{fig:1}
\end{figure}

For many decades novae have been used as rulers of distance outside our Galaxy \citep{Pritchet1987,DellaValle1995,Shara2018a}, so it appears somehow ironic that the erroneous interpretation of S And in terms of nova explosion has significantly delayed the correct measurements of the distance to M31 and more generally  the correct interpretation of the spiral Nebulae as systems external to the Milky Way. However, this conundrum was taken by \citet{Lundmark1923} the other way around, he assumed that spiral Nebulae were extragalactic systems and in a series of paper published between 1920 and 1923 he realized first that \emph{``\dots Spiral Nebulae were extragalactic objects \dots some Nova got very bright\dots''}, and therefore \emph{``\dots a division into two magnitude classes is not impossible.''} (see Fig.~\ref{fig:1}). 

The brightest class of explosive variables was labelled in the following decades in different ways. \citet{Lundmark1927} called them upper-class novae, while \citet{Hubble1929} calls S Andromedae {\sl exceptional case} and finally \citet{Baade1934} proposed the definition currently adopted of {\sl Super-novae}. In the following years the hyphen has disappeared. \citet{Lundmark1935} pointed out the existence, at the faint tail of the luminosity distribution of novae, of two other groups of ``temporary stars'': the U Gem or Dwarf Novae and nova-like objects. He also provided the first classification in terms of luminosity at maximum for the whole family of explosive variables: 

\begin{itemize}
\item[A.] The upper-class novae or super-novae, having an absolute magnitude at maximum ($M_{\max}$)   around $-15$ mag;
\item[B.] The middle-class novae, or ordinary novae, with an absolute magnitude at maximum very close to $-7$ mag;
\item[C.] The lower-class novae, or the dwarf novae, having an absolute magnitude at maximum around +3 mag or +4 mag.
\end{itemize}

Lundmark also provided the first estimate of their frequency of occurrence in the Milky Way: 1 in 50 years for super-novae and about 50 ordinary novae per year.  \emph{``Thus the ordinary Novae are at least 2500 times more frequent than the super-Novae.''} We note that current estimates of this ratio for spirals galaxies range between $2 \times 10^3$--$2 \times 10^4$ \citep{DellaValle1994}.

\subsection{First theories} 

Since the times of Riccioli (1596--1671), Hevelius (1611--1687) and Fabricious (1564--1617) many theories were developed to explain the nature of the nova phenomenon.  Gian Battista Riccioli,  in the ninth book of its encyclopedic work ``\emph{Almagestum Novum}'', dedicated to ``Comets and New Stars'', listed a dozen of  theories to explain the nature of these transients phenomena. At that time the physical understanding of nova explosions was completely inaccessible, so it is not surprising that the theories presented in the Riccioli's work appear today as naive and extravagant. For example, it was assumed that these ``temporary'' stars were not ``new'' but in fact they were ``old'' and that they were hidden in the depths of the ether, invisible to us, because of their vast distance. These stars were supposed to be in motion and when they approached the Earth, they became visible and when they moved away they became invisible again. This idea was initially conceived to explain the behaviour of Tycho's new star, but it was also used to explain the appearance and disappearance of periodic stars as Mira Ceti.  Many other theories were devised in the following centuries and in some cases they were characterized by more physical insight with respect to the oldest ones. For example, \citet{Maupertuis1732} suggested that the appearance and disappearance of new stars could be originated by the unbalanced action of planets around stars that were flattened by their very high angular momentum. When the geometrical alignment between the flattened star and the observer turned from edge-on to face-on, the observer could measure a dramatic change in the brightness of the star.  According to \citet{Newton1729}, novae were old stars that had exhausted their reserve of light and vapors and therefore were not anymore visible in the sky. They could have returned to ``new splendor'' after being fed with fresh supply of gas provided by comets that fell upon them.  In this case the idea that a nova was an ``old'' star that could be rejuvenated or reinvigorated by gas coming from another external celestial body, appears remarkable and not distant, at least in its general principle, from modern ideas, in which the fuel source are not comets but a companion star. Finally, we note the very imaginative hypothesis elaborated by \citet{Monck1885} to explain the nature of S Andromedae:

\begin{quote}
``Gaseous nebulae sometimes occupy degrees in the sky \dots The rush of a star through a gaseous nebula is therefore no improbable event and it is certain that if occurred the star would be highly heated, and its light greatly intensified. An idea that was borrowed by shooting stars events that are known to be dark bodies rendered luminous for a short time by rushing through our atmosphere \dots ''
\end{quote}

\subsection{Modern times}\label{sec:1.2}

We had to wait for the 20th century, particularly the 1950s and 1960s to come to understand the nova phenomenon in its general lines. The beginning of ``modern times'' might be set by the work of \citet{Schatzman1951} who highlighted the role of $^3$He in triggering ``\emph{une onde de detonation}'' that might be at the origin of the nova phenomenon. The following steps, i.e., the discovery that DQ Her was a binary system \citep{Walker1954} and that the binary condition was common to all cataclysmic variables including novae \citep{Kraft1964}, laid  the physical basis to conclude that nova outbursts  are related to mass accretion phenomena onto the surface of degenerate hot companion \citep{Kraft1965,Paczynski1965}. All previous items, complemented with the first hydrodynamic studies of nova ejections \citep{Sparks1969},  provided the pillars on which the modern studies on nova explosions have then progressed \citep{Starrfield1975,Prialnik1978,Nariai1980}. 

The current framework for Classical Novae  establishes that they are explosive events caused by the ignition of the thermonuclear runaway (TNR) on the surface of a white dwarf that is accreting material, under degenerate conditions, from a low main sequence companion \citep[e.g.][]{Gallagher1978}. Most novae are semi-detached binary star systems where a late-type main sequence star, which has filled its Roche-lobe, transfers mass through the inner L$_1$ Lagrangian point, to the white dwarf companion. When the gas falls onto the compact object, because of its angular momentum with respect to the accreting star, it will not fall directly onto the white dwarf but will start orbiting around it. Gas particles, having different orbits will collide with each other and, shortly after their bulk motion will be circularized through friction and viscosity processes, the gas will eventually form an accretion disk before falling onto the compact object. In passing we note that this process can be partially or totally inhibited by the presence of strong magnetic fields \citep{Wickramasinghe2014}. If the mass accretion rate onto the white dwarf is sufficiently low the accreted gas becomes degenerate, and both pressure and temperature in the accreted envelope will dramatic increase. Finally, TNR will ignite at the base of the accreted envelope leading to a nova eruption. The explosion is not destructive for the white dwarf, therefore nova explosions are ``recurrent'' and the period of time between two explosions can be as short as a few months \citep{Kato2014} and as long as $\sim 10^5$ years. The length of this time scale marks the difference from the so called ``Recurrent Novae'', (RNe) i.e., novae for which at least two outbursts have been recorded (typically RNe show more explosions over the lifetime of an astronomer) and Classical Novae (CNe) characterized by much longer intervals of time between two explosions.  

In a very simplified scheme a white dwarf composed of either carbon-oxygen (CO) or oxygen-neon-magnesium (ONeMg) accretes hydrogen rich material from a companion star at a rate of $10^{-9\pm1}\,M_{\odot}/{\rm yr}$ and as the matter piles up onto its surface, it becomes dense and hot. It is heated at its bottom mainly by gravitational compression. When the pressure and the temperature at the bottom of the accreted layer reaches the critical values of 
\begin{equation}
    P_{\rm crit} = \frac{G M_{\rm WD} \Delta m}{4 \pi R^4},
\end{equation}
of $\approx 10^{19}$--$10^{20}$ dyne cm$^{-2}$ \citep{Fujimoto1982,Truran1986} and T $\sim 10^7$K, the nuclear reactions ignite. In fact theoretical calculation for realistic nova outbursts  show that $P_{\rm crit}$ also depends on other quantities such as $\dot{M}$ and $\Delta M_{\rm ign}$ \citep{Kato2014}. These reactions produce short-lived (about 100s) $\beta^+$ radioactive nuclei that can be transported by convection in the outer envelope. This is efficiently heated up, leading to higher temperatures, which in turn causes more nuclear burning. The temperature eventually rises and causes the softening of the degeneracy of the layers above the nuclear burning and a TNR flash occurs then expelling the ejecta. At the critical pressure, it corresponds a critical ignition mass, which is a function of the radius of WD and its mass:
\begin{equation}
    \Delta M_{\rm ign} \approx \frac{4 \pi P_{\rm ign}}{G}\frac{R_{\rm WD}^4}{M_{\rm WD}}.
\end{equation}

Therefore, for a given $\dot{M}$, the recurrence time:
\begin{equation}
\Delta T_{\rm rec} \sim \Delta M_{\rm ign}/ \dot{M} \sim  R^4/M_{\rm WD}
\end{equation}
is proportional to $R^4/M_{\rm WD}$. 

From the radius of the WD expressed as a function of its mass \citep{Eggleton1983}:
\begin{equation}
    R_{\rm WD} \approx 8.5 \times 10^8 \left[ 1.286 \left( \frac{M_{\rm WD}}{M_{\odot}}\right)^{-2/3} - 0.777 \left( \frac{M_{\rm WD}}{M_{\odot}} \right)^{2/3} \right]^{1/2}{\rm cm} \,,
\end{equation}
we derive that the more massive WDs, e.g., $1.35\,M_{\odot}$, are characterized by recurrence time shorter by a factor 282 with respect to light WDs with masses of $0.6\,M_{\odot}$. This first order computation is in good agreement with the recurrence times scale obtained by several authors \citep{Truran1986,Politano1990,Ritter1991}, see also Table \ref{tab:1}, for which $\Delta T_{0.6}/\Delta T_{1.35} = 322$. 

\begin{table}[]
    \caption{Nova recurrence time scale. Table reproduced with permission from \citet{Truran1986}.}
    \label{tab:1}
    \centering
    \begin{tabular}{c|c}
    \hline\hline
     Mass WD    & Interval \\
      $(M_{\odot})$   & $(10^5)$ years \\
         \hline
       0.60  & 12.9 \\
       0.70  & 7.3 \\
       0.80  & 4.2 \\
       0.90  & 2.4 \\
       1.00  & 1.2 \\
       1.10  & 0.64 \\
       1.20  & 0.28 \\
       1.30  & 0.09 \\
       1.35  & 0.04 \\
       \hline
    \end{tabular}
\end{table}

\begin{figure}
    \centering
    \includegraphics[width=9cm]{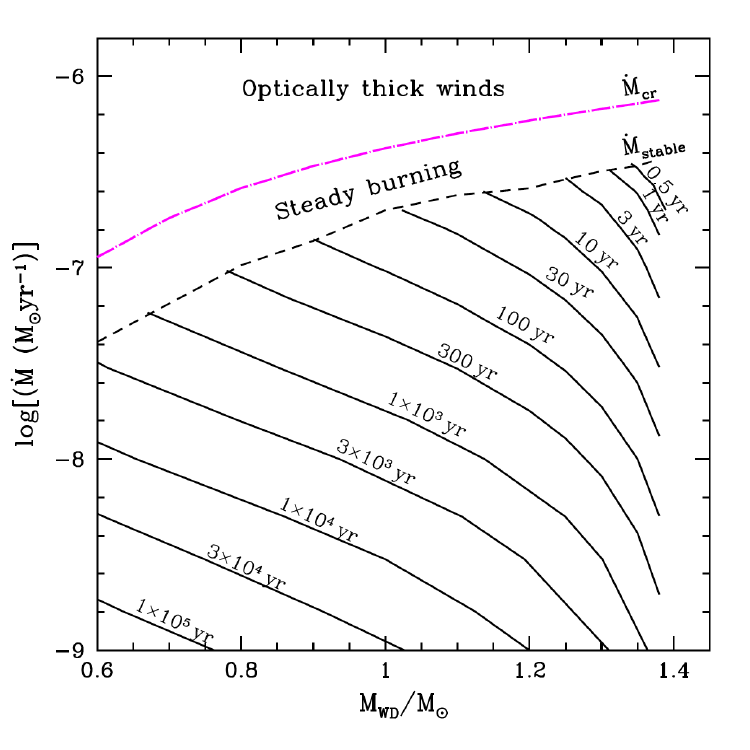}
    \caption{The distribution of recurrence periods of classical novae compared with the accretion rate and the mass of the primary white dwarf, according to the simulations by \citet{Kato2014}.}
    \label{fig:Kato2014}
\end{figure}

\citet{Yaron2005} have computed an extended grid of multi-cycle nova evolution models and obtained recurrence times at varying  the mass of and temperature of the WD and $\dot{M}$. For a typical value of $\dot{M} = 10^{-9} M_\odot/{\rm yr}$ we consistently derive a ratio of 390. Accurate grids provided by \citet{Kato2014} computed for a range  $10^{-6} < \dot{M}< 10^{-9} M_\odot/yr$ give consistent results within a factor about two for $\dot{M}\approx 10^{-9} M_\odot/yr$, see Fig.~\ref{fig:Kato2014}. Now we can estimate on the basis of simple arguments the duration of the nova duty cycle. A system formed by a low main sequence companion of 0.5$M_{\odot}$ can experience about 0.5 $M_{\odot} \times \alpha / M_{\rm env}$ outbursts. The parameter $\alpha$ is a coefficient that accounts for the efficiency of the accretion through the inner Lagrangian point, $L_1$, which is estimated to be of the order of 20\% (see Sect. 12). For $M_{\rm env}\sim 10^{-5}\,M_{\odot}$ and $\sim 10^{-4}\,M_{\odot}$ for fast and slow novae (see Sect. 10) we derive about $10^{4\div 3}$ outbursts during a nova cycle, consistently with the ``classical'' estimates \citep{Bath1978}. 

\begin{figure}
    \centering
    \includegraphics[width=9cm]{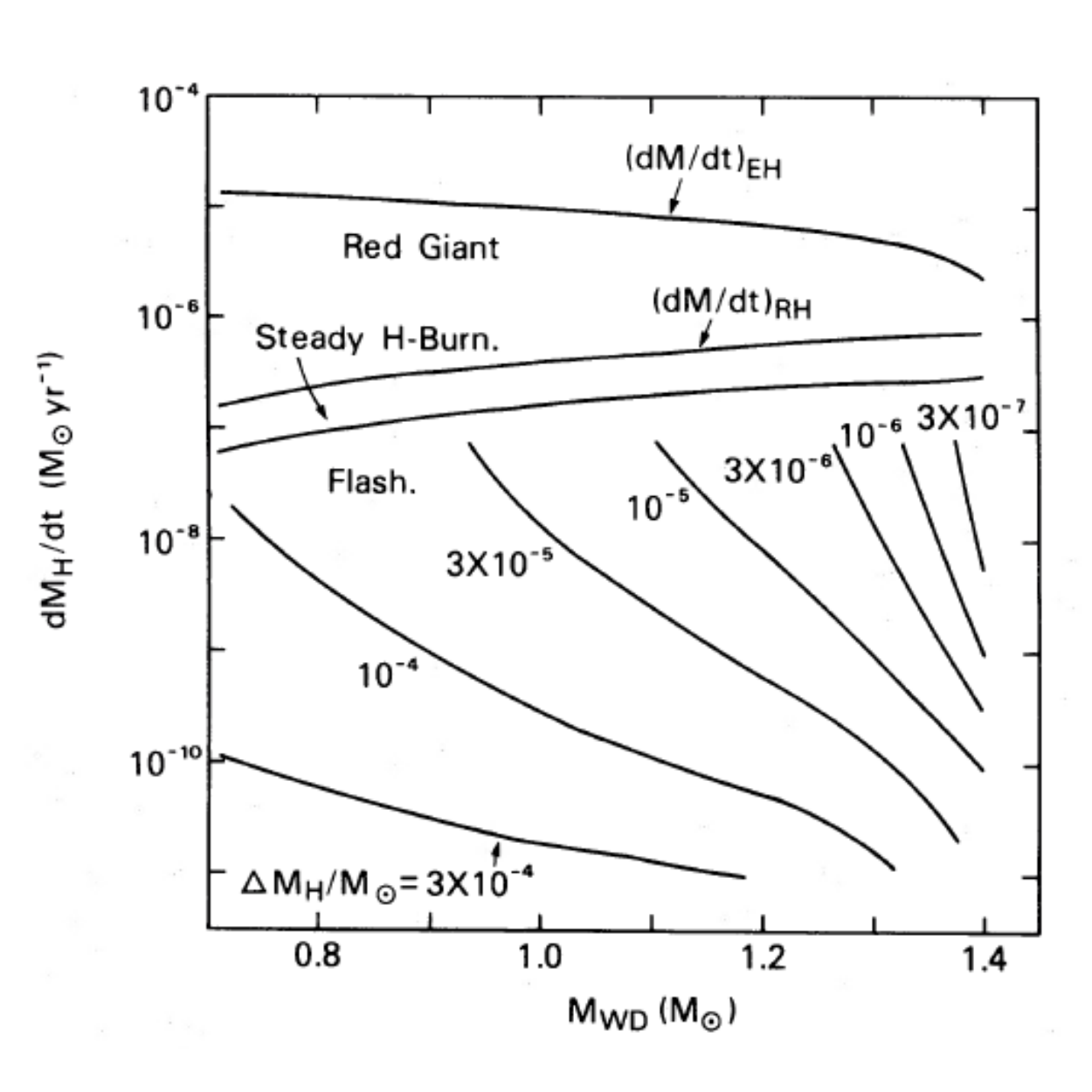}
    \caption{Masses of the accreted envelopes as a function of the mass of the WD and $\dot{M}$. Image reproduced with permission from \citet{Nomoto1982a}.}
    \label{fig:2}
\end{figure}

\begin{figure}
    \centering
    \includegraphics[width=9cm]{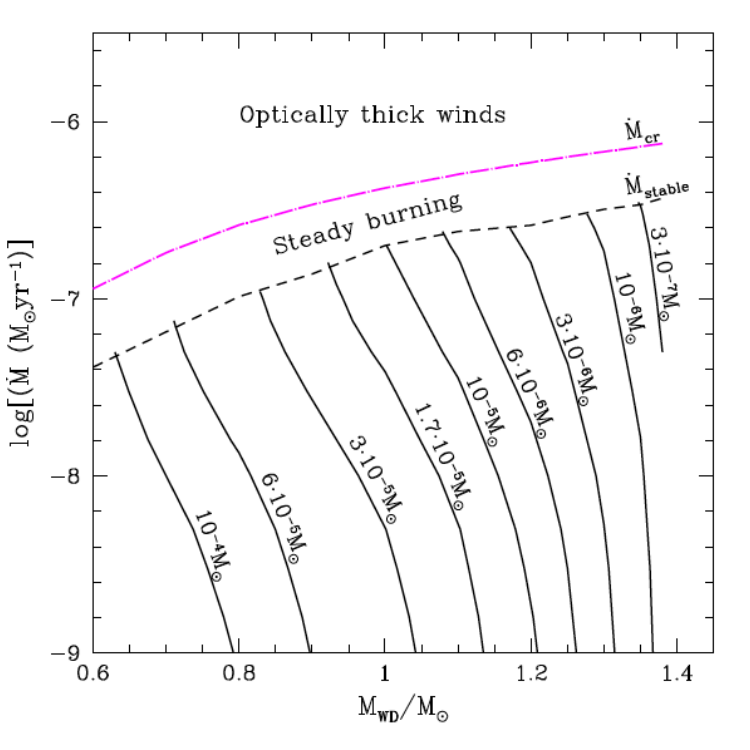}
    \caption{The distribution of the ignition mass as a function of the accretion rate and the mass of the primary white dwarf, according to the simulations by \citet{Kato2014}.}
    \label{fig:Kato2}
\end{figure}

Thus, the mean lifetime for nova systems turns out to be $10^{4\div 3} \times \Delta T_{\rm rec}$, therefore $\sim  10^7$ years and $ \sim 10^9$ years for massive and light nova progenitors, respectively.  The connection among $M_{\rm WD}$, $\dot{M}$, and M$_{\rm env}$ was explained by several authors e.g. \citet{Nomoto1982a} (Fig.~\ref{fig:2}) and revised in recent years by \citet{Kato2014} (Fig.~\ref{fig:Kato2}). The more massive WDs expel the lightest envelopes with the highest ejection velocities, up to 4000--5000 km/s in the most extreme cases. In many post-novae a ``quiet'' H-burning continues in a kind of steady state then originating the so called SuperSoft X-ray phase \citep{Ogelman1984,Orio2001,Henze2010}, see also Sect. 12.2.

The physical processes regulating the ejection of the gas are still matter of debate. The entire evolution of a nova outburst can be explained in terms of a continuum radiation wind \citep{Kato1994} or through an homologous ballistic ejection of a discrete shell \citep{Shore1996}. More recently, the detection of high-energy radiation from novae combined with radio and optical high-resolution  observations led to the proposal of a ``multi-phase'' ejection mechanisms for novae. The current evidence is consistent with an early time ballistic ejection, with a toroidal geometry distribution and a late time bi-polar faster wind \citep{Mukai2019}. Their interaction would give rise to shocks and then to the observed high-energy radiation \citep{Metzger2015}.

\subsection{Classical novae as cataclysmic variables}

Classical Novae fall in the faint tail of stellar explosions that occur in galaxies.  With an emitted radiant and kinetic energies of about $10^{45}$ erg \citep{Gallagher1978}.  They are many orders of magnitude less energetic than gamma-ray bursts \citep{Amati2006}, Kilonovae \citep{Abbott2017} and Supernovae \citep{Gal-Yam2012}.  In terms of magnitude at maximum these eruptions can reach an absolute magnitude as bright as nova LMC 1991, $M_V \approx$ -10 mag \citep{DellaValle1991,Schwarz2001}, approximately one order of magnitude less luminous than Luminous Red Novae \citep{Pastorello2019}. 

Novae belong to the very heterogeneous family of semi-detached binaries and different phenomena are observed according to the nature of the receiving star. If it is a main sequence (MS) star, we have an Algol-type eclipse system characterized by periods of hours/days. If it is a collapsed object, as a white dwarf (WD), we enter in the heterogeneous realm of Cataclysmic Variables  (CVs), normally characterized by periods of few hours (but see GK Per, \citealt{Bianchini1981}). With the limits, that are inevitable when we proceed to a schematization, we can broadly divide the CVs family in different sub-classes according to masses of the companion stars and the amount of mass transfer rates. Decades of studies on CV properties effectively summarized by \citet{Warner2003} converge toward the following ``outline'':  

\begin{itemize}
\item if the donor star is a low mass MS star and the receiving star is a WD $(0.7\,M_{\odot})$ and $\dot{M}\sim 10^{-11}\,M_{\odot}/{\rm yr}$, the binary system is labeled as Dwarf Nova (DN); 
\item if the donor star  is a low mass MS star and the receiving star is a high mass WD $(0.7\,M_{\odot} < M < 1.1 M_{\odot})$ and $\dot{M} \sim 10^{-9}$--$10^{-10}\,M_{\odot}/{\rm yr}$, the binary system is a  Classical Nova (CN);
\item if the donor star  is a MS star (or a giant) and the receiving star is a high mass WD ($\sim 1.3\,M_{\odot}$ ) and $\dot{M}$ $\sim$ $10^{-7}$--$10^{-8}\,M_{\odot}/{\rm yr}$ the system is a Recurrent Nova (RN);
\item if the donor star is a low mass MS star and the receiving star is a  NS/BH, then, the system is a Low Mass X-ray Binary (LMXB);
\item if the donor star is a high mass MS (e.g., blue giant) star and the receiving star is a  NS/BH, the system is a High Mass X-ray Binary (HMXB);
\item if the donor star  is a low MS star and the receiving star is a very high mass WD ($> 1.3\,M_{\odot}$ ) the transfer can bring the WD across the Chandrasekhar limit and the WD explodes as SN~Ia \citep{Whelan1973,Nomoto1982a,Nomoto1982b,Starrfield2019};
\item if the donor star is another WD, the coalescence can bring the binary system across the Chandrasekhar limit and the double degenerate (DD) system will explode as SN~Ia \citep{Iben1984,Iben1985}. As an alternative the coalescence of two WDs may not result in a SN-Ia explosion, but eventually producing an \emph{``accretion-induced-collapse''} to a neutron star \citep{Nomoto1991}.
\item if the donor star  transfer Helium onto the WD it can explode as SN~Ia by igniting off-center at sub-Chandrasekhar masses \citep{Yungelson1998},
or feeding the WD in a Helium-rich supersoft X-ray system (\citealt{Kato1989}, \citealt{Hachisu1999}) and typified by the Recurrent Nova U Sco,  in which the WD grows in mass to the Chandrasekhar limit and may explode as SN Ia \citep{Kato2012}.

In the 1980s, it was proposed \citep{Shara1986} an unified scenario in which Classical Novae and Dwarf Novae were the same systems observed at different evolutionary stages (see Sect. 3.3).

With ``nova-like'' we label all poorly studied cataclysmic variable stars which resemble novae due to their photometric behaviour at maximum or minimum or due to their spectral development. It is not infrequent that as soon as more observations are collected it is possible to reclassify some ``nova-like'' objects as belonging to one of the above reported classes.

\end{itemize}

\section{Introduction}

\begin{figure}
    \centering
    \includegraphics[width=10cm]{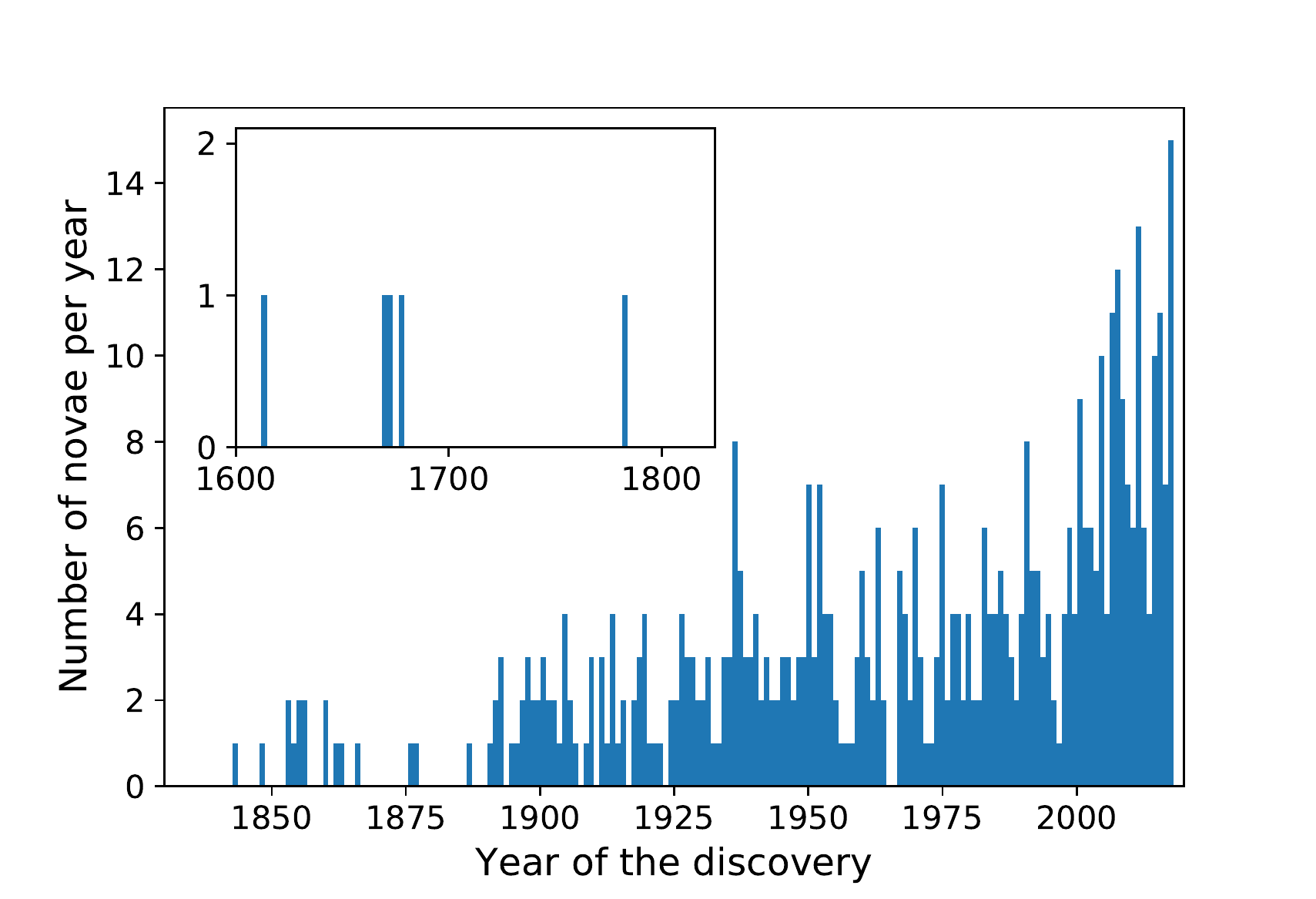}
    \caption{The number of novae vs.\ year of discovery. The total number of novae plotted in this figure can be found at the \emph{Project Pluto} website (\url{https://projectpluto.com/galnovae/galnovae.txt} maintained by Bill Gray).}
    \label{fig:3}
\end{figure}

About 400 novae have been observed in the Milky Way since CK Vul 1670 and WY Sge 1783, but only a fraction of them has been observed into some details sufficient to carry out interesting analysis (Fig.~\ref{fig:3}).

Collections of nova data are periodically updated, reviewed and enriched, see for example \citet{Cecchini1942,Duerbeck1981,Duerbeck1987,Ozdonmez2018,Hachisu2018,Schaefer2018} -- S18 hereafter -- and \citet{Selvelli2019} -- SG19 hereafter -- and some data from these papers have been used in this work. Reviews on novae have not been very popular since the exhaustive paper by \citet{Gallagher1978}. However, we notice there has been an increased interest in this field: the books ``Cataclysmic Variable Stars'' by \citet{Warner2003} and ``Classical Novae'' by \citet{Bode2008}, and the more recent reviews by \citet{Woodward2011}, \citet{Poggiani2018}, \citet{Darnley2019} and \citet{Shafter2019}. 

Spectroscopic and photometric data for historical novae have been extracted from the classical books ``\emph{The Galactic Novae}'' by \citet{Payne1957} and ``\emph{Classical Novae}'' by \citet{Bode2008}, while more recent data have been directly derived from literature (e.g., \citealt{Ribeiro2013,Munari2017}). However, both historical and new records have been revised according to the Tololo Nova Classification in the spectroscopic ``types'' Fe II and He/N Novae \citep{Williams1991,Williams1992,Williams1994a,Williams1994b}.  The modern spectroscopic classification, usually based on low-intermediate resolution, was subsequently integrated by high resolution observations (R $\sim$ 50,000) carried out at ESO-La Silla at the end of 1990s for Milky Way and Magellanic Clouds Novae \citep{DellaValle2002,Mason2005}. These high resolution data set have provided the observational basis for the detections in the nova ejecta of the so called Transient Heavy Element Absorptions (THEAs, \citealt{Williams2008}) and $^7$Li/$^7$Be \citep{Izzo2015,Tajitsu2015,Molaro2016,Izzo2018,Molaro2020}. Lithium and Berillium were ones of the most important and puzzling missing pieces of chemical evolution of the Milky Way \citep{Romano2003} and  their detection have significantly contributed to assess the role of novae in the Galactic Nucleosynthesis.

Systematic surveys to search for novae in external galaxies have a long historical record. Pioneer surveys on photographic plates were mostly devoted to study the nova population in M31 \citep{Hubble1929,Arp1956,Rosino1956,Rosino1964,Rosino1973,Rosino1989}, M33 \citep{Rosino1973b}, and in the LMC \citep{Graham1979}. We note that the former surveys were often supplemented by more fragmentary observations \citep{SharoV1979} or by ``new concept'' surveys, as the one carried out on the M31 bulge in H$\alpha$ narrow band filter with one-meter class telescopes equipped with CCD detectors \citep{Ciardullo1987}. Photographic  data obtained in the 1950s Palomar campaign for the discovery of classical novae in M81 and in the early 1970s at Asiago Observatory for studying novae in M33, were published only in the 1990s \citep{Shara1999,DellaValle1994b}. The modern search for novae in extra-galactic systems has received a significant boost of interest after the seminal paper by \citet{Pritchet1987} dedicated to study novae in the Virgo Cluster. The paper was a ``classical'' search for novae in B filter, carried out with a four-meter class telescope (the Canada-France-Hawaii Telescope, CFHT) equipped with CCDs. A work that, despite of only 9 novae discovered in 15 half-nights run,  has considerably influenced the subsequent studies in the extragalactic nova topics including the distance scale debate \citep{Jacoby1992,vandenBergh1996}.  In the following years surveys for nova searches in extragalactic systems (corroborated by some spectroscopy) have been carried out in M31 \citep{Tomaney1992,Darnley2004,Darnley2006,Shafter2001,Shafter2011,Kasliwal2011,Cao2012}, M51 and M101  \citep{Shafter2000}, in NGC 1316 \citep{DellaValle2002b}, M49 \citep{Ferrarese2003,Curtin2015}, M33 \citep{Shafter2012}, M87 \citep{Shafter2000,Madrid2007,Curtin2015,Shara2016}, M84 \citep{Curtin2015}, NGC 2403 \citep{Franck2012}. Finally, we point out the first homogeneous survey on novae in the Magellanic Clouds \citep{Mroz2016} since \citet{Graham1979}. Sporadic observations carried out on M32 and NGC 205 \citep{Neill2005} have discovered one or possibly two nova outbursts. 

The structure of this review is as follows: Sect. 3, 4 and 5 discuss the properties of the Galactic nova populations, their photometric and spectroscopic behaviours, their locations inside the Milky Way, their rate of occurrence and the Maximum Magnitude vs. Rate of decline Relationship (MMRD). Sect. 6 makes a summary of the main nova surveys in the Local Group galaxies and beyond, up to Virgo and Fornax clusters and discusses their frequency of occurrence. Sect. 7 is dedicated to estimate the nova rate in the Globular Cluster systems. Sect. 8 compares the MMRDs in different Hubble type galaxies and discusses the outliers of the MMRD. In Sect. 9 we calibrate the LMC, M31 and Virgo MMRD relations with GAIA Data Release 2 (DR2 -- \citealt{GAIA2018}) data to determine the distances toward the respective hosts. Sect. 10 dwells on the contribute of novae to the chemical evolution of the host galaxy. Sect. 11 discusses the role of CNe as possible progenitors of SNe-Ia. In Sect. 12 we briefly report on the observations of novae at different wavelengths (radio, X- and $\gamma$-rays) and, finally, in Sect. 13 we give our conclusions.    

\section{Nova populations in the Milky Way}

\citet{Baade1944,Baade1957} introduced the concept into astronomy that different kinds of stellar populations have different spatial distribution within the galaxies, therefore the starting point to search for a link between speed classes and properties of the nova progenitors is to establish the spatial distribution of novae inside the respective parent galaxies. Due to their intrinsic luminosity, novae can be easily identified in the Milky Way and in external galaxies, thus novae are ideal tracers of their parent stellar population. However, in spite of these favorable conditions, the nova parent population is still an open question.  Historical data on galactic novae \citep{Hubble1929,McLaughlin1942,McLaughlin1945a,McLaughlin1946,Payne1957} have received discrepant interpretations. According to Hubble:

\begin{quote}
``In general, the distribution of novae follows the distribution of luminosity in the nebula. The concentration in the nuclear region is conspicuous.''
\end{quote}

\begin{figure}[htb]
    \centering
    \includegraphics[width=10cm]{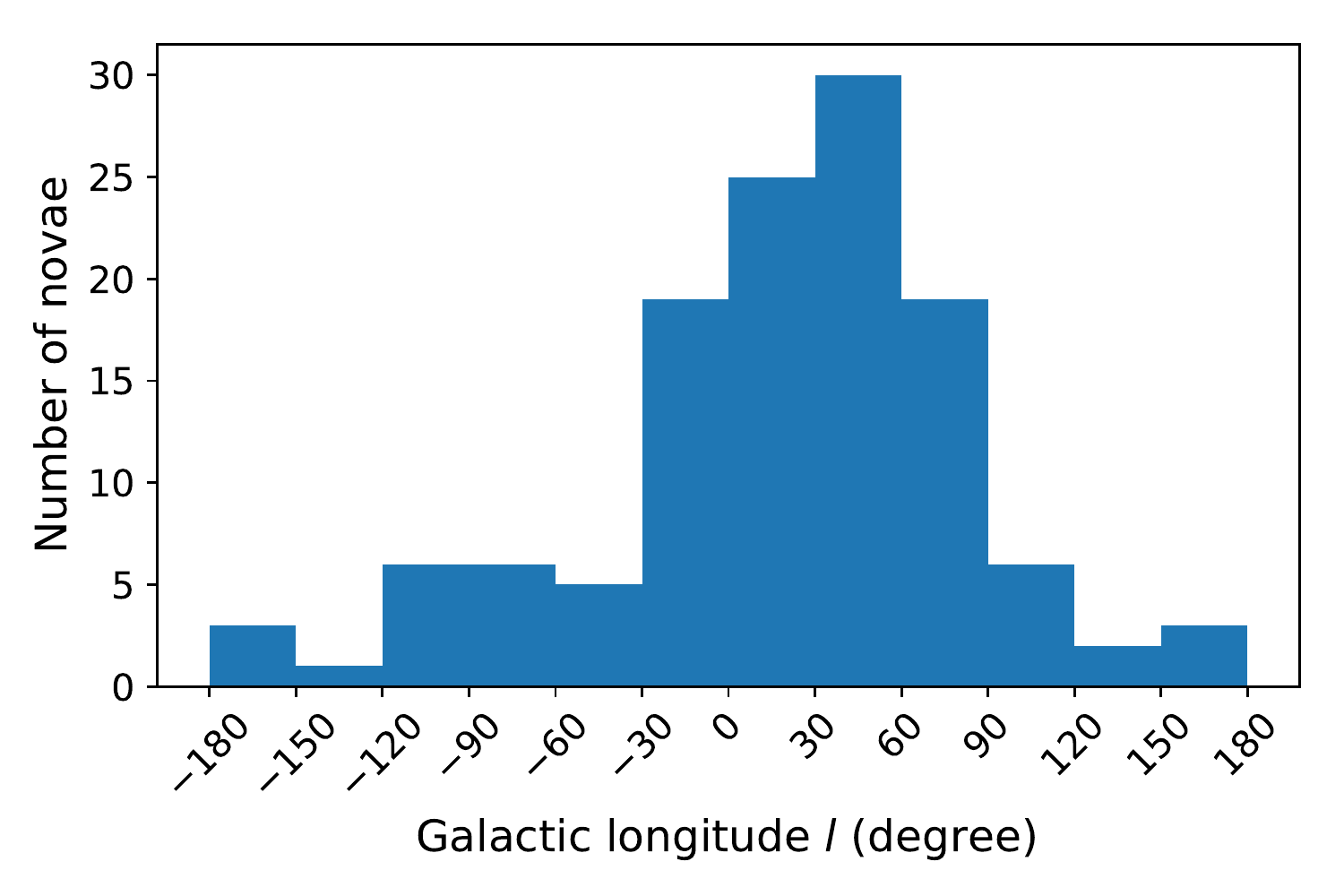}
    \caption{Distribution of the galactic longitudes of novae inside the Milky Way. Data from \citet{Ozdonmez2018}}
    \label{fig:4}
\end{figure}

\begin{figure}[htb]
    \centering
    \includegraphics[width=10cm]{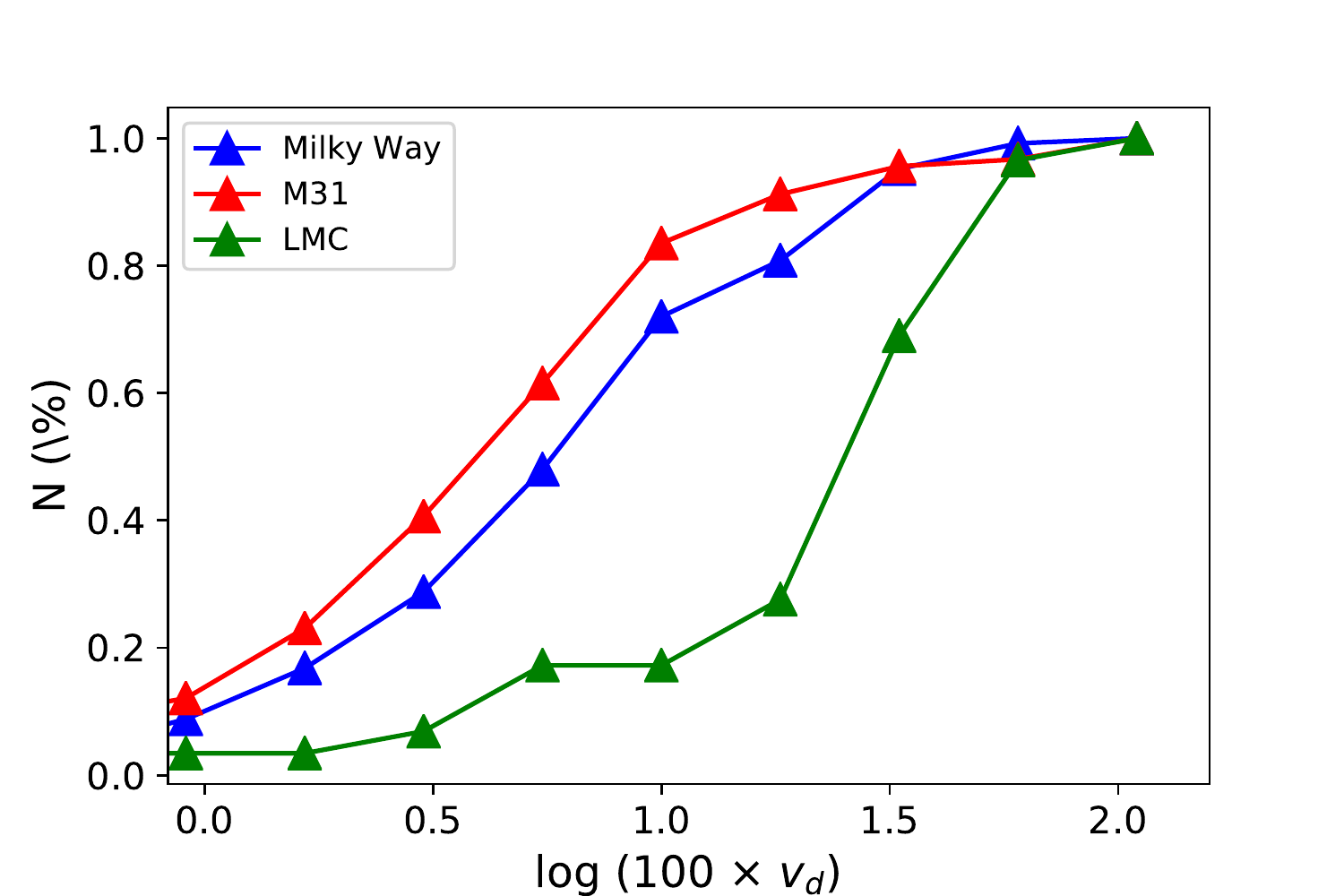}
    \caption{Distributions of the rates of decline for Milky Way, M31 and LMC. MW data \citep{Ozdonmez2018}, M31 data \citep{DellaValle1995}, LMC data \citep{DellaValle1995,Shafter2013}.}
    \label{fig:5}
\end{figure}

Modern data confirm the strong concentration of novae in the direction of the galactic bulge (Fig.~\ref{fig:4}). \citet{Kukarkin1949,KopyloV1955} and \citet{Plaut1965} pointed out the existence of a thickening of novae towards the galactic plane and the galactic center, and the latter author classified novae as belonging to the `disk population'. \citet{Minkowski1948} and \citet{Payne1957} showed that the galactic longitudes of novae and planetary nebulae (PNe) have similar distributions and therefore novae, like PNe, should belong to Pop II stellar population. \citet{Baade1958} assigned novae to Pop II stellar population because of the occurrence of a few ones, like T Sco 1860, in very old stellar population systems of the Globular Clusters M80. \citet{Iwanowska1961} suggested that novae are a mixture of Pop I and Pop II objects. \citet{Patterson1984} pointed out that novae belong to an 'old disk' population. Observations of novae in extra-galactic systems have helped to clarify the picture: \citet{DellaValle1993} compared the cumulative distributions of the rates of decline for M31, LMC and Milky Way nova populations and found that galactic and M31 distributions are indistinguishable, whereas M31 and LMC distributions do not come from the same population at $\gtrsim 99\%$ significance level, see Fig.~\ref{fig:5}.  

Since the speed class of a nova is related to some of the  most important properties characterizing a nova system, like the mass of the WD and the amount of the ejected matter (see Sect. 3.2), the lack of systematic differences in the distributions of the rates of decline between Milky Way and M31 novae would suggest that they come from similar parent stellar populations, which are different from the one in which LMC novae are produced. On the other hand LMC is a bulge-less galaxy and, therefore, novae in the LMC originate from disk population by definition. Therefore, it seems logical to mainly associate novae in M31 and in the Milky Way to a bulge stellar populations. This conclusions is strengthen by spectroscopic observations of ``disk'' novae in the M33 and LMC \citep{Shafter2012,Shafter2013}, which show remarkable differences compared with ``bulge'' M31 novae \citep{Shafter2011} (see also Sect. 6).

\subsection{Nova lightcurves: the speed classes}

\begin{figure}[htb]
    \centering
    \includegraphics[width=10cm]{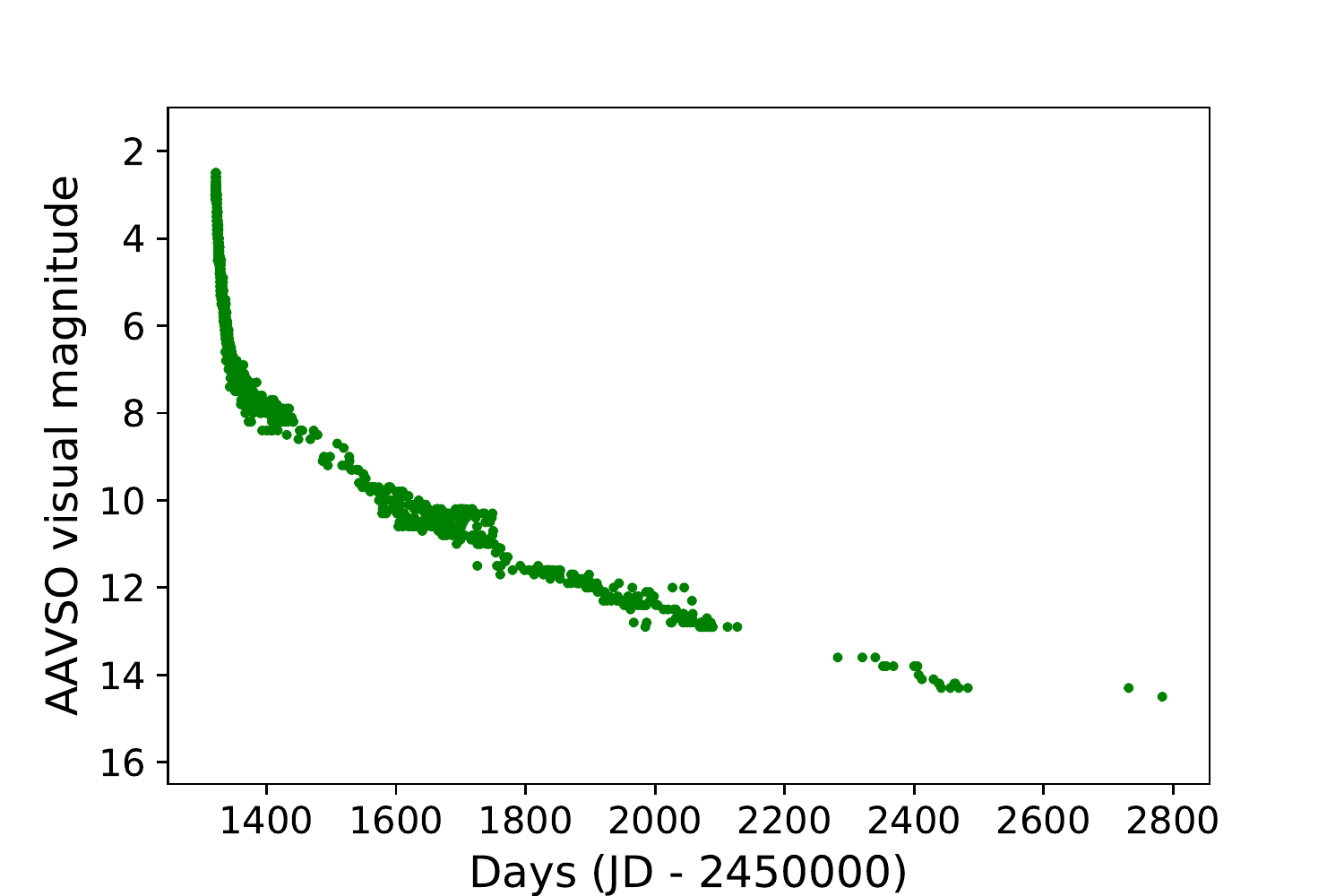}
    \caption{Lightcurve of V382 Vel (Nova Velorum 1999) characterized by a smooth decline. Data obtained from the AAVSO Data Archive.}
    \label{fig:6}
\end{figure}

\begin{figure}[htb]
    \centering
    \includegraphics[width=10cm]{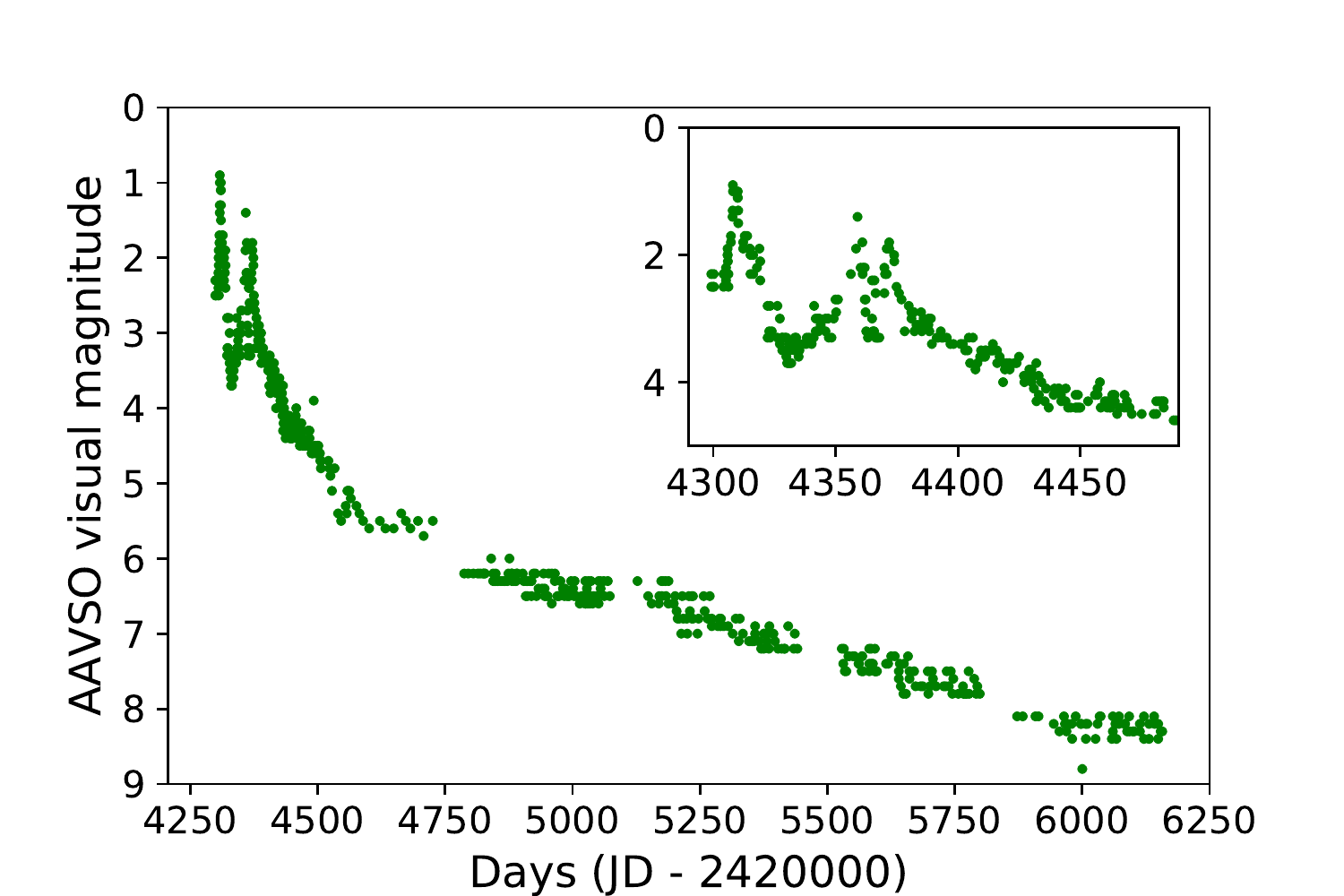}
    \caption{Lightcurve of RR Pic (Nova Pictoris 1925) characterized by a slow rising to maximum light and slow decline exhibiting secondary maxima. Data obtained from the AAVSO Data Archive.}
    \label{fig:6b}
\end{figure}

\begin{figure}[htb]
    \centering
    \includegraphics[width=10cm]{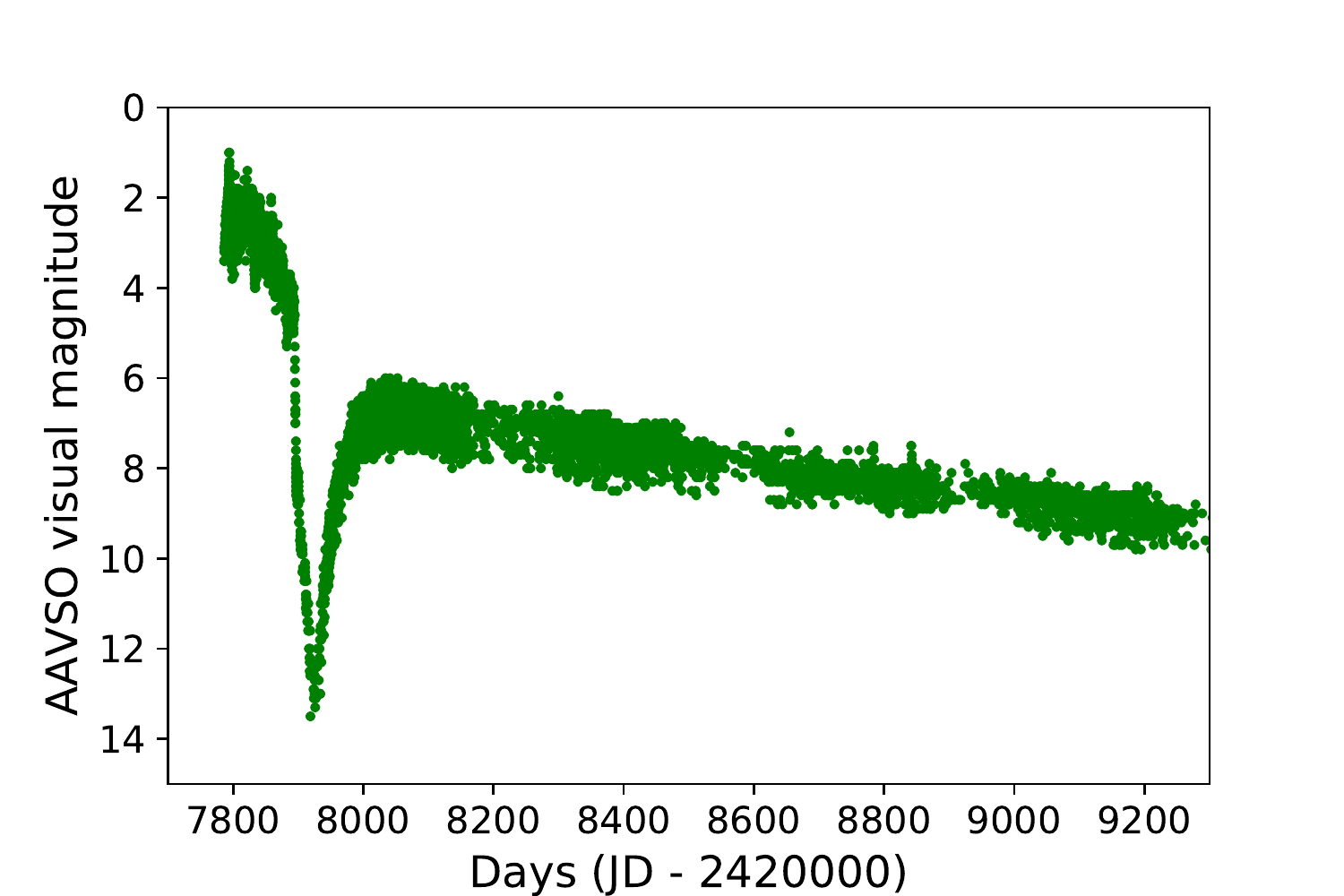}
    \caption{Lightcurve of DQ Her (Nova Herculis 1934) characterized by a slow decline with oscillations, followed by a deep trough due to dust formation. Data obtained from the AAVSO Data Archive.}
    \label{fig:8}
\end{figure}

\begin{figure}[htbp]
    \centering
    \includegraphics[width=9cm]{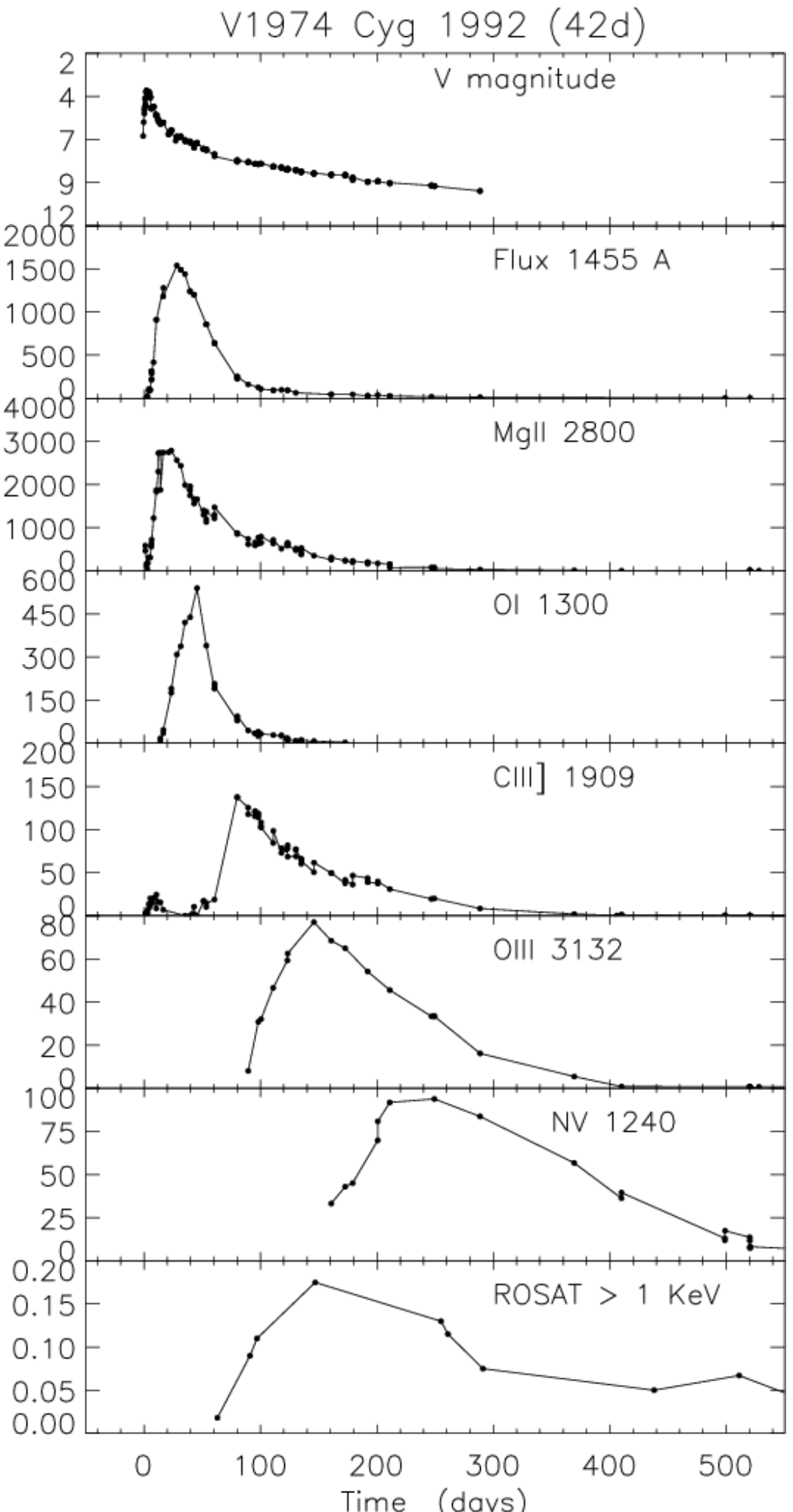}
    \caption{Lightcurve of V1974 Cyg (Nova Cygni 1992) as observed in the optical (upper panel), in the 1455\,{\AA} UV continuum emission (second panel from the top) and in several UV emission line fluxes (third to seventh panels from the top). The last panel shows the evolution in X-rays as observed from ROSAT \citep{Cassatella2004}.}
    \label{fig:V1974Cyg}
\end{figure}

The systematic studies of novae aimed at measuring their distances and luminosity at maximum, date back to the 1920s. \citet{Lundmark1922} derived an average absolute magnitude at maximum of $M_V = -6.2$ mag, but characterized by a broad dispersion, typified by the cases of Nova Lac 1910 and T Sco 1860, with  $M_V =-1.1$ mag and $M_V =-9.1$ mag, respectively. In a following paper, \citet{Lundmark1923} revised the previous value by averaging the distances obtained for about two dozen of novae with four different methods and he obtained $\langle M_V \rangle =-7.2$ mag, not far from $\langle M_V \rangle = -7.5$ mag found in modern studies (SG19). Novae normally show a fast rise to maximum light of the order of few hours, but in some case may last up to several days (e.g., RR Pictoris). The timescale and the morphology of lightcurves during the decline exhibit a very high degree of individuality (see Figs.~\ref{fig:6}, \ref{fig:6b} and \ref{fig:8}). In recent years, the possibility of using multi-wavelength observations led to important developments in understanding the evolution of nova outbursts. Novae lightcurves observed at different wavelengths (optical, UV and X-ray) exhibit different phenomenology. The ionisation of the ejecta generally increases with time and the peak of the emission shifts from optical to UV and finally to X-rays. A clear example is V1974 Cyg in Fig.~\ref{fig:V1974Cyg}, which reports the evolution of the emission as observed in different wavelengths \citep{Cassatella2004}.

In the attempt to understand the link between timescale and morphology, several authors have tried to classify the nova light curve on the basis of either the rates of decline, expressed in magnitude per day \citep{McLaughlin1939,McLaughlin1942,Payne1957} or by grouping novae in several families \citep{Duerbeck1981} according to common behaviors exhibited by lightcurves after maximum light. For example, class A of \citet{Duerbeck1981} was defined as ``smooth fast decline'', class C was defined as ``extended maximum, deep minimum'' and class E, defined as ``extremely slow nova with irregular lightcurve''. The classification in speed class, reported here (see Table~\ref{tab:2}) is the one adopted by \citet{Payne1957}.

\begin{table}
    \caption{The nova speed classification as reported in \citet{Payne1957}. }
    \label{tab:2}
    \centering
    \begin{tabular}{c|c|c}
    \hline\hline
     speed class &  $t_2$ & mag/day \\
         \hline
        very fast           &  $10^d$ & $>0.2$ \\
        fast                &  $11^d-25^d$ & 0.18 -0.08  \\
        moderately fast     & $26^d-80^d$  & 0.07-0.025   \\
        Slow                & $81^d-150^d$  & 0.024-0.013  \\
        very slow           & $151^d-250^d$  & 0.013 - 0.008  \\
       \hline
    \end{tabular}
\end{table}

The rate of decline are expressed in terms of mag/day or $t_2$ and $t_3$, i.e., the times that a nova takes to decay by 2 or 3 magnitudes from maximum light. 
Multiple-peaks are often observed in slow novae, thus, this ambiguity might hamper the correct use of the MMRD for this nova subclass. This situation is well represented by the light curves of many galactic novae \citep{Duerbeck1981} such as NQ Vul 1976 or RR Pictoris 1925, for example. If we compute the rate of decline after the first maximum we would obtain $t_3$ of a few days rather than 65 and 130 days as actually measured after averaging and smoothing the secondary maxima.

This classification is very popular and still adopted (see, e.g., \citealt{Warner2003}) and it is certainly useful for quickly providing the observers with some important pieces of information necessary to trigger and plan the photometric and spectroscopic follow-up of an outbursting nova. However, in the last years it emerged the need to move from a simple grouping of novae, according to their speedy class or morphology of the lightcurves, toward a physical classification of nova outbursts. The idea that we develop in this paper is to relate the morphological classification of Payne-Gaposchkin with: i) the ``disk'' and ``bulge'' Galaxy environments by measuring the galactic latitude and longitude of nova systems and their height above the galactic plane \citep{DellaValle1992,DellaValle1998}; ii) the spectral classification of novae in Fe II and He/N classes \citep{Williams1991,Williams1992}. 
Recently, \citet{Darnley2012} have proposed a CN classification in three classes, based on the evolutionary stage of the respective secondary stars: main sequence nova, sub-giant nova, and red-giant nova.

\subsection{Disk and bulge novae}

\begin{figure}[htb]
    \centering
    \includegraphics[width=10cm]{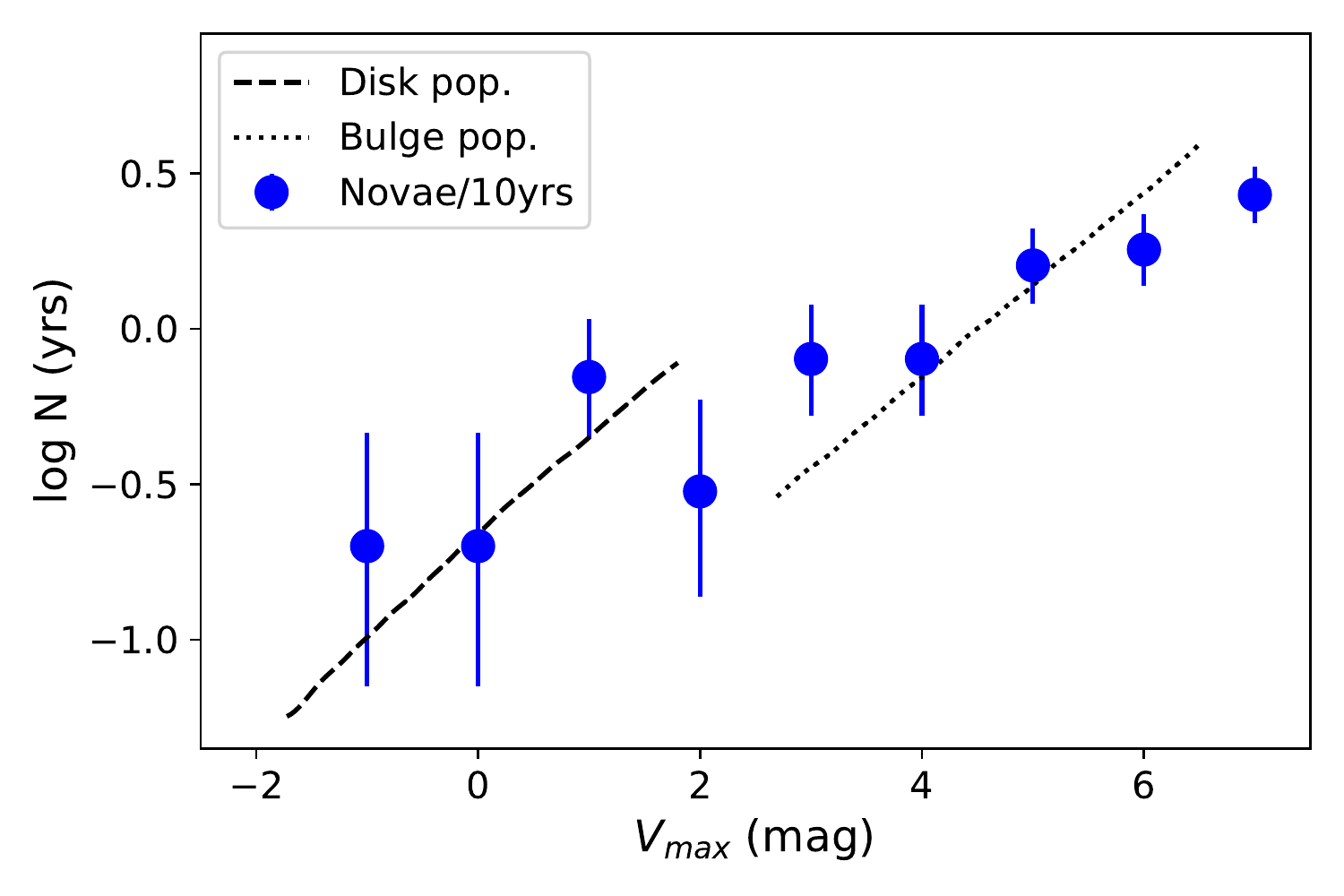}
    \caption{Adapted and updated from Fig.~2 of \citet{Duerbeck1990}. Galactic nova counts vs. apparent magnitude at maximum.  The distribution shows that up to mag $\sim 2$ we mostly observe bright ``disk'' novae. As we proceed toward fainter magnitudes the number of bright novae is vanishing and the distribution starts to increase due to the observations of faint bulge novae.} 
    \label{fig:10}
\end{figure}

\begin{figure}[htb]
    \centering
    \includegraphics[width=10cm]{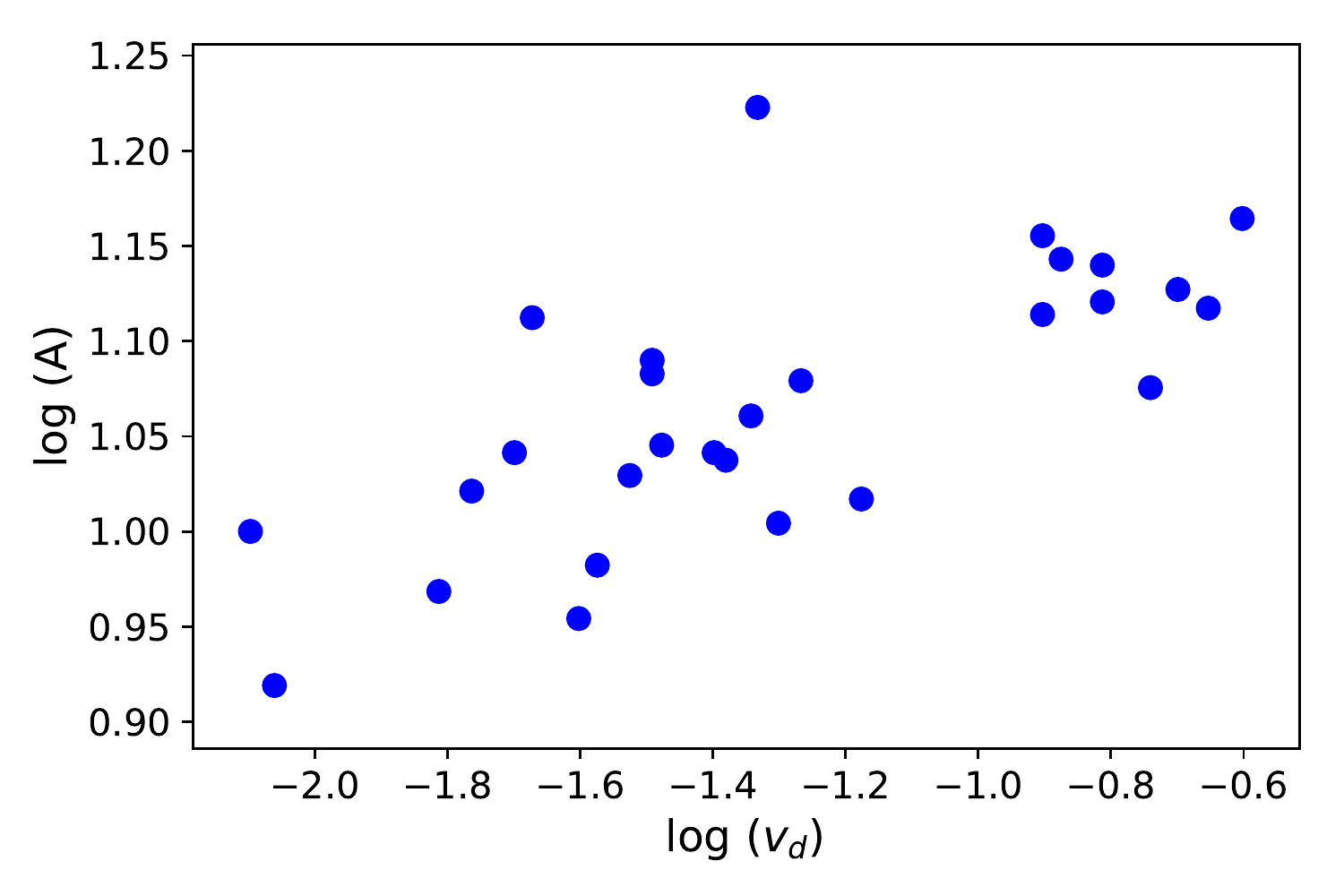}
    \caption{Distributions of the outburst amplitude plotted against the rate of decline (expressed as $v_d$ = $3/t_3$). Data from S19 and SS19, see also Table \ref{tab:app1}.}
    \label{fig:11}
\end{figure}

The quantitative characterization of the concept of nova populations into two classes of objects, i.e., fast and bright \emph{`disk novae'} and slow and faint \emph{`bulge novae'}, has been independently elaborated by \citet{Duerbeck1990}, and by \citet{DellaValle1992}, \citet{DellaValle1994}, \citet{DellaValle1998}. The former demonstrated that nova counts in the Milky Way do not follow a unique distribution (see Fig.~\ref{fig:10}), the latter authors showed that the rates of decline, which traces at first order some of the parameters driving the strength of the outburst, correlates with both the spatial distributions of novae inside the Milky Way and their spectroscopic types.  Theoretical calculations suggest that the strength of a nova outburst is a strong function of the mass of the underlying WD \citep{Shara1981,Livio1992}: the brighter is the nova at maximum, the larger is the amplitude of the outburst (see Fig.~\ref{fig:11}) the more massive is the underlying WD.

\begin{figure}[htb]
    \centering
    \includegraphics[width=10cm]{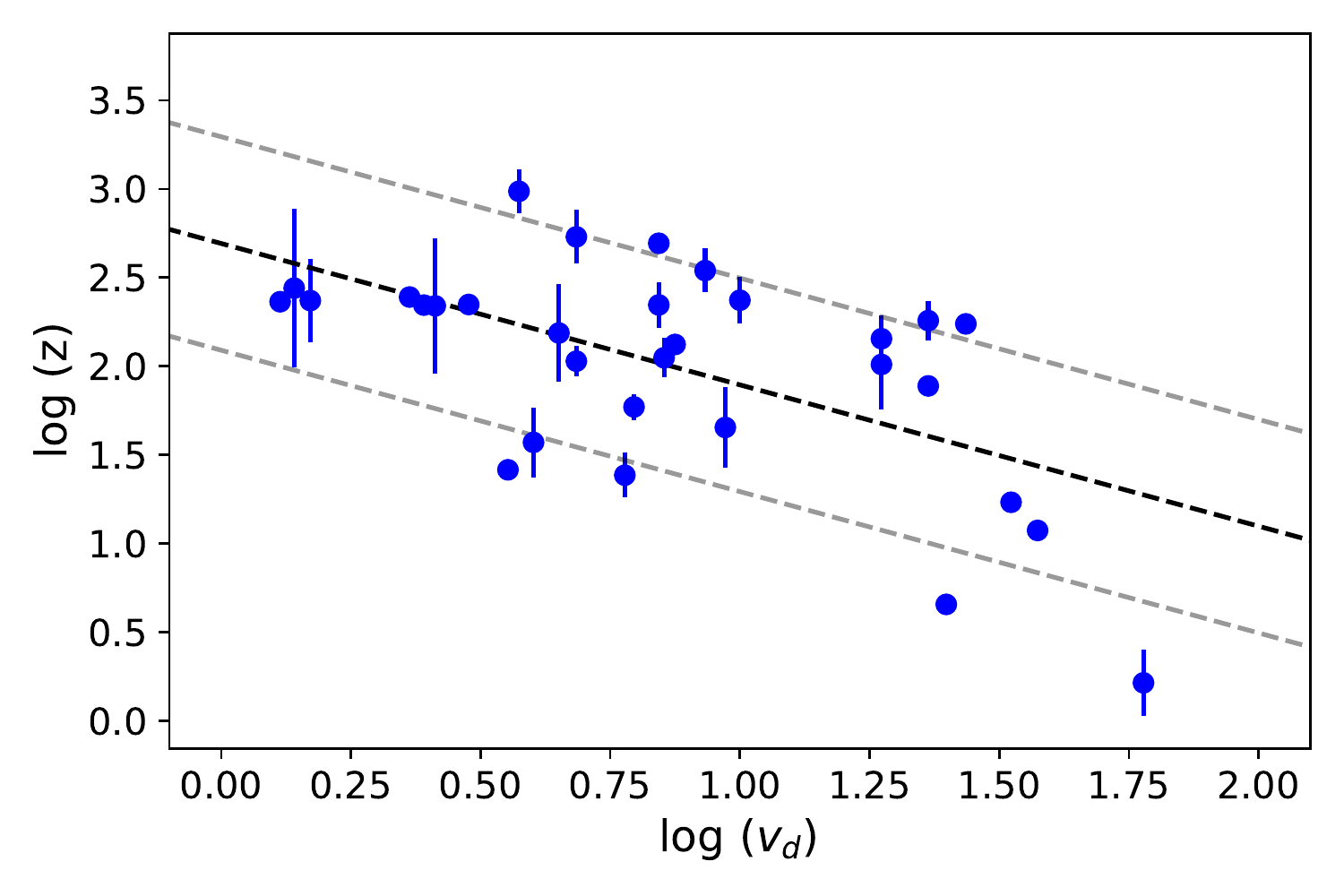}
    \caption{Relation between the height z above the galactic plane and the rate of decline obtained from novae which distances could be determined with GAIA DR2. The gray dashed lines mark the $3 \sigma$ error strips. Data from S18 and SG19, see also Table~\ref{tab:app1}.}
    \label{fig:13}
\end{figure}

Since the luminosity at maximum of a nova correlates with the rate of decline, the distribution of the rates of decline traces the distribution of the masses of the WDs in nova systems. On the other hand, there exists a strong relation between the final mass of the WD and the initial mass of the progenitor star on the main sequence (MS): the higher the initial mass, the higher will be the mass of the WD.  For example, \citet{Cummings2018} found -- admittedly with some degree of uncertainty -- that a star of $7.5\,M_{\odot}$ produces a WD of $\sim 1.3\,M_{\odot}$, while stars of initial mass $\sim 1\,M_{\odot}$ produces a WD of $\sim 0.6\,M_{\odot}$. Galactic novae then should be characterised by different scale heights above the galactic plane according to the initial mass of the progenitor star.  Novae originating from the more massive WDs should be located in the thin disk of the Galaxy, while novae from less massive WDs should occur in the thick disk, in the bulge or in the Galactic halo. Therefore, we should observe a trend between the rates of decline and the their heights above the galactic plane. To test this hypothesis we use the GAIA DR2 distances reported in the golden sample of S18 and SG19, to derive the height above the galactic plane:
\begin{equation}
z=10^{0.2\times (m-M+5-A)}\times \sin b \,,
\end{equation}
where $b$ is the galactic latitude of each nova. The S18 and SG19 samples include 15 novae that are in common to both samples. To test our prediction we have merged the two data sets into a ``bona fide'' sample (Table~\ref{tab:app1})  complemented by the recent GAIA DR2 measurement of the distance to V1500 Cyg (Nova Cyg 1975)\footnote{We derived the GAIA DR2 distance of V1500 Cyg using the parallax provided in the GAIA DR2 database and using the formula given in \citet{Bailer-Jones2018}, see also \citet{Muraveva2018}. The GAIA DR2 archive is available at \url{https://gea.esac.esa.int/archive/}}. Figure~\ref{fig:13} shows strong evidence for the existence of a trend between the rate of decline and the height above the galactic plane. The correlation 
\begin{equation}
\log z = -0.797 (\pm 0.199) \times v_d + 2.692 (\pm 0.206) 
\end{equation}
is characterized by a confidence level of $\gtrsim 3\sigma$, according to the Student's t-distribution.

The trend illustrated in Fig.~\ref{fig:13}  indicates that the fastest nova systems, which contain the most massive WDs, are concentrated close to the galactic plane. It is difficult to understand how such a distribution could be due to selection effects, since there is no obvious mechanism to prevent the discovery of fast and bright novae at high $z$. Obviously, we cannot exclude that some heavily absorbed slow and faint nova at small $z$ can have been overlooked.

\begin{figure}[htb]
    \centering
    \includegraphics[width=10cm]{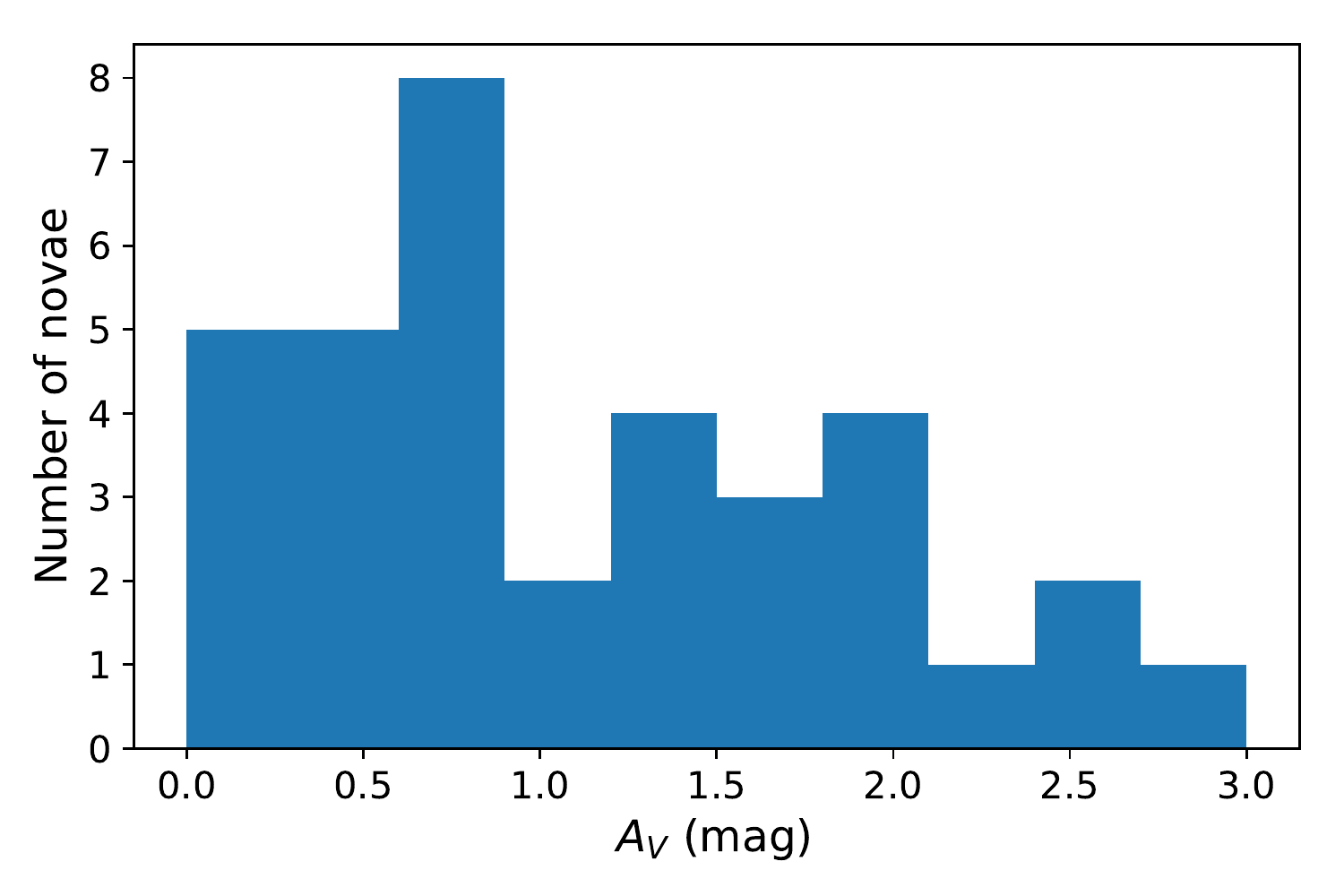}
    \caption{The distribution of the absorptions $A_V$ (mag) as given in \citet{Downes2000}, S18 and SG19 (see also Table \ref{tab:app2})}
    \label{fig:16}
\end{figure}

However, we note that the observed distribution (Fig.~\ref{fig:16}) of nova absorptions as estimated by different authors peaks at A$_V$ $\sim 1$ mag and extends up to $\sim 3$ mag. Therefore, the faint tail of the luminosity function of novae at maximum light, characterized by an absolute magnitude at maximum of $\sim -6$ mag, is bright enough to be detected, with modern surveys, at several kpc distance toward the galactic center even under more severe absorption constraints. 

We note that the attached error bars to each data point, due to the exquisite accuracy of the GAIA DR2 distances, are very small. However, the scatter of the relation is intrinsically large. This fact implies that other parameters in addition to $M_{\rm WD}$,  such as the $\dot{M}$ \citep{Prialnik1982,Townsley2004,Kato2014}, the envelope mass at ignition \citep{Hachisu2010}, the temperature of the WD \citep{Prialnik1995,Yaron2005} or the chemical composition of the accreted matter \citep{Kovetz1985,Starrfield1999} affect the luminosity at maximum, the rate of decline, the ejection velocity and the mass of the expelled layers. 

In the following, we label \emph{`disk'} novae the objects characterized by $z\lesssim 150$ pc.  They are mostly `fast' declining objects characterised by a bright peak at maximum $M_V \sim -9$ mag,  $t_2 \lesssim 12^d$ and $t_3 \lesssim 20^d$ and mainly belongs to the He/N class. Slow novae occur at higher z, up to $1$ kpc, and are characterized by $M_V \sim -7.5$ mag at maximum and $t_2$ and $t_3$  $\gtrsim 12^d$ and $20^d$, respectively. Hereafter, we label these high $z$ slow declining objects \emph{`bulge'} novae \citep{DellaValle1998}.

\subsection{Post-novae}

\begin{figure}[htb]
    \centering
    \includegraphics[width=10cm]{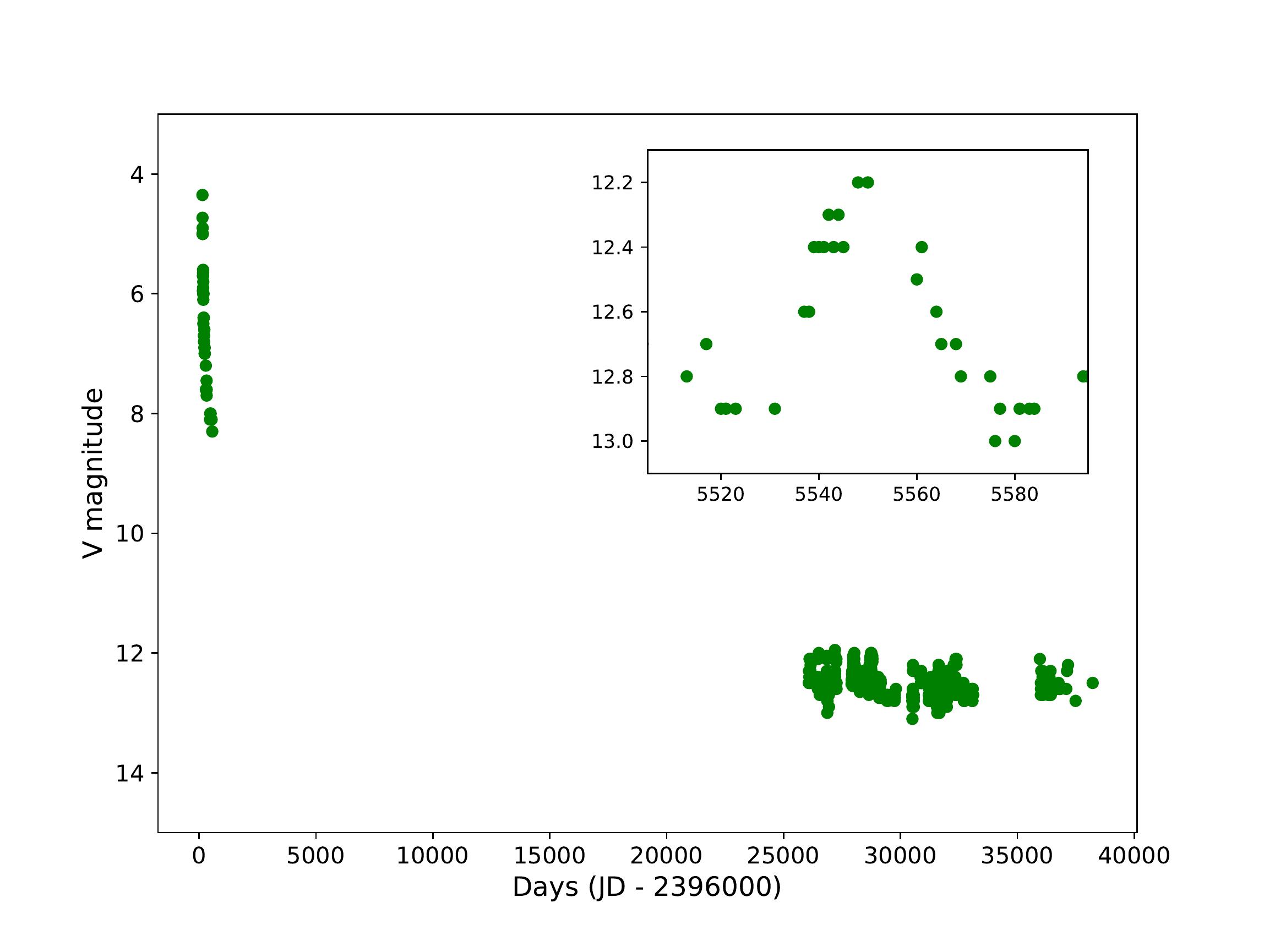}
    \caption{v841 Oph lightcurve. Inset: an example of DN-like outburst occurred during the post-outburst stage.}
    \label{fig:19}
\end{figure}

\begin{figure}[htb]
    \centering
    \includegraphics[width=12cm]{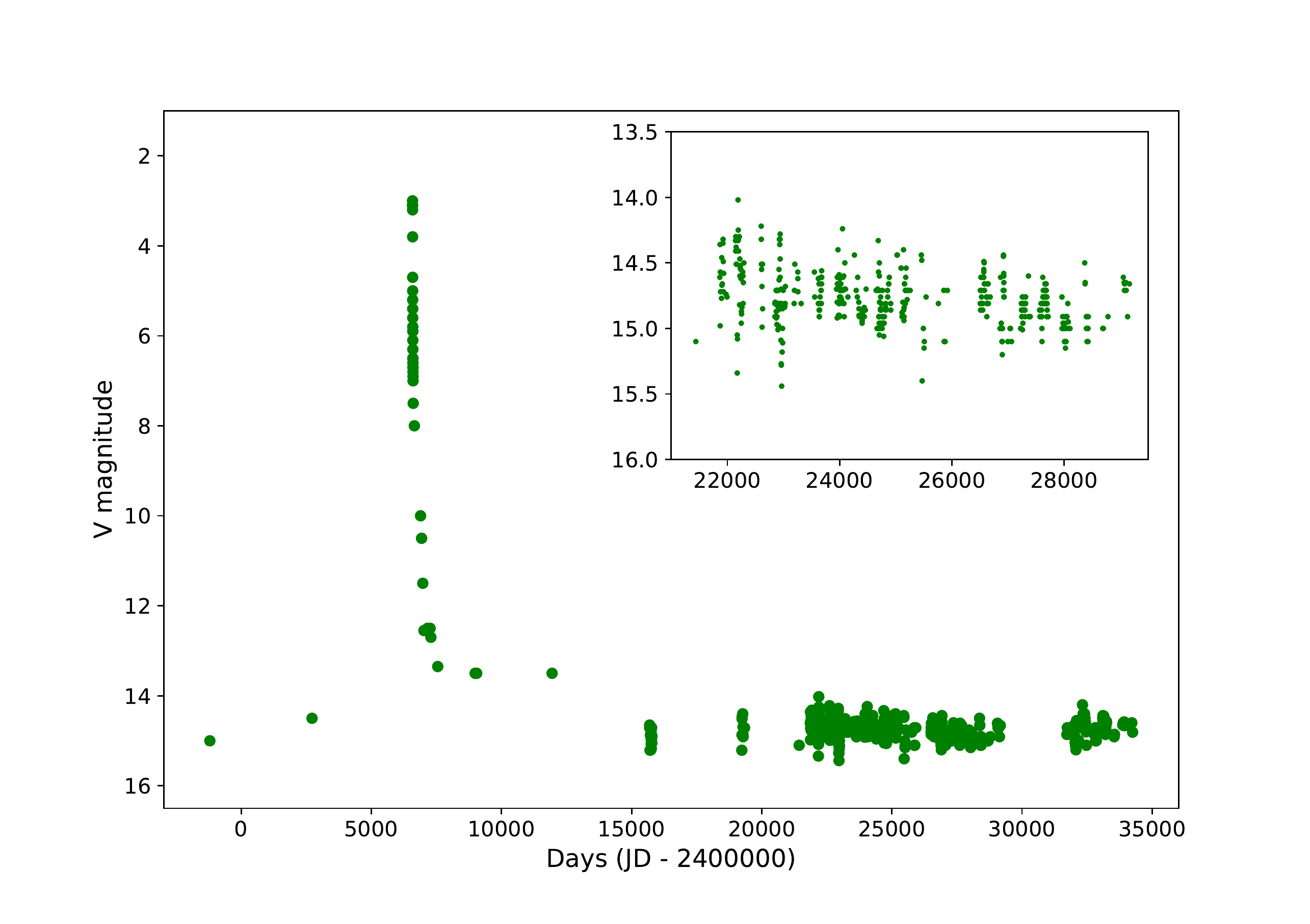}
    \caption{Q Cyg lightcurve. Inset: an example of modulation of the post-outburst lightcurve on time scale of hundreds days.}
    \label{fig:18}
\end{figure}

\begin{figure}[htb]
    \centering
    \includegraphics[width=12cm]{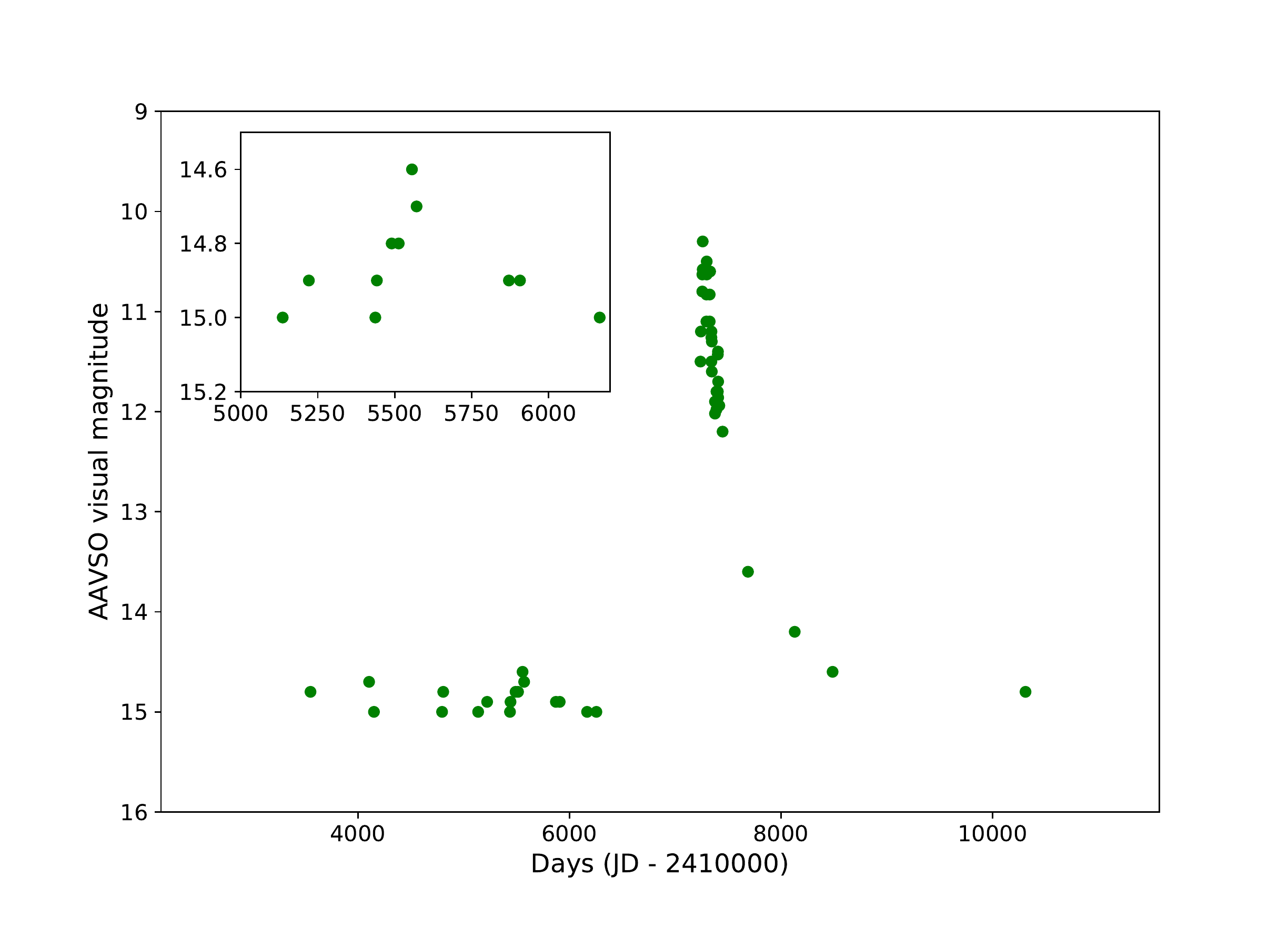}
    \caption{AR Cir lightcurve. Inset: example of DN activity during the pre-outburst stage.}
    \label{fig:21}
\end{figure}

We can estimate as $\Omega \sim 1.5\times GM^2 / R \sim 10^{51}$ erg the binding energy of a $1.4\,M_{\odot}$ WD. On the other hand, the energy budget associated with a nova outburst is $\sim 10^{45}$ erg. This simple orders of magnitude comparison provides a clear indication that the nova outburst is not a catastrophic event for the WD, unlike from SN-Ia explosions. As a consequence, some decades after the outbursts the systems go back to quiescence and the post-nova magnitudes become very similar to those observed during the pre-nova stage \citep{Robinson1975}. The return to the minimum light is not as quick as the luminosity increasing to maximum.  This fact can be quantified by studying the lightcurves of the oldest galactic novae for which a significant record of observations since the maximum epoch do exist: Nova Oph 1848 (v841 Oph), Nova Cyg 1872 (Q Cyg) and Nova Cir 1906 (AR Cir). The light curves reported in Figs. \ref{fig:19}, \ref{fig:18}, \ref{fig:21} have been obtained from the Steavenson's observations published between 1920 and 1950.  An inspection of the data shows that v841 Oph and Q Cyg have returned to the quiescence state, at V $\sim$ 12.5 mag and V $\sim$ 15 mag in about 80 and 50 years, respectively.

Post-nova stages observations are not very frequent. When they are available, they are characterized by variation of the magnitude at minimum as large as $\sim$1--2 mag. These brightness variation are due to multiple factors such as: brightness modulations possibly related to orbital period \citep{Woudt2004}, solar-type cycles activity of the secondary \citep{Bianchini1987,Warner2003} on time scale of the orders of months/years (see inset in Fig.~\ref{fig:18}) but also to dwarf nova activity on time scales of 50--270 days \citep{DellaValle1987,DellaValle1990}.

\begin{figure}[htb]
    \centering
    \includegraphics[width=10cm]{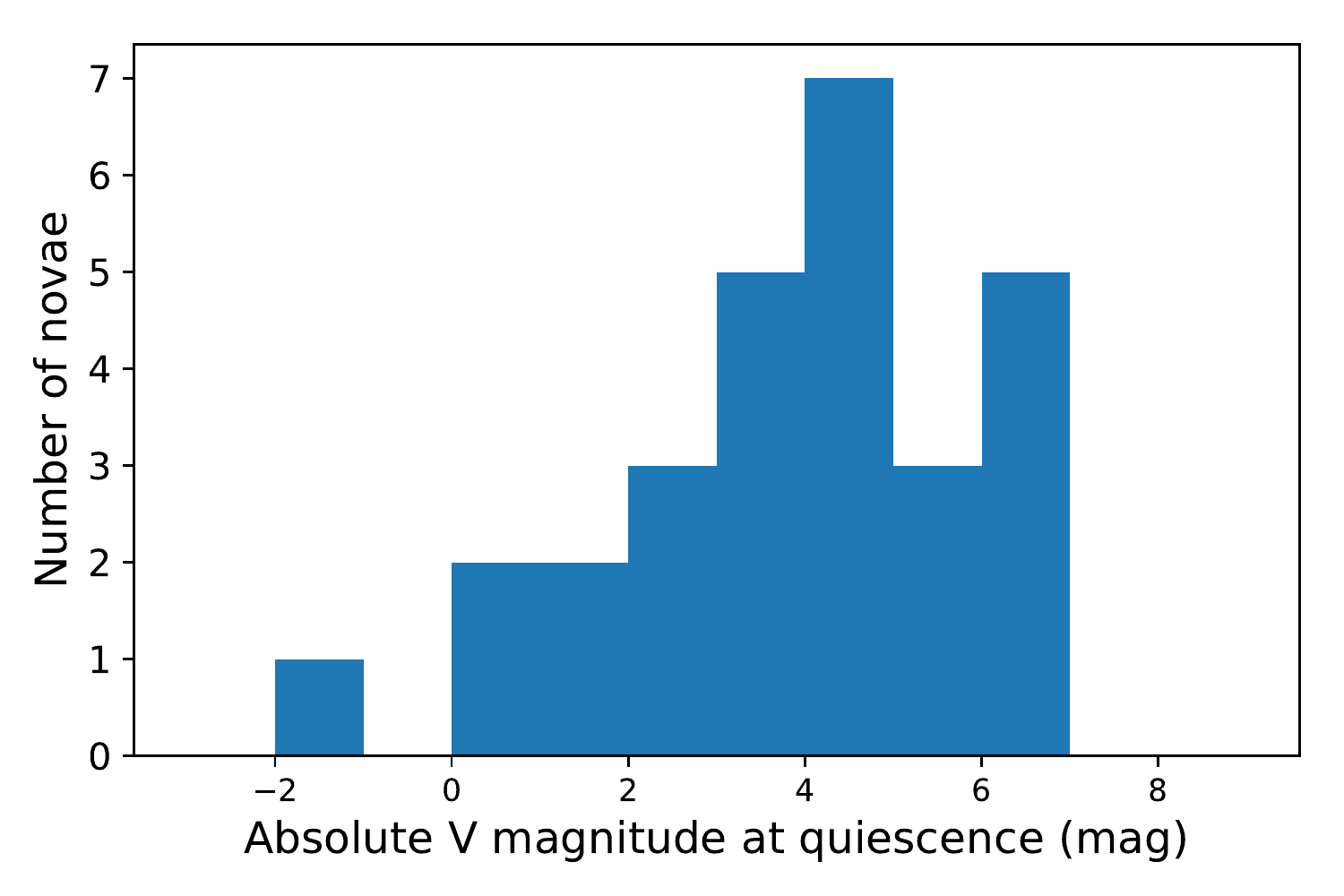}
    \caption{The distribution of absolute V magnitudes of GAIA DR2 novae at quiescence.}
    \label{fig:q}
\end{figure}

The distribution of the absolute magnitude at minimum, see Fig.~\ref{fig:q}, is a distribution characterised by a mean brightness of M$_V=3.8\pm$0.24 mag. Such distribution of luminosities requires a mass transfer rate between $\sim 10^{-8}$  and $10^{-9}\,M_{\odot}/{\rm yr}$ (see Fig.~5 in SG19). The brightness at minimum of the fastest declining novae implies mass transfer rates that are a full order of magnitude below the average value for most novae, close to the DNe values \citep{Vogt1990}. 

The link between DNe and CNe has been hypothesized long ago in the framework of the so called ``Hibernation'' scenario \citep{Shara1986} (see also \citealt{Hillman2020}) which was originally elaborated to explain an apparent two orders of magnitude discrepancy between the density of classical novae inferred in the solar neighborhoods and the density deduced from nova surveys in M31, nova theory and U, B photometric surveys. Intrinsically low CN space density would imply very short recurrence times between the outbursts, which in turn would imply high values of the $\dot{M}$, which are not observed.   According to these authors the space density of CNe is much higher but novae would spend most time between two consecutive outbursts in a ``hibernation'' state.  This stage would be characterised by low or very low mass transfer rates ($10^{-10}$--$10^{-12}\,M_{\odot}/{\rm yr}$) as result of the increasing separation between the companion stars after each outburst, as consequence of the conservation of the angular momentum of the binary system. The canonical mass transfer rate of $10^{-8}$--$10^{-9}\,M_{\odot}/{\rm yr}$ would be recovered as soon as the  magnetic braking and/or gravitational radiation emission decrease the separation between the companion stars.  During this intermediate stage in which $\dot{M}$ increases from hibernation to the canonical nova values, the system would exhibit DN activity.  This scenario has received in the years some observational support.

First of all, the dwarf nova activity was detected not only after the nova outbursts, but it was also observed decades before the outbursts in V446 Her or IV Cep \citep{DellaValle1990} and AR Cir (see inset in Fig.~\ref{fig:21}). \citet{Duerbeck1992} could measure an average decline of 10$\pm$ 3 millimag per year  in the post outbursts decline of nine old novae, admittedly with large individual scatter. However, these data allowed Duerbeck to conclude that the observed trend was \emph{``compatible with the prediction of the hibernation hypothesis''}. Finally in more recent years, \citet{Shara2007,Shara2012} discovered nova shells around two dwarf novae: Z Cam and AT Cnc.  

On the other hand the ``Hibernation'' scenario was introduced on the basis of a supposed existence of a two orders of magnitude discrepancy between the  density of CVs occurring close to the Sun and the rate of CVs that was believed to be as high as $\rho\sim 10^{-4}\,{\rm pc}^{-3}$ \citep{Shara1990}.  This density value of CVs has been the subject of many revisions in the last two decades that we have summarized in Table~\ref{tab:6}.

\begin{table}[htb]
    \caption{Collection of measurements of CVs density close to the Sun.}
    \label{tab:6}
    \centering
    \begin{tabular}{l|l}
    \hline\hline
     Density (pc$^{-3}$) & Authors\\
         \hline
$10^{-5} $	& \citet{Patterson1998}\\
$3\times10^{-6} $ & \citet{Warner2001}\\
$3 \times 10^{-5}$ & \citet{Schwope2002}\\
$< 5 \times 10^{-7} $ & \citet{Cieslinski2003}\\
$0.3-3.3 \times 10^{-5}$ & \citet{Pretorious2007}\\
$< 2 \times 10^{-5} $ & \citet{Hernandez2018}\\
$< 5 \times 10^{-6} $ & \citet{Schwope2018}\\
       \hline
    \end{tabular}
\end{table}

These values are 1--3 orders of magnitude smaller than the $\rho \sim 10^{-4}\,{\rm pc}^3$, which originated the quest for the existence of a hibernation stage for CVs. \citet{DellaValle1993} found a density of 
nova outbursts in the galactic disk of: 
\begin{equation}
\rho=0.30 \pm 0.15 \times 10^{-10}\,{\rm pc}^{-3}\,{\rm yr}^{-1}
\end{equation}
consistent  with previous estimates of $\rho=0.5 \times 10^{-10}\,{\rm pc}^{-3}\,{\rm yr}^{-1}$ \citep{Duerbeck1984,Patterson1984}, $\rho=0.2$--$0.9 \times 10^{-10}\,{\rm pc}^{-3}\,{\rm yr}^{-1}$ \citep{Naylor1992} once we have removed from their original values the contribute of bulge novae. 
The next step is to derive the density of CNe from the density of nova outbursts by multiplying the observed density of nova outbursts for the recurrence time between the outbursts. Disk novae that are mostly fast and bright originate from massive WDs, say $1.1$--$1.3\,M_{\odot}$. According to Sect.~1.2, these novae have recurrence times of the order of $2 \times 10^4$ years, which gives a density of Classical Nova progenitors of: 

\begin{equation}
\rho_{\rm CNe}=0.6 \times 10^{-6}\,{\rm pc}^{-3}. 
\end{equation}

This result is consistent with the lower density values reported in Table~\ref{tab:6}, which were derived with a broad variety of approaches including OGLE survey \citep{Cieslinski2003}, ROSAT observations \citep{Schwope2002,Pretorious2007,Schwope2018} and SDSS survey \citep{Hernandez2018}. The still large range of uncertainties in the knowledge of the CV density does not exclude  that there is some room for the existence of a ``hibernation'' stage in a ``mild'' version, i.e., it should be active over a shorter interval of time than originally hypothesized. 

\subsection{The interaction of nova with the interstellar medium}

The interaction of the expanding shells with the interstellar medium was discussed by \citet{Oort1926} and \citet{Ambartsumian1952}. These models were developed on simple assumptions: the conservation of energy, average expansion velocities of the nova shell and ejecta masses of $\sim 10^{-5}\,M_{\odot}$.  After assuming a range of interstellar densities, Oort could estimate: i) the ``half-lifetimes'' that is the time after which the expansion velocities of the ejecta have decreased to 50\% of its initial value, to be 50--100 years and ii) 10,000 years the time needed to velocities of nova ejecta to become indistinguishable from the chaotic motions of the circum-burst interstellar medium. Attempts to measure the deceleration of the nova ejecta \citep{Duerbeck1987} were hampered by the the inhomogeneities  affecting the comparison between images of post-nova nebulae obtained decades apart. Also one should considers the possibility that the circumburst medium could have been swept away by previous outburst episodes. \citet{Downes2000} in their survey on nova shells, which included also the \citet{Slavin1995} data, decided to derive the distances via nebular parallaxes toward each nova by assuming no deceleration for the ejecta. The correctness of this assumption has been recently verified by \citet{Santamaria2020}. These authors compared new narrow-band images of nebulae around old novae: FH Ser, V533 Her, DQ Her, V476 Cyg, T Aur (aged 50 to 130 years) with images from archives and found no evidence for deceleration.  They thus concluded that nova shells can continue to expand for several hundred years without undergoing a significant deceleration.

\subsection{The galactic nova rate}

The knowledge of the galactic nova rate is important for at least two reasons: i) some nova-like systems are believed to be progenitors of SNe-Ia (see sect. 11) therefore the correct measurement of the nova rate is an essential ingredient to be compared with the relatively well known SN-Ia rate in the Milky Way \citep{Cappellaro1999,Cappellaro2015}; ii) the role of novae as contributors to the Galactic nucleosynthesis has been well established on theoretical grounds \citep{D'Antona1991,Romano2001,Travaglio2001,Romano2003,Prantzos2012}. Recently  observations of $^7$Be \citep{Tajitsu2015} and $^7$Li \citep{Izzo2015} have provided significant empirical support to these theoretical predictions.  However, despite of the importance of these issues the correct value of the galactic nova rate is still poorly known (see Table \ref{tab:4}).  Just to give an idea here below we report a short summary of the estimates available in literature. \citet{Lundmark1935} derived $\sim 50$ novae/yr. \citet{Allen1954} inferred from his paper on the whole-sky statistics of celestial object about $R=100\div 200$ novae/yr, which is similar to the estimate obtained later on by \citet{SharoV1972} of $R=260$ novae/yr and at variance with $R\sim 50$ novae/yr estimated by \citet{KopyloV1955}. In more recent epochs \citet{Liller1987} have corrected the ``observed'' rate of 3.7 novae/yr for incompleteness and selection effects and they derived 74 novae/yr $\pm 24$. \citet{Hatano1997} have revised \citet{Liller1987} and assumed that most novae are produced in the disk and found $R=41\pm 20$. \citet{vandenBergh1991} derived about $R=16$ novae/yr by scaling the nova rate of M31 to the GCs population ratio  Galaxy/M31=0.56. \citet{Ciardullo1990,DellaValle1994,Darnley2006} found $R=11\div 46$ novae/yr  and $R=15\div 24$ and $R=34^{+15}_{-12}$ respectively after scaling the nova rates measured in extragalactic systems to the luminosity of the Milky Way. \citet{Shafter1997} found $R=35\pm 11$ by studying the distribution of novae inside the Milky Way and applying a number of correction factors for incompleteness and observational bias. More recently, the same author \citep{Shafter2017}  has increased his previous estimate to $R=50^{+31}_{-23}$ by revising the correction factors to the galactic nova counts and by assuming uniform properties between bulge and disk novae.  

\begin{table}[htb]
    \caption{One century of galactic nova rate measurements.}
    \label{tab:4}
    \centering
    \begin{tabular}{l|l}
    \hline\hline
     Nova/yr  &  author \\
         \hline
 50 & \citet{Lundmark1935}\\
 100--200 & \citet{Allen1954}  \\
 50 &  \citet{KopyloV1955}\\ 
 260 &  \citet{SharoV1972}\\
 74 & \citet{Liller1987}\\
 11--46 & \citet{Ciardullo1990}\\
 16 &  \citet{vandenBergh1991} \\
 $20\pm 5$ &  \citet{DellaValle1994}\\
 $41\pm 20$ & \citet{Hatano1997}\\
 $35\pm 11$& \citet{Shafter1997}\\
 $34^{+15}_{-12}$   & \citet{Darnley2006} \\
 $50^{+31}_{-23}$  & \citet{Shafter2017} \\
       \hline
    \end{tabular}
\end{table}


\begin{figure}[htb]
    \centering
    \includegraphics[width=10cm]{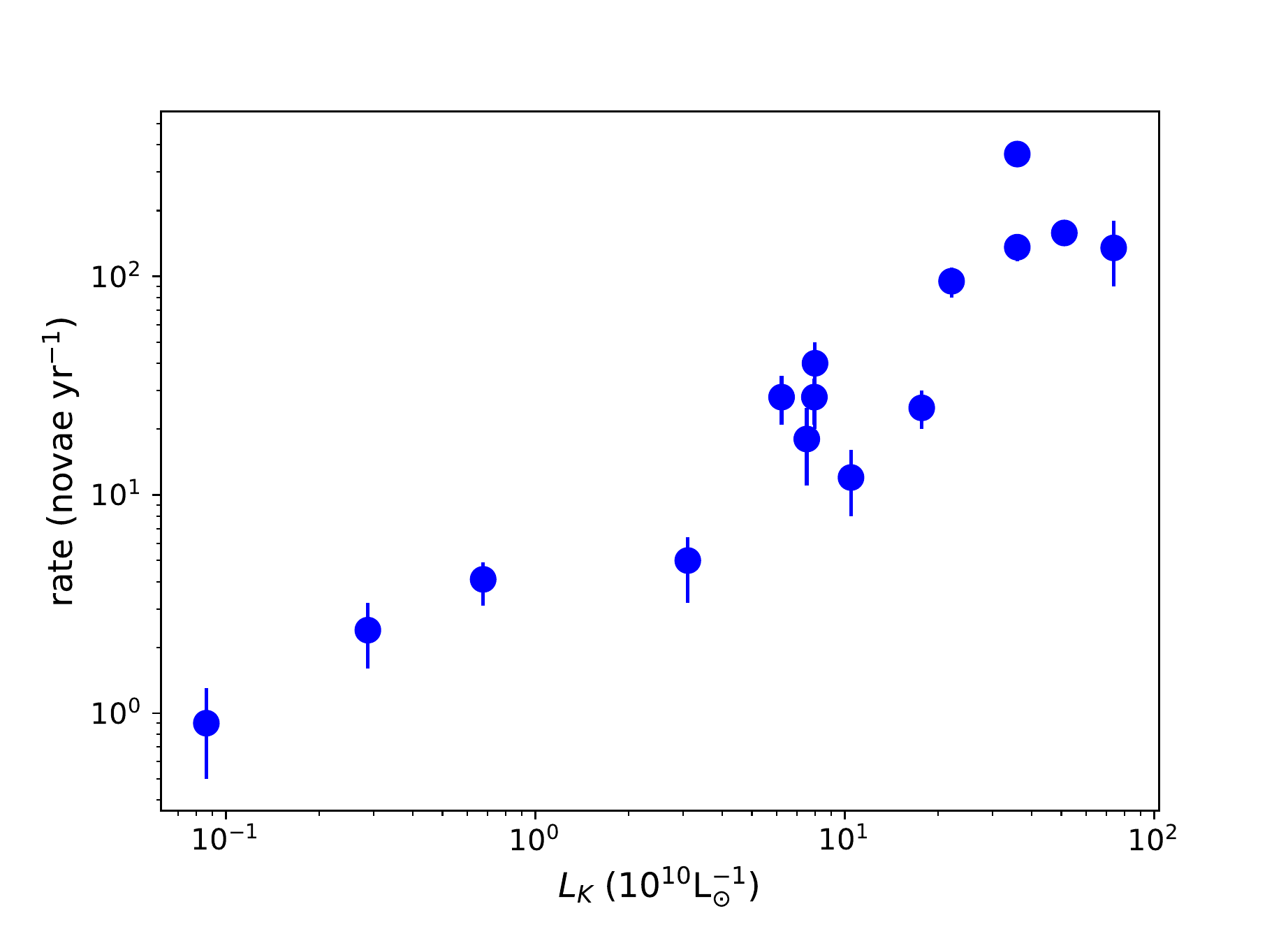}
    \caption{Extragalactic nova rates plotted agains the $K$-band luminosity (data from Table \ref{tab:sample} in Sect. 6.6).}
    \label{fig:ratesLK}
\end{figure}

Figure~\ref{fig:ratesLK} shows the relationship between the nova rate and the $K$ luminosity of the parent galaxy. A least squares fit to the data points gives: 
\begin{equation} 
\log N \approx 0.779 \times \log L_{K} - 7.09
\end{equation}
 
After assuming for the Milky Way $L_{B} \approx 2\times 10^{10}B_{\odot}$ \citep{vandenBergh1988}, and $(B-K)=3.50$, corresponding to Sb Hubble type, we find $\approx 22$ novae/yr, admittedly with a large attached error. However, this result is consistent with the recent ``direct'' measurement of \citet{Mroz2015}. These authors have measured the bulge nova rate of the Milky Way with the data of the OGLE survey.  They find a rate for the Galactic bulge of $13.8\pm 2.6$ novae/yr that according to these authors:  \emph{``\dots is much more accurate than any previous measurement of this kind, thanks to many years monitoring of the bulge by OGLE survey''}.  A ``rounded'' rate of about $14 \pm3$ novae per year in a galaxy like the Milky Way, where the nova production is bulge dominated, implies a global rate that hardly can exceed 30 novae/yr in good agreement with the frequencies found in the 1990s \citep{Ciardullo1990,DellaValle1994,Shafter1997}. Only if incompleteness affects dramatically the galactic nova counts in the disk we can relax the assumption that novae in the Milky Way are mainly produced in the bulge and therefore values of the galactic nova rate  as large as $\sim 50$ novae/yr are still possible \citep{Shafter2017}. Presently the existence of this bias is not supported by the observations.  

\section{The spectral evolution of novae} 

Many contributions have been devoted to describe and to summarize the spectroscopic evolution of nova outbursts, mostly dedicated to identify the different nova stages, their absorption and emission lines and measuring the expansion velocities of the ejecta \citep{Payne1957,McLaughlin1960}. More recently, \citet{Shore2012} has provided a general physical description of this sequence of stages that helped a lot to enlighten the physics underlying the spectroscopic evolution of novae. A similar description based on the color-color evolution of nova outbursts has been also provided by \citet{Hachisu2014}. 

\begin{figure*}
    \centering
    \includegraphics[width=10cm]{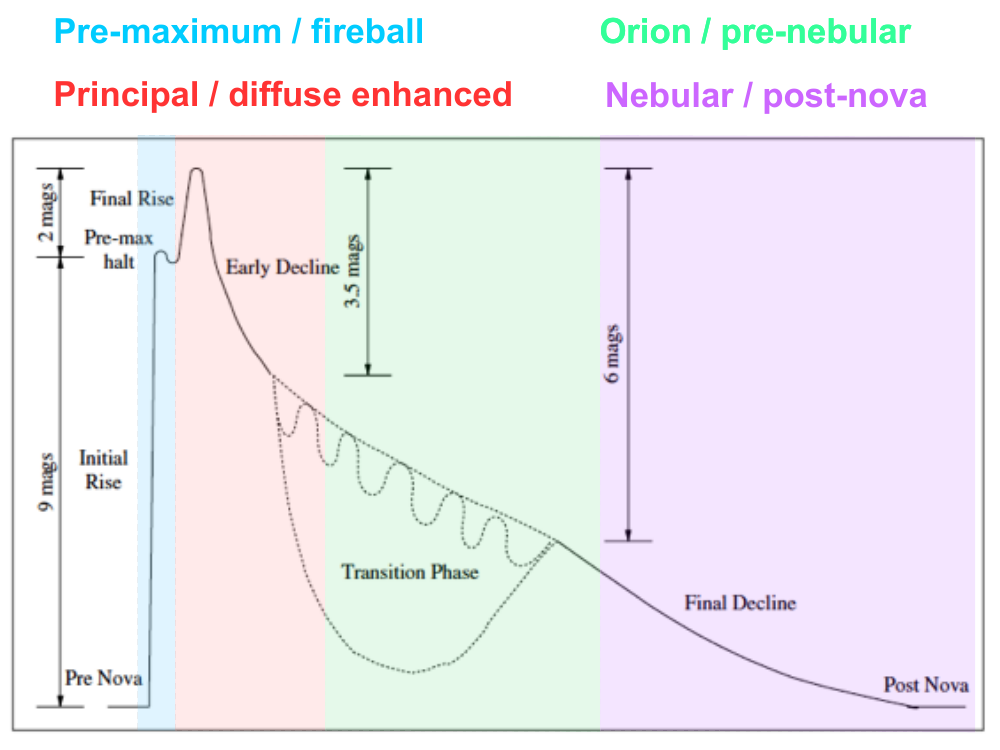}
    \caption{The average lightcurve morphology of a nova. We have added in different colors the description of the main spectroscopic phases. Image reproduce with permission from \citet{Warner2003}}
    \label{fig:22}
\end{figure*}

In the following, we have linked the ``classical'' photometric stages with the ``old'' and ``new'' spectroscopic descriptions of nova phases (see Fig.~\ref{fig:22}). The spectral evolution of novae is basically driven by two physical parameters: the mass density of the ejecta and their expansion velocities.In spite of the broad phenomenology observed in nova spectra six spectral stages have been defined and in the following we provide an essential description for each of them.

\begin{figure}[htb]
    \centering
    \includegraphics[width=10cm]{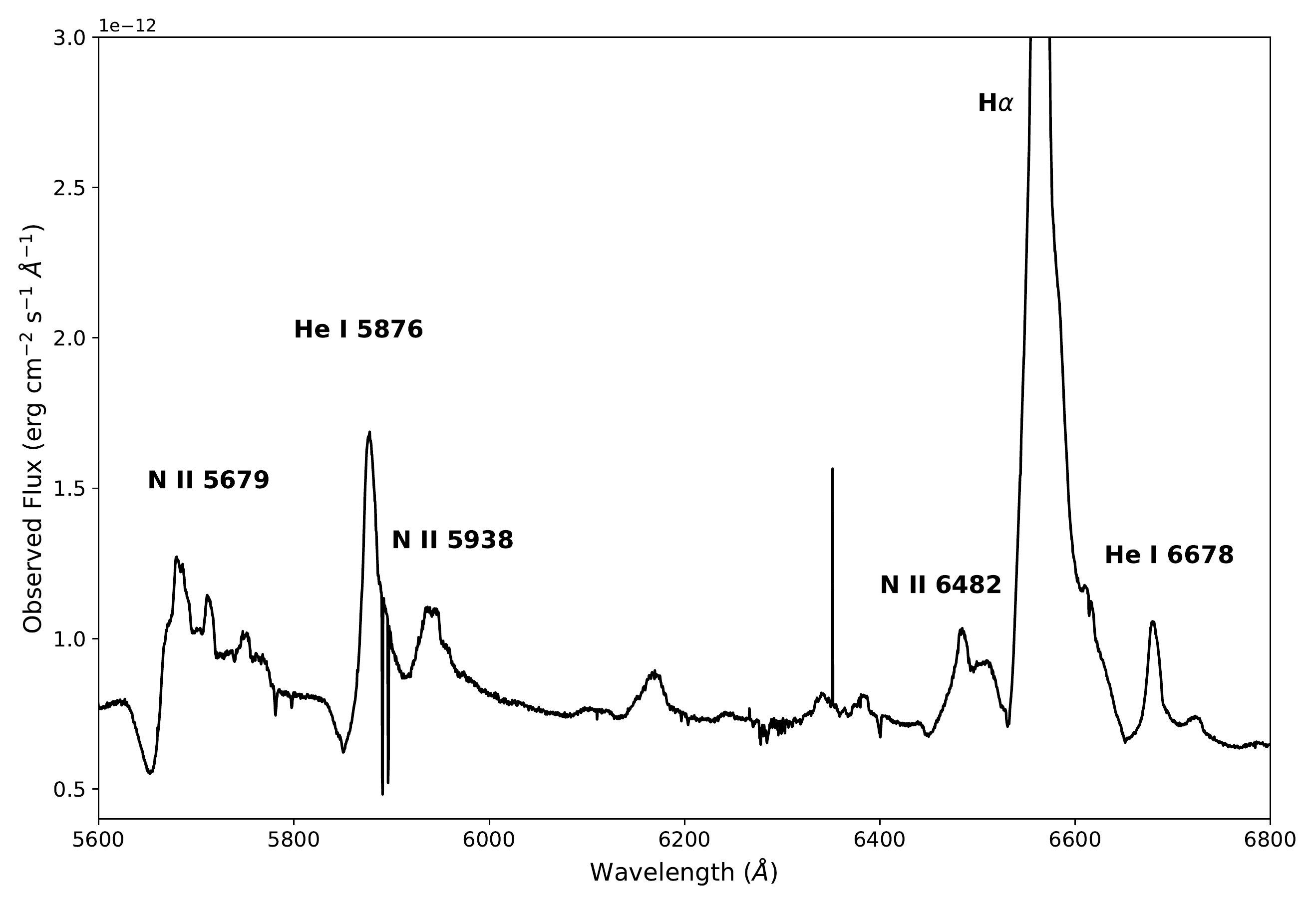}
    \caption{Initial rise/pre-maximum spectrum of T Pyx \citep[adapted from ][]{Izzo2012}.}
    \label{fig:23}
\end{figure}

\textbf{Initial rise -- Pre-maximum spectrum:} it corresponds to the  adiabatic free expansion of the ejecta that is rapidly cooling since the expansion regime is much shorter than the characteristic cooling time of the ejecta. This phase is also referred as 
``hot fireball'' spectrum and it is observed very seldom because it implies: i) the discovery of a nova during its rise at maximum, that normally lasts from a few hours to few days and ii) the immediate availability of telescope time, two conditions that normally do not occur frequently.  Some novae show a pre-halt maximum (the so-called shoulder phase) in the light curve, which is characterized by bluer colours and that anticipates the final rise to the maximum brightness. This feature in the light-curve has been caught in very few novae, such as T Pyx \citep{Shore2011} and V1974 Cyg \citep{Shore1993}.  The spectra bear some similarity to F supergiants \citep{Kodaira1970}. Balmer and non-Balmer lines are characterized by high expansion velocities (2000--3000 km/s) as measured from the minimum of the respective P-Cygni profiles. These features are produced during the initial expansion of the optically thick envelope ejected by the nova. (see Fig.~\ref{fig:23}). 

\begin{figure}[htb]
    \centering
    \includegraphics[width=10cm]{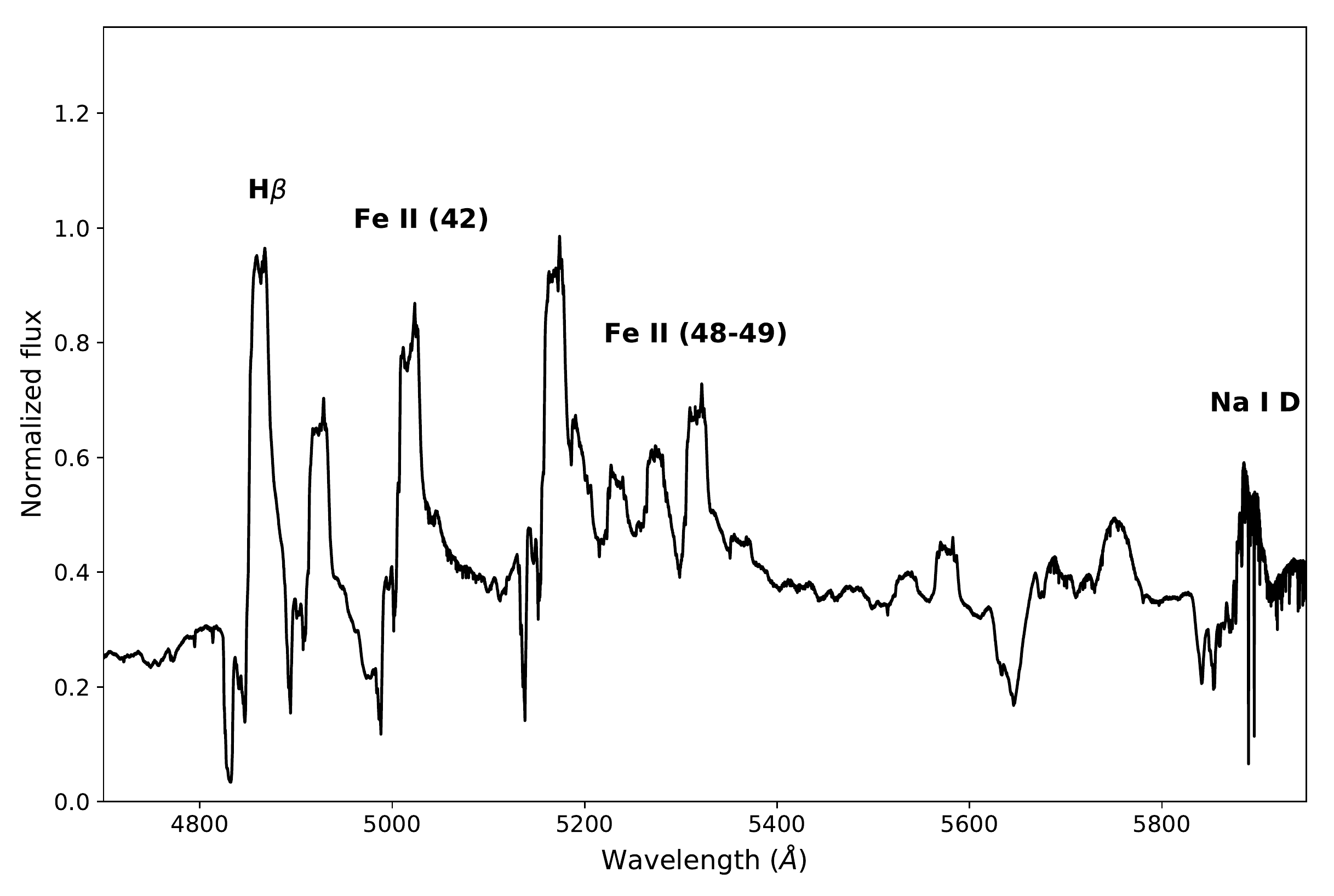}
    \caption{Principal spectrum phase of V1369 Cen, showing the presence of diffuse enhanced absorptions in the P-Cygni of the brightest emission lines.}
    \label{fig:24}
\end{figure}

\textbf{Maximum light -- Principal spectrum:} this stage corresponds to the rise to the maximum brightness in V and R magnitudes and a contemporaneous decrease in luminosity at blue and UV wavelengths. This is the time when the fireball reaches the critical cooling temperature, the pseudo-photosphere starts to recede and the spectrum is being dominated by the free-free emission from thinner plasma outside the photosphere \citep{Hachisu2006}. Consequently, the nova ejecta itself acts now as a passive media, by converting the UV flux into longer wavelengths emission and then showing low-ionization transitions in the spectra, such as the Fe II group species (hence the name \emph{``iron-curtain''}), Na I D and Ca II. These lines are flanked by broad P-Cyg profiles with blue-shifted velocities generally lower than those observed during the fireball phase (see Fig.~\ref{fig:24}). At the same time, the optically-thick stage in UV bands leads also to the presence of ``pumped'' transitions, like the O I 8446\,{\AA}, whose upper level is over-populated by the H Ly-$\beta$ 1025.72\,{\AA} transition, which is almost coincident with the transition O I 1025.76\,{\AA} \citep{Bowen1947,Williams2012}. High-resolution spectroscopic observations have also revealed the presence of narrow and low-ionization heavy element absorptions (THEA - \citet{Williams2008}, see Sec. \ref{sec:42}), which are detected during the brightest epochs of several nova outbursts with expanding velocities relatively low ($\sim 500$ km/s or less). 

During the principal phase, the so-called \textbf{Diffuse Enhanced} absorption systems appear during the very early decline. The only feature that differentiates the diffuse enhanced from the principal spectrum is the presence of multiple P-Cyg profiles (Fig.~\ref{fig:24}).  All these absorption components show an apparent acceleration of the ejecta with time, as inferred from the P-Cygni absorptions. However, this acceleration is not real, it rather originates in the downturn density of the ejecta and the contemporaneous recession of the hot pseudo-photosphere. Indeed, a decreasing density implies a more optically thin ejecta, which implies a fragmentation of the broad P-Cygni absorption lines into several weaker structures. Very recently, it has been proposed that the appearance of new and faster absorption systems in the P-Cygni profiles of Balmer lines correlates with the presence of rebrightenings, or flares, in the light curves \citep{Li2017,Aydi2019}. These new systems would be the result of fast wind ejection from the underlying WD and their interaction with the pre-existing ejecta will lead to strong shocks, originating the high-energy emission observed in some novae \citep{Metzger2015,Cheung2016,Mukai2019,Aydi2020}, see also Sect. 12. Another explanation for the presence of the rebrightenings is that nova ejecta is actually composed by strong density inhomogeneities created during the TNR \citep{Williams2016}. 

\begin{figure}[htb]
    \centering
    \includegraphics[width=10cm]{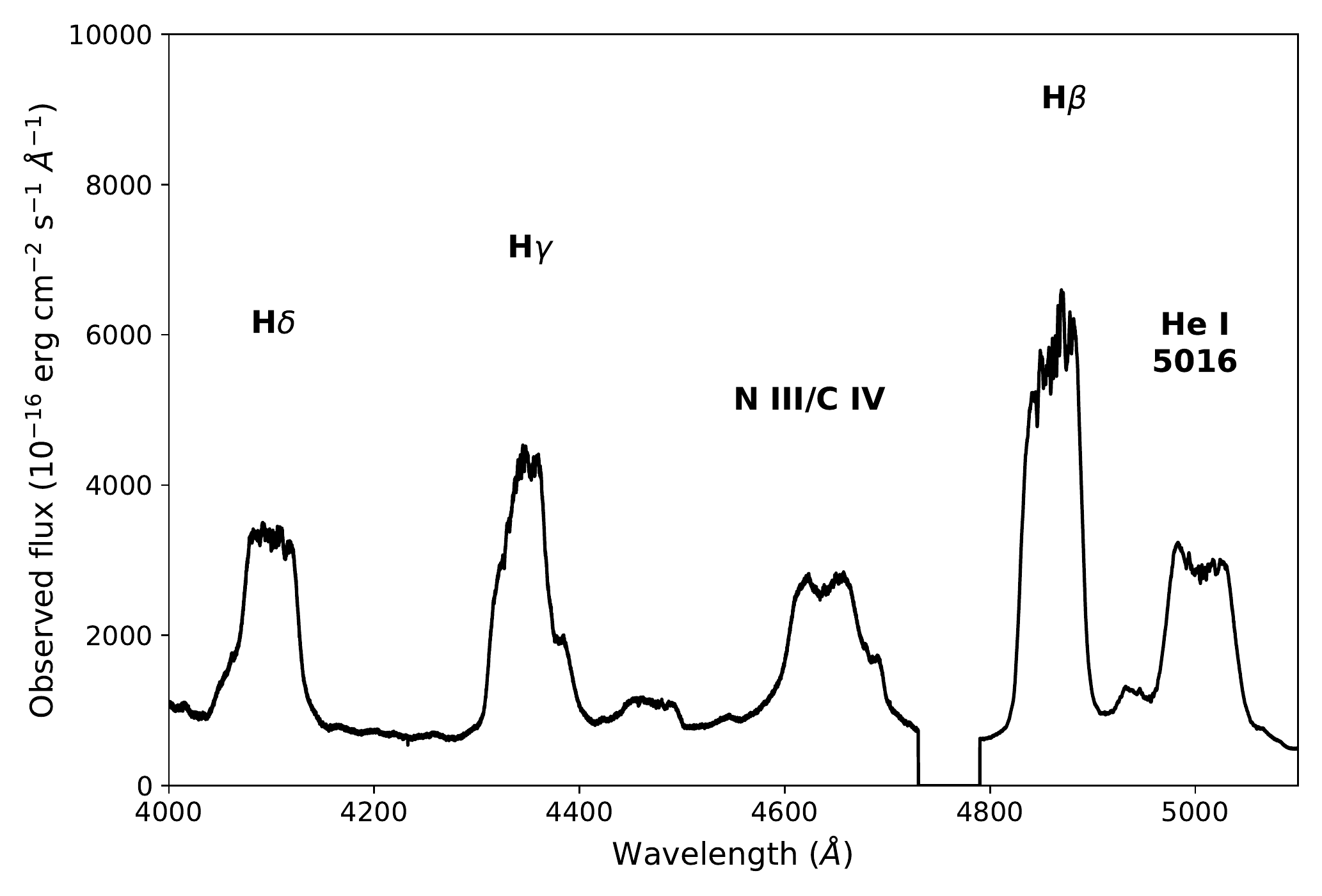}
    \caption{Orion phase spectrum observed in ASASSN-16kt.}
    \label{fig:25}
\end{figure}

Toward the end of the early decline and at the beginning of the transition stage (that is the transition from optically thick to optically thin condition) another spectral system appears, the so called \textbf{Orion spectrum}. The optical spectra now show features due to high-ionization lines (He I, O II, N II), and the CIII/NIII lines Bowen blend (``4640 band'') see Fig.~\ref{fig:25}. Their appearance in the spectrum is due the low density of the ejecta combined with the intense UV flux. An important characteristics of this stage is that the expansion velocities measured from these transitions do not generally coincide with the ones inferred during the iron-curtain phase at optical wavelengths. As the density further decreases  the spectrum develops nebular lines which increase their brightness with time \citep{Osterbrock2006}. During this stage the absorption systems disappear.

\begin{figure}[htb]
    \centering
    \includegraphics[width=10cm]{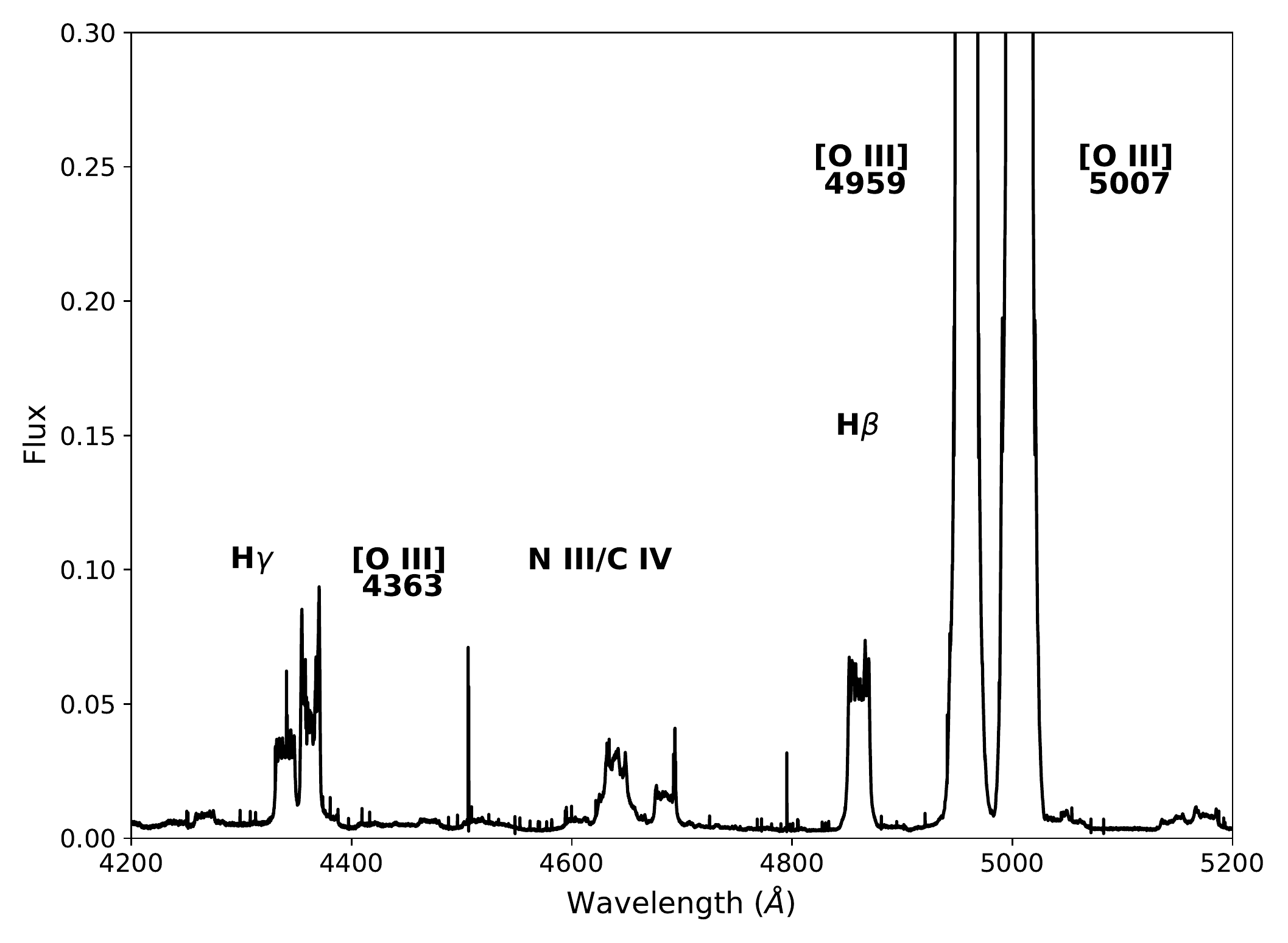}
    \caption{Nebular spectrum observed in V1369 Cen \citep{Izzo2015,Mason2018}.}
    \label{fig:26}
\end{figure}

The enhancement in flux of the He II marks the end of the optically thick phase (e.g., \citealt{Munari2014}) and the beginning of the \textbf{Nebular phase} followed by the increase in luminosity of forbidden transitions
\citep{Rafanelli1978} of oxygen auroral and nebular [O III] 4363 and  4959, 5007, respectively and nitrogen [N II] 5755, 6548, 6584 at optical wavelengths, see Fig.~\ref{fig:26}.  In the spectra of ONeMg novae we also observe bright [Ne III] 3869, 3968 and [Ne V] 3346, 3426.  It is well known \citep{McLaughlin1943,Bappu1954,Payne1957} that spectroscopic observations obtained during the Nebular stage, when the ejecta are optically thin, are useful to study the geometrical and ionization structure of the ejecta (e.g., \citealt{Shore1996,Shore2013, Vanlandingham2001,Iijima2012,DeGennaroAquino2014,DeGennaroAquino2015,Mason2018}). Nebular lines show flat-top or saddle shape or split profiles, often characterized by fine structures on their top, then suggesting both the presence of density variations in the ejecta and in the geometric distributions as polar caps or equatorial rings. 
During the nebular phase, the peak of the emission shifts to UV wavelengths, and finally to X-rays. The UV wavelength range was studied in very details thanks to the International Ultraviolet Explorer (IUE) satellite during the 1980s and the first half of the 1990s (e.g., SG19). In this phase emission lines of ionised helium and high ionisation of carbon, nitrogen, oxygen and neon  freshly-produced elements are observed \citep[e.g.,][and references therein]{Cassatella1979,Stickland1981,Shore1996,Cassatella2002,Cassatella2004}. Optical observations combined with UV spectra obtained at these epochs,  allowed the determination of the ejecta abundances \citep{Vanlandingham1996,Vanlandingham1997,Schwarz2002,Shore2003,Schwarz2007}. These results have definitely proven that element abundances in novae are  generally super-solar, up to 1--2 orders of magnitude (\citealt{Livio1994,Gehrz1998,Gehrz2008}, see also Jos\'e, Shore and Gehrz in \citealt{Bode2008}). During the nebular stage, novae can also achieve their highest degree of excitation, when the ejecta are completely optically thin and the underlying hot WD surface is revealed as a bright source in soft X-rays \citep{Oegelman1993} due to the burning of residual hydrogen-rich material near its surface (see also Sect. 12). In these conditions soft X-rays and the strong UV photon flux are capable to photo-ionize heavy elements, but the low density of the nova environment at these epochs will avoid their complete recombination, resulting in the presence of highly-excited coronal lines in the spectra such as [Ni\,{\sc viii}] 4446, 4493; [Ni\,{\sc ix}] 4332, 4404;  [Fe\,{\sc vii}] 6086; [Fe\,{\sc x}] 6376; [Fe\,{\sc xiv}] 5303; and [Ar\,{\sc x}] 5535 \citep{Rosino1989}. 

Finally, few years after the outburst the object develops the \textbf{Post-nova spectrum} that is characterized by the gradual disappearance of coronal and then forbidden lines, due to the decrease of the ionization level (see \citealt{Tappert2014} and references therein). Residual emission lines remain strong or at least still detectable, e.g.,  H$\alpha$+[N II],  NIII 4640 and  He II 4686 emission line originating from the accretion disc around the WD. The latter two lines, for example, were still visible in the spectra of FH Ser 10 years after the outburst \citep{Rosino1986}.  

\subsection{The Tololo spectroscopic taxonomy}  

\begin{figure}[htbp]
    \centering
    \includegraphics[width=10cm]{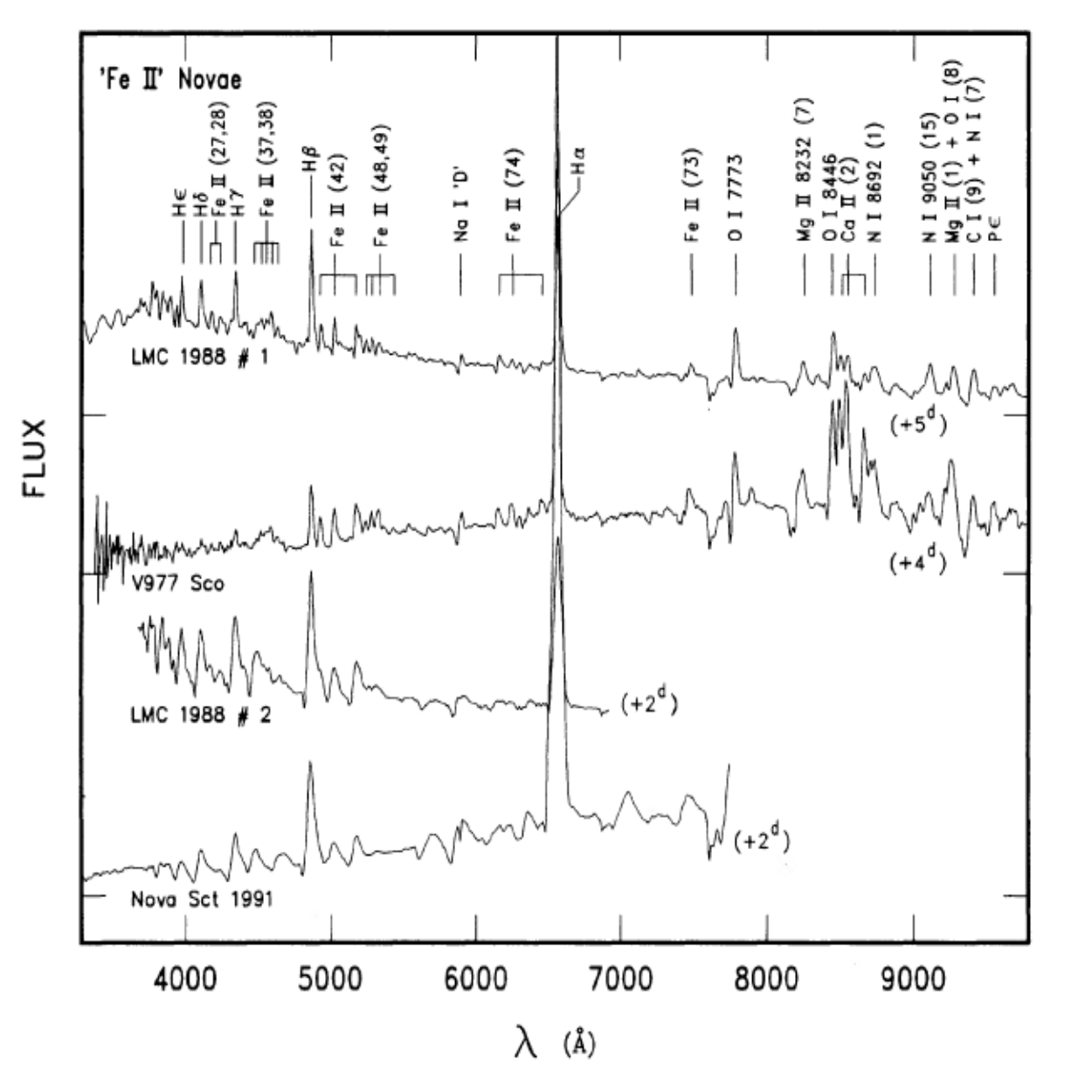}
    \includegraphics[width=10cm]{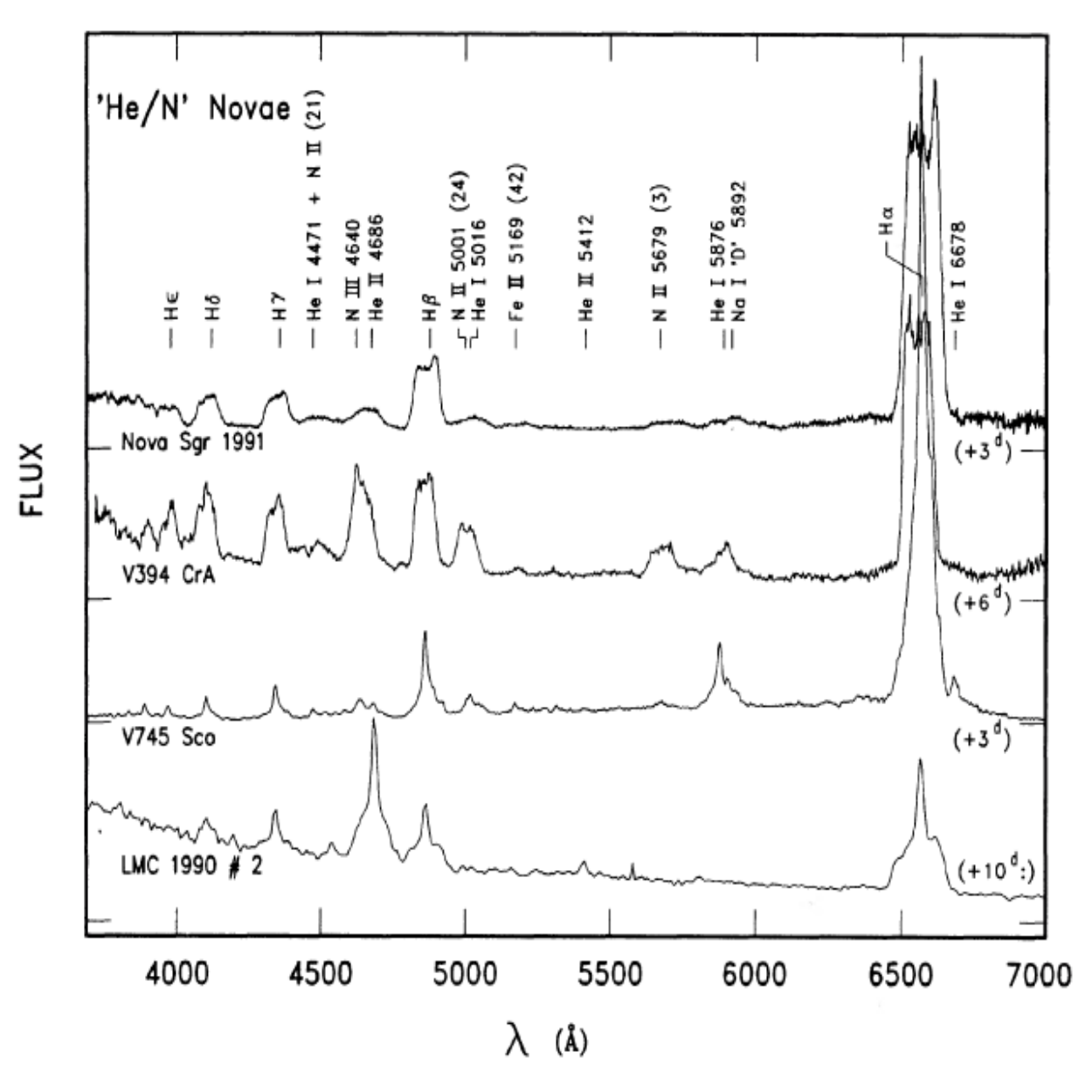}
    \caption{Typical ``Fe-II'' nova spectra obtained close to maximum light (top panel). ``He/N'' spectra of novae observed a few days past maximum (bottom panel). Image reproduced with permission from \citet{Williams1992}.}
    \label{fig:27}
\end{figure}

\citet{Williams1992,Williams1994a,Williams1994b} after studying two dozen of galactic and Magellanic novae concluded that novae can be broadly divided into two spectroscopic classes: the Fe II and He/N novae. The former are characterized by slow spectroscopic evolution with expansion velocities $\lesssim 2500$ km/s (FWZI) and the presence of Fe II lines in the early emission spectrum as the strongest non-Balmer lines. The latter are fast spectroscopically evolving novae, characterized by high expansion velocities ejecta $\gtrsim 2500$ km/s (FWZI) with He and N lines being the strongest non Balmer lines in the emission spectrum near maximum. Hybrid objects (e.g., V1500 Cyg), which evolve from Fe II to He/N, are classified as Fe II-b (b=broad) and are physically related to the He/N rather than to Fe II class (see Fig.~\ref{fig:27}).

\begin{figure}[htb]
    \centering
    \includegraphics[width=10cm]{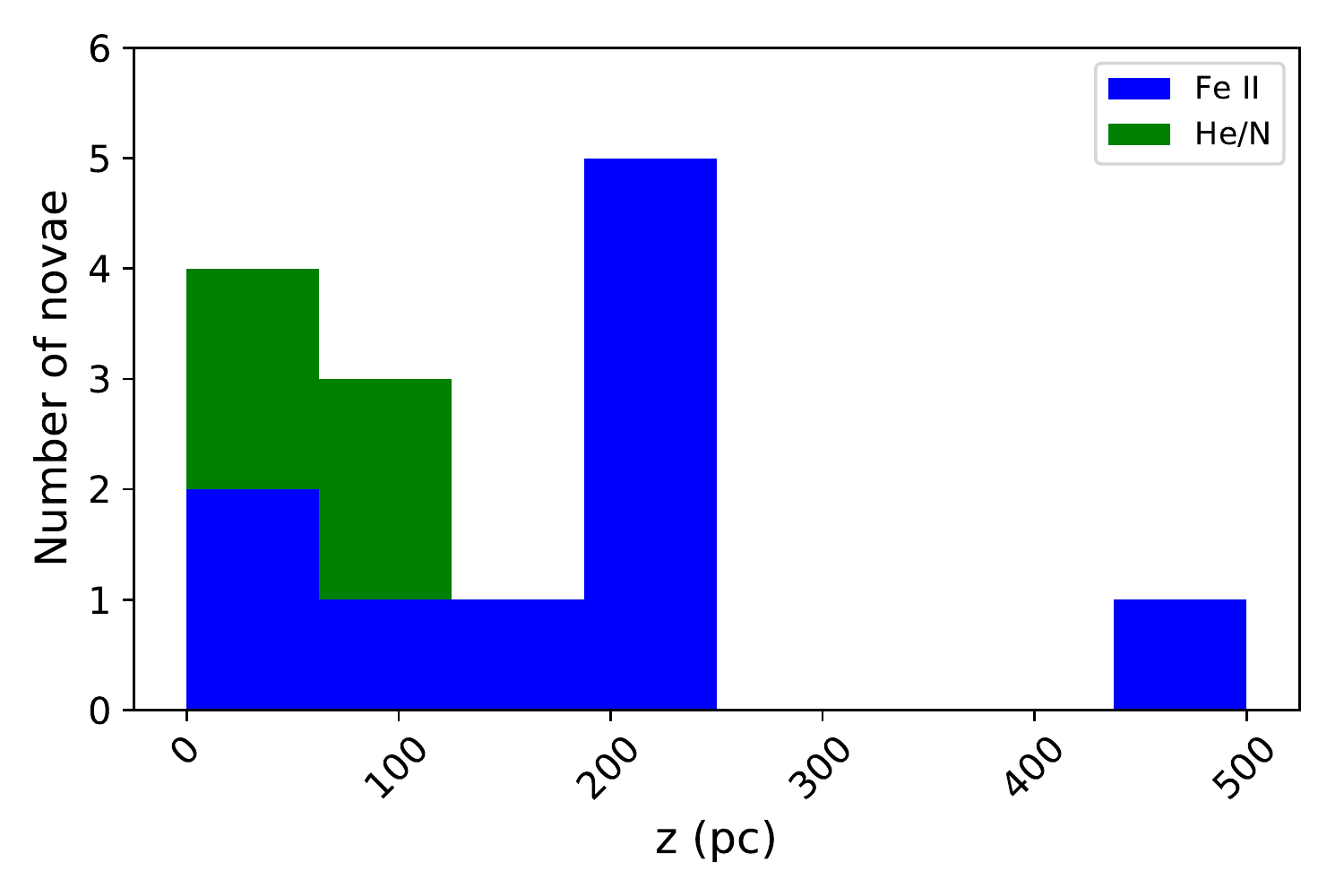}
    \caption{Histogram of the heights of novae above the galactic plane according the Tololo spectroscopic classification. Data obtained from Gaia DR2 distances published in S18 and SG19, while spectral type obtained from \citet{DellaValle1998}}.
    \label{fig:28}
\end{figure}

\begin{figure}[htb]
    \centering
    \includegraphics[width=10cm]{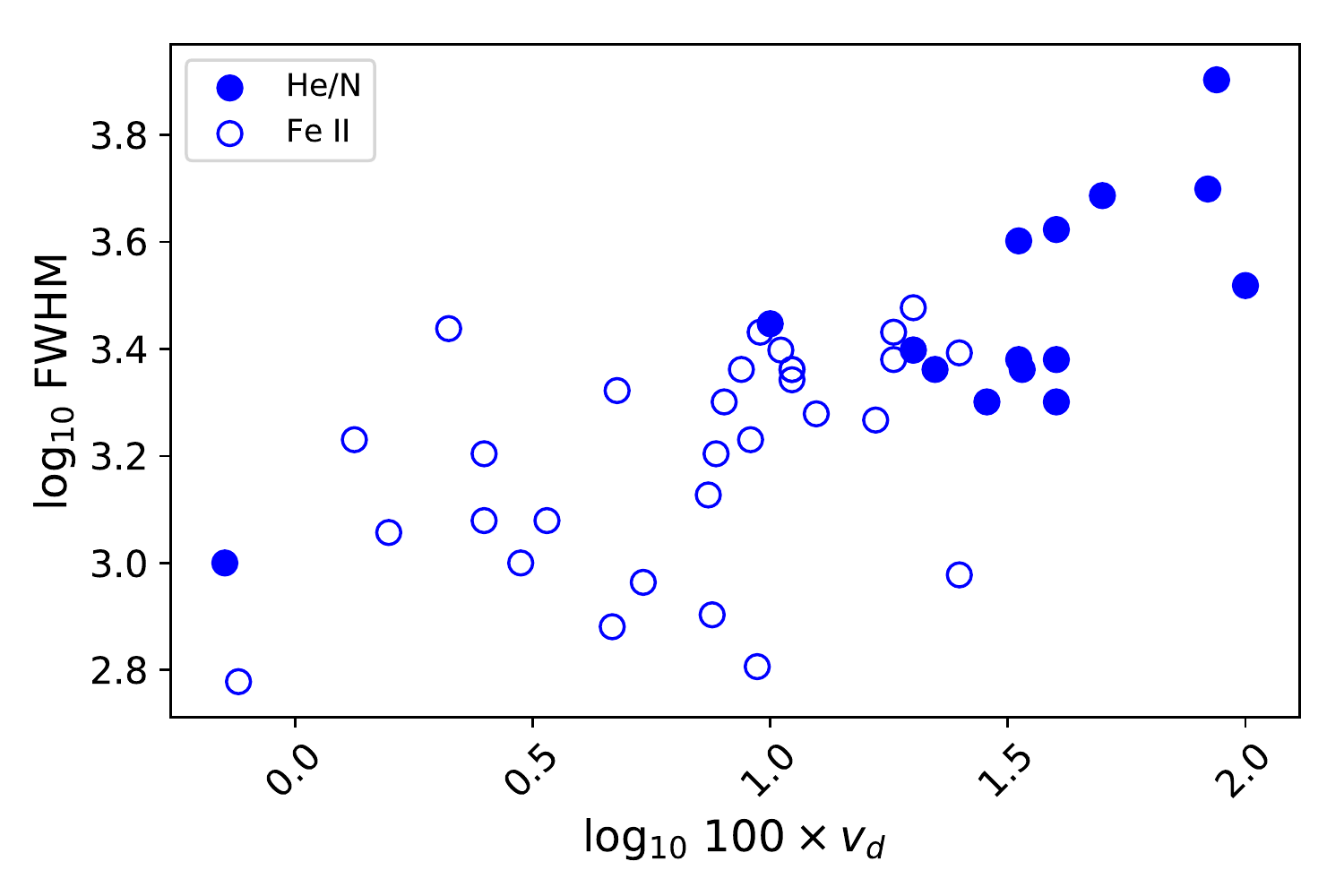}
    \caption{Relationship between the rate of decline and average expansion velocity (obtained from the FWHM) at early stages. Filled dots refer to the He/N and Fe II-broad (hybrid novae) classes. Light dots to Fe II novae. The filled dot at small rate of decline is V5558 Sgr a Fe II-b nova. Data from S18, SG19, \citep{DellaValle1998} and \citep{Ozdonmez2018}}.
    \label{fig:MDV1998}
\end{figure}

In Fig.~\ref{fig:28} we report  the heights above the galactic plane of the novae of our bona fide sample (S18 + SG19) after being classified according the Williams' criteria. The plot shows that novae belonging to the He/N and Fe II-b classes tend to concentrate close to the Galactic plane with a typical scale height $\sim 100$ pc, whereas Fe II novae are distributed more homogeneously up to $z \sim 300$ pc and possibly beyond \citep{DellaValle1998}. Figure~\ref{fig:MDV1998} shows an upgraded version of the relationship between the rate of decline and the average expansion velocity of the nova ejecta measured at early stages, first observed by \citet{McLaughlin1940}. Filled dots refer to the He/N novae and they are characterized by the largest expansion velocities. On the basis of these plots we can infer that objects previously classified as `disk' and `bulge/thick-disk' novae tend to correspond to the  He/N (+ Fe II-b) and Fe II spectroscopic classes. To explain this behavior one should consider the following argument: the more massive the WD (for a given $\dot M$ and T$_{\rm WD}$) the smaller is the mass of the accreted envelope (in view of $\Delta M_{\rm acc} \propto R^4_{\rm WD}/M_{\rm WD}$) the more violent is the outburst and the larger the fraction $\Delta m_{\rm shock}/\Delta m_{\rm wind}$ where $\Delta m_{\rm shock}$ is the shell mass ejected at the maximum and $\Delta m_{\rm wind}$ is the fraction of the shell mass ejected in the subsequent continuous optically-thick wind phase. Since He/N spectra are formed in the shell ejected at maximum light it is very likely that He/N novae are generally associated with massive WDs, which eject the less massive envelopes (see Sect. 1.2). In the framework of this interpretation the distributions reported in Fig.~\ref{fig:28} and Fig.~\ref{fig:MDV1998} are  simple consequences of the trend exhibited by the rate of decline vs.\ height above the galactic plane (see Fig.~\ref{fig:13} ).

\subsection{The THEAs: Transient Heavy Elements Absorptions}\label{sec:42}

The previous spectroscopic classifications were significantly enriched by high resolution observations of novae (R $\sim 50,000$) carried out at ESO La Silla Observatory at the end of 1990s \citep{DellaValle2002,Mason2005,Ederoclite2006} and at other facilities \citep{Sadakane2010}. These observations have revealed \citep{Williams2008} the presence around maximum light of a new spectroscopic system formed by transient heavy element absorptions (THEAs). These authors were able to identify more than 300 absorption lines of single ionized elements with $E_{\rm ion} < 6$ eV belonging to the iron group, such as: Sc, Ti, V, Cr, Fe, Sr, Y, Zr. The FWHMs of THEAs show a broad range of velocities between 30--300 km/s. They all progressively weaken and disappear over timescales of days or weeks.  

\begin{figure}[htb]
    \centering
    \includegraphics[width=\textwidth]{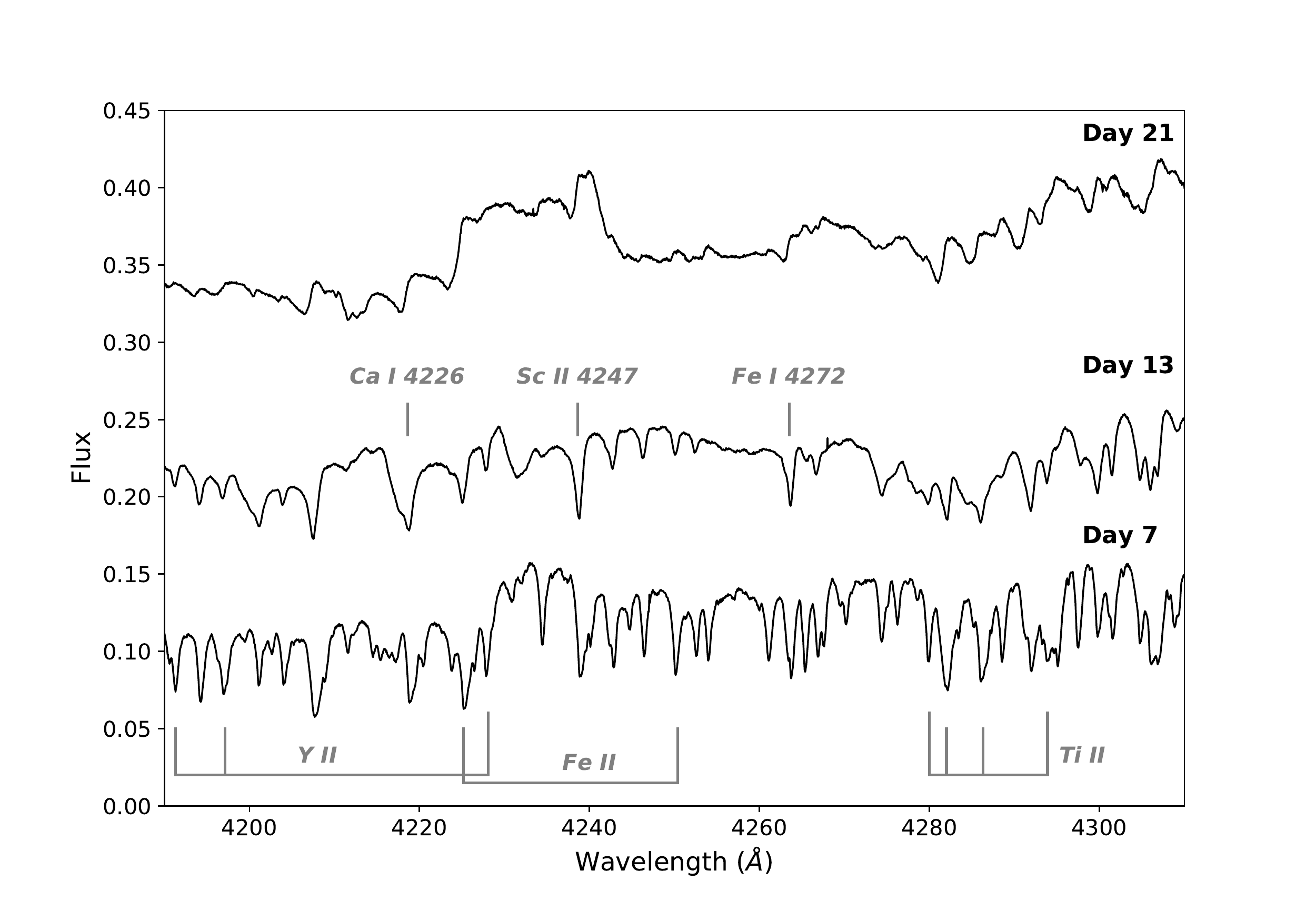}
    \caption{A selected wavelength range showing the optical spectral evolution in V1369 Cen within the first three weeks from the nova outburst. The THEA narrow absorptions are clearly detected within the first two weeks, when the ejecta are still optically thick. The ejecta density decreases with time and then the faint and narrow absoptions are the first features to disappear. Image reproduced with permission from \citet{Izzo2015}}
    \label{fig:29}
\end{figure}

The observations show that THEAs are superimposed on the principal absorption systems of novae and this implies that THEA systems originate outside the expanding photosphere, so they are not a direct product of the nova outburst. After days or weeks when the nova ejecta reach the THEAs, the transient systems are not observed any longer. Another property of THEA lines is that the time at which these lines are not observed depends on the rate of decline of novae. The three fastest novae in the sample of \citet{Williams2008} do not show THEA, indeed, while the slowest objects can exhibit THEAs up to 1--2 months past maximum. The excellent temporal coverage suggests as most plausible explanation that the faster ejecta, which are associated with the fastest declining novae, have reached the THEA absorptions very few days past maximum, before our observations could started. As an alternative one should figure out a mechanism for which the THEA are not developed around fast declining novae. 
The origin of this material can be ascribed to activity of the companion  star. The chemical species observed in THEA systems and the contemporaneous lack of H would seem to exclude that the observed material is a residual of previous outbursts. In the paper by \citet{Williams2008}, it was proposed that they originate from a continuous mass ejection from the secondary star and some empirical estimates suggest that the involved amount of gas is $\sim 10^{-5}\,M{_\odot}$ located at 10--100 AU from the binary system. 



\section{The maximum magnitude vs.\ rate of decline relationship in the Milky Way (MMRD)}

\begin{figure}[htb]
    \centering
    \includegraphics[width=10cm]{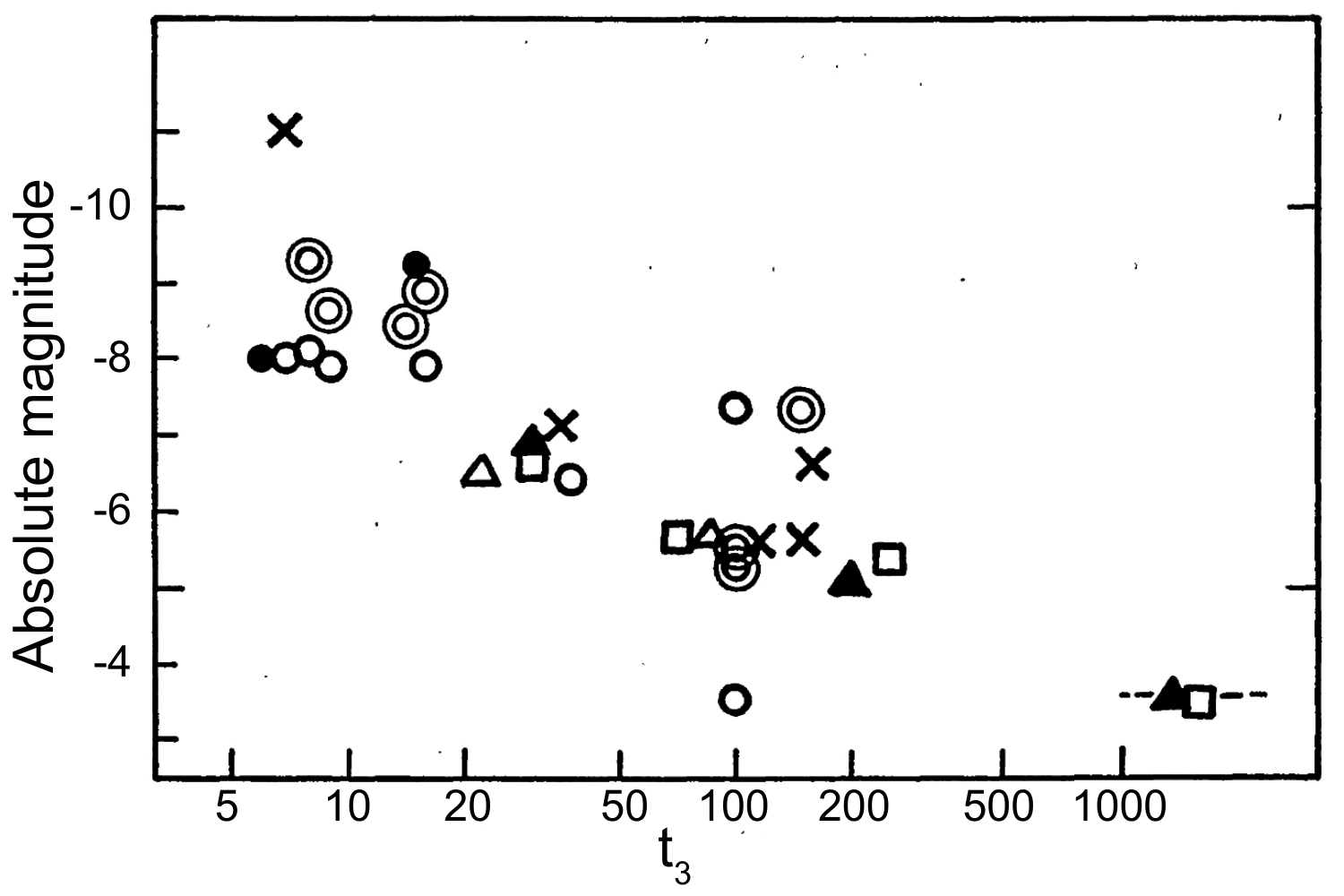}
    \caption{The maximum magnitude vs. rate of decline relation calibrated by \citet{McLaughlin1945c}. Double circles: nebular expansion; open circles: intensity of the interstellar Calcium; solid circles: T Coronae and T Scorpii; crosses: galactic rotation; solid triangles: novae in Andromeda; open triangles: novae in the LMC; squares: novae in Sagittarius.}
    \label{fig:MMRDMcLau}
\end{figure}

The existence for this relation was first suggested by \citet{Zwicky1936}, who named it as \emph{``life-luminosity''} relation. It was expressed as: $M_{\max} = -5\times \log \tau_{\Delta m}$ + const, where $\tau_{\Delta m}$ is the time necessary to decrease $\Delta m$ magnitudes from maximum. Unfortunately he calibrated the relation by using two galactic novae, Nova Per 1901 and Nova Aql 1918, and three Supernovae, SN 1936A in NGC 4273, SN 1926A in NGC 4303 and SN 1885A in M31,  so the results were misleading. The first calibration of the MMRD relation entirely based on nova events was carried out by \citet{McLaughlin1945c}. He used: i)  13 Galactic Novae whose distances were determined with the nebular parallaxes, the intensities of interstellar absorption lines or the residual velocities from interstellar lines interpreted as due to galactic rotation, ii) novae coming from the Hubble survey in M31 and iii)  three novae from LMC (Fig.~\ref{fig:MMRDMcLau}).  

In the following decades a number of relations based on different nova samples, calibrators and assumptions on the galactic absorption in the interval 0.8 mag/kpc \citep{McLaughlin1945a} up to 3.5 mag/kpc \citep{KopyloV1952} have been provided by many authors:

\begin{itemize}
\item $M_o = 2.0\times \log t_3$ - 10.1 \citep{Vorontsov-VelyaminoV1947}; 
\item $M_o = 3.7\times \log t_3$ - 13.8 \citep{KopyloV1952};  
\item $M_{pg} = 2.5\times \log t_3$ - 11.8 \citep{Schmidt1957};  
\item $M_{pg} = 2.5\times \log t_3$ - 11.5 \citep{Schmidt1957}; 
\item $M_V = 2.5\times \log t_3$ -11.75 \citep{Schmidt1957}; 
\item $M_B = 1.8\times \log t_2$ - 11.5 \citep{Pfau1976}; 
\item $M_{pg} = 2.4\times \log t_3$ - 11.3 \citep{deVaucouleurs1978}.
\end{itemize}

\begin{figure}[htb]
\centering
\includegraphics[width=10cm]{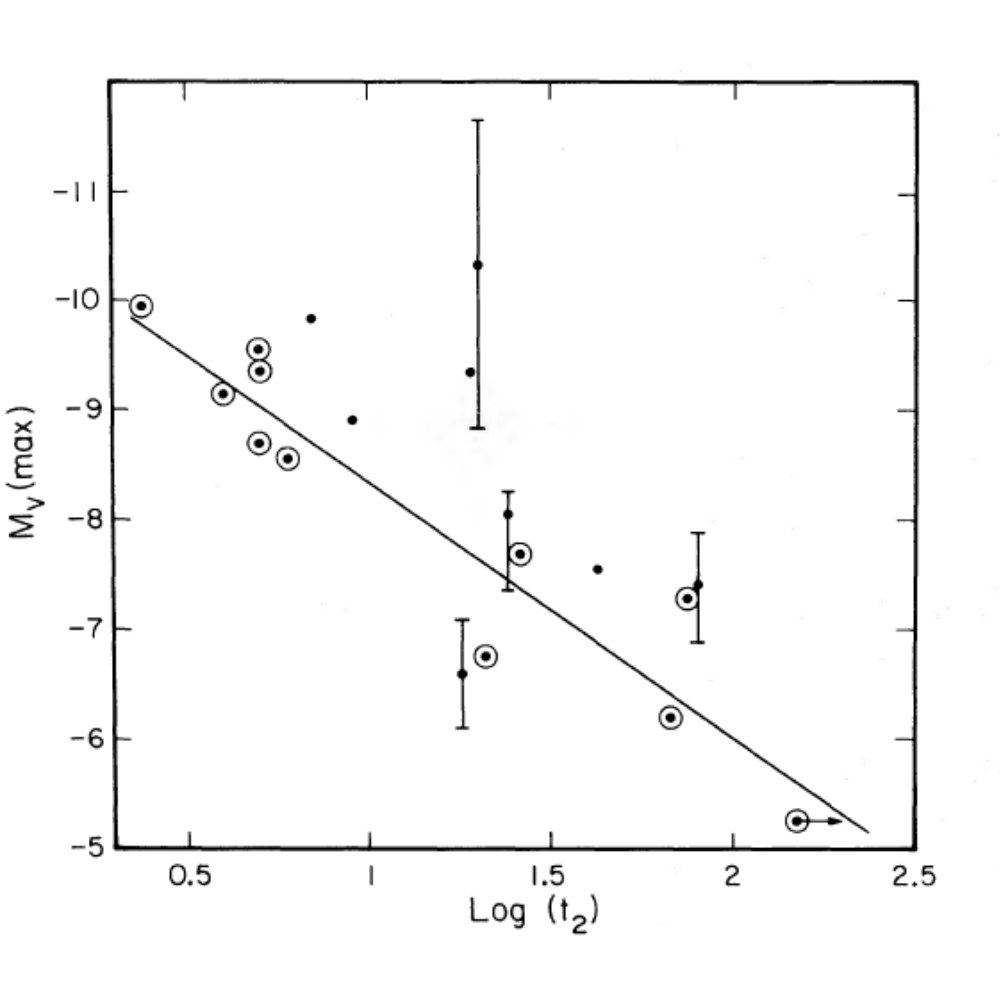}
\caption{The MMRD relation obtained via nebular parallaxes of 19 novae. Image from \citet{Cohen1985}.}
\label{fig:cohen}
\end{figure}

In the 1980s, the calibration of the MMRD relation received its first breakthrough. \citet{Cohen1983} and \citet{Cohen1985} were able to measure the size of 19 shells around post-novae. These observations, together with the i) expansion velocities measured from the spectra, ii) the assumption of spherical symmetry in the expansion of the ejecta and iii) individual estimates of the galactic absorption toward each nova, provided for the first time a set of nova distances, which were homogeneously measured, and a new calibration of the MMRD relationship.  The result, reported in Fig.~\ref{fig:cohen}, yielded:

\begin{equation}
M_V = 2.41\pm 0.23\times \log t_2 - 10.70\pm 0.30.
\end{equation}

\citet{Downes2000} have increased this sample to 30 objects and improved the quality of ``old'' measurements of nova shells surrounding nine post-novae with ground-based and Hubble Space Telescope (HST) data. Their linear fits gives:

\begin{equation}
M_V = 2.54 \pm 0.35\times \log t_3 -11.99\pm 0.56
\end{equation}
\begin{equation}
M_V = 2.55 \pm 0.32 \times \log t_2 -11.32\pm 0.44
\end{equation}
 
\begin{figure}[htb]
    \centering
    \includegraphics[width=10cm]{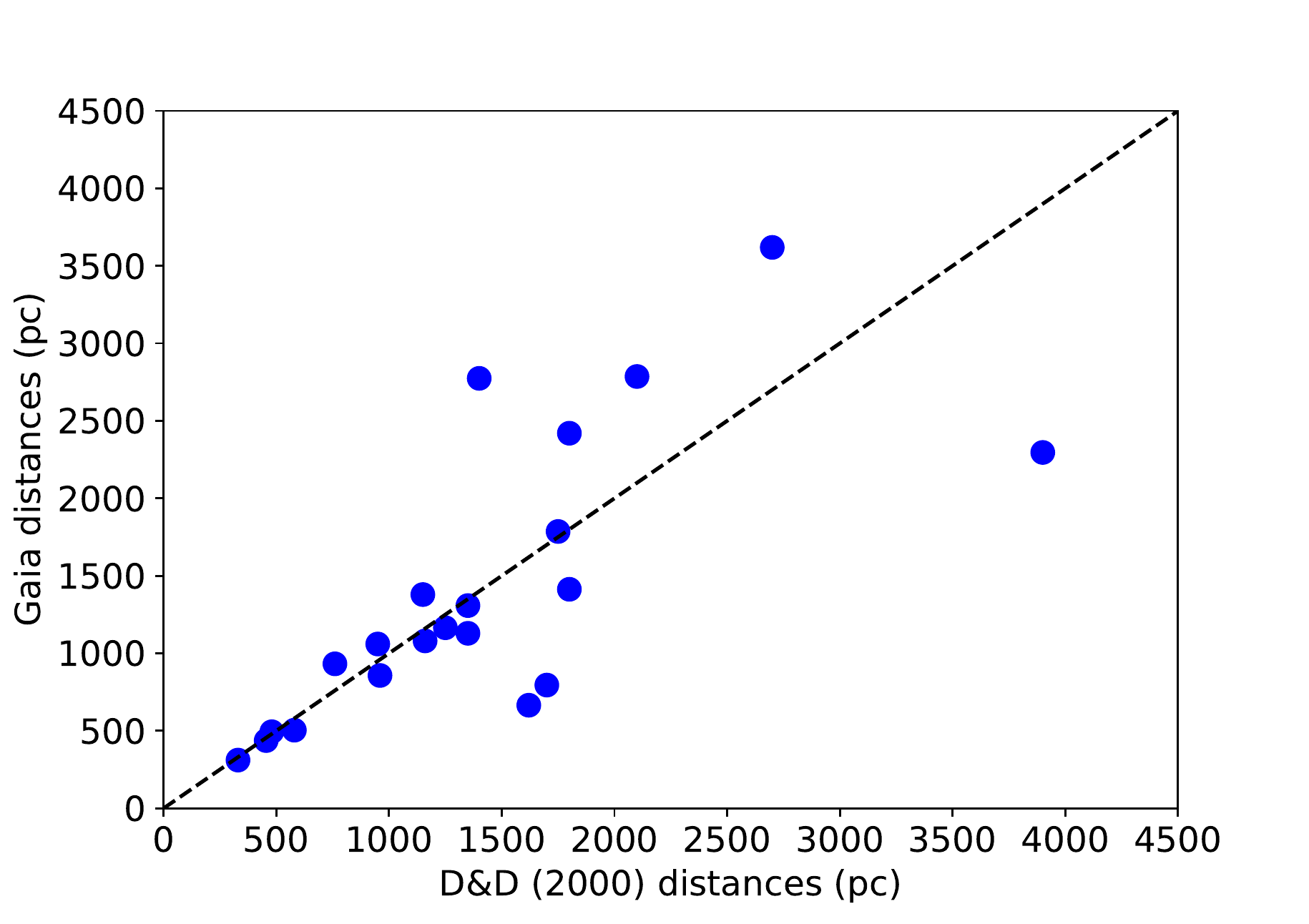}
    \caption{Distances obtained via nebular parallaxes (data from Downes \& Duerbeck 2000) vs. GAIA DR2 distances (data from S18 and  SG19).}
    \label{fig:31}
\end{figure}

The GAIA DR2 distances \citep{GAIA2018} give the opportunity to check the reliability of nova distances measure via nebular parallaxes. Fig.~\ref{fig:31} shows that nebular parallaxes are certainly consistent with GAIA DR2 measurements below 1.5 kpc. The scattered points at distances larger than 1.5 kpc is likely due to failures on the assumptions of the nebular parallaxes methodology (e.g., spherical symmetry expansion) and in particular to the poor quality of the A$_V$ estimates. 
 
\medskip 

\begin{figure}[htb]
    \centering
    \includegraphics[width=10cm]{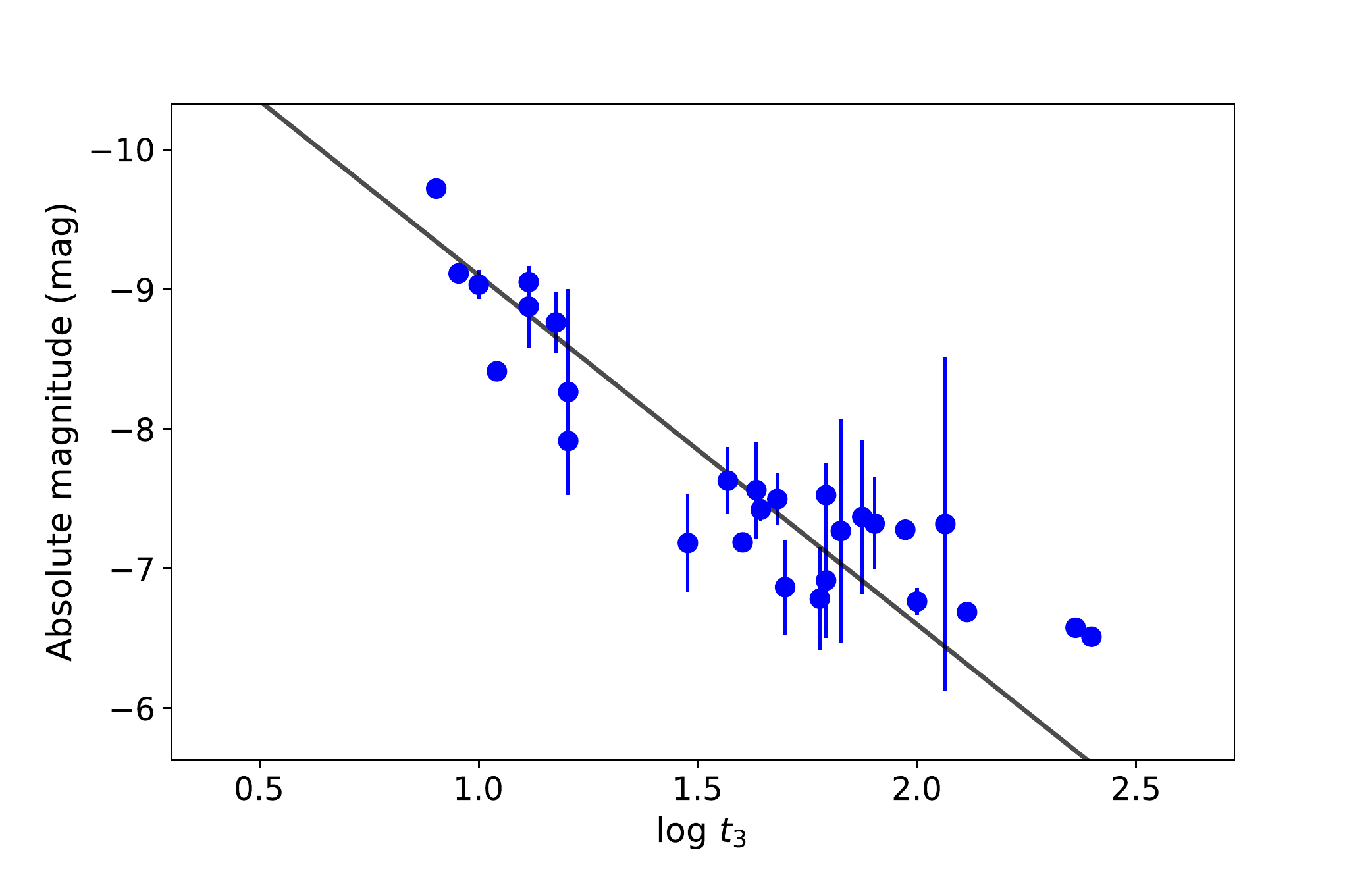}
    \caption{The absolute V magnitude plotted against the $t_3$ value for the SG19 and S18 nova sample. The solid line corresponds to the best-fit found by \citet{Hachisu2009}}
    \label{fig:Jap}
\end{figure}

The existence of the MMRD can be broadly outlined from simple physical arguments \citep{Bath1978,Kantharia2017} or theoretically modelled \citep{Hachisu2015}.  At first order one can estimate the luminosity at maximum of a nova as: $L=4\pi R^2 \sigma T^4$, where $R \sim v\times t$ with $v$ being the average expansion velocity of the ejecta and $t$ the time from the outburst.  This simplified approach shows in very intuitive way why novae characterized by the largest expansion velocities can achieved the highest luminosity at maximum light. For this reason their envelopes become optically thin very soon and their decline in luminosity proceed faster than in intrinsically fainter (and slowly expanding) novae. The tacit assumption in this argument is that the masses of the ejecta have roughly the same mass for all CNe: and this is true within a factor about ten, between $10^{-4}$ and $10^{-5}\,M_\odot$. For example, some RNe or \emph{ultra-fast} novae \citep{Yaron2005,Shara2017b} that expel envelopes, about 100--1000 times lighter than observed in CNe,  achieve the transparency condition ($\tau<<1$) in the ejecta very soon, therefore, their maxima are fainter and also they decay much more quickly than CNe. The MMRD relation was analytically  derived by \citet{Shara1981} and more recently computed from the theoretical modeling of different lightcurves calculated for a large variety of white dwarf chemical composition and masses \citep{Hachisu2006,Hachisu2007,Hachisu2009,Hachisu2019,Hachisu2008,Kato2009}. After normalizing the lightcurves for an appropriate time scale factor, the latter authors derive a ``universal'' behavior that can be described as : $M_V = 2.5\times \log t_3 - 11.6$,  which fit the GAIA data within a sufficient degree of approximation (see Fig.~\ref{fig:Jap}). 

However, even after the papers by \citet{Cohen1985} and \citet{Downes2000} the large scatter exhibited by the MMRD, about 3 magnitudes (see Fig.~\ref{fig:cohen}) has hampered the use of novae as effective distance indicators. This scatter is essentially produced by three factors. The first one is connected to the nebular parallax methodology that assumes symmetric expansion for the nova ejecta. By neglecting the role of opening angles and inclination of the ejecta it can bring into the distance measurements significant errors \citep {Ford1988, Shore2012, Wade2000} at least for more distant and old novae. The second element is the lack of a method to measure with the necessary accuracy the visual extinction toward the nova. We observe novae from inside our Galaxy and the estimates of the reddening toward each nova are very crude (see Sect. 5.2).  They are based on empirical relations based on the intensity of interstellar lines \citep{Munari1997} and from the comparison of predicted and observed emission lines ratios  \citep{Robbins1968a,Robbins1968b}. For example, unblended H and He recombination lines can be used for  this  goal if  they  are  optically  thin and not affected by collisions \citep{DellaValle2002}. Yet the attached errors are large and the final estimate of absorptions very uncertain. Finally on the top of this we should add the intrinsic scatter of the MMRD relationship.

\subsection{The shape of the MMRD relationship} 

\begin{figure}[htb]
    \centering
    \includegraphics[width=10cm]{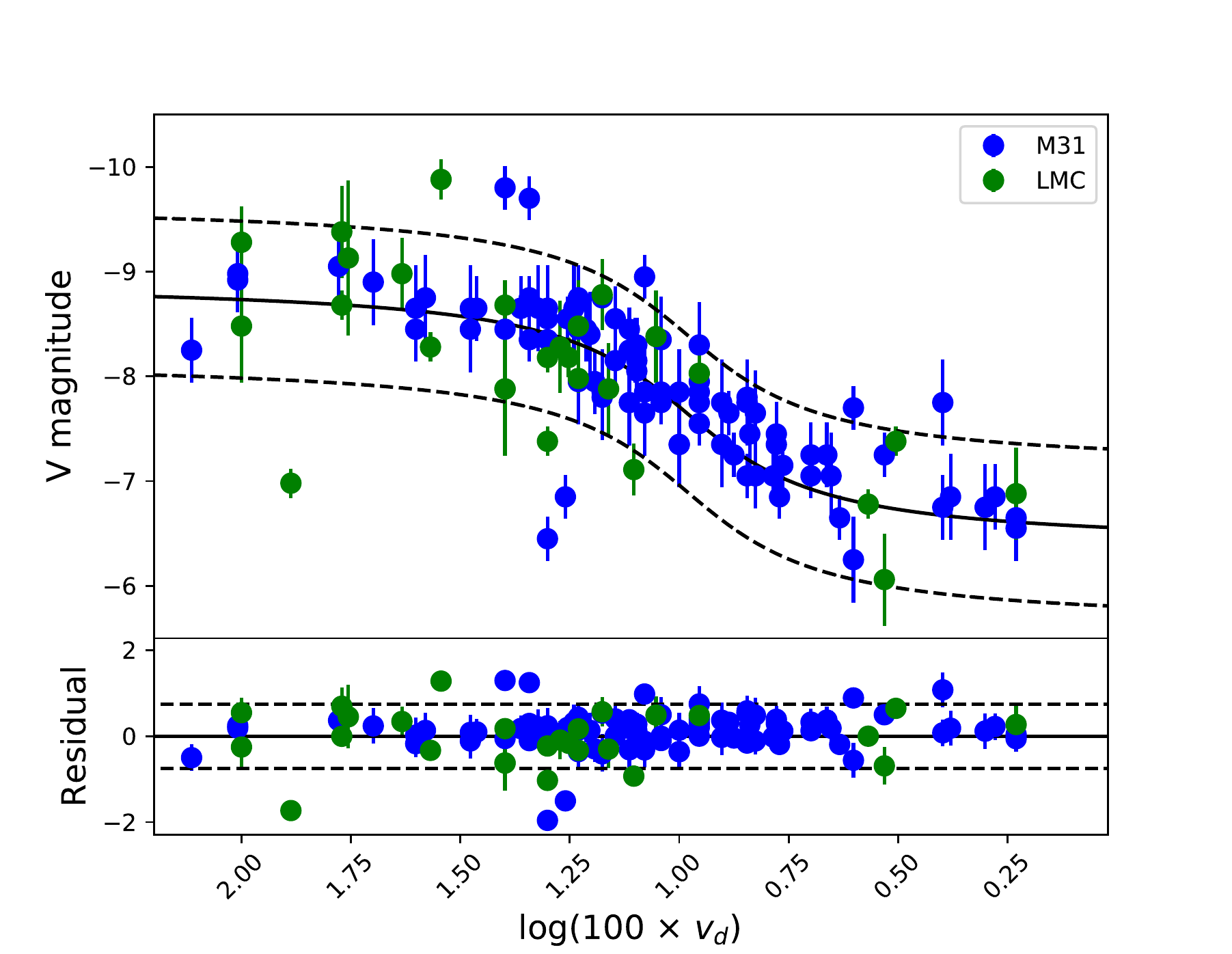}
    \caption{M31 and LMC novae in the MMRD plane (data from \citealt{DellaValle1998,Shafter2013}). The best fit is the solid line described by Eq.~(\ref{eq:MMRD_MW}). With black and grey dashed lines we report the $\pm 3\sigma$ strips, with $1\sigma=0.25$. A $\tilde{\chi}^2$ test indicates that the M31 and LMC nova data samples are better fitted by an S-shape formulation than a linear function: $\tilde{\chi}^2_{S} = 1.70$ while for the linear model $\tilde{\chi}^2_{\rm lin} = 1.87$. }
    \label{fig:MMRDM31LMC}
\end{figure}

Local effects as the reddening toward each nova and the difficulty of determining distances accurately suggest to look at other galaxies to calibrate and to determine the ``true'' shape of the MMRD relationship. These drawbacks are well detectable by comparing Fig.~\ref{fig:cohen} with Fig.~\ref{fig:MMRDM31LMC}.

The former plot is the MMRD derived through nebular parallaxes by \citet{Cohen1985}. Here the uncertainties on the distances and absorptions, increase the intrinsic scatter of the MMRD, which can well fitted by a linear regression. Fig.~\ref{fig:MMRDM31LMC} represents the MMRD derived by studying the nova population in M31 and LMC (data from \citealt{DellaValle1995,Shafter2013}, all data have been collected in Table~\ref{tab:app5} and Table~\ref{tab:app6}). This plot shows a linear trend only between  $0.04 < v_d < 0.2$ ($10 < t_2 < 50$ days). For the entire range of $v_d (=2/t_2$) the analytic representation is a reverse S-shaped function \citep{Capaccioli1989,DellaValle1995}. A $\tilde{\chi}^2$ test indicates that M31 and LMC nova data are better fitted by a S-shape function (see Eq.~\ref{eq:MMRD_MW} in Sect.~5.2) than a simple linear fit: $\tilde{\chi}^2_{S} = 1.70$ while for the linear model $\tilde{\chi}^2_{\rm lin} = 1.87$. The linear fits performed on novae in the Milky Way are essentially due to scanty statistics combined with the large  uncertainties associated to the measurements of the interstellar absorption along the line of sight to each single nova.

\begin{figure}[htbp]
    \centering
    \includegraphics[width=10cm]{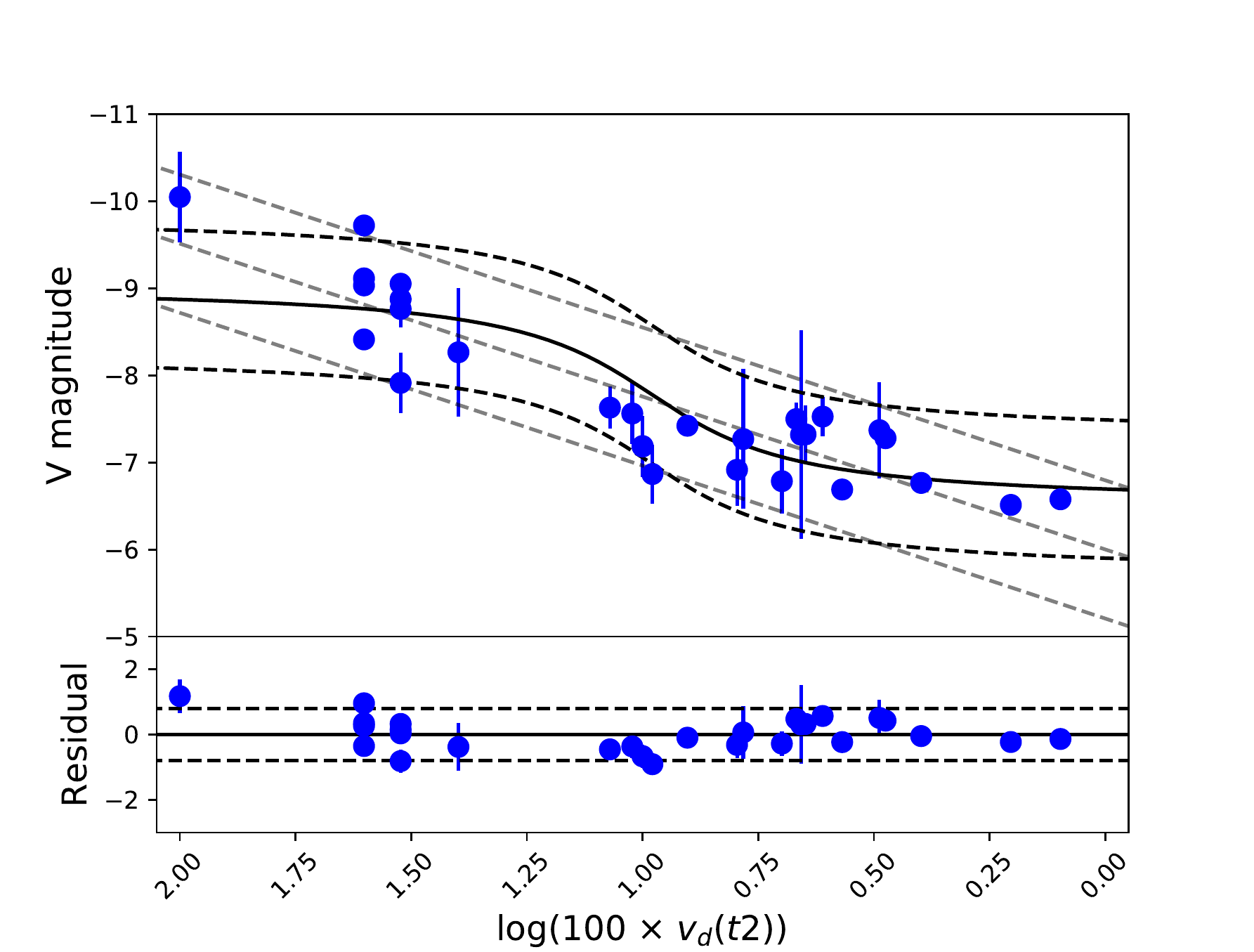}\\
    \includegraphics[width=10cm]{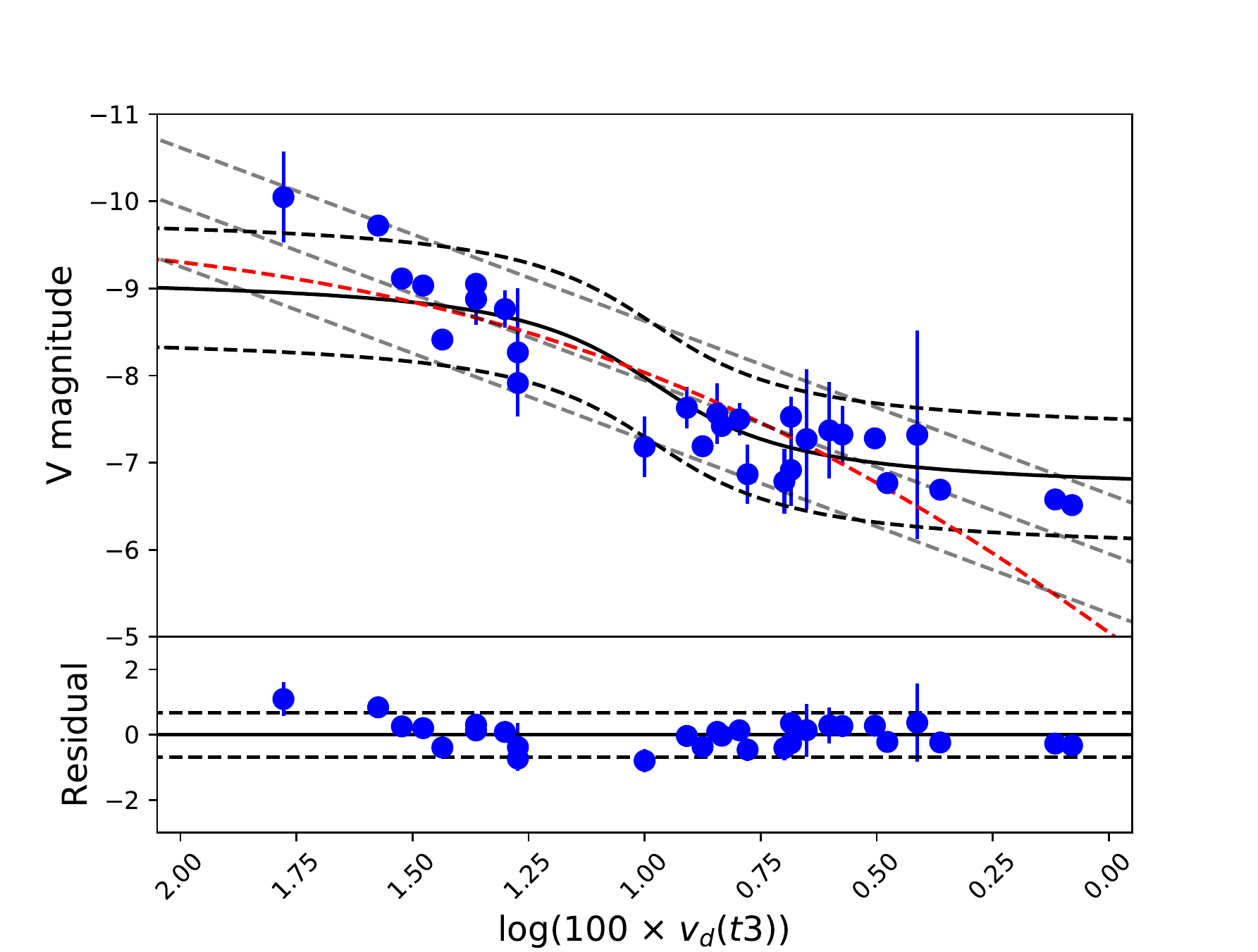}
    \caption{(Upper panel) Absolute magnitudes derived from GAIA DR2 distances and rate of decline for the $t_2$ parameter, fitted by the S-shape reverse relation. The solid line is the best fit Eq.~(\ref{eq:MMRD_MW}), where the two dashed lines identify the $\pm 3\sigma$ strip, with $\sigma = 0.26$ mag. The dashed gray lines correspond to the fit obtained with a linear model. The magnitude of the intercept of the MMRD with the y-axis is $M_V=-8.85$ mag for the $t_2$ parameterization. (Lower panel) Same analysis, but for the $t_3$ parametrization. The two dashed lines identify the $\pm 3\sigma$ strip, with $\sigma = 0.23$ mag. The dashed red line is the theoretical track (Eq.~\ref{eq:14}). The magnitude of the intercept of the MMRD with the y-axis is $M_V=-9.02$ mag. Data have been obtained from Table~\ref{tab:app1} and Table~\ref{tab:app2}. $\tilde{\chi^2}_{\rm S-shape}$ = 1.02, $\tilde{\chi^2}_{\rm linear}$ = 1.05.}
    \label{fig:GOLDt3}
\end{figure}

Theoretical attempts to explain the MMRD relation also provide evidence that a linear fit to the nova data only provides a rough first order approximation. \citet{Hartwick1978,Shara1981,Livio1992} were able to derive simple relationships between the absolute magnitude at maximum and the mass of the underlying WD. For example, \citet{Livio1992} derives:
\begin{equation}
M_{\max} \approx -8.3 -10\times \log M_{\rm WD}/M_\odot
\end{equation}
and in turn a theoretical MMRD relation of the form:
\begin{equation}\label{eq:14}
t_3= 51.3 \times 10^{0.2\times (M_B +9.76)} \times [10^{0.067\times (M_B + 9.76)} - 10^{-0.067 \times (M_B + 9.76)}]^{3/2}  
\end{equation}
that is expressed in days. The red dashed line in Fig.~\ref{fig:GOLDt3} shows the theoretical relation, superimposed on the GAIA MW data, which is expressed in absolute magnitudes. 
The flattening of the observed distribution at high luminosities is real and can be interpreted as a physical threshold due to the fact that the WDs in nova systems which result in super-Eddington novae are approaching the $\sim 1.4\,M_{\odot}$ Chandrasekhar limit. The flattening at faint level of luminosity, characterized by m$_{pg}$ $\sim$ 19 mag, is relatively near to the photographic detection limit of the M31 surveys by Rosino, and may well represent the bright wing of Eddington novae which are populating the bottom of the MMRD relation.  Therefore, we cannot exclude that the faint flattening might be the result of an observational bias. However, we note that the nature of the outburst for very low mass WD has not been fully explored and, therefore, it is still possible that the observed flattening is a consequence of the physics of the outburst.

\subsection{How good is the MMRD?}

Recently, by taking advantage from nova distances determined by GAIA DR2, the MMRD has been tested by S18 and SG19 and they have achieved very different conclusions. The former author concluded that \emph{``\dots the MMRD should no longer be used''} whereas the latter authors reached opposite conclusions. These conflicting results are even more puzzling after considering that the two samples share 15 novae and the parameters used to compute the MMRD are very similar. Then we have formed a ``bona fide'' sample of 29 objects to test the MMRD relation. This sample  includes the ``golden sample'' of S18, the SG19 novae and V1500 Cyg, whose parallax measured by GAIA DR2 is $p = 0.777 \pm 0.187$ mas (see Table \ref{tab:app1}). Two novae have been removed from S18 ``golden sample'' sample: BT Mon and V1330 Cyg, whose magnitudes at maximum and rates of decline reported by the author are incorrect.  The maximum light of BT Mon was certainly much brighter than the value of $m(\max) = 8.5$ mag reported by S18. This conclusion can be reached after noting that spectroscopic observations of BT Mon \citep{McLaughlin1941} revealed that this nova was in the so called ``Orion'' phase when it was observed at $m=8.5$ mag. Since Orion phase is subsequent to the principle and diffuse enhanced spectra (see Sect. 4), this magnitude value cannot correspond to the magnitude at maximum of the nova. On the basis of these arguments, \citet{Smith1998} suggested that the maximum of nova BT Mon could be as bright as m$_v\approx 4$ (see SG19 for a deep discussion of the BT Mon case). Also the estimate of the magnitude at maximum of V1330 Cyg, $m(\max)=9.9$ mag, reported by S18 appears incorrect. According to \citet{Ciatti1974}:

\begin{quote}
``the maximum was lost. At the time of the discovery the nova was just at the end of the early decline, in the transition phase, showing the typical fluctuations of a moderately fast nova\dots. Since in general the transition takes place when the nova has fallen about 3.5 mag below maximum \dots it is likely that the maximum was attained near May 15 with 7.5.''
\end{quote}

The same authors estimate $t_2=12^d$, which is much shorter than $t_2=161^d$ reported by S18. In view of these uncertainties we have exclude from our ``bona fide'' sample both BT Mon and V1330 Cyg.   

After applying the reverse S-shape function, derived from M31 and LMC novae, to our ``bona fide'' sample, we get the following absolute calibration for the MMRD of the MW:
\begin{equation}\label{eq:MMRD_MW}
M_V = -7.78(\pm 0.22) - 0.81 \times \arctan ((1.32 - \log t_2)/0.23)
\end{equation}

\begin{figure}[htb]
    \centering
    \includegraphics[width=\textwidth]{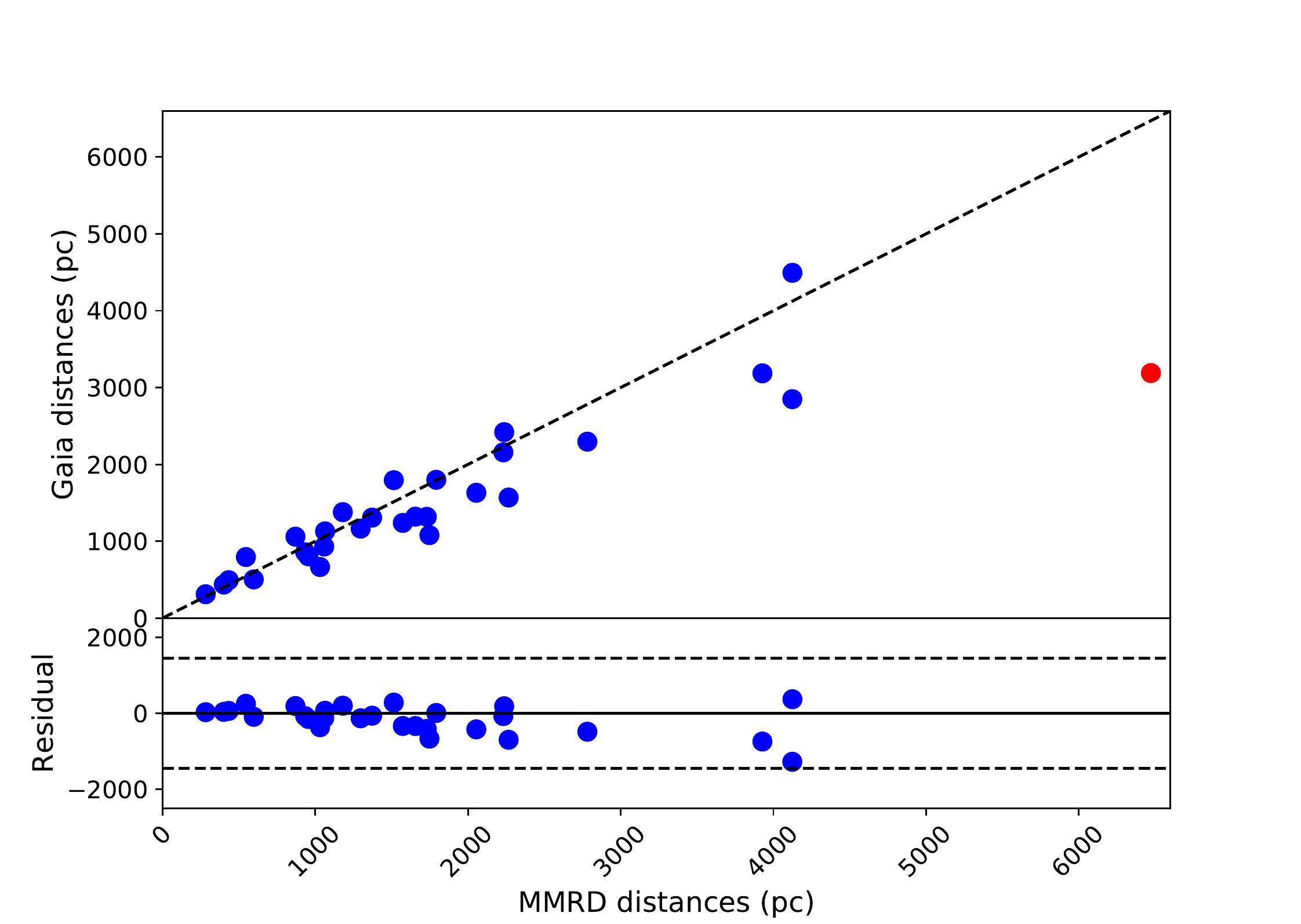}
    \caption{GAIA DR2 distances vs.\ MMRD distances. The red point marks CI Aql. MMRD distances are computed using the MMRD formulation of Eq.~(\ref{eq:MMRD_MW}). Data from Tables \ref{tab:app1}.}
    \label{fig:GAIAMMRD}
\end{figure}

In Fig.~\ref{fig:GAIAMMRD} we compare the distances obtained via GAIA DR2 with the distances obtained via MMRD. The plot shows a good agreement between GAIA DR2 and MMRD distances. Basically, the most deviating object is the Recurrent Nova CI Aql, which is characterized by one of the highest correction for absorption ($A_V=2.6$ mag). The inspection of the residuals indicates that above $\sim 2$ kpc the MMRD tends to overestimate the distances up to $\sim 30\%$. Unless to figure out some bizarre  mechanism for which the MMRD relation is followed only by novae nearby the Sun, the most simple explanation of this behavior is not a failure of the MMRD but a generalized underestimation of the measure of $A_V$ toward the more distant novae. Our analysis support the SG19 results, contrary to S18 conclusions.

\section{Nova populations in external galaxies} 

The use of galactic data to study the properties of the Milky Way nova populations has been often questioned because of the observational bias due to both interstellar absorption in the galactic disk and our position within the Galaxy. These effects can be largely minimized by studying the spatial distribution of nova populations in external galaxies where novae are all at essentially the same distance.   

\subsection{Novae in M31}\label{sec:6.1}

\begin{figure}[htb]
    \centering
    \includegraphics[width=10cm]{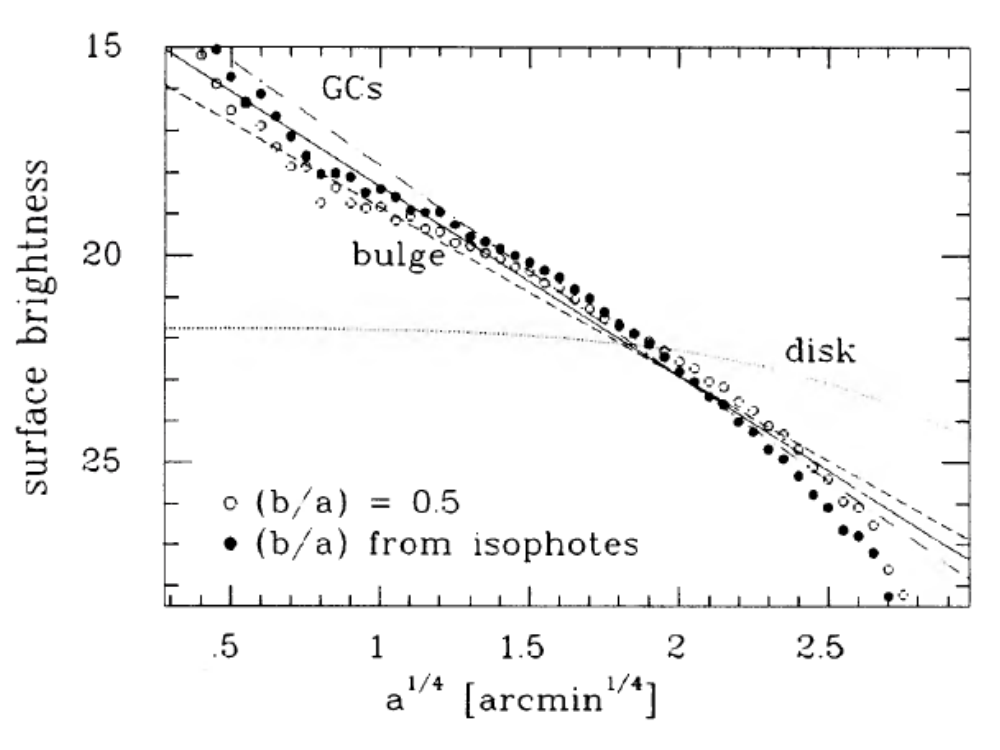}
    \caption{Surface density profiles of M31 nova population (light and filled dots) as a function of the semimajor-axis distance from the center. The solid and dashed lines represents the photometric $r^{(1/4)}$ models of the bulge \citep{Walterbos1986,deVaucouleurs1958}. The long dashed is the density profile of GC systems. The nova data (circles) follows the light profile of the bulge. Image reproduced with permission from \citet{Capaccioli1989}.}
    \label{fig:Capaccioli}
\end{figure}

Due to its closeness and the relatively high frequency of nova events, the M31 nova population has been the only one extensively studied. However, a simple skimming through past literature reveals the existence of different ideas on the nova population assignment: intermediate populations \citep{Arp1956,Rosino1964}, halo population \citep{Wenzel1978}, mainly bulge population \citep{Ciardullo1987,Capaccioli1989,Shafter2001},  mostly disk population \citep{Hatano1997}. Although it is possible that a fraction of M31 novae assigned to the bulge by \citet{Ciardullo1987} and \citet{Capaccioli1989}  are in fact physically related to the disk and their allotment to the bulge was a simple consequence of neglecting the geometrical projection effects \citep{Hatano1997}, there are two arguments that strongly suggest that most novae in M31 mainly belong to the bulge. First of all, \citet{Capaccioli1989} show a close similarity between the projected density distribution of the nova system with the decomposed light profile of M31 into the two main components: bulge and disk (see Fig.~\ref{fig:Capaccioli}). In this plot, novae follow the solid line, i.e., the $r^{1/4}$ model for the bulge along the major axis resulting from the decomposition in bulge/disk components of the B band luminosity profile \citep{Walterbos1986}, while the dotted line is the exponential model of the major axis B light profile of the disk from the same author. It is apparent that the nova surface density follows the light profile of the bulge and unless severe selection effects, due to dust for example prevent the discovery of disk novae in M31,
one can conclude that novae ignore the trend of the light of the disk. We note that \citet{Kaur2016} results on M31 seem to exclude that dust can induce such a selection effect.  

\citet{Shafter2001} have reappraised the Hatano et al.'s conclusions, after comparing the spatial distribution of novae with the background light of M31. They indeed confirm that most novae in M31 originate in the bulge: likely about 70\% and not less than 50\%. Some interesting hints come also from spectroscopic observations \citep{Shafter2012}. These authors find that about $\sim$80\% of M31 novae belong to the Fe II spectral class. In view of what we discussed in Sect. 4.1, we can conclude that most novae in M31  are ``bulge'' novae.  

The first estimate of nova rate in M31 dates back to \citet{Hubble1929} who derived from a sample of about eighty novae an observed rate of 21.5/yr. After applying a number of correction factors he concluded: \emph{``that thirty per year is a reasonable estimate of the frequency of novae in M31''}. Not far from this result appears the estimate by \citet{Arp1956}. He obtained from an almost continuous 1.5 year survey carried out with the 60-inch telescope at Mount Wilson an observed rate of 19.85 per year, which he corrected to a ``real'' frequency of $26\pm 4$ novae per year to account for systematic bias. By taking advantage of two nova surveys carried out at Asiago Observatory with the 1.22-m \citep{Rosino1964,Rosino1973} and 1.82-m \citep{Rosino1989} telescopes, \citet{Capaccioli1989} derived from a global sample of 142 novae, an observed rate of $29\pm 4$ novae per year. Later analysis of \citet{Shafter2001} based on 53 new discovered novae found a rate slightly larger than previous estimates: $37^{+12}_{-8}$ novae per year, and about 70\% of them  produced in the bulge. \citet{Darnley2006} reported in the framework of POINT-AGAPE microlensing survey a nova rate in M31 of $65^{+16}_{-15}$ novae/yr from an ``observed'' sample of 20. This is a full factor two larger than previously estimated. All nova rate data are summarized in Table~\ref{tab:3}. 

\begin{figure}[htb]
    \centering
    \includegraphics[width=10cm]{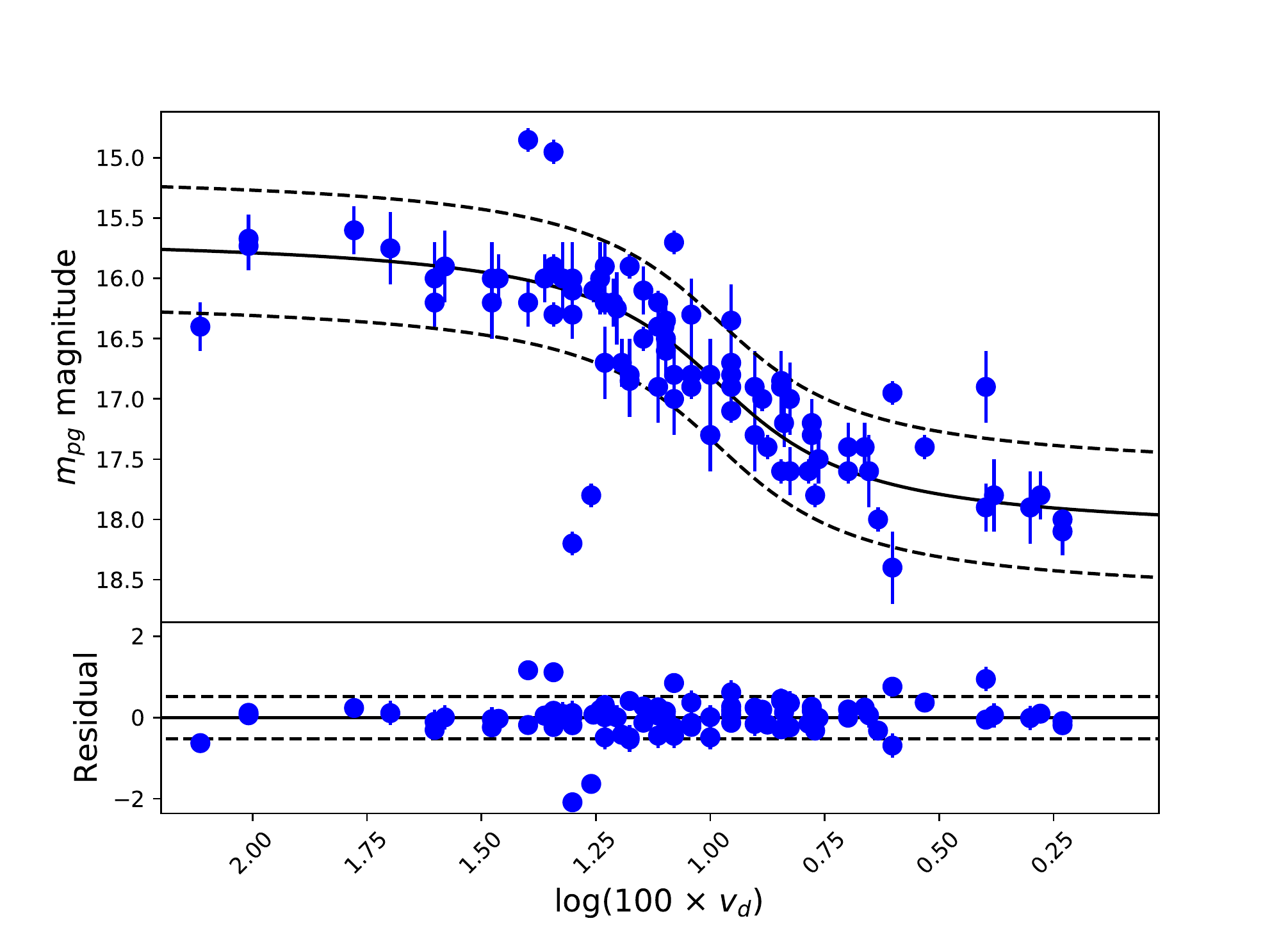}
    \caption{The MMRD relation for 91 M31 novae. The two dashed lines mark the $\pm 3\sigma$ strips, with $\sigma = 0.17$ mag strip. The magnitude of the intercept of the MMRD with the y-axis is $M_{pg}=15.76$ mag. Data have been obtained from Table \ref{tab:app5}}.
    \label{fig:CapaccioliM31}
\end{figure}

\begin{table}[htb]
    \caption{Collection of nova rate measurements for M31.}
    \label{tab:3}
    \centering
    \begin{tabular}{c|c|c|c}
    \hline\hline
     Novae&rate $(yr^{-1})$&  author & method\\
         \hline
 85 & 30 & \citet{Hubble1929}& observations\\
 30 & 26$\pm$4 & \citet{Arp1956}& observations\\
 142 & 29 $\pm$ 4 & \citet{Capaccioli1989}& observations\\
 72 & 37+12 -8 & \citet{Shafter2001}& observations\\
 22 & 65 +16-15 & \citet{Darnley2006}& observations\\
 - & 97 & \citet{Chen2016}& Theory\\
 -& 106 & \citet{Soraisam2016} & Theory+ analysis of the bias\\
       \hline
    \end{tabular}
\end{table}

A technical analysis of the completeness factors and criteria adopted by \citet{Darnley2006} to derive the latter result is outside the scope of this review. Nevertheless, we point out the following: according to \citet{Shafter2001} the nova rate in M31 can be as high as $\sim 50$ novae/yr \emph{``if the luminosity-specific disk and bulge nova rates are comparable''}.  On the other hand, \citet{Darnley2006} concluded that \emph{``M31 bulge CN eruption rate per unit $r$ flux is more than five times greater than that of the disc.''}. It looks like that the two conclusions are inconsistent with each other. A possible way out is that the nova rate in M31 is significantly higher than reported above. For example, \citet{Chen2016} by using different star formation histories such as a starburst case  for early type galaxies, a constant star formation rate case for spiral galaxies and a composite case for ``early'' spirals such as M31, estimated a new rate of 97 novae per year for Andromeda galaxy. These authors also found other interesting results. The nova rate at 10 Gyr in an elliptical galaxy is up to 20 times smaller than a spiral galaxy with the same mass.  Therefore the specific nova rate per unit of luminosity (or mass) in the ``Ellipticals'' peaks around 1 Gyr and declines by 2 orders of magnitude at 10 Gyr. In ``Spirals'' the specific nova rate peaks around 2 Gyr and declines by a factor of 4 after 10 Gyr. The specific nova rate for elliptical galaxies at 10 Gyr is $1\div 2$ per $10^{10}\,M_{\odot}\,{\rm yr}^{-1}$, which is consistent with previous predictions \citep{Matteucci2003} and observations (see Fig.~\ref{fig:Hubblenova}).  The normalized nova rate for Spiral at 10 Gyr is $\sim$20--40 per $10^{10}\,M_{\odot}\,{\rm yr}^{-1}$, significantly higher than measured in early type galaxies. This result is qualitatively in agreement with observations \citep{DellaValle1994,DellaValle2002,Alis2014} although the ratio is much larger than observed. According to \citet{Chen2016}  this mismatch may be due to either the incompleteness of past surveys or incorrect assumptions on the SF history in their models. By applying the normalized nova rates in spirals derived by \citet{Chen2016}, to the LMC, one should observe between 6 and 12 novae/yr, which are unrealistic values for the nova rate in our satellite galaxy (see Sec. \ref{sec:6.3}). Therefore we tend to downsize the case for incompleteness and to agree with the authors that the mismatch is possibly due to some incorrect assumptions on the SF history, although the trend of the normalized nova rates exhibited by early-type to late-type hosts is qualitatively consistent with observations. 

Similar high frequency have been obtained by \citet{Soraisam2016} after assuming that specific nova rates in the bulge and disc are the same and by correcting the Arp's and Darnley et al.'s rate for an hypothetical bias against the discovery of fast novae due to the poor observing cadence of these surveys. These authors estimated 106 novae yr$^{-1}$. We note that Arp survey consisted of about 1000 plates taken on 290 nights distributed over $\sim$18 months, June 1953--Jan 1955. These data imply about one observing night every due days and about three plates per night. With such a ``dense'' sampling it is rather difficult to think that Arp might have missed a significant number of ``fast'' and ``bright'' novae. Arp indeed concluded: \emph{``Consequently of the 30 novae discovered, for only 5 were the maxima missed by more than a day''}. Therefore the paucity of detection of fast and bright novae in M31 does not seem the result of an observational bias rather being due to the poor efficiency of old stellar populations to produce fast novae, which originate by massive WDs. These novae are characterized by lifetimes of the order of a few $\times 10^7$ years (see Sect. 1.2) and therefore we do not expect that they are predominant in $\sim 10^9$ years old bulge stellar population. In the following, we adopt $R_{\rm M31} = 40^{+20}_{-10}$ novae/yr as an ``educated guess'' for the frequency of occurrence of novae in M31. 

\subsection{Novae in M33}

Until the 1990s the nova rate in M33 was based on the \citet{Hubble1953},  \citet{Arp1956} and \citet{SharoV1993} estimates, ranging between 1 and 0.5 novae/yr.  However, in the 1960s a nova survey on M33 started at Asiago Observatory, although sparser than the M31 survey carried out in the same years. Those observations allowed to discover 5 novae and to conclude that the nova rate in M33 was not \emph{``so low as previously assumed''} \citep{Rosino1973b}.  A new surveys was started a few years later at the 1.82-m telescope and five more novae were discovered \citep{DellaValle1988}. Altogether these observations have provided the database to measure a rate of $R_{\rm M33}=4.6 \pm 0.9$ novae/yr \citep{DellaValle1994b}.  A few years later \citet{Williams2004} started an H$_\alpha$ survey which led to the discovery of four novae brighter than $m_{\rm H\alpha}=18$ mag, and to derive a nova rate of $2.5\pm0.7$ novae/yr. In Table~\ref{tab:M33}, we summarize the results reported by these three independent surveys. A grand total of 14 novae have been discovered during a total control-time of 5.5 yr. After extrapolating the observed number of novae to the whole M33 field, according to the correction factors provided by the different authors we find: ${\rm R}_{\rm M33}=3.5^{+0.95}_{-0.8}$ ($1\sigma$ poissonian error).
          
\begin{table}[h!]
\caption{Summary of nova surveys in M33. Col.~2: control time of the
survey; Col.~3: ``observed'' number of novae; Col~.4: correction factor to the whole galaxy field; Col.~5: ``observable'' number of novae. Data from \citet{Rosino1973b,DellaValle1988,Williams2004}.}
	\label{tab:M33}
	\begin{center}
	\begin{tabular}{lcccc}	
	\hline \hline
	Nova Survey & C-Time & Novae & Corr.& Novae \\
	Telescope   & yr     & obs   & field&      \\
	\hline
	1.22m	    & 0.85   & 5     &  1   &  5   \\
    1.82m       & 1.9    & 5     &  1.4 &  7   \\
	1.00m	    & 2.75   & 4     &  1.75&  7   \\   
\hline	
    		    & 5.5    & 14    &      & 19   \\
\hline \hline
	\end{tabular}
	\end{center}
\end{table}

\begin{figure}[htb]
    \centering
    \includegraphics[width=10cm]{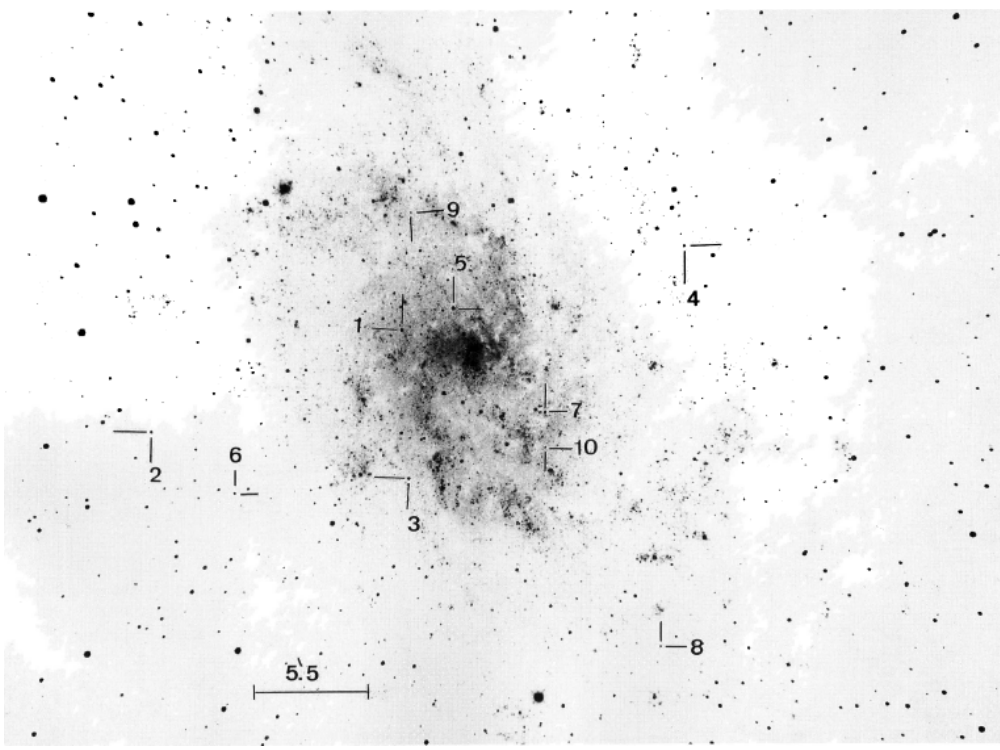}
    \caption{Novae in M33 galaxy \citep{DellaValle1994b}. }
    \label{fig:MDV1994}
\end{figure}

We note that to compute the control time of their survey \citet{Williams2004} assumed a set of template $H_\alpha$ lightcurves based on the peak mag and rate of decline derived for M31 novae. The authors are well aware of the possible drawbacks of their approach and indeed they warn the readers: 

\begin{quote}
  ``in our analysis is the assumption that the lightcurve properties  of novae in M31 which were used in the Monte Carlo and mean H$\alpha$ lifetime nova rate calculations are representative of novae in M33. If M33 contains a higher percentage of ``fast'' novae compared with M31, then we may be overestimating the mean nova lifetime in M33 and thus underestimating the true nova rate.''
\end{quote}

Their assumption is fully justified if one compares nova populations having similar properties (e.g., novae of the bulge of M31 with novae in an elliptical galaxy in Virgo), but it is not applicable here because M33 is an almost bulgeless galaxy \citep{Bothun1992} and therefore its nova production necessarily originates in the disk, in particular, Fig.~\ref{fig:MDV1994} shows that novae appear superimposed on the arms of the parent galaxy. \citet{Shafter2012} were able to carry out a spectroscopic survey on novae in M33, a notable piece of work given the small number of novae that can be discovered during a campaign. They found that only 2 out of 8 discovered objects were members of the Fe II  Williams' spectroscopic class. This result is different (at a confidence level of $\sim 95\%$) from what expected on the basis of Milky Way \citep{DellaValle1998} and M31 \citep{Shafter2011} results, where most nova populations ($\sim 80\%$) are Fe II objects. Therefore the assumption that the properties of novae in M33 were similar to those of novae in M31 is not anymore justified and the rate of 2.5 novae/yr should be regarded as a robust lower limit for the nova rate in M33. Fast Novae close to maximum light evolve more quickly than slow novae also if observed in $H_\alpha$ light,  although at slower rate than exhibited in $m_{pg}$ light, possibly a factor $\sim 1.5$ rather than $\sim 5$. As a consequence the control time of \citet{Williams2004} is likely to be overestimated by $\sim 30\%$ and the nova rate in M33 underestimated by an equal amount.  Therefore the nova rate in M33 is larger than previously estimated, we assume hereinafter: ${\rm R}_{\rm M33}=4.1^{+1}_{-0.8}$.

\begin{figure}[htb]
    \centering
    \includegraphics[width=10cm]{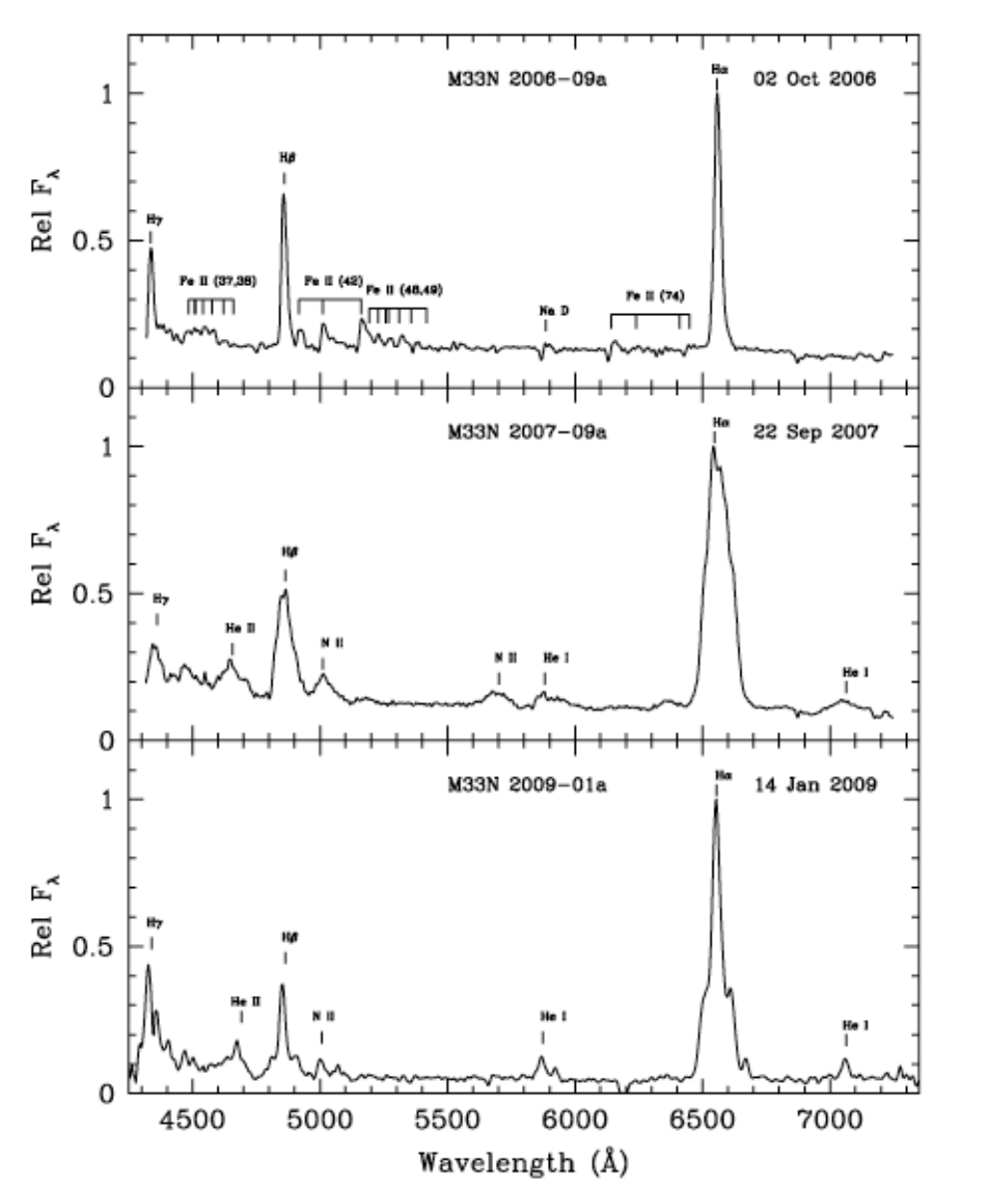}
    \caption{Spectra of M33 novae. M33N 2006-09a appears to be a Fe II nova, while 2007-09a and 2009-01a are He/N systems. Image reproduced with permission from \citet{Shafter2012}.}
    \label{fig:Shafter2012}
\end{figure}
 
\subsection{Novae in the Magellanic Clouds}\label{sec:6.3}

The properties of the outbursts of LMC novae and their frequency of occurrence have been well established only recently. Until the 1990s only 20 objects -- the first one observed in 1926 \citep{Buscombe1955} -- were available and a reliable estimate of the nova rate was jeopardized by the fragmentary nature of observations, in some cases even one discovery every ten years.  A first summary of the available data until Nova LMC 1988 has been provided by \citet{Capaccioli1990}. In spite of the paucity of the data these authors could compile the MMRD relation that showed a remarkable percentage excess of fast novae compared to M31.  

\begin{figure}[htb]
    \centering
    \includegraphics[width=10cm]{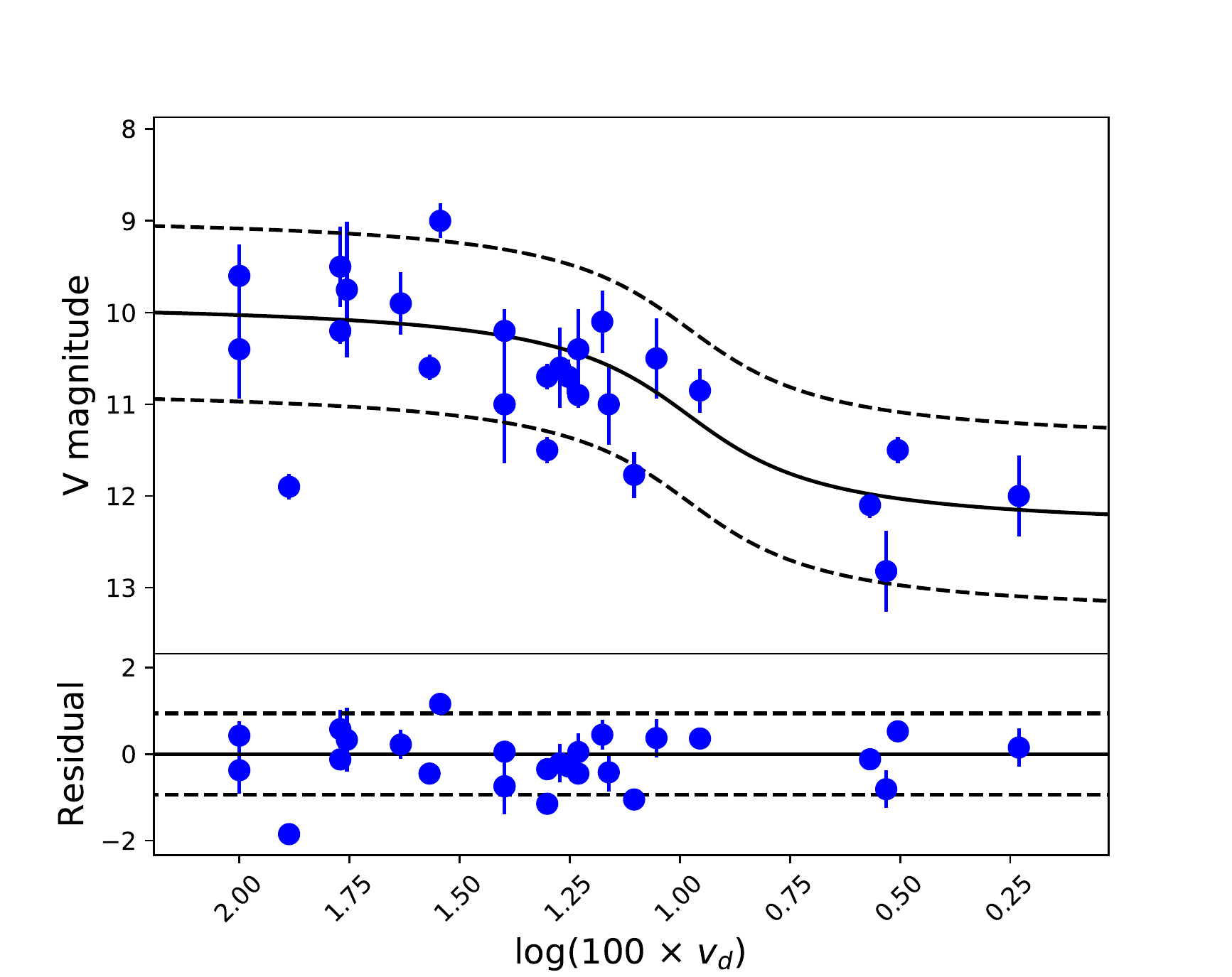}
    \caption{The MMRD relation for 27 LMC. The two dashed lines mark the $\pm 3\sigma$ strip. $1 \sigma$ corresponds to = 0.32 mag. The magnitude of the intercept of the MMRD with the y-axis is $V=10.03$ mag. Data have been obtained from Table~\ref{tab:app6}}.
    \label{fig:MMRDLMC}
\end{figure}

The data could be fitted with the same MMRD derived from M31 novae, obviously after applying a different zero point.  In other words the MMRD relation seems universal in terms of relation between peak magnitude and rate of decline but not as it concerns the relative percentage of the bright and fast novae with respect to faint and slow. This piece of observation has an important consequence. We have previously commented about the rates of decline from maximum light as a tracer for differences in the parameters of nova outbursts (Sect. 3.2), particularly for the mass of the underlying WD. Most novae in the LMC are bright and fast declining then suggesting a membership to disk nova population which originates from more massive WDs. However, other parameters are at play. For example \citet{Starrfield2000} explain this behavior on the basis of the metallicity of the accreted material. \citet{Kato2013} after studying the effects of low Z on the properties of nova outbursts achieved similar conclusions. These authors concluded that LMC novae should be brighter and eject more material than Galactic novae of the same type since the LMC metallicity is about one-third that of the Galaxy. The age of the parent stellar populations of fast and slow novae were also studied in details by \citet{Subramaniam2002} on a sample of 15 LMC novae.  They were able to prove an upper and lower limits, for the age of parent stellar population of fast and moderately fast novae of 3.2 and 1 Gyr,  unlike the stellar regions around slow novae mostly in the range 4--10 Gyrs. 

More recently \citet{Shafter2013} revised and collected data for 43 novae and through a detailed analysis of 29 lightcurves and 18 spectra, was able to confirm that LMC novae:  i) have faster rates of decline than novae in the Milky Way and in M31; ii) show a high percentage of He/N and Fe IIb novae compared with M31. Therefore bright He/N disk-novae dominate the LMC nova production unlike M31 where faint Fe II bulge novae are predominating.  

The measurement of the nova rate in the LMC is a recent story. The first estimate was provided by \citet{Graham1979} at the end of a $\sim 10$ years survey which allowed to discovery 10 novae. He suggested a rate between 2 and 3 novae/yr, and today we can say that he was right. \citet{Capaccioli1990} found $2\pm 1$ novae/yr, \citet{DellaValle2002} found $2.5\pm 0.5$ novae/yr and recently \citet{Mroz2016} from the OGLE survey derived $2.4\pm 0.8$ novae/yr. 

The estimate of the rate in the SMC is a more critical issue because a continuous monitoring of this galaxy for nova searches has never been carried out if not very recently thanks to the OGLE surveys \citep{Mroz2016}. Here below, we describe four different approaches:
\medskip                                                        

i) we can derive a first order estimate after assuming that the nova rate scales as the K-band luminosity of the host.  After assuming $K_{\rm LMC}=-2.17$ and $K_{\rm SMC}=-0.43$, we derive a nova rate of $\sim 0.5$ novae/yr. 

ii) A second measurement derives from the discovery of 4 novae in the SMC during 1999 \citep{Welch1999a,Glicenstein1999a,Glicenstein1999b,Welch1999b}. Although none of the objects has been spectroscopically confirmed, their lightcurves show sudden changes of the brigtness by more than $\sim 7$ mag on time scales of days/weeks. This behaviour is typical for novae, although dwarf-nova bursts originated by the rare sub-class of WZ Sge \citep{Ortolani1980} show a similar photometric behaviour. However, their intrinsic luminosity excludes this possibility. After assuming that these 4 events are novae, a simple application of poissonian statistic \citep{Gehrels1986} implies that the nova rate in the SMC has to be larger than 0.7 novae/yr \citep{DellaValle2002} at $\sim 99\%$ confidence level.  

iii) A third approach is based on the intensive monitoring of the SMC operated in the early 2000s, for example by Bill Liller, which led to the discovery of 6 novae.  We derive an ``observed'' rate of ${\rm R}_{\rm SMC}=0.75$ novae yr$^{-1}$ which should be regarded as a robust lower limit.   

iv) Finally, \citet{Mroz2016} further extended the SMC nova database and derive a rate of ${\rm R}_{\rm SMC}=0.9 \pm 0.4$ novae yr$^{-1}$ fully consistent with the rates derived above. This is the rate that will be adopted hereinafter. 

\subsection{Novae in M81}
 
The analysis of 79 plates obtained in the 1950--55 Palomar campaign for the discovery of Cepheids and novae in M81 has been published many years later by \citet{Shara1999}. These authors concluded that the spatial distribution of 23 discovered novae is \emph{consistent with the two populations divisions as argued previously}. Particularly, they estimate that the fraction of novae belonging to the bulge is not larger than $\sim 60\%$. From the data reported in the paper it has been possible to derive the control time of the survey and a rate of $24\pm 8$ novae/yr in agreement within the errors with the rate estimated by \citet{Neill2004} of $33^{+13}_{-8}$ novae per year. The average value that we adopt hereinafter is: $28^{+7}_{-6}$ novae per year.

\subsection{Novae in Virgo and Fornax ellipticals}

First observations of novae in Virgo galaxies were reported by Hubble \citep{Bowen1954} but only after \citet{Pritchet1987} the study of novae outside the Local Group of galaxies changed from sporadic interest into a mature field of research. These authors estimated on the basis of 9 detections in the B band a rate of $\sim 160\pm 57$ novae/yr, which almost corresponds to the rate in M49 because 8 out of the 9 novae were discovered in this galaxy. \citet{Shafter2000} used the Kitt Peak 4-m telescope to monitor M87 in H$_\alpha$ filter and they discovered 9 novae from which a rate of $91\pm 34$ novae/yr was estimated. \citet{Ferrarese2003} reported after a 55 days campaign carried out with HST/WFPC2 in V and I colors the discovery of 9 novae in M49, from which they derived a rate of $100^{+35}_{-30}$ novae/yr. \citet{Curtin2015} have monitored in H$_\alpha$ M87, M49, and M84 with the Canada-France-Hawaii 3.6-m telescope over about one year of control time and they estimated $154^{+23}_{-19}$, $189^{+26}_{-22}$ and $95^{+15}_{-14}$, respectively. By averaging these measurements, we find ${\rm R}_{\rm M49}=158^{+20}_{-17}$ and  ${\rm R}_{\rm M87}=136^{+19}_{-17}$ for M87. These rates are consistent with the frequency of occurrence of $\sim$90--180 novae/yr estimated by \citet{DellaValle2002b} for the giant elliptical NGC 1316 in Fornax cluster. Recently, \citet{Shara2016} have derived from a new survey on novae in M87 a higher rate of $363^{+33}_{-45}$  (see next paragraph). 

\subsection{Nova rates as function of the Hubble type of the hosts}\label{sec:6.6}

Several authors \citep{DellaValle1994,Shafter2000} have studied the trend of the nova rates as a function of the total and normalized luminosity of the parent galaxy. They suggested, though with different views that there may exist a systematic difference in the nova rate per unit of $K$ luminosity in galaxies of different colors (Hubble types) and/or classes of luminosity (see their Fig.~3 and 4). One result on which both teams seem to converge is that ``blue'' galaxies of the sample, characterized by low luminosity classes are more prolific nova producers than early-type spirals and ellipticals by a factor $\sim 3$. The current available data are summarized in Table~\ref{tab:sample} with obvious meaning of the acronyms. 

\begin{table*}
  \caption{Extragalactic Nova Data. References: \citet{Arp1956} (1) \citet{Pritchet1987} (2), \citet{Capaccioli1989} (3) \citet{Ciardullo1990} (4), \citet{Capaccioli1990} (5), \citet{DellaValle1994} (6),  \citet{Ferrarese1996} (7), \citet{Shara1999} (8) \citet{Shafter2000} (9) \citet{Shafter2001} (10), \citet{DellaValle2002} (11),  \citet{DellaValle2002b} (12), \citet{Ferrarese2003} (13), \citet{Williams2004} (14), \citet{Neill2004} (15), \citet{Darnley2006} (16),  \citet{Coelho2008} (17), \citet{Guth2010} (18), \citet{Curtin2015} (19), \citet{Mroz2016} (20), \citet{Shara2016} (21).}
  \label{tab:sample}
	\begin{center}
	\begin{tabular}{lcccccllll}
	\hline \hline
    Galaxy     &     B       & A$_{\rm gal}$ & A$_{\rm int}$  &  (B-K)   &  ($m-M$)    &   Rate &  $L_K/L_{\odot}$   &  $\nu_K$  & Ref.\\
          &     (mag)       & (mag) & (mag)  &  (mag)   &  (mag)    &   ($\#$/yr) &  $(10^{10})$   &  (*)  & \\
   \hline                                                             
       M31   &    4.36$\pm$0.02 & 0.46 & 0.2  & 3.85    &  24.38    &  40$^{+20}_{-10}$     &   8.0   & 5.0  & 1,3,10,16 \\
       N5128 &    7.84$\pm$0.06 & 0.50 & 0.0  & 3.38    &  28.12    &  28$^{+7}_{-7}$       &   6.2   &  4.5 & 4 \\
       M33   &    6.27$\pm$0.03 & 0.18 & 0.40 & 2.87    &  24.64    &  4.1$^{+1}_{-0.8}$    &   0.7   &  6.1 & 6,14 \\
       LMC   &    0.91$\pm$0.05 & 0.26 & 0.13 & 2.74    &  18.50    &  2.4$^{+0.8}_{-0.8}$  &   0.3  &  8.4 & 5,20 \\
       SMC   &    2.70$\pm$0.10 & 0.20 & 0.41 & 2.71    &  18.99    &  0.9$^{+0.4}_{-0.4}$  &   0.1 &  10.4  & 11,20\\
       M49   &    9.37$\pm$0.06 & 0.10 & 0.00 & 4.30    &  31.06    & 158$^{+20}_{-17}$     &  51.1   &  3.1 & 2,13,19 \\
       M100  &   10.05$\pm$0.08 & 0.11 & 0.06 & 3.84    &  31.04    &  25$^{+5}_{-5}$        &  17.7   &  1.4 & 7 \\
       M87   &    9.59$\pm$0.04 & 0.10 & 0.00 & 4.17    &  31.03    &  136$^{+19}_{-17}$    &  36.0   &  3.8 & 9,19 \\
             &               &      &      &         &           &  363$^{+33}_{-45}$   &   &  10.1  & 21 \\
       M51   &    8.96$\pm$0.06 & 0.15 & 0.12 & 3.43    &  29.42    &  18$^{+7}_{-7}$       &   7.5   &  2.4 & 9 \\
       M101  &    8.31$\pm$0.09 & 0.04 & 0.03 & 3.24    &  29.34    &  12$^{+4}_{-4}$       &   10.4   &  1.1 & 9,17 \\
       M81   &    7.89$\pm$0.03 & 0.35 & 0.37 & 3.99    &  27.80    &  28$^{+7}_{-6}$       &   7.9   &  3.5 & 8,15 \\
       N1316 &    9.42$\pm$0.08 & 0.09 & 0.00 & 4.15    &  31.66    &  135$^{+45}_{-45}$    &  73.8   &  1.8 & 12 \\
       M84   &   10.01$\pm$0.12 & 0.18 & 0.00 & 3.76	 &  31.32    &  95$^{+15}_{-15}$     &  22.1   &    4.3   & 19\\
       M94   &    8.99$\pm$0.05 & 0.08 & 0.09 & 3.72    &  28.21    &  5$^{+1.8}_{-1.4}$    &  3.1    &    1.6    & 18 \\
	\hline \hline
	\end{tabular}
	\end{center}
\end{table*}

\begin{figure}[htb]
    \centering
    \includegraphics[width=10cm]{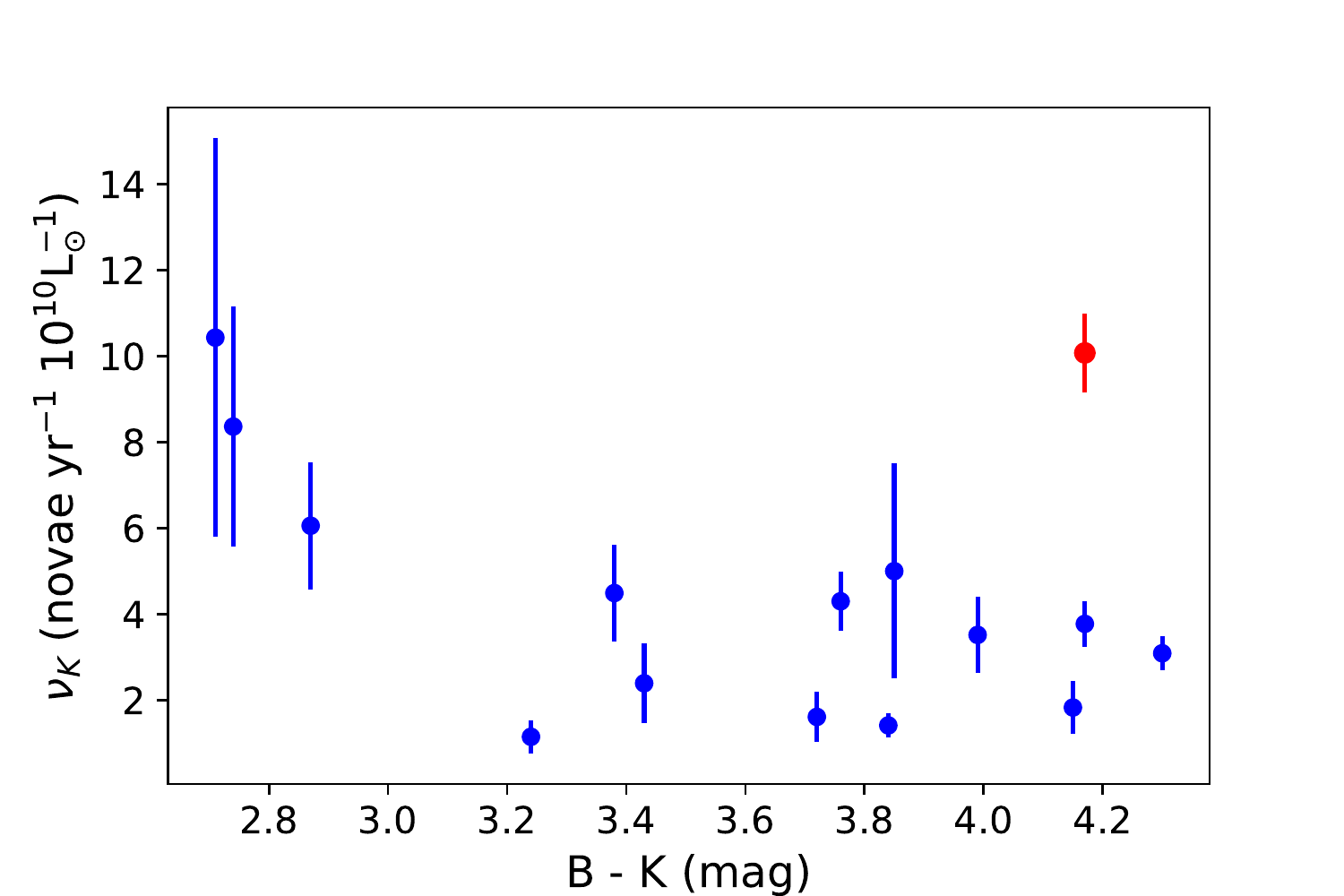}
    \caption{The specific nova rate (nova rate per unit of luminosity of the host) plotted as a function of galaxy color. The red circle corresponds to the nova rate value in M87 as inferred by \citet{Shara2016}. Data from Table \ref{tab:sample}.}
    \label{fig:Hubblenova}
\end{figure}

Analysis of the data reported in Fig.~\ref{fig:Hubblenova} confirms that ``red'' galaxies, say, redder than (B--K)$\sim $ 3.2 show a constant trend of the normalized nova rate, close to $\sim 2$  while blue systems, belonging to low luminosity class galaxies, exhibit a normalized luminosity a factor at least $\sim 3$ times larger.

The existence of an overproduction of nova outbursts in the LMC and SMC and to a smaller extent in M33 might be a consequence of the so called `selection effects' on the nova frequency, investigated by \citet{Truran1986} and \citet{Ritter1991}.  It is due to the different recurrence times and number of nova progenitors per unit of luminosity in the respective hosts.  The argument is the following: from \citet{DellaValle1998} we derive an average magnitude at maximum for a typical nova originating in an old stellar population (``bulge nova'') of M$_B\sim -7$ mag. The same authors derive M$_B\lesssim -9$ mag for the average nova produced in a star forming stellar population (``disk nova''). After using the relation obtained by \citet{Livio1992} between the magnitude at maximum and the mass of the underlying white dwarf, we obtain: $M_{\rm WD}\sim 0.8\,M_\odot$ and $M_{\rm WD}\sim 1.2$--$1.3\,M_\odot$ respectively (see also SG19).  We know that nova systems in disk dominated galaxies result in more frequent nova events due to the shorter nova recurrence time associated with massive WDs.  For example, from \citet{Yaron2005} we obtain for novae characterized by the above reported masses,  recurrence times of $\sim 4.5\times10^4$ years (by interpolating data of their Table~3 for $M=0.65$ and $M=1\,M_{\odot}$, $\dot{M}=10^{-9}$ and $T=50,000$ K) and $3\times10^3$ yr for a $M=1.25\,M_{\odot}$, $\dot{M}=10^{-9}$ and $T=50,000$ K. On the other hand, according to \citet{Cummings2018} and assuming a Salpeter IMF the number of nova progenitors characterized by $\sim 0.8\,M_{\odot}$ white dwarf is about 5 times larger than the number of novae  with $\sim 1.2/1.3\,M_{\odot}$ white dwarf. Therefore the observable outburst ratio between  disk and bulge novae becomes: 

\begin{equation}
\rho_{\rm disk}/\rho_{\rm bulge} \sim \Delta~T_{\rm bulge}/\Delta~T_{\rm disk} \times N^{prog}_{\rm disk}/N^{prog}_{\rm bulge} \approx 15/5 \sim 3 ,
\end{equation}
in good agreement with observations. 

Recently, \citet{Shara2016} derived from 41 novae discovered in a HST survey on M87 a global rate of $363^{+33}_{-45}$ novae/yr. If confirmed this high rate implies a specific (normalized) nova rate of about $7.9^{+2.3}_{-2.6}$ which is comparable to those exhibited by Magellanic Clouds (see Fig.~\ref{fig:Hubblenova}). Therefore the density of nova outbursts in giant elliptical would be similar to that observed in dwarf irregular star forming galaxies. In turn this implies that the recurrence times of novae harbored in binary systems including light WDs -- the only ones active in the old stellar populations of ellipticals -- should be considerably shorter than estimated by several authors \citep{Yaron2005,Starrfield2019}, or that the number of potential progenitors of CNe vastly exceed the value expected from stellar evolution arguments.   

One possible explanation suggested by \citet{Shara2016} is that all previous ground-based surveys for novae in external galaxies have missed most novae, especially the ones belonging to the hypothetical class of ``fast'' and ``faint''.  There are two arguments against this hypothesis: i) \citet{Pritchet1987} survey on Virgo ellipticals was densely sampled with 11 observing runs distributed on 15 nights. None faint and fast novae were discovered but seven faint and slow declining objects. This behavior is reminiscent of the Arp's result described in Sect. \ref{sec:6.1} and it is suggestive that old bulge stellar population are inefficient producers of bright and fast novae due to the lack of active nova systems containing massive WDs; ii) \citet{Ferrarese2003} survey was carried out with HST as well, so it is difficult to call for a massive observational bias affecting this survey. Recently, \citet{Shafter2017} through an independent analysis of the HST data resize the result of \citet{Shara2016} and  concluded that nova rates as low as ~200 per year remain plausible.

A speculative possibility is represented by a further develop of an interesting result obtained by \citet{Livio2002}. After the serendipitous discovery of a SN Ia near the jet of the active galaxy 3C 78, these authors have examined the issue of whether jets can enhance accretion onto white dwarfs via Bondi accretion process. They also present evidence for observations of 11 novae in close proximity to the jet.  From a simple application of the  \citet{Bondi1952} formula for accretion on a star, we derive that this process is efficient  only for very low stellar velocities (a few km/s). If we assume typical star velocities of the order of 10--100 km s$^{-1}$ as measured by \citet{Williams1994b} for nova systems in the Milky Way and the jet has a width of about 0.1 kpc, the transit time is less than 100 Myr during which the accreted mass onto massive WDs would be only $10^{-5}$-$10^{-7}\,M_{\odot}$ similar to the mass of nova accreted envelopes. This makes the accretion process from the ISM likely inefficient to trigger SN Ia explosions while it may play an important role to trigger nova explosions.  If true, this result may account for the high frequency of nova events observed 
near the nucleus/jet of M87 \citep{Madrid2007} that might be considerably higher than observed in the rest of the galaxy.

\subsection{Nova outbursts as a function of the metallicity content and class of luminosity of the hosts}

The observed trend of the normalized nova rate might be also interpreted in terms of differences in the metallicity of the parent stellar population. For example \citet{Starrfield1998} have discovered, on theoretical grounds, that novae originating in low metallicity stellar environments are more brighter at maximum light, than novae originating in populations characterized by solar metallicity and  this fact is observed in the LMC, which is characterized by a metallicity that is about 1/3 of that of the Milky Way. 
The presence of a nova overproduction in blue galaxies with respect to red galaxies may also hint for the existence of a link between star formation and the frequency of occurrence of nova outbursts. For example, \citet{Yungelson1997} have suggested that in irregular and late dwarf spiral galaxies the nova rates is mainly determined by the current star formation rate in the disk, while the nova rate in elliptical galaxies mainly depends on the mass of the galaxy. Since the star formation rate in giant spirals was, several Gyr ago, much higher than the present one \citep{Gavazzi1996} it turns out that most stars in the disk of giant spirals are old and their nova population characterized by a frequency of occurrence of nova systems not too different from that occurs in their bulges. In addition since the present SFR decreases moving from late to early spirals, it is not unexpected to find that the integrated colors of giant spirals should be redder than the integrated colors of dwarf spirals where the SFR is still active. This scenario provides a simple explanation of the fact that giant late spirals, as M51 and M101 have redder colors than dwarf galaxies of comparable Hubble type and a nova frequency of occurrence similar to that of the bulges. 

\section{Novae in globular clusters}

The researches for novae in globular clusters (GCs) have always given a very limited number of detections. Two novae have been so far discovered in Milky Way GCs:  M80 and M14. The former, T Sco 1860, occurred in M80 \citep{Sawyer1938} and it was a moderately fast nova with a t$_3$ of about three weeks.  The latter was a nova discovered by \citet{Sawyer1964}  on photographic plates taken with the 72 inches telescope at the Dominion Observatory in 1938.  M14 has been imaged on 124 nights from 1932 to 1963 and the nova was detected during its final decline in all eight plates obtained in 1938. The post-nova candidate was possibly identified by \citet{Shara1986}. The whole collection was formed by 2500 plates that have imaged 50 GCs. In a zero order approximation we can set a lower limit for the rate of  novae in Milky Way GCs,  after considering an effective survey time of about 30 years on 50 GCs and a single detection. By combining these data the rate of novae should be larger than $6.6\times 10^{-4}$ novae ${\rm GC}^{-1}{\rm yr}^{-1}$. In modern times nova outbursts have been unsuccessfully searched by \citet{Ciardullo1990}
in 54  globular clusters of M31 on a mean effective survey time of about 2 years.  Similar results were obtained by \citet{Tomaney1992} that carried out a survey on 
200 M31 GCs over 1 year of effective surveillance time.  The null  result  for  both surveys can provide some constraints on the nova rate in GCs. In particular, from simple Poisson statistics we find that the probability of obtaining a null result are 95\% and 5\% when the expected values  are  0.05 and  3.0, respectively.   Therefore one should expect that the nova rate in GCs may be larger than the 95\% probability value, i.e., $1.6\times 10^{-4}$ novae ${\rm GC}^{-1}{\rm yr}^{-1}$ and hardly higher than the  5\%  probability  value, i.e., $10^{-2}$ novae ${\rm GC}^{-1}{\rm yr}^{-1}$. The lower limit is consistent with the result of \citet{Shara2004}, who  estimated a rate for the M87 of $\sim 4 \times 10^{-3}$ novae GC$^{-1}{\rm yr}^{-1}$. The results based on X-ray surveys on M31 GCs \citep{Henze2013} of: $1.5\times 10^{-2} {\rm GC}^{-1}{\rm yr}^{-1}$ matches the upper limit on the GCs nova rates obtained via optical surveys.  Finally, \citet{Curtin2015} and \citet{Doyle2019} find a GC nova rate of: $\sim 4 \times 10^{-4}$ and $5\times 10^{-4}$ novae ${\rm GC}^{-1}{\rm yr}^{-1}$, respectively. 

The scanty statistics results in a global rate affected by almost two orders magnitude uncertainty: between a few $\times 10^{-4}$ and $10^{-2}$ novae GC$^{-1} yr^{-1}$. 

During their nova survey on M49, M87 and M84, \citet{Curtin2015} found that two out of 84 novae were discovered in the GC systems of these galaxies. The GC systems correspond to 2.4$\%$ of detected novae, while the mass of GCs comprise only the 0.23$\%$ of the combined mass of the galaxies. Thus, these authors speculate that the nova rate in GCs might be enhanced by a factor 10 relatively to the nova rates measured in the host galaxies. Such a high nova rate in GCs might receive some support by the study of \citep{Kato2013} who found on theoretical grounds that the nova population in GCs should be mainly formed by fast novae. These novae are easily missed during low-cadence surveys that characterized the search for CNe in such stellar systems. On the other hand, \citet{Cao2012} by studying the M31 globular clusters system concluded that the M31 GC specific nova rate is not significantly higher than the Andromeda specific nova rate. We should take both conclusions with some degree of  caution because they are still dominated by small number statistics.    

\section{The MMRD relationship in galaxies of different Hubble types} 

\begin{figure}[htb]
    \centering
    \includegraphics[width=15cm]{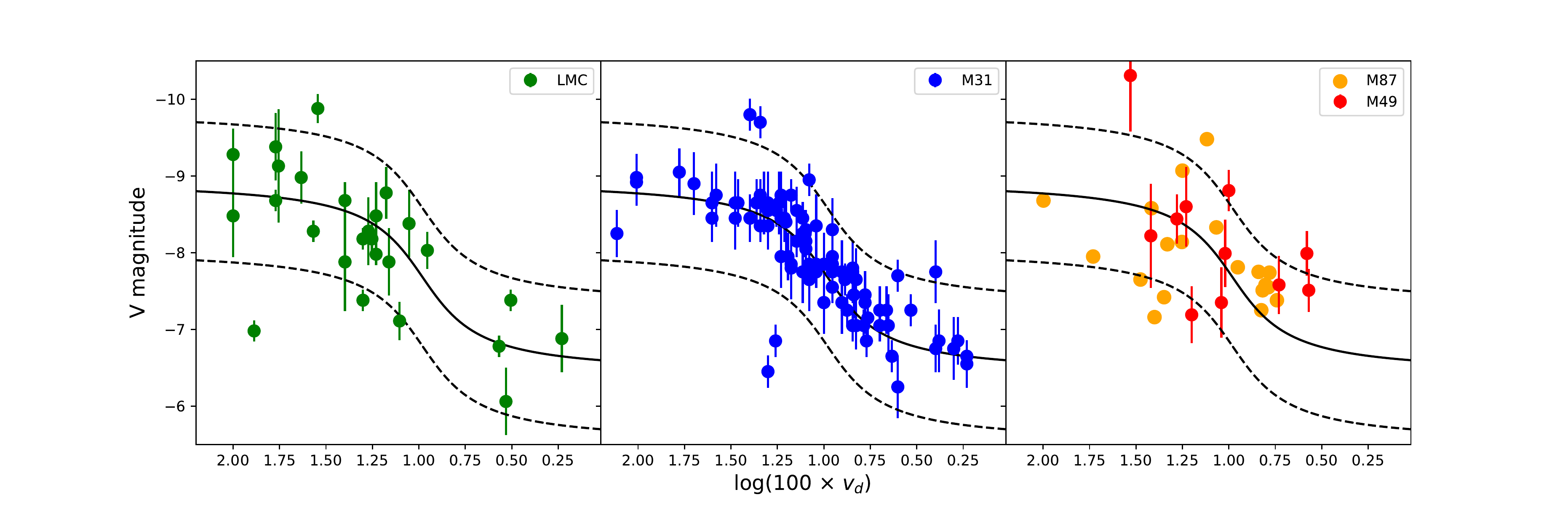}
    \caption{The comparison among the MMRD relationships for LMC, M31 and Virgo novae. The best fit is the solid line described by Eq.~(\ref{eq:MMRD_MW}). The two dashed lines mark the $\pm3\sigma$ strips. }
    \label{fig:comp}
\end{figure}

Figure~\ref{fig:comp} shows the maximum magnitude vs. rate of decline relationship for LMC, M31 and Virgo novae. This plot indicates that the nova production in the LMC is clearly biased towards fast and bright novae, whereas the M31 nova population exhibits a prominent `slow' component. By the same token the trend exhibited by novae discovered in Virgo. These trends are confirmed by a Kolmogorov--Smirnov test carried out on the  rates of decline distributions of the respective nova populations.  We find that the lightcurves of M31 novae mimic well the trend exhibited by the Virgo nova population (KS Probability = 0.7), but deviate strongly from the behaviour of the LMC novae (KS Probability = 0.005) which are on average faster. It is not obvious to explain this behavior in terms of an observational bias: indeed the brightest novae are detected in the nearest galaxy (LMC) and they would be missed in the more distant ones (Virgo), while faint novae would be detected in 15--20 Mpc away systems and missed in nearby galaxies, such as the MCs. The apparent magnitudes at maximum of novae belonging to the faint tail of the LMC, reach V=12-13 mag which are well within the capability of surveys currently in progress \citep{Mroz2016}, therefore faint and slow novae in MCs are intrinsically rare objects. Obviously it is possible that some fast nova is missed in poor sampled surveys \citep{Rosino1989} but this doesn't seem the case for the \citet{Arp1956}, \citet{Pritchet1987} and \citet{Ferrarese2003} surveys which observations were carried out -- on average -- every other night.  

Since the luminosity at maximum and the speed class of a nova depends on the physical parameters that determines the strength of the outburst \citep{Yaron2005} the rate of decline should be regarded as a valuable tracer of intrinsic differences in the nova populations. The differences in the MMRDs find a simple explanation in the framework of the two nova populations scenario: nova systems in disk dominated galaxies are associated with more massive WDs then resulting in faster, intrinsically brighter and more frequent nova events (see Sect. \ref{sec:6.6}). Given the relatively short lifetime of the nova phase for these systems (see Sect. \ref{sec:1.2}) they can be detected in galaxies where the star formation is still active, while the nova population in old stellar systems is mainly formed by ``old'' nova systems that contain light WDs and are characterized by nova lifetime of the order of $\sim 10^9$ years.  Therefore the lack of fast and bright novae in old stellar systems as elliptical galaxies appears to be real. 

\subsection{Outliers of the MMRD}

\begin{figure}[htb]
    \centering
    \includegraphics[width=10cm]{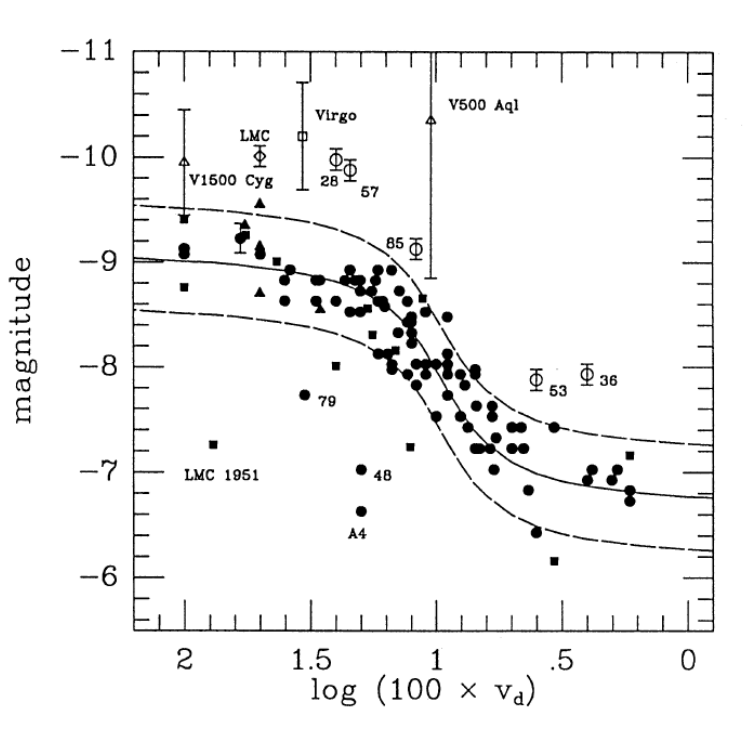}
    \caption{The MMRD relation for the LMC reported in \citet{DellaValle1991} showing the bright outliers of the relation, as open circles, and the faint ones, the filled circles.}
    \label{fig:MDV1991}
\end{figure}

The existence of outliers of the MMRD was first reported by \citet{Arp1956}, see his Fig.~32. Arp himself suggested that nova number 4 was a heavy absorbed object, ``obscured'' by 2.2 mag. \citet{Rosino1973} found two more deviating objects Nova 48 and Nova 66 which were found to be spatially coincident with nova 79 and nova 81, therefore he classified them as RNe. Both novae have fallen below the MMRD, then providing the first suggestion that some RN might be outlier of the MMRD. \citet{DellaValle1991} found another faint outlier of the LMC MMRD: Nova LMC 1951 which was about two magnitudes dimmed than expected from the measured rate of decline. He also found evidence for the existence of a family of superbright novae (see for example Fig.~\ref{fig:MDV1991}) brighter than expected up to $\sim 1$  mag above the $+3\sigma$ strip of the MMRD: Nova LMC 1990-1 \citep{Schwarz2001}, Rosino's novae in M31, nr.~28, 57, 53, 36, 85 and one object in Virgo. In the following we have assumed as outliers the points that fall outside the $\pm 3\sigma$ strip. Faint and fast novae are rare objects, so they may well be patchy absorbed objects or intrinsically faint novae (such as RNe, for example) which eject very light envelopes. More difficult is to explain the existence of ``super-bright'' novae, because it would imply the existence of an extra-energy source other than nuclear burning. The spectroscopic and photometric evolution of nova LMC 1991 excludes its membership to the class of ``red luminous novae'' \citep{Rich1989,Mould1990,Martini1999}. According to \citet{Iben1993} one possible explanation is the dissipation of the orbital energy motion of the secondary engulfed in the expanding nova ejecta during the nova common envelope phase. An alternative may be the relatively rare super-bright nova explosions which occur in CV-like systems formed by a cold low-mass degenerate dwarf accreting matter from a close low-mass companion \citep{Iben1992}. 

\begin{figure}[htb]
    \centering
    \includegraphics[width=10cm]{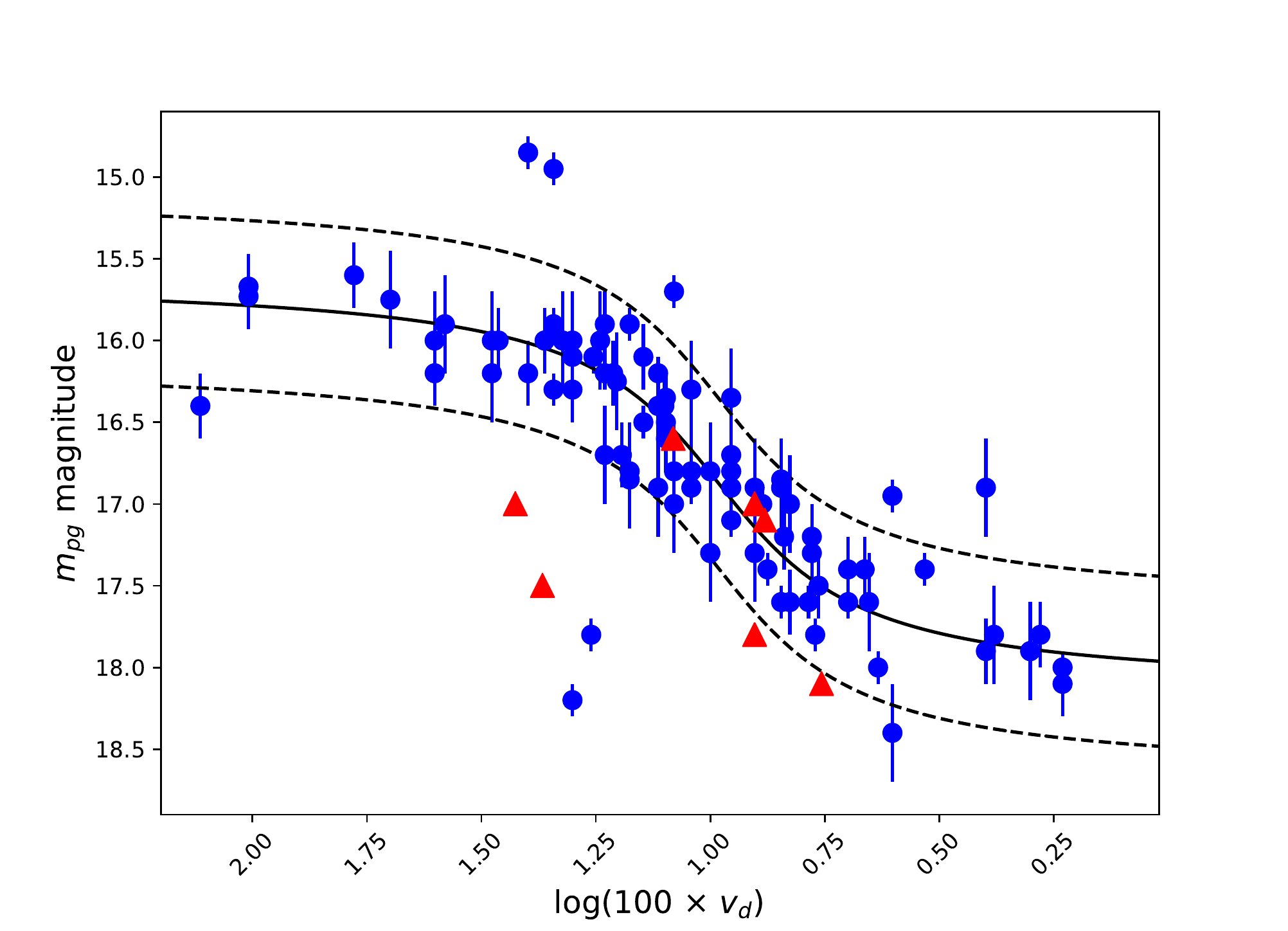}
    \caption{Blue dots: M31 novae from \citet{Arp1956} and \citet{Rosino1964,Rosino1973,Rosino1989}. Red triangles are \citet{Kasliwal2011} novae for which the MMRD parameters were computed with sufficient accuracy. The two dashed lines mark the $\pm3\sigma$ strips. Data from Table~\ref{tab:app6b}.}
    \label{fig:MMRDMansi}
\end{figure}

More recently a paper by \citet{Kasliwal2011} claim on the basis of 20 novae discovered in M31 (10), M81 (6), M82 (3) and NGC 2403 (1) that this sample is inconsistent with the canonical MMRD. A close inspection of their Fig.~11,  reveals that only 2 objects out of 17 (for which was possible to determine the MMRD parameters) match the MMRD. Other 4 objects are borderline and consistent with the relation within the errors. In other words between 65\% and 90\% of novae in \citet{Kasliwal2011} sample appear to be outliers of the relation. This is a surprising result if compared with the percentage of outliers discussed above. Taking these figures at their face values the MMRD should be a very poor relation, but this seems out of question after the recent GAIA DR2 results (see SG19 and fig. \ref{fig:GAIAMMRD} in this paper). The robustness of the relation is confirmed by the fact that the ``canonical'' MMRD is based on 119 M31 and LMC nova lightcurves, see Fig.~\ref{fig:MMRDM31LMC}, and only $\sim 5\%$  of them are outliers.  The M31 data come from three different surveys obtained with three different telescopes (60-inch, mount Wilson and 1.22 and 1.82 cm at Asiago Observatory) from two experienced observers as \citet{Arp1956} and \citet{Rosino1964,Rosino1973,Rosino1989}. 
LMC data are a collection of relatively sparse observations, but in spite of this,  the MMRD relation for the LMC, see Fig.~\ref{fig:MMRDLMC}, shows only two (7\%) obvious outliers. Therefore it is difficult to understand where the \citet{Kasliwal2011} high percentage of outliers is produced. The daily cadence of the Kasliwal et al.'s survey -- which is not very different from the Arp one -- does not seem at the origin of this high percentage of outliers. \citet{Kasliwal2011} finds $t_2$ values not unusual for nova systems and much longer than the $t_2$ of a few hours hypothesized for the ultra-fast nova population.   
In the attempt to clarify this point, in Fig.~\ref{fig:MMRDMansi} we have added to the Arp and Rosino novae the eight M31 Kasliwal et al.'s objects for which the authors reported the magnitude at maximum and the $t_2$ rate of decline (see their Table~5). The use of apparent magnitudes rather than absolute magnitudes has the advantage to minimize the corrections to the original observations by avoiding the magnitude revision for background and foregrounds absorptions and removing the systematics due to the adoption of the distances toward each galaxy. The $g$ mag have been transformed to $m_{pg}$ through $m_{pg}=g-0.05$ derived from the photometric corrections reported in \citet{Capaccioli1989} and \citet{Jordi2006}.  

The lack of errorbars do not allow to be more quantitative but we think that five objects match the relation and another one might be consistent within the uncertainties with it. The two faintest novae might well be patchy absorbed novae and/or RNe caught in outburst. Admittedly on scanty statistics the fraction of outliers  objects in M31 (2 out of eight) turns out to be consistent with previous findings.   

\begin{figure}[htb]
    \centering
    \includegraphics[width=10cm]{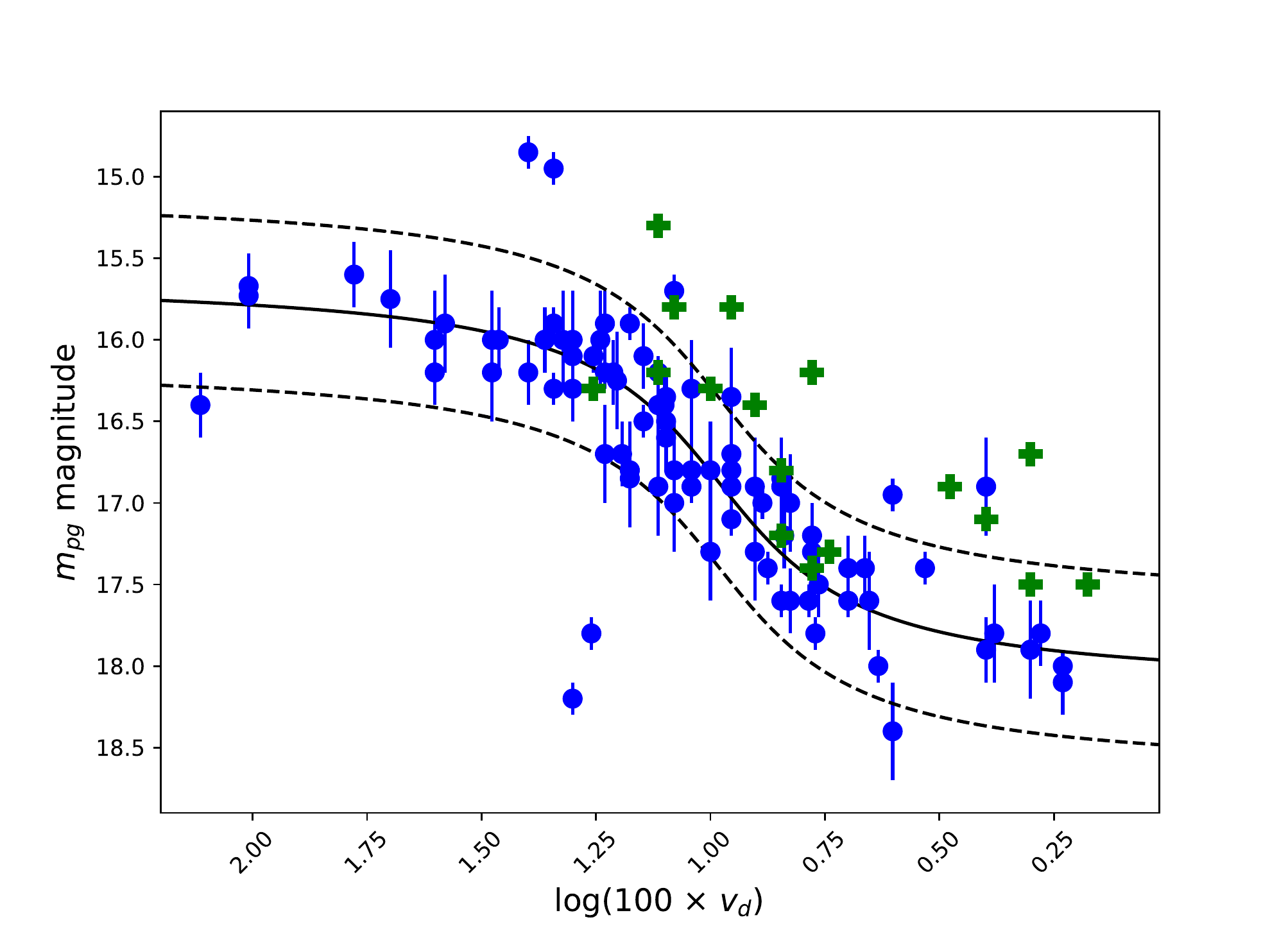}
    \caption{Blue dots: M31 novae from \citet{Arp1956} and \citet{Rosino1964,Rosino1973,Rosino1989}. Green crosses: M31 novae from \citet{Hubble1929} whose magnitudes turned out to be systematically too bright by $\sim 0.7$ mag. The two dashed lines mark the $\pm3\sigma$ strips. Data from Table~\ref{tab:app6c}.}
    \label{fig:M31Hubble}
\end{figure}

Finally we note the following: the discovery of a new photometric sub-class of ``faint'' and ``fast'' classical novae by \citet{Kasliwal2011} was essentially based on their Fig.~11, where it appears that most of newly discovered novae fall into the faint and moderately fast region of the MMRD after revision of their apparent magnitudes for internal and foreground absorptions toward each galaxy and by assuming the corresponding distance moduli.   Systematic behaviours exhibited by data can hide pitfalls and other explanations. For example a similar systematic behaviour, even if observed in a different region of the MMRD plane, is exhibited by \citet{Hubble1929} novae. In Fig.~\ref{fig:M31Hubble}  we have added the Hubble data to Arp and Rosino novae (Hubble data from Table~VI of \citealt{Capaccioli1989}). 

Most Hubble novae are systematically above the $+3\sigma$ strip of the MMRD. They seem to identify a sub-class of ``bright'' and ``slow'' CNe. In fact these data points simply suffer of a $\sim -0.7$ mag offset in the zero point of the photometric calibration. 

\begin{figure}[htb]
    \centering
    \includegraphics[width=10cm]{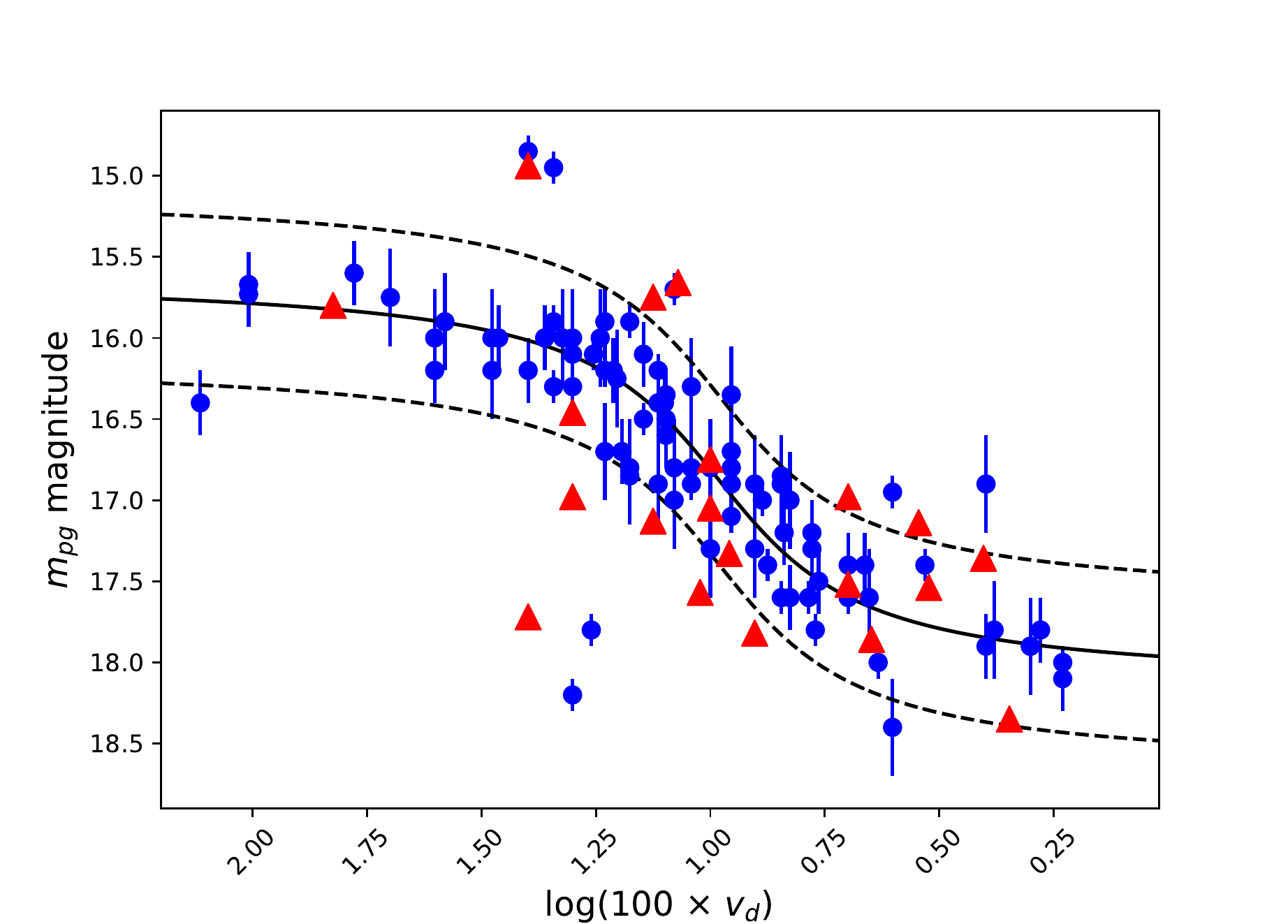}
    \caption{Blue dots: M31 novae from \citet{Arp1956} and \citet{Rosino1964,Rosino1973,Rosino1989}. Red triangles: M31 novae from \citet{Cao2012} data. With the exception of three obvious outliers, Cao et al.'s data match the MMRD within $\pm 3\sigma$ (dashed lines). Data from Table~\ref{tab:app6b}.}
    \label{fig:MMRDCao}
\end{figure}

However, the observational evidence for the existence of a new photometric sub-class of ``faint'' and ``fast'' classical novae has vanished after \citet{Cao2012} have published the data of their new survey on M31 novae. Fig.~\ref{fig:MMRDCao} has been obtained, by adding the \citet{Cao2012} dataset to the Arp and Rosino data, by using the photometric correction term $m_{pg}=R+0.1$. We confirm all parameters reported by \citet{Cao2012} with one exception: the $t_2$ value for nova 2010-06b. The first maximum of this ``slow nova'' is followed by a sharp and fast decline after which we can observe secondary maxima followed by strong fluctuations in brightness and finally a more gentle decline. To compute the correct rate of decline we have applied the smoothing procedure described in Sect. 3.1. We wish to point out that this ``procedure'' is standard (i.e., not ``ad hoc'') for measuring $t_2$ or $t_3$ in nova lightcurves. This is the case for nova 2010-06b (that shows some resemblance to the nova n. 36 of Rosino) for which a $t_2=25^d$ is a more realistic choice. The MMRD parameters for the fair quality data of the Cao's sample for which only lower limits to the $t_2$ were provided by the authors, have been directly derived from the published lightcurves (see Table~8 in \citealt{Cao2012}) and our results are fully consistent with the original estimates of \citet{Cao2012}. All these data points match satisfactorily the MMRD relation and there are only two obvious outliers. The very bright nova 2009-10b, which achieved a magnitude at maximum of $14.9$ mag (almost $-10$ mag in terms of absolute magnitude at maximum). Such bright novae are unusual in M31 but not unique events. Nova 2009-10b is a ``twin'' of novae nr.~28 \citep{Rosino1964} and 57 \citep{Rosino1973} indeed. The other one is nova 2010-06b that falls in the region of patchy absorbed CNe or RNe.

\begin{table}[htb]
  \caption{Fair quality data from the \citet{Cao2012} sample and revised estimates.}
\label{tab:5}
    \centering
    \begin{tabular}{c|l|c|c}
    \hline\hline
    Nova  & $t_{\rm 2,Cao}$ &  $t_{\rm 2,revised}$ & V$_{\max}$ \\
\hline
2009-08b & $>35$ & 45 & 17.76 \\
2009-11b & $>37$ & 90 & 18.25 \\
2009-11d & $>17$ & 25 & 17.72 \\ 
2010-06a & $>31$ & 40 & 16.88 \\
2010-10a & $>9 $ & 15 & 17.03 \\
2010-10b & $>41$ & 60 & 17.44\\
2010-12b & $>13$ & 17 & 15.56 \\
\hline
\end{tabular}
\end{table}

The third paper claiming the existence of a significant fraction of outliers in the MMRD relation is by \citet{Shara2017b}. These authors based their claim on the \citet{Kasliwal2011} results and on their own survey on M87 novae \citep{Shara2016}. They discovered 41 novae and determine the MMRD parameters for 21 lightcurves. A quantitative inspection of each lightcurve confirms the parameters derived by \citet{Shara2016} with the exception of nova 8 for which we find $t_2=7$ rather than 3.7 and novae 18 and 22 for which is impossible to derive the respective $t_2$ due to the incompleteness of the lightcurves.

 \begin{figure}[htb]
    \centering
    \includegraphics[width=10cm]{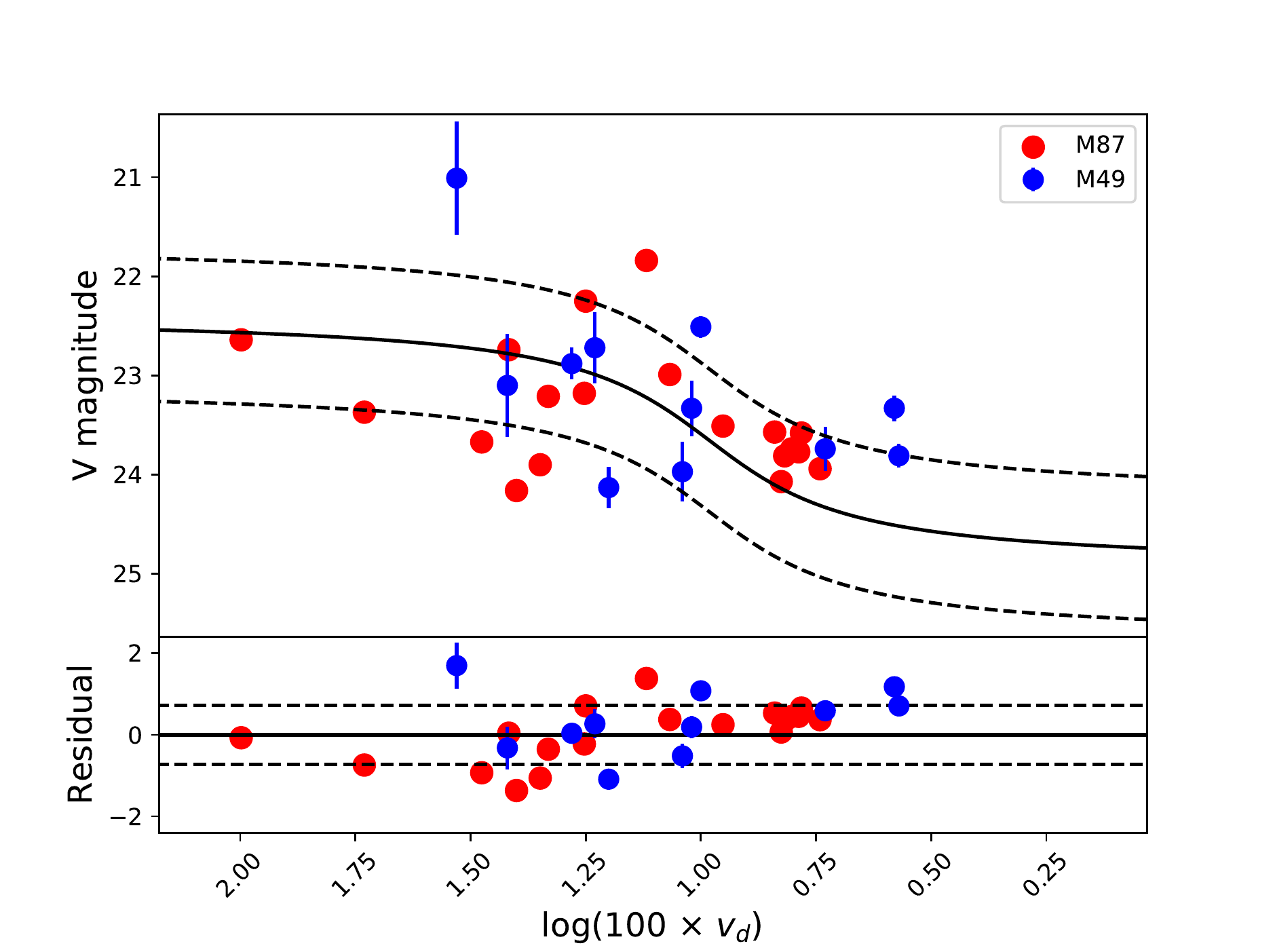}
    \caption{Blue dots: ten M49 and one NGC 4365 novae from \citet{Pritchet1987} and \citet{Ferrarese2003}. Red dots: nineteen M87 novae from \citet{Shara2016}. The magnitude of the intercept of the MMRD with the y-axis is $V = 22.54$ mag. Dashed lines mark  the $\pm 3\sigma$ strip, with $\sigma = 0.36$ mag.}
    \label{fig:VIRGO}
\end{figure}
 
Figure~\ref{fig:VIRGO} shows the positions of novae discovered in Virgo ellipticals \citep{Pritchet1987,Ferrarese2003,Shara2016} into the MMRD plane after assuming that M49 and M87 are basically at the same distance: $(m-M)= 31.10$ mag and 31.12 mag, respectively.\footnote{\url{http://leda.univ-lyon1.fr/}} Virgo novae are still consistent with the MMRD, but their distribution is characterized by a larger dispersion than it is observed in M31 and LMC. Indeed out of 30 objects (see Table~\ref{tab:app7}) about 20\% are outliers at level of $\pm 3\sigma$. The larger dispersion is likely due to the presence of systematics related to photometric measurements at very faint luminosity levels in three different galaxies carried out by three different teams \citep{Pritchet1987,Ferrarese2003,Shara2016}. 

The distribution of the data points is consistent with the MMRD relation, particularly the HST data do not show the existence of any faint and fast subclass of CNe, but possibly the existence of a superbright outlier similar to nova 2 in NGC 4472 discovered by \citet{Pritchet1987}. Obviously, due to the time sampling of the lightcurves we cannot exclude that these surveys have missed the ultra-fast novae with $t_2$ as short as a few hours \citep{Yaron2005,Shara2017b}. However, due to the tiny masses of the accreted envelopes, $10^{-8}\,M_{\odot}$ \citep{Shara2017b} their peak magnitudes should be considerably dim with respect to CNe and so they are rather hard to discover in such distant galaxies.  

An important result obtained by \citet{Shara2016} is the discovery of a conspicuous faint and slow nova population, which is typical of an old parent stellar populations. From their published lightcurves we can infer that novae from n.~25 through n.~41 peak between $V\sim 24$ mag and $V\sim 26$ mag and are characterized by $t_2$ values that unfortunately can be only roughly estimated, but they very likely range in the interval $40^d$--$200^d$. These 17 data points would consistently populate the faint and slow region of the M87 MMRD in the range of rates of decline corresponds to 0.7--0.01, in terms of $\log (100 \times v_d)$ units.

\section {Novae as distance indicators}

There are several reasons for which novae were believed to be good distance indicators. They can be summarized as follows:

i)  Novae are bright, they can reach M$_B \sim - 10$ mag, about 3 magnitudes more luminous than Cepheids of longer periods.

ii) Unlike Cepheids that can be detected only in spirals, novae can be ``easily'' recognized also in ellipticals.

iii)  Intrinsic differences in the outburst properties, existing between novae associated with the bulge or spiral arms stellar populations do not affect the zero point of the MMRD relationship, but only change the relative percentage of fast and bright novae. Particularly, we note that the luminous plateau of the MMRD relationship, characterized by an absolute magnitude of M$_V=-8.85\pm 0.10$ mag (calibrated with GAIA DR2 nova distances), seems to be a ``standard candle'', because it is constant along the different Hubble types of the host galaxies.  

iv) There exists a good theoretical understanding of the MMRD, i.e., of the tool that it is used.

On the other hand, there are also some drawbacks for using novae as distance indicators: 

i) The cosmic scatter of the MMRD relationship is relatively small, $\sigma$ = 0.25 mag, which brings about 12\% uncertainty in the distance. However, it is not so small as it would be required in the 1--2\% cosmology accuracy era \citep{Riess2019}.  

ii) Novae are transient objects so the time of their occurrence in the sky is unpredictable and therefore their observations are not easy to be scheduled. This aspect is not considered with favor by the various Time Allocation Committees.

\subsection{The distance to M31, LMC and Virgo}

To test the reliability of novae as distance indicators we measure the distances to M31, LMC and Virgo ellipticals by comparing the respective MMRD relations with the Galactic MMRD (Eq.~\ref{eq:MMRD_MW} and Fig.~\ref{fig:GOLDt3}) obtained using GAIA DR2 distance measurements, i.e., the ``bona fide'' sample described in Sect. 5.2 with the additional case of V1500 Cyg.\footnote{The GAIA DR2 distance of V1500 Cyg used in this paper, $d = 1.29 \pm 0.31$, can be derived using the parallex measurement as provided in the GAIA DR2 data archive: \url{https://gea.esac.esa.int/archive/}} Particularly, we compare the absolute magnitude of the high luminosity plateau of the Galactic MMRD: $M_V=-8.85$ (see Fig.~\ref{fig:GOLDt3}) with the apparent magnitudes of the high luminosity plateau of M31, LMC and Virgo MMRDs, which are $m_{pg}=15.76$ mag, $M_V=10.03$ mag and $M_V=22.54$ mag, respectively.  The distance modulus for M31 can be derived by combining the new data with Eq.~(4) from \citet{Capaccioli1989}:
\begin{align}
(V-M_v)_{\circ} +A_B &= [m_{pg}-M_V]- 0.26 \nonumber\\
&= \,[15.76-(-8.85)] -0.26 = 24.35\pm 0.31 (1\sigma)  
\end{align}
which is in excellent agreement with the Cepheids measurement of $24.38$ mag obtained by \citet{Riess2012}. Obviously, novae due to the dispersion of the MMRD relation, provide the distance to M31 with an accuracy that is three times less than that provided by \citet{Riess2012}. 

The comparison between Galactic and LMC MMRD relations finds a distance modulus:
\begin{equation}
(V-M_v)_\circ = 10.03-(-8.85)- 3.16\times E(B-V)=18.44 \pm 0.41 (1\sigma)  \end{equation}
The color excess $E(B-V)=0.14\pm 0.05$ mag has been derived from 6 novae reported by \citet{Shafter2013} and it implies $A_V=0.44$ mag. Despite of scanty statistic this value is in good agreement with $A_V=0.3\pm 0.1$ mag derived by \citet{Capaccioli1990} from observations of reddened LMC Cepheids. The distance modulus of $(m-M)=18.44$ mag is in excellent agreement with the eclipsing binary distance of $(m-M)=18.477\pm 0.004$ (statistical)$\pm 0.026$ (systematic) mag, obtained by \citet{Pietrzyski2019} by studying 20 eclipsing binaries in the LMC. Novae exhibit an uncertainty on the distance to the LMC of $\sim 19\%$ which is much larger than the one derived by \citet{Pietrzyski2019}. Finally the comparison between the Galactic and Virgo MMRDs finds, after assuming $A_V=0.07$ mag: 
\begin{equation}
(V-M_v)_\circ = 22.54-(-8.85)-0.07=31.32 \pm 0.44(1\sigma) 
\end{equation}
in agreement within $\sim 1\sigma$ with recent measurements of Virgo cluster ellipticals of $(m-M)_{\rm M49}=31.10$ mag, and $(m-M)_{\rm M87}=31.12$ mag.\footnote{see Hyperleda archive: \url{http://leda.univ-lyon1.fr}}.

\subsection{Novae as standard candles}

The concepts of ``distance indicators'' and ``standard candles'' are often interchangeably used. In fact, they are different. The observation of a SN-Ia (standard candle) allows, at least in principle, to immediately compare the apparent and absolute magnitudes, and then to compute the distance to the host galaxy by observing a single object. On the other hand if we observe a single Cepheid (distance indicator) this is not enough to make a direct distance measurement of the distance to the host galaxy. We need first to observe a number of Cepheids with different periods in such a way to determine the ``P-L'' or the ``P-L-C'' relations, and then we use these relations to determine the distance. In this respect when we use the MMRD relation, novae behave as distance indicators.   

The most popular method for using novae as ``standard candles'' is the so called \citet{Buscombe1955} law. They  first pointed out that all novae, irrespective of their rates of decline, have about the same absolute magnitude 15 days past maximum. This is a simple consequence of the MMRD relationship for which the most luminous novae decline faster and viceversa. As a consequence there is an epoch in which the lightcurves of fast and intrinsically bright novae cross the lightcurves of intrinsically faint and slowly declining objects. There are several calibration of the $\langle M_{15} \rangle$ magnitude: $\langle M_{V15} \rangle$ = $-5.24 \pm 0.15$ mag \citep{vandenBergh1987}, $\langle M_{V15} \rangle$ = $-5.60\pm 0.43$ mag \citep{Cohen1985} $\langle M_{V15} \rangle$= $-5.69\pm  0.42$ mag \citep{Capaccioli1989}. \citet{Downes2000} measured the distances of a dozen of galactic novae via nebular parallaxes and obtained $\langle M_{V15} \rangle$= $-6.05\pm 0.44$ mag. \citet{Darnley2006} found $M_{I15}=-6.3\pm 1$ mag. We know that a nova at maximum light has a color index of $(B-V)_{\circ} =0.23$ mag \citep{vandenBergh1987} that corresponds to a spectral type between A7 and F0, which in turn it is characterized by $(V-I)_{\circ} \approx 0.4$ \citep{Allen1973}. This color index implies $M_{V15}\approx -5.66\pm 1.00$ mag, after correcting for $A_V=0.23$ mag derived from $A_B=0.31$ mag \citep{Burstein1984}. 
SG19 from a sample of 17 galactic novae have recalibrated the $M_{V15}$ with GAIA DR2 distances and used a homogeneous treatment of the reddening correction \citep{Selvelli2013}. They find a weighted mean of $M_{V15}=-5.71\pm 0.10$ mag. 

From the Arp's high quality sample of 23 M31 novae we compute $m_{pg15}=18.98\pm 0.58$ mag.  The distance modulus can be obtained by the usual formula (see above) and $A_B=0.31$ mag \citep{Burstein1984}: 

\begin{equation}
(V-M_v)_{\circ} + A_B= -(m_{pg15}- M_{V15}) - 0.03 = 24.35\pm 0.59   
\end{equation}

which agrees with the MMRD distance within a few hundredths of magnitude, admittedly with a lower accuracy.  For the LMC we assume $M_{V15}=13.05\pm 0.25$ mag \citep{Capaccioli1990} and $A_V=0.44$  mag (see above) and we find: 

\begin{equation}
(m-M)=18.32\pm 0.27
\end{equation}

in agreement within about a tenth of magnitude with the MMRD distance. 
\medskip 

\citet{Ferrarese2003} by using 5 novae discovered in M49 find a $M_{V15}=24.77\pm 0.19$ mag, which would implies $M_{V15}=-6.26\pm 0.19$ mag, which is about $2.5\sigma$ brighter than the SG19 calibration. This result is borderline but it is not totally off the wall after considering the small number (five) of objects in the Ferrarese et al.'s  sample.

\section{The role of novae in the Galactic Nucleosynthesis}

Novae may significantly contribute to the galactic nucleosynthesis \citep{Gehrz1998} because they produce considerable concentrations of $\beta^+$ isotopes such as $^{13}C$ $^{15}N$ \citep{Sparks1978,Peimbert1984,Williams1985} $^{22}$N, $^{26}$Al \citep{Hillebrandt1982,Kolb1997}, $^{22}$Ne \citep{Livio1994}, $^7$Li \citep{Arnould1975,Starrfield1978,D'Antona1991,Izzo2015} and $^7$Be \citep{Tajitsu2015,Molaro2016}. The abundances of these elements in nova ejecta are significantly enhanced with respect to solar concentrations. The quantitative assessment of the contribution of novae to the chemical evolution of the Milky Way needs three quantities: i) the galactic nova rate; ii) the mass of the ejected envelopes; iii) the abundances of heavy elements in the nova ejecta. The Galactic nova rate has been extensively discussed in Sect. 3.5 and in the following we adopt a frequency of occurrence for novae in the Milky Way of about 25 novae/yr.

\subsection{The mass of the ejecta} 

\begin{figure}[htb]
    \centering
    \includegraphics[width=10cm]{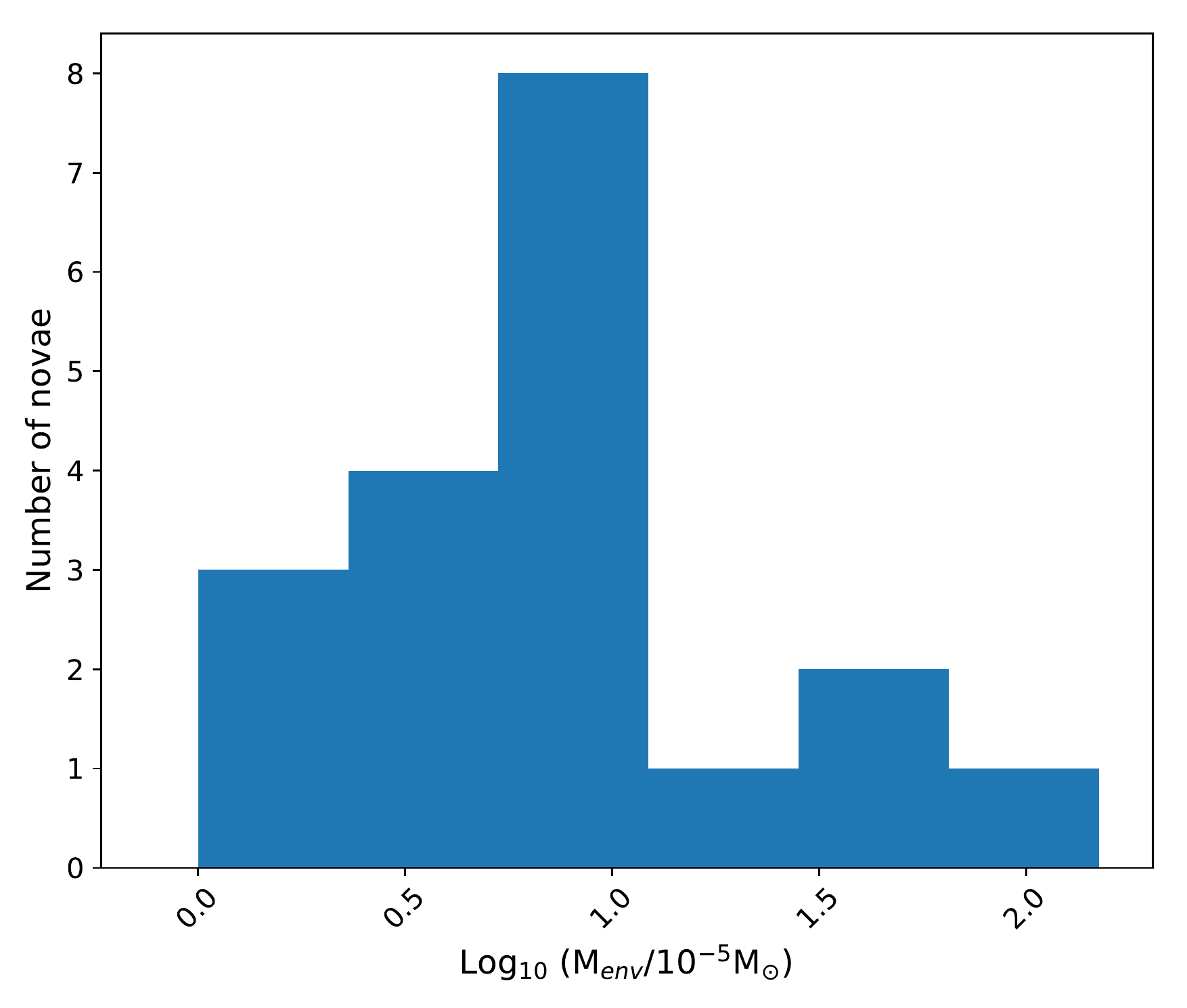}
    \caption{Histogram of the nova mass envelopes. Data from \citet{Warner2003,DellaValle1997,DellaValle2002,Mason2005,Ederoclite2006,Izzo2015,Molaro2016}.}
    \label{fig:masshisto}
\end{figure}

\begin{figure}[htb]
    \centering
    \includegraphics[width=10cm]{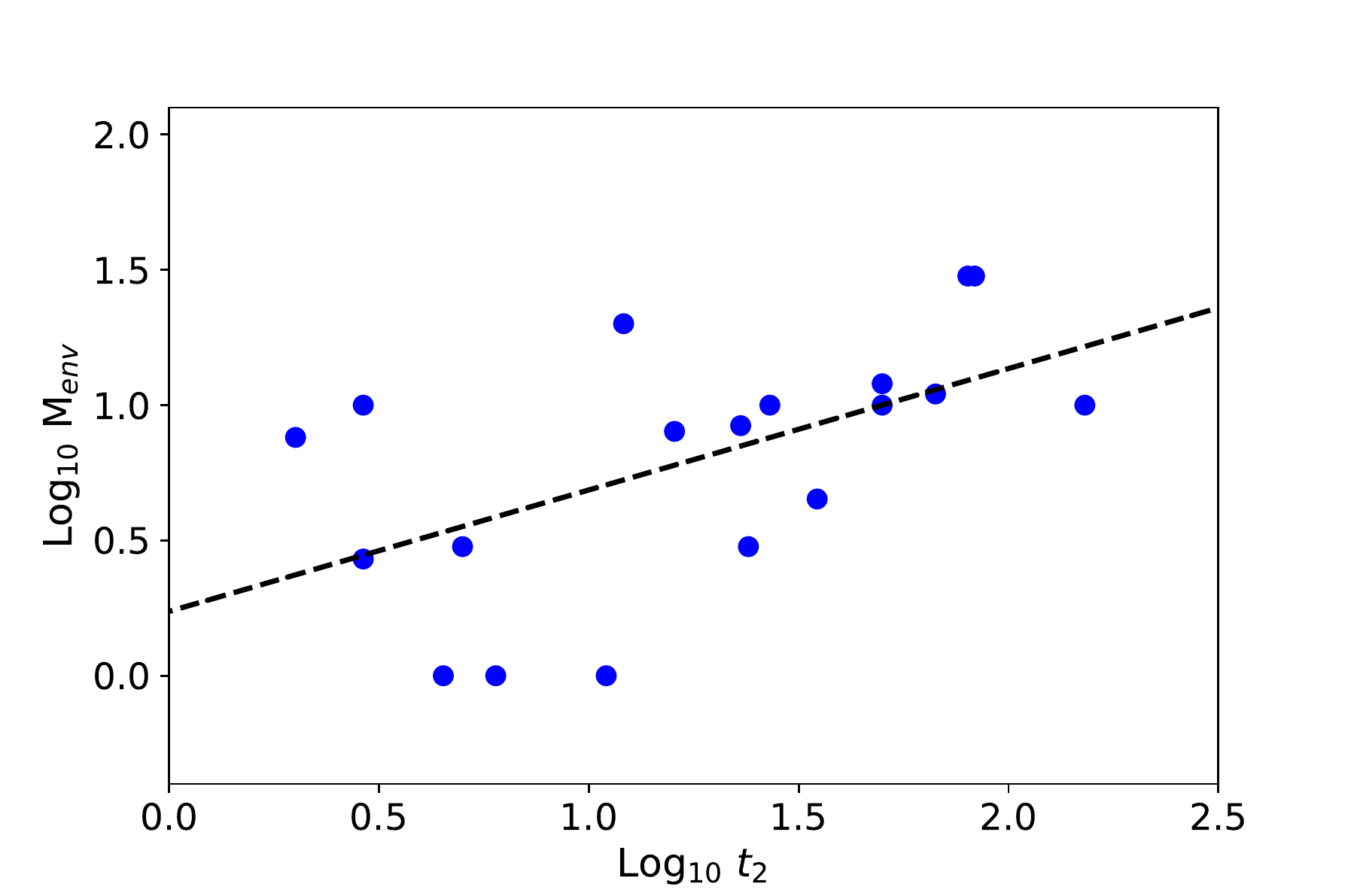}
    \caption{The correlation between the nova envelopes mass vs.\ rate of decline.  Mass of the envelopes in units of $\log (10^{-5})\,M_\odot$. Data from \citet{Gehrz1998} and \citet{Warner2003}.}
    \label{fig:masst2}
\end{figure}

We present in Fig.~\ref{fig:masshisto} the estimates of the mass distribution of the ejecta collected from literature. The data distribution peak at $\times 10^{-4}\,M_\odot$. 
Fig.~\ref{fig:masst2} shows that the  ejected mass marginally correlates (95\% c.l.) with the class of speed of the nova, according to: 

\begin{equation}
\log M_{\rm env} = 0.45 \times  \log t_2 + 0.24
\end{equation}

This result is not unexpected. Since the more massive white dwarfs need, on average, to accrete less material to achieve the conditions to trigger the outburst, they should eject less massive envelopes. Now we are in the position to assess -- at least as order of magnitude -- the contribute of a nova system to the chemical enrichment of the host galaxy. We derive that about $0.25\,M_\odot$ are processed through the nova ejecta over 100 years which is the characteristic time scale for Supernova explosions in the Milky Way. This figure has to be compared with $\sim 10\div 20\,M_\odot$ ejected by a typical core-collapse SN. It follows that only nuclei enriched, in the nova ejecta, by a  factor $\sim 50$ can contribute significantly to the galactic nucleosynthesis. In old and passive stellar systems like the elliptical galaxies where the formation of new stars has ceased long ago and therefore the rate of core-collapse is practically zero, the contribute of novae may be more significant. In this systems only SNe-Ia occur and these SN explosions are the result of the disruption of a CO WDs that release about $1.3\,M_\odot$ into the interstellar gas of the parent galaxy every 100 years or so.

\subsection{The abundances in the nova ejecta} 

\begin{figure}[htb]
    \centering
    \includegraphics[width=10cm]{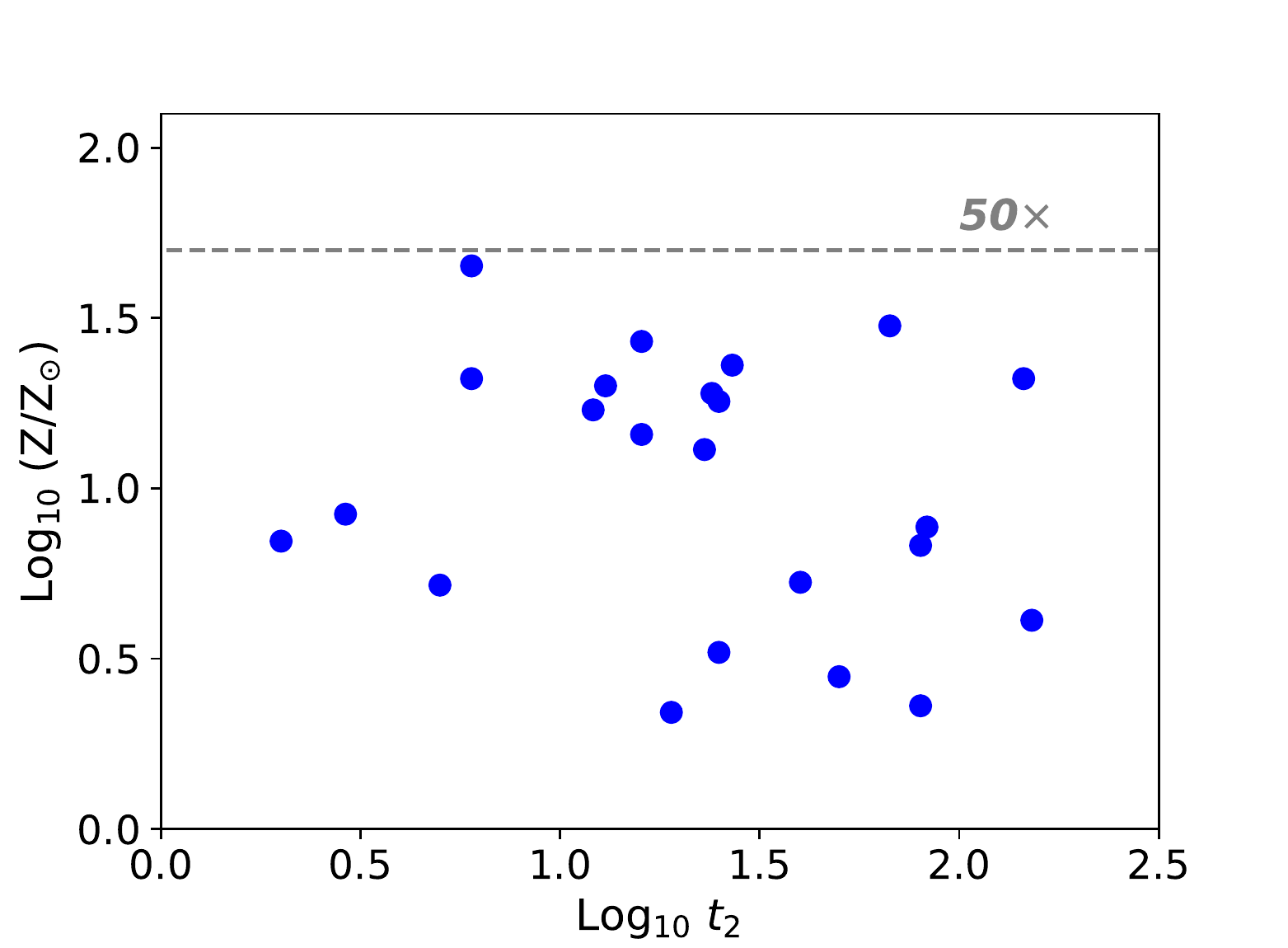}
    \caption{Abundances in the nova ejecta vs.\ rate of decline. The data do not show a significant correlation. The grey horizontal line represents an over abundance of 50 times the solar one. Data from \citet{Gehrz1998} and \citet{Warner2003}.}
    \label{fig:abut2}
\end{figure}

Recognition of unusual chemical abundances in the nova ejecta dates back many years \citep{Payne1957,Williams1977,Gallagher1978}. This topic has been reviewed several times \citep{Livio1994,Gehrz1998,Jose1998} and more recently by \citet{Jose2012} and \citep{Hernanz2015}.  One point stands clearly: all novae, for  which reasonable data are available (even admittedly characterized by large uncertainties), appear to be enriched in either He or/and heavy elements. This result is not unexpected because TNR proceed in the accreted H-rich envelope by means of the CNO burning chain. Fig.~\ref{fig:abut2} illustrates the average trend of the heavy elements abundance in the nova ejecta as a function of the rate of decline. Two facts emerges: 1) the available data suggest that the enrichment into the nova envelopes is normally below the $\sim 50$ time threshold 2) the heavy elements abundance in the ejecta do not seem to correlate with the speed class of the novae. 

\subsection{Novae as lithium factories}

During the TNR phase preceding the nova outburst, the synthesis of light chemical elements, and some of their isotopes, takes place. As soon as the the matter piles up onto the WD and the degeneracy pressure is reached
at the bottom of the accreted layer the thermonuclear reactions (TNRs) start. The first reaction to be ignited is the proton-proton chain. As the temperature increases, convective motions bring unburned elements from the interior of the WD into the burning region, allowing the ignition of the CNO cycle. This nuclear reaction increases the production of $\beta$-unstable isotopes like $^{13}$N, $^{15}$O and $^{18}$F, that will finally lead to the outburst and matter ejection \citep{Sparks1977,Starrfield2009}. During these processes,  helium can also produce $^7$Be, via the $^3$He -- $^4$He reaction channel, an isotope which decays into the stable $^7$Li after a half-life time decay of $\sim$ 53 days \citep{Cameron1971}.

Abundance studies of some of the best observed novae led to the evidence that CNO elements are very abundant in nova ejecta \citep{Gehrz1998}, in particular of the order of ten times the Solar value for the nitrogen, oxygen, neon and magnesium \citep{Vanlandingham1996,Schwarz2002}. On the other hand, we had little information about the lithium production. In classical novae, convective motions can transport a large quantity of freshly-produced $^7$Be in its most external regions. In these external layers, lithium can survive and then ejected once the nova outburst is triggered by the TNR. An immediate consequence is that CNe may represent one of the most efficient astrophysical sources of lithium  \citep{Starrfield1978}. However, long-term spectroscopic observations of CNe in outburst have never revealed the presence of Li in the nova ejecta \citep{Friedjung1979}. 

\begin{figure}[htb]
    \centering
    \includegraphics[width=10cm]{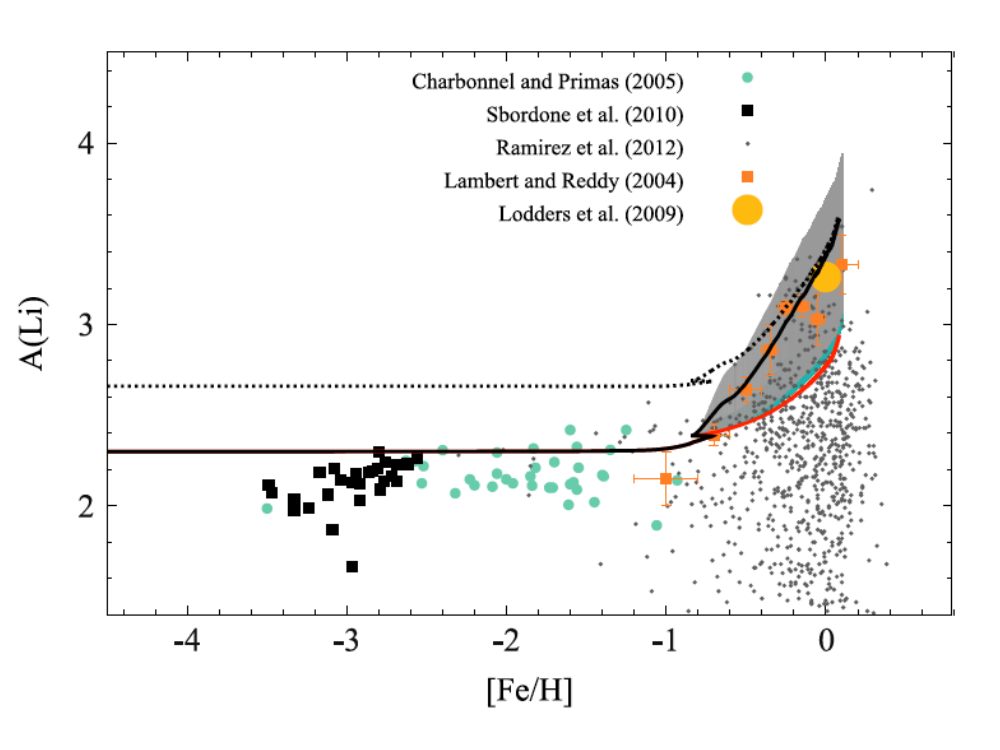}
    \caption{The distribution of the abundance of lithium ($A({\rm Li})$) as measured for stars with different metallicities ($[Fe/H]$) in the Galaxy \citep{Spite1982,Spite1990}. The black continuous line corresponds to the predictions of the chemical evolution models \citep{Romano2001} assuming the average yield of lithium provided by novae of $M_{\rm Li} = 2.55 \times 10^{-10}\,M_{\odot}$ and with a primordial abundance of $A({\rm Li}) = 2.3$. The grey dashed region marks the upper and lower limit considering the uncertainties on the lithium yield. The red line represent the predictions of the chemical evolution model when neglecting the novae lithium yield. Dashed black line represents the evolution of the lithium if we start from a primordial abundance as observed by Planck of $A({\rm Li}) = 2.7$ \citep{Coc2014}. Data from \citet{Charbonnel2005,Lambert2009,Lodders2009,Sbordone2010,Ramirez2012}. Image reproduced with permission from \citet{Izzo2015}.}
    \label{fig:spite}
\end{figure}

Lithium is one of the most fragile element in the Universe, given its low binding energy per nucleon \citep{Conti1968}. Consequently, lithium is easily destroyed in many astrophysical environments, mainly via proton collision, at a temperature of $\sim 2 \times 10^6$ K. Lithium is also the only stable ``metal'' formed during the Big Bang Nucleosynthesis, given the absence of a stable element with atomic number $A=8$ (the only possible one, $^8$Be, has a very short life, $t \approx 10^{-17}$ s).  Within the standard BBN theory, it is possible to predict the primordial lithium abundance from the estimate of the baryon-to-photon ratio parameter \citep{Dunkley2009,Fields2011,Coc2013}, which shows a discrepancy with the lithium abundance inferred from old stars observed in the Galactic halo, the so-called ``cosmological lithium problem'' \citep{Spite1982,Sbordone2010}. Similarly, the lithium abundance inferred from younger Galactic stellar populations is much higher than the abundance estimated from the BBN  \citep{Spite1982,Spite1990}, then suggesting the presence of unknown lithium factories in the Galaxy, this is the so-called:  ``Galactic Lithium problem'' \citep{D'Antona1991,Matteucci1995}. In recent years, different astrophysical sources have been proposed for the lithium nucleosynthesis such as: the cosmic-ray spallation \citep{Reeves1990}, the hot bottom burning in low-mass red giant branch (RGB) stars \citep{Sackmann1999}, intermediate asymptotic giant branch (AGB) stars \citep{Sackmann1992,Travaglio2001}, carbon stars \citep{Abia2001} but it turned out that they were unable to produce the full amount of lithium observed in the Milky MilKy \citep{Romano2001}. 

\subsection{``The Hunt for Red Lithium''} 
 
The idea that novae could be efficient lithium producers has been predicted long ago  \citep{Arnould1975,Starrfield1978} but it has never been confirmed (e.g., \citealt{Friedjung1979, DellaValle2002}). The typical signature of lithium in the spectra of CNe is the resonance transition of its neutral state, which is a very tight absorption doublet centered at $^7$Li I 6707.8\,{\AA}. Although \citet{Friedjung1979} did not detect any absorption line corresponding to $^7$Li I, he was able to put upper limits on the lithium abundance in nova ejecta. Lithium was extensively and unsuccessfully searched in the following years. A possible evidence for the presence of a transient $^7$Li emission line was made for the case of Nova Vel 1999 \citep{DellaValle2002}, but further analysis using UV spectral data from Hubble Space Telescope showed its origin as due to neutral nitrogen \citep{Shore2003}.

\begin{figure}[htb]
    \centering
    \includegraphics[width=10cm]{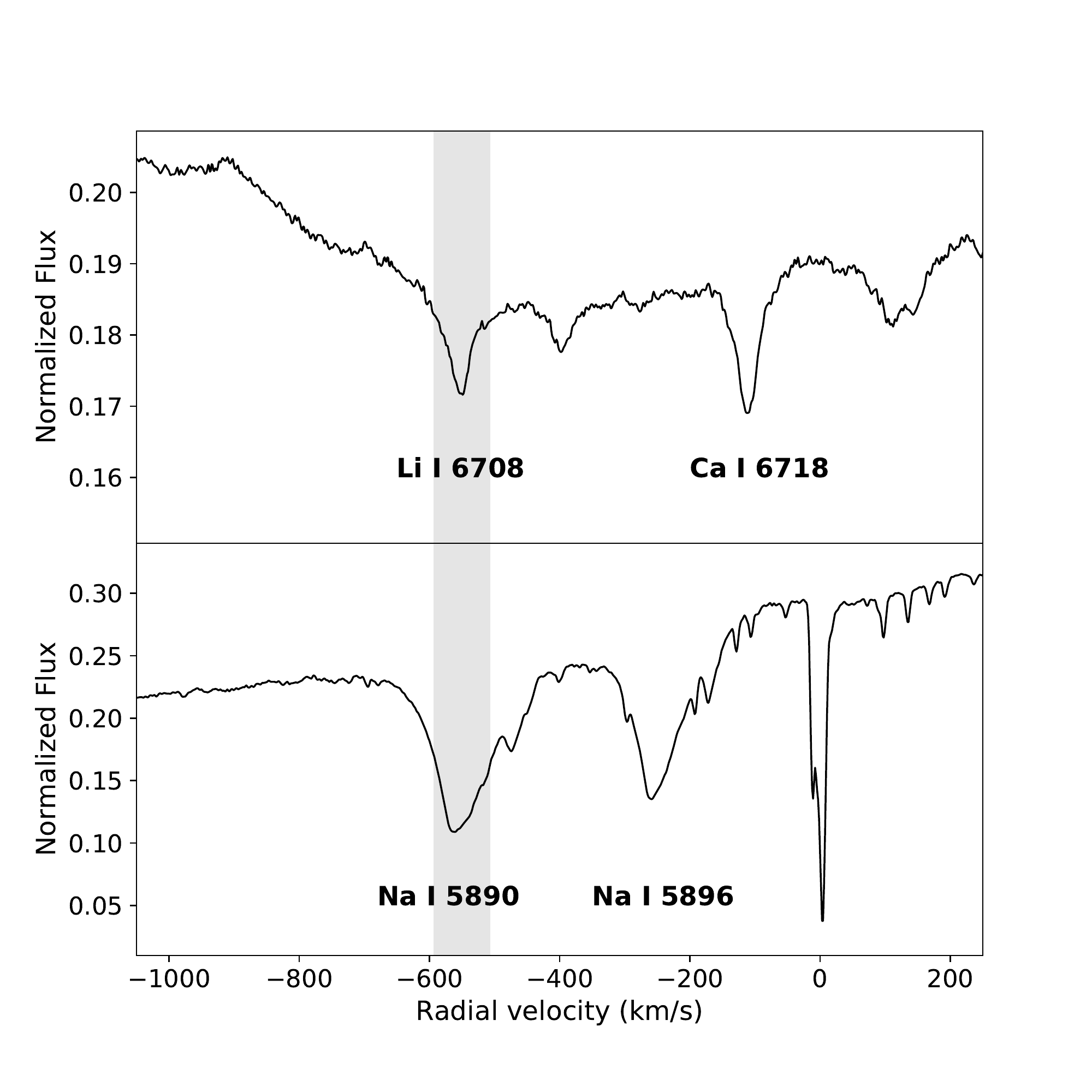}
    \caption{The spectrum of V1369 Cen observed 6 days from the discovery with FEROS. The upper panel shows the spectrum centered at the lithium resonance transition wavelength of Li I 6708\,{\AA}, while the lower panel shows the spectrum centered ad the Na ID2 5890\,{\AA}. The presence of Li I 6708 is confirmed by its oserved ejecta velocity, which is the same of Na I D and other absorption features in the spectrum \citep{Izzo2015}.}
    \label{fig:LiCen}
\end{figure}

The presence of Lithium in CNe remained a mistery until 2013.  Two very bright novae were discovered and then observed with high-resolution spectrographs by two distinct groups: V339 Del and V1369 Cen. The presence of blue-shifted absorption lines due to $^7$Be II was reported in the late (more than 50 days from the outburst) spectra of V339 Del \citep{Tajitsu2015}, while early spectra of V1369 Cen revealed the presence of an absorption feature that was confirmed to be due to the resonance transition of neutral Lithium at the expanding velocity of $V_{\rm exp} = -550$ km/s \citep{Izzo2015}. For this latter case, by using an approach similar to the one used by \citet{Friedjung1979}, the total mass of lithium ejected was computed to be $M_{\rm Li} = (0.3-4.8) \times 10^{-10}\,M_{\odot}$. After assuming this amount of lithium as ejected by all slow novae in the Galaxy, and using accurate numerical codes developed to compute the chemical evolution of the Galaxy \citep{Romano2001}, it was proved that novae can explain the observed over-abundance of lithium observed in young stellar populations \citep{Spite1982,Fields2014,Izzo2015}.
The detection of Li I in the early spectra of V1369 Cen does not represent the unique case so far. In 2015 an other bright slow nova was discovered in the direction of the Sagittarius constellation, V5668 Sgr and similarly to V1369 Cen, early spectra showed the presence of a faint absorption line that corresponds to the Li I $\lambda$6708 transition blue-shifted at a velocity of $v_{\rm exp} \sim 500$ km/s \citep{Wagner2018}. The same velocity is displayed by other typical transition observed in novae spectra as Na I D2 $\lambda$5890 and Fe II $\lambda$5169. This nova also showed the presence of blue-shifted absorptions related to the resonance doublet of $^7$Be II  $\lambda\lambda$3130/31 \citep{Molaro2016}. By following a similar approach \citet{Molaro2016} have estimated the ejected mass of $^7$Be to be $M_{\rm Be,Sgr} = 7 \times 10^{-9}\,M_{\odot}$. In passing, we note that \citet{Prantzos2012} in a detailed review on possible lithium factories in the Galaxy, explicitly mentioned that:

\begin{quote}
``if novae provide $\sim$50--65\% of solar lithium, they ought to have a Li yield of $\sim 10^{-9}\,M_{\odot}$.''
\end{quote}

A prediction in excellent agreement with the result derived from observations. After assuming a typical nova life-time of 10$^{10}$ yrs, \citet{Molaro2016} suggest that a dozen of novae/yr like V5668 Sgr can produce the total amount of $^7$Li observed in the Galaxy.

\begin{figure}[htb]
    \centering
    \includegraphics[width=6cm]{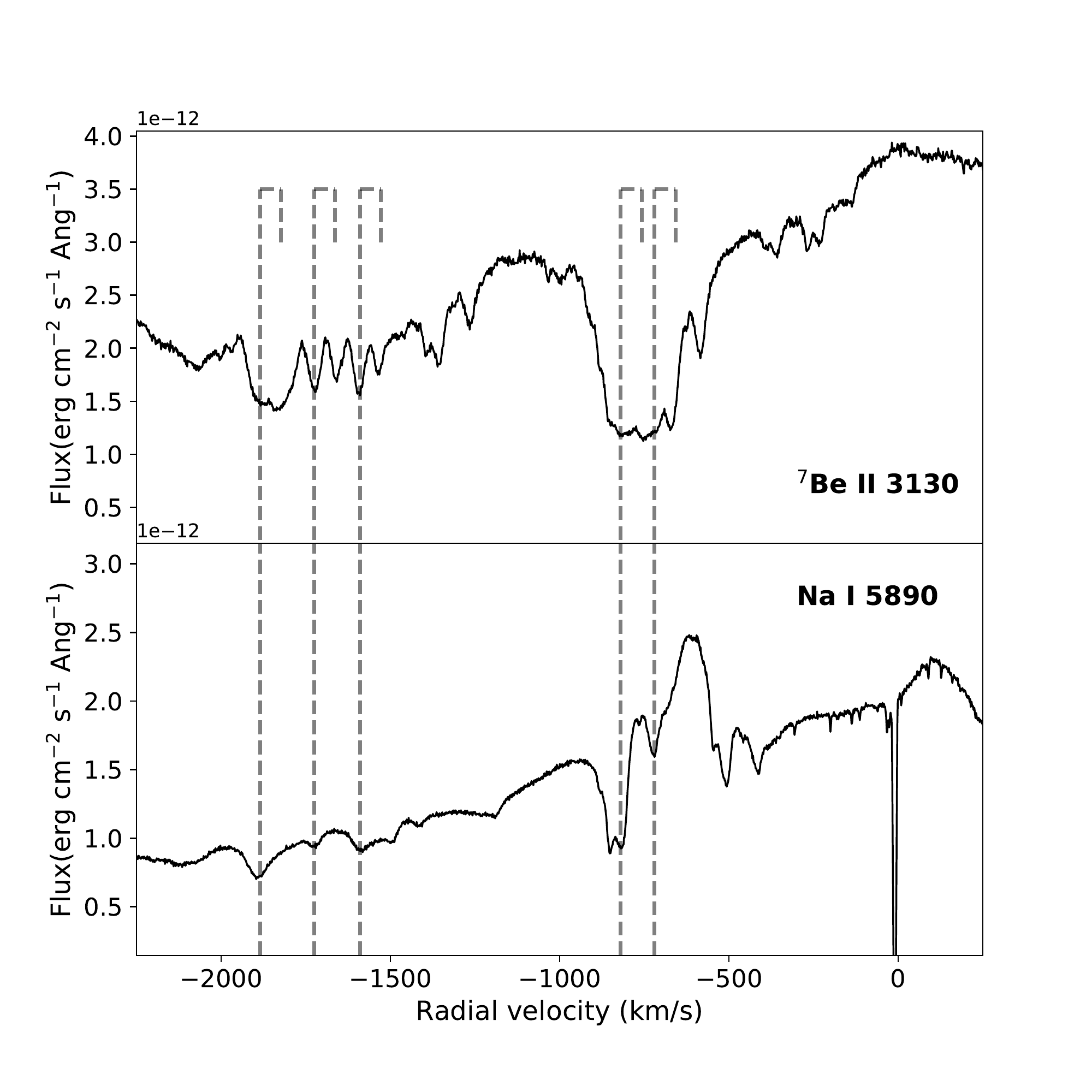}
    \includegraphics[width=5.7cm]{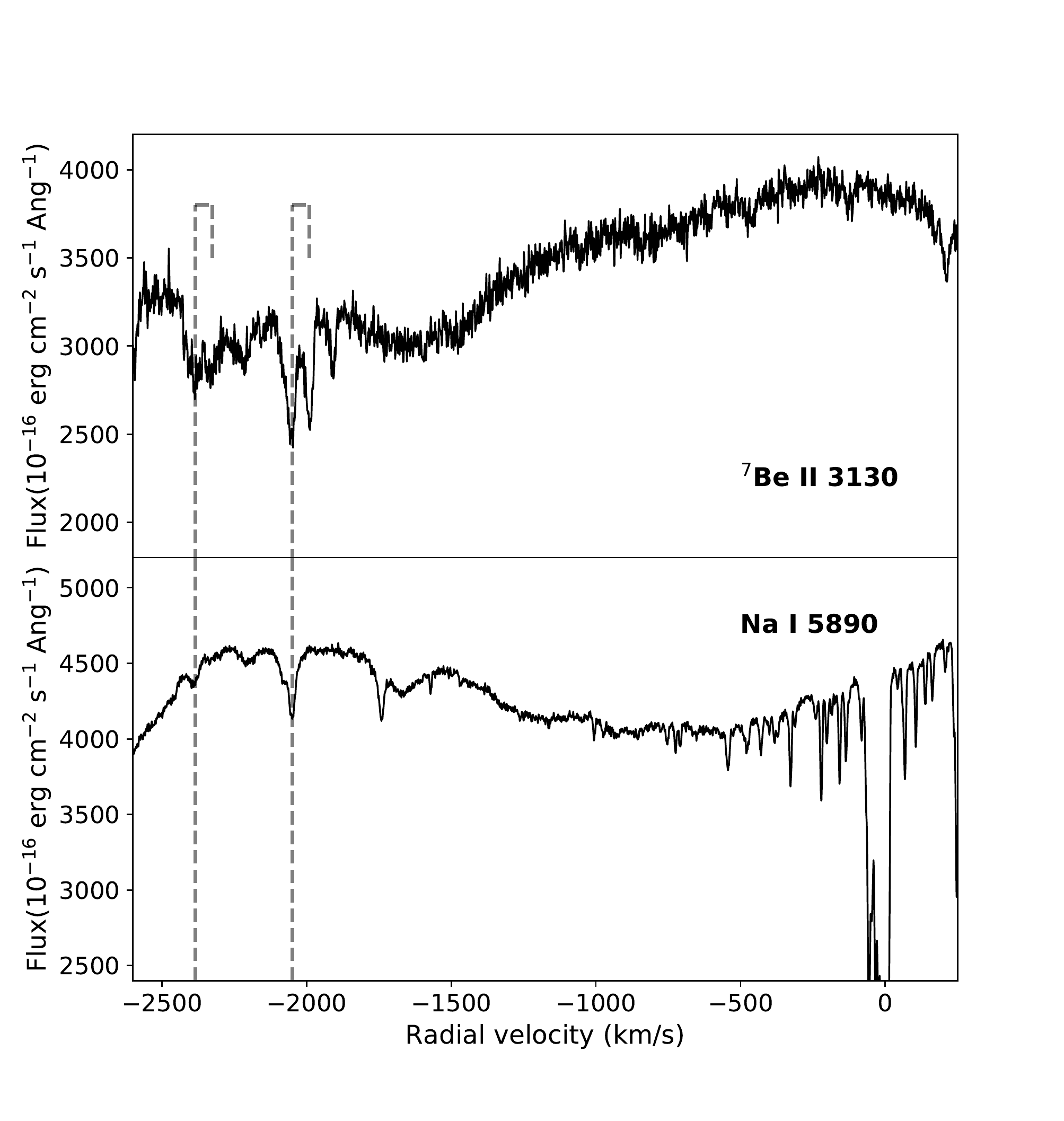}
    \caption{The spectrum of V5668 Sgr (left panel) and V407 Lup (right panel) observed 88 days and 8 days, respectively, after the nova discovery \citep{Molaro2016,Izzo2018}. The upper panel shows the spectrum centered at the $^7$Be resonance transition wavelength of 3130.583\,{\AA}, while the lower panel shows the spectrum centered ad the Na ID2 5889.93\,{\AA}. The presence of $^7$Be is clearly marked by the combined identification of the two components of the doublet, at the same expanding velocity than other features, in this case of Na I 5890.}
    \label{fig:Be}
\end{figure}

In recent years, additional evidences of $^7$Be have been found in other novae \citep{Tajitsu2016, Izzo2018, Selvelli2018, Molaro2020}. 

\section{Novae as Supernova-Ia progenitors}

SNe-Ia are commonly referred as standard candles even if it would be more appropriate to describe these objects as ``standardizable'' candles because their high luminosity at maximum correlates with their rate of decline. Differently from CNe the intrinsically brighter SNe-Ia are characterized by the slowest rates of decline \citep{Phillips1993}. The process of standardization works well on empirical basis \citep{Perlmutter1998,Riess1998,Schmidt1998} but still it is far from being completely understood \citep{Mazzali2001,Wygoda2019} and the uncertain nature of the binary systems does not help to clarify this issue. SNe-Ia basics that can be directly derived from observations can be summarized as the following:

\begin{itemize}
\item SNe-Ia occur in early and late Hubble types;
\item Occurrence in early type galaxies with SFR less or much less than  a few $10^{-1}\,M_\odot/{\rm yr}$ implies that SNe-Ia are not caused by the collapse of the core of stars with $M > 8\,M_\odot$;
\item The lack of H and He in the ejecta points in the direction of highly evolved progenitors;
\item The energy for unit of mass, i.e., $1/2\times (10^4\,{\rm km/s})^2$ is similar to the one obtained from the conversion of C (7.7 Mev b.e. per nucleon) into Fe (8.8 Mev b.e. per nucleon).  
\end{itemize}

On the basis of these arguments, the common wisdom strongly suggest that SNe-Ia are the thermonuclear disruption of mass-accreting WDs in binary systems. Historically two scenarios are commonly accepted for producing SNe-Ia: the so called double degenerate \citep{Webbink1984,Iben1984,Tornambe1986} where two CO WDs in a binary systems make coalescence as result of the lost of orbital energy for GWs  and the single degenerate \citep{Whelan1973,Nomoto1982a} in which a WD accretes material from a MS star (see \citealt{Maoz2014,Livio2018} for an update  review on this topic). In the latter framework a SN-Ia can be produced by the explosions of  WDs in recurrent- or symbiotic- nova systems, when the respective WDs  approach the Chandrasekhar limit after having accreted material from the companion stars.  Because these Cataclysmic Systems typically contain massive WDs characterized by a high accretion rates, they are regarded as possible SN-Ia progenitors. Recently \citet{Starrfield2019} have found on the basis of 1-D hydrodinamical simulations that all C-O WDs in the range of masses $0.6$--$1.35\,M_\odot$, which accrete material from companions within a narrow interval of  $\dot{M} \approx 1.6\times 10^{-10}\,M_{\odot}$/yr would grow in mass up to 80\% of the accreted mass after each nova cycle. These authors conclude that novae can be an important channel for type Ia progenitors. These results are still preliminary and they should be taken with some degree of caution. We note for example that even if WDs are growing in mass because the ejected/accreted mass is smaller than 1, the Chandra mass might not be reached. This possibility depends among other things on the mass of the secondary and on the efficiency of the accretion onto the WD that is often defined as $\eta = \Delta m_{\rm growth}/\Delta m_{\rm trans}$. 
\citet{Kato1989} computed $\eta$ for different sizes of the Roche lobes and they never found, even under the most favorable conditions (high value of $M_{\odot}$), a value larger than 0.3--0.4. For example a $1\,M_\odot$ WD which accretes material at $\dot{M} \sim 10^{-10}\,M_\odot$ yr$^{-1}$ and it is characterized by $\eta \lesssim 0.1$  requires a secondary star with a mass much larger than that of the primary,  to reach the Chandrasekhar limit of the order of 
$M_2 \sim 0.35/\eta \approx 3.5\,M_\odot$. The CV configuration where $M_2$ is $> M_1$ might be not characterized by stable mass transfer \citep{Warner2003}.
 
Symbiotics as progenitors of SNe-Ia had some popularity in the 1990s thanks to \citep{Munari1992}, which showed that the density of progenitors  of Symbiotic novae was vastly consistent with the rate of SN-Ia explosions.  A few years later this scenario was questioned by radio observations of SN 1986G in Centaurus A \citep{Boffi1995} that failed to reveal the radio synchrotron emission \citep{Chevalier1988} that should have been produced during the interaction between the SN blast wave and the circumburst material. The impact of this negative result was partially soften by the fact that SN 1986G was somehow peculiar object and therefore it was not representative of the main bulk of SNe-Ia. The result from radio observations of 27 SNe-Ia obtained with the Very Large Array two decades later by \citet{Panagia2006} was much more sharp. There was not a single positive detection and the lack of radio signal implied the existence of a very low circumburst density. Thus \citet{Panagia2006} were able to to set an upper limit of $\dot{M} \lesssim 3\times 10^{-8}\,M_\odot/{\rm yr}$  on the mass loss rate of the secondary star, which is 1--2 orders of magnitudes smaller than the $\dot{M}$ expected from red giant donor. This upper limit was further constraint up to $\dot{M} \lesssim 5\times 10^{-9}\,M_\odot/{\rm yr}$  by \citet{Chomiuk2016}.  

The most direct pieces of observation, to date,  which might rule out Symbiotics to form the main bulk of SN-Ia progenitors are the deep pre-explosion images, obtained with HST, of the  site  of  the ``standard'' SN-Ia  2011fe. \citet{Li2011} did not find a suitable candidate progenitor at the SN location and by taking advantage of the proximity of the M101 host galaxy, they could set a firm upper limit on the magnitude of the progenitor system. They could conclude that the progenitor was: ``\emph{$10\div 100$ times fainter than previous limits on other SN-Ia progenitors. This directly rules out luminous red giants and the vast majority of helium stars as the mass donating companion. These observations favour a scenario where the exploding WD of SN 2011fe accretes matter either from another WD, or by Roche-lobe overflow from a subgiant or main-sequence companion star}''.  

Finally, we point out the vain search for a surviving companion in the SN-Ia remnant SNR 0509-67.5 by \citet{Pagnotta2014}. These authors could concluded that: \emph{``This lack of any ex-companion star to deep limits ($M_V=8.4$ mag) rules out all published single-degenerate models for this supernova. The only remaining possibility is that the progenitor of this particular type Ia supernova was a double-degenerate system.''}

Then the question becomes whether or not some room is left for the existence of the single degenerate scenario or is it completely ruled out? 

Probably the most convincing piece of evidence that a single degenerate system can explode as SN-Ia is given by the detection of circumburst material around SN-Ia 2006X \citep{Patat2007a}. The analysis of the evolution in the spectroscopic features of NaI-D doublet induced by the change of the ionization conditions of the SN radiation field have proven that this material was located at distances from the SN of about $D \sim 10^{16}$ cm and it was expelled by the Supernova progenitor about 50 yr before the explosion. The existence and the fast spectroscopic evolution of this circumburst material is  reminiscent of the THEA absorptions observed by \citet{Williams2008}, around novae shortly after the maximum light. According to \citet{Patat2007a} these erratic or recurrent in time emission episodes rule out a double degenerate scenario progenitor for SN 2006X. In passing we also note that a similar study carried out by (almost) the same team \citep{Patat2007b} on SN-Ia 2000cx was not able to detect a behavior similar to that exhibited by SN 2006X.  Taking at their face values these results may suggest that both scenarios are at play. If this fact is true a second question arises: what is the nature of the single degenerate systems? According to general believe some of these systems might be Recurrent Novae (e.g. \citealt{Kato2012}), such as for example T Pyx \citep{Godon2014}, but see also \citet{Selvelli2008b} for a different view.  Out of about 400 nova outbursts in the Galaxy, Fig.~\ref{fig:3}, only ten of them has been identified as RNe. The rigorous correction of these numbers for incompleteness is very questionable procedure because many assumptions are involved. However, we can apply at first order approximation the following argument. From Sect. 3.5  we assume a galactic nova rate of 25 novae/yr, therefore in the last 156 years (from U Sco in 1863) in the Galaxy have occurred about 4000 nova explosions. It means that, on average, the probability to detect in this period of time a nova outburst was about 10\%. Since to identify a Recurrent Nova we need to observe at least two outbursts from the same system, the probability to have observed a RN is at first order only $\sim 1\%$. Since we have observed in the MW about a dozen of RNe, their actual number has to be $\sim$ 10 times as larger. This simple order of magnitude calculation would indicate a ``true'' ratio RNe/CNe of $\sim 30\%$ for our Galaxy cf. \citep{Pagnotta2014}.  We note that \citet{DellaValle1996} have estimated for M31 and LMC a fraction RNe/CNe of 10\% and 30\%  respectively and they concluded that \emph{``RN-type systems can account for at most a few percent of the SN-Ia rate''}. \citet{Shafter2015} consistently estimated that \emph{`` as many as one in three nova eruptions observed in M31 arise from progenitor systems having recurrence times $< 100\,yr$\dots''} and they concluded that it is very unlikely that RNe could play a major role as SNe-Ia progenitors in a galaxy like M31. 
 
These last results about the contribute of RNe to the SNe-Ia rates seem to relegate the SD to a completely secondary role with respect to double degenerate. However, this story does not seem to be over yet. \citet{DellaValle1994} and \citet{Mannucci2005} have found that the  SNe-Ia rates,  per unit of luminosity of the respective host galaxies, are a full order of magnitude higher in late type/\emph{blue} galaxies than observed in E-S0/\emph{red} galaxies. This result can be explained by assuming that a significant fraction of Ia events in late spirals/irregulars originates in a relatively young stellar component  characterized by evolutionary times shorter than $\sim 100$ Myr \citep{Mannucci2006}, while the SNe-Ia in passive elliptical galaxies have a time evolution which is described by a much wider distribution, well characterized by an exponential function with a decay time of about 3 Gyr. Recently \citep{Ilkiewicz2019} have studied a subset of Symbiotic binaries, which contain a Mira donor that is characterized by orbital periods of dozens of years. Obviously these symbiotic stars are not normally considered suitable SN-Ia progenitors because of their large separation (however see \citealt{Hachisu1999} for a discussion of the symbiotic scenario). This team has analysed the evolution of V407 Cyg
and they find that there is 90\% probability that this system can explode as SN-Ia and this probability increases to 97\% for WD with $\sim 1.35\,M_{\odot}$, which incidentally coincides with the value of the mass suggested by \citet{Hachisu2012}. \citet{Ilkiewicz2019} predict that the white dwarf in V407 Cyg will reach the Chandrasekhar limit in 40--200 Myr and eventually will explode as SN-Ia. This evolutionary time is consistent with the ``prompt'' SN-Ia population identified by \citet{Mannucci2006} (see also \citealt{Greggio2010}). Very recently \citet{Darnley2006,Darnley2019} have provided convincing evidence that the progenitor of the yearly erupting RN M31N 2008-12a is a single degenerate system with a red giant donor (see also \citealt{Williams2016}). \citet{Kato2015} have determined that the central WD is of CO type and that its mass is $\sim 1.38\,M_{\odot}$. Moreover, the WD is quickly growing in mass, by ejecting only $\sim$40\% of the matter accreted during each nova outburst cycle (which corresponds to $\sim 10^{-7}\,M_{\odot}$). Therefore, it should  pass the Chandrasekhar limit in about 100,000 years and it might terminate its life with a SN-Ia eruption.

\section{Multi-wavelength novae}

The possibility to investigate nova outbursts in other frequencies than the optical and near-infrared electromagnetic range has become a standard procedure in the last decade, thanks to a new-generation of instruments and detectors operating at radio, infrared (IR) and high-energies as X- and $\gamma$-rays. These observations have revealed that novae undergo a complex evolution characterized by several emission phases and a non-spherical geometry for the nova ejecta. In the following we briefly review  some of the most interesting results obtained in recent years.

\subsection{Novae in the infrared}

Abundances from novae can also be inferred from IR observations, in particular for Ne, Mg and Al \citep{Gehrz2008,Gehrz2015}. IR observations have also demonstrated that CNe, particularly of the CO class, are dust grain factories \citep{Geisel1970,Hyland1970,Clayton1979,Gehrz1980,Gehrz1980b,Gehrz1990,Gehrz1995a,Mason1996,Mason1998,Woodward2011,Chesneau2012}. In these objects the free-free emission that dominates the optical peak is often followed by a dust-dominated phase, while in ONe novae an Orion phase immediately follows the Principal stage \citep{Gehrz1995b,Woodward1995,Woodward1997}. 
NIR observations of novae have also provided convincing evidence for bi-polar ejection of nova shells \citep{Chesneau2007,Chesneau2011,Chesneau2012}. For example NIR-interferometric observations of V339 Del \citep{Schaefer2014} obtained during the earliest phases have detected from the second day an aspherical distribution of the light that can be explained by assuming prolate or bipolar expanding structures. We address the reader to the papers by \citet{Chesneau2012b} and by \citet{Gehrz2008} for a deep insight into IR emission from novae.

\subsection{X-ray novae}

Classical Novae have been detected in X-rays since the first observations by the ROSAT satellite in the 1980s \citep{Krautter1996,Orio2001}. During the bright outburst phases, X-ray radiation is emitted as a consequence of three physical processe: 1) a thermal emission from the underlying WD, which has been observed in several novae \citep{Krautter1996,Mukai2001,Balman2001,Schwarz2011,Page2015,Osborne2015} when the nova ejecta starts to become optically thin, what is called also as  supersoft X-ray source phase (SSSs); 2) a hard X-ray component, originated in the shocks between the ejecta with material surrounding the WD progenitor, such as previously ejected shells or a dense interstellar medium \citep{Mukai2001,Metzger2015} 3) emission from the re-established accretion during the quiescent phase \citep{Takei2011}. ROSAT observed the supersoft emission in a few objects, among them V1974 Cyg one of the most extensively studied Galactic novae by ROSAT \citep{Balman1998}. The breakthrough in this field occurred with the commissioning of the Neil Gehrels Swift satellite \citep{Gehrels2004}. Thanks to  its capability to re-point a target within minutes/hours, this satellite is particularly suitable for observations of targets of opportunity such as nova stars. For example, one of the most consistently investigated cases was the 2006 outburst of the recurrent nova RS Oph \citep{Osborne2011}. Today we have observed X-rays emission from the very earliest phases through the whole visibility stage for
about half of the 80 novae observed in X-rays  \citep{Ness2007,Osborne2015}, including some objects that have occurred in M31 and M33 and MCs \citep{Aydi2018}.

The results of a systematic survey on novae in the Milky Way and M31 with Swift has provided a number of important results: i) there exists an early phase characterised by hard X-ray emission due to the presence of shocks within the different components of nova ejecta \citep{Chomiuk2014a} and with pre-existent circum-burst medium; ii) the onset of the supersoft phase begins soon in novae characterised by small  $t_2$ values, which are suggestive of low mass ejecta \citep{Henze2014}; iii) the duration of the supersoft phase is an indicator of the WD mass: the higher is the mass the shorter is the supersoft phase \citep{Hachisu2006,Hachisu2010}; iv) the temperature of the thermal component observed during the supersoft phase, modeled with a simple blackbody function, is directly related to the mass of the underlying WD; v) X-ray photons rising from the supersoft phase can destroy the dust produced at later phases \citep{Gehrz2018}, but see also \citet{Shore2018}; vi) some novae show a modulation in X-rays, whose characteristic period varies from few dozens of seconds like RS Oph, KT Eri and V5668 Sgr \citep{Osborne2011,Ness2015} up to few days like Nova LMC 2009A. We address the reader to the paper on Swift nova observations by \citet{Osborne2015}.

In the last years, 13 novae belonging to the Milky Way \citep{Orio2005,Mukai2008}, M31 \citep{Orio2006,Orio2010,Henze2010,Henze2011,Henze2012,Henze2014} and SMC \citep{Orio2018} have been monitored with XMM-Newton and Chandra. The main goal was the study of supersoft X-ray sources at high-resolution, aimed at both determining the duration of their SSS phase and the physical parameters of the observed atmosphere, including abundance studies \citep{Ness2009,Pinto2012}. Some theoretical implications have been discussed by \citet{Hernanz2010}. During these surveys was possible to discover several RNe \citep{Henze2018} and remarkably enough some very rare events such as a nova in outburst in a globular cluster of M31 \citep{Henze2013}. 

\subsection{Gamma-ray novae}

\begin{figure}[htb]
    \centering
    \includegraphics[width=9cm]{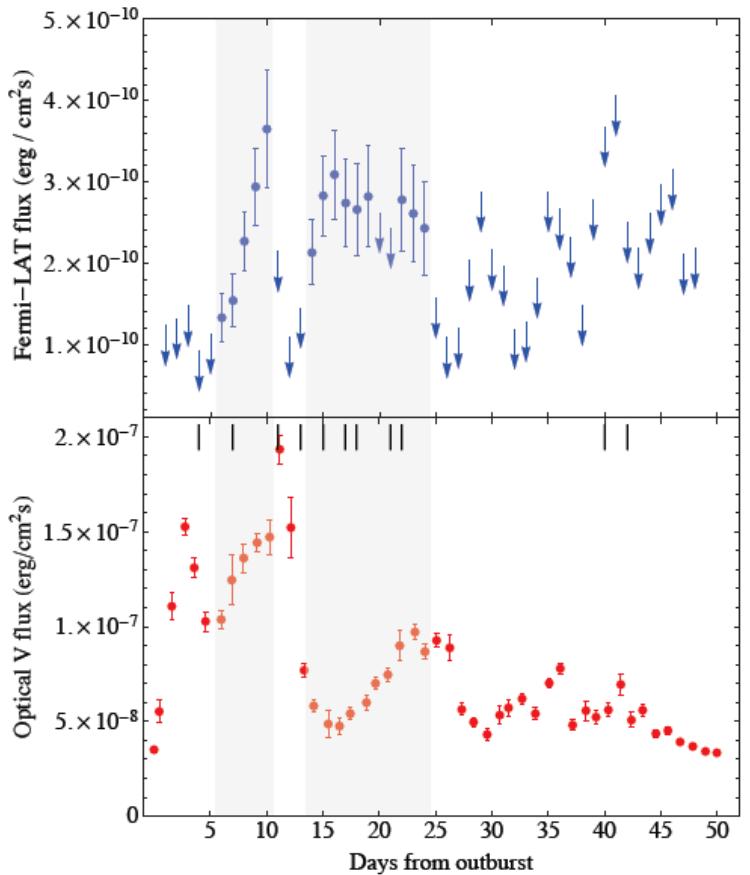}
    \caption{The optical emission of V1369 Cen (lower panel) compared with the emission detected by the Fermi-LAT detector (upper panel, courtesy M. Kovacevic). Grey regions mark the time intervals corresponding to a positive detection of Fermi-LAT.}
    \label{fig:LAT}
\end{figure}

In 2014, Classical Novae, were definitely revealed as a new class of gamma-ray sources \citep{Ackermann2014}, although the first nova-like observed in gamma-rays was the symbiotic nova V407 Cyg \citep{Abdo2010}. To date about a dozen of novae and symbiotic systems have been detected in gamma-rays at energy larger than 100 MeV, but below few GeV. The physical origin of this emission is not yet clear, but it is likely related to strong shocks driven by both the dense nova ejecta when colliding with the tenuous interstellar medium or the wind from the secondary \citep{Martin2013}. As an alternative the internal shocks are produced by the interaction of a fast wind radiatively driven by the nuclear burnings on the white dwarf \citep{Martin2018} with inhomogeneities formed into the expanding shells of the nova ejecta (see the case of V1369 Cen \citealt{Mason2018}). However, we cannot exclude that a combination of these two mechanisms is at play. Although these inhomogeneities are often observed as knots in some historical novae, such as T Pyx \citep{Shara2015}, only $\sim 15\%$ of the objects optically discovered show gamma emission. This rises  the question whether or not gamma-ray novae are intrinsically rare among galactic novae or this small percentage results from a distance bias \citep{Morris2017,Franckowiak2018}, which prevents the discovery of high energy emission in most CNe because of their distances.    

The high-energy radiation observed in novae does not generally show a direct link with the optical emission \citep{Franckowiak2018} with the exception of bright LAT novae such as:  ASAS-SN 16ma \citep{Li2017}, TCP J1734475-240942 \citep{Munari2017}, V906 Car \citep{Aydi2020} and V1369 Cen (see Fig.~\ref{fig:LAT}) where a correlation between the observed optical light curve and the $\gamma$-ray flux was observed, then implying that a considerable fraction of the optical radiation comes from the re-processing of the gamma-ray radiation at longer wavelengths. This evidence also suggests that the optical radiation observed in the nova phenomenon is not only originating from the thermal energy released in the TNR \citep{Metzger2015}, but it is partly due to the re-processing of high-energy radiation coming from acceleration of particles likely originating by magnetic field amplification in the shocks \citep{Li2017} or radiation processes in leptonic and/or hadronic scenarios \citep{Vurm2018}. 

After connecting the nova high-energy radiation with the multiple absorption of the principal spectra (see Sect. 4), the modelling  of nearby nova remnants \citep{Slavin1995,Bode2002,Moraes2009,Ribeiro2009,Ribeiro2013,Ribeiro2013b}, IR interferometry \citep{Chesneau2012b,Schaefer2014}, and the delayed emission from radio observations (see next subsection) a more clear scenario of nova phenomenon emerges. In this framework immediately after the ignition of the TNR, a first component with an oblated/prolate matter geometry distribution expands with velocities of few hundreds of km/s. At later times, a faster ($\sim$1000--5000 km/s) wind, is emitted from the poles of the central WD, with relatively high opening angles, and then it catches up with the dense earlier component. The interaction between these two ejecta can lead to particle acceleration that give origin to the observed gamma-ray radiation. The structure of the remnant will show years after the outburst the consequences of this interaction. For example, it was observed in FH Ser 1970 an ellipsoid structure characterized by denser ring-like component along the minor axis \citep{DellaValle1997,Gill2000}. 

Novae are also factories of isotopes, like $^7$Be and $^{22}$Na, which decay via electron capture or through the $\beta^+$-process emitting very high-energy photons at specific energies (see \citealt{Hernanz2012} for a review). For example, $^7$Be decays into $^7$Li emitting a photon with an energy of 478 keV. The recent discovery of a consistent amount of beryllium produced in novae implies that we should be able to detect this line with space-based detectors like INTEGRAL. However, a detection of this emission line is still missing, likely due to the large distance of the novae for which beryllium and/or lithium have been observed \citep[see, e.g.,][]{Siegert2018}. The detection of this 478 keV line would represent the conclusive confirmation that novae are one of the main factories of lithium in the Milky Way.
 
\subsection{Novae at radio frequencies}

\begin{figure}[htb]
    \centering
    \includegraphics[width=9cm]{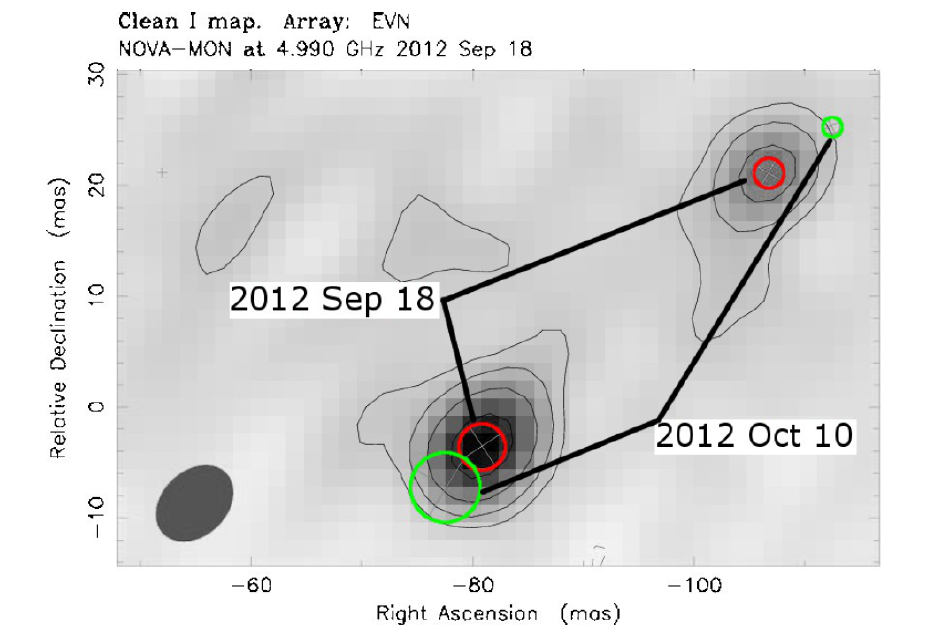}
    \caption{The European VLBI Network (EVN) radio image of the remnant of the V959 Mon as obtained on September 18, 2012. The bi-polar shape of the remnant is clearly shown by the darker blobs, where the red cicles mark the centered position on September 18 while the green circles mark the observed positions $\sim$ 22 days later \citep{Chomiuk2014b}.}
    \label{fig:nova}
\end{figure}

One of the most efficient methods to measure the mass ejected in novae consists in the analysis of radio emission emitted by the thermal bremsstrahlung free-free emission, in particular for the case of an impulsive ballistic shell ejection \citep[see, e.g.,][and Seaquist \& Bode in \citealt{Bode2008}]{Seaquist1977}. Novae have been observed in radio since the 1970s, with FH Ser being one of the first studied case \citep{Hjellming1970,Wade1971}, where \citet{Hjellming1979} succeeded in reproducing the phenomenology observed in radio with a spherical shell linear gradient model characterised by only three variable parameters. Other novae, like HR Del and V1500 Cyg can be very well explained by this ionized shell model \citep{Ennis1977}. Spatially-resolved radio observations have also allowed to resolve the ejecta which are characterised by a non-spherical emission and in several cases by a bi-polar geometry and lobes \citep{Taylor1988,Hjellming1995}. The first nova that was resolved in radio was V1974 Cyg \citep{Eyres1996}, which showed the presence of an evolving remnant only $\sim$80 days after the outburst. At these epochs, radio images already showed deviation from spherical symmetry. These observations have contributed to improve the theoretical models that were used to reproduce the observed shape of the nova remnants \citep{Lloyd1996}.  

Recent observations show a more complex structure. While a non-spherical geometry characterized by bi-polar ejecta, is observed in almost all cases (\citealt{Rupen2008,Sokoloski2008,Chomiuk2014b}, see also Fig.~\ref{fig:nova}) with some exceptions like v351 Pup \citep{Wendeln2017}, additional pieces of this jigsaw have been discovered. First, the presence of a non-thermal emission in few novae like RS Oph \citep{Rupen2008} V1723 Aql \citep{Weston2016a} and v5589 Sgr \citep{Weston2016b} and in the gamma-ray nova V1324 Sco \citep{Finzell2018}. These observations were explained as synchrotron radiation rising from shocks originating in the interaction between a faster wind with a oblated/toroidal slower components (\citealt{Vlasov2016}). Another interesting result consists in the presence of a delayed (with respect to the initial optical outburst) event of mass ejection observed in the recurrent nova T Pyx \citep{Chomiuk2014a}, characterised by an amount of mass larger than expected from recurrent novae \citep{Nelson2014}. This result may suggest that the impulsive ejection model of novae needs some adjustments. On the other side it confirms that radio observations are more efficient in measuring the masses of the nova ejecta than optical observations (e.g., the case of v458 Vul, \citealt{Roy2012a}). We address the reader to \citet{Roy2012b} for a paper which summarize the radio studies on novae. 
The future for radio studies on nova stars is very promising in view of the new facilities that will be come operational in the next decades such as: SKA \citep{O'Brien2015} or the next generation Very LArge Array \citep[ngVLA][]{Linford2018}. 

\section{Conclusions}

The new distances to galactic novae provided by GAIA DR2 have allowed us to re-analyze the numerous collections of data for novae in our Galaxy and in external systems. From our analysis the following results emerge: 

1) The analysis of the nova counts, their heights above the galactic plane and rates of decline of novae inside the Milky Way has allowed to derive, on more robust empirical basis, the notion of disk and bulge nova populations. Fig.~\ref{fig:13} shows strong evidence ($>3\sigma$) for the existence of a correlation between the rate of decline and the height above the galactic plane. This trend indicates that the fastest nova systems, which contain the most massive WDs, are concentrated close to the galactic plane. We then derive that the typical \emph{disk nova} is a fast evolving object whose lightcurve exhibits a bright peak at maximum M$_V\sim -9$ ($t_2\lesssim 13^d$ or $t_3\lesssim 20^d$) a smooth early decline and preferentially belongs to the He/N (or Fe II broad) spectroscopic class. The progenitor is preferentially located at small heights above the galactic plane ($\lesssim 150$pc) and the associated WD is rather massive, $M_{\rm WD}\gtrsim 1\,M_\odot$. The typical \emph{bulge nova} is a slow evolving object whose lightcurve exhibits a fainter peak at maximum, M$_V\sim -7.2$ mag ($t_2 \gtrsim 13^d$ or $t_3 \gtrsim 20^d$), often multiple maxima, dust formation, maximum standstill and belongs to the Fe II spectroscopic class. The progenitors extend up to 500 pc and beyond from the galactic plane and are likely related to a Pop II stellar population of the galactic thick-disk/bulge. Therefore they are associated (on average) with less massive WDs, $M_{\rm WD}\lesssim 1\,M_\odot$. The characteristic recurrence times between outbursts for nova systems formed by massive WDs are $\Delta T_{\rm rec} \lesssim 10^{3\div 4}$ years and their duty cycle lasts $\approx$ a few $\times 10^7$ years. For nova systems formed by light WDs $\Delta T_{\rm rec} \lesssim 10^{5\div 6}$ years and their lifetime activity lasts $\approx \times 10^9$ years.

2) Analysis of the MMRD relationship for LMC, M31 and Virgo novae confirm the existence of systematic differences in the distributions of the rates of decline of the respective nova populations. At least 70\% of novae in the LMC are bright and fast and therefore associated with massive WDs while fast novae in M31 are $\lesssim 25$\%. The two distributions are significantly different (K-S gives $\gtrsim$99\%).  The analysis of the spatial distributions suggests that nova populations in M31 and M81 are a mixture of disk and bulge novae, $\sim$30\%--70\% and 40\%--60\% respectively. Novae in the LMC and M33 originate from disk population whereas novae in M87 and NGC 1316 are obviously related to an old stellar population.

3) Analysis of the distributions of the rates of decline as a function of the absolute magnitude at maximum of LMC, M31 and Virgo, reveals the existence of a scant group of deviating objects. This sub-sample includes super-bright novae which extra energy input might be due to dissipation of orbital energy motion of the secondary engulfed in the nova ejecta during the nova common envelope stage. Also there are objects which fall below the $3\sigma$ strip. They can well be RNe that eject very light envelopes ($10^{-7}$--$10^{-8}\,M_{\odot}$),  hundreds or thousands times lighter than envelopes associated with CNe or patchy absorbed novae.  Outliers of the M31 MMRD found by \citet{Kasliwal2011} might well be classified into the previous classes of deviating objects without calling for the existence of a sub-class of faint and fast CNe that at the moment appear only predicted. \citet{Cao2012} observations of novae in M31 are fully consistent with the MMRD relation. If we consider novae observed in the nearest galaxies, such as M31 and LMC, we have a total number of 144 data points (see Tables~\ref{tab:app5}, \ref{tab:app6}, \ref{tab:app6b}, \ref{tab:app6c}), 12 of which (8\%) are obvious outliers of the MMRD at $3\sigma$ level (see Fig.~\ref{fig:MMRDMansi} and Fig \ref{fig:MMRDCao}). Finally we note that some outliers could be due to the observations of intracluster or intragalaxies novae. Recently \citet{Darnley2020} have convincingly demonstrated the existence of two novae belonging to the Andromeda giant stellar flow. These novae certainly belong to the M31 `` environment '' but they could possibly be located at slightly different distance from that which characterizes the main bulk of the M31 novae.

4) Comparison between the distances obtained for galactic novae with the MMRD relation and GAIA DR2 agree to better (or much better) than 30\%  then confirming that the MMRD relation can be used to estimate the distance to novae in the Milky Way with an interesting accuracy, contrary to recent claims. Most of the error (still) associated to distance measurements obtained via MMRD comes from the uncertainty in estimating: i) the absorption A$_V$ toward each nova; ii) the correct computation of the MMRD parameters, such as the observed magnitude at maximum and the rate of decline; iii) misuse of the MMRD to intrinsically faint objects such as RNe.  

5) GAIA DR2 distances have allowed a new absolute calibration of the MMRD relation. The comparison between the absolute and observed MMRD relations of M31, LMC and Virgo finds: $(V-M_v)_{\rm M31}=24.35\pm 0.31$ mag, $(V-M_v)_{\rm LMC}=18.44\pm 0.41$ mag, and $(V-M_v)_{\rm Virgo}=31.32\pm 0.44$ mag, which are in excellent agreement with the current distance measurements for Local Group and Virgo galaxies, respectively. Obviously, given the intrinsic dispersion of the MMRD it appears hard to measure the distances to extragalactic systems to better than $\sim$10\% which make novae not longer competitive distance indicators in the era of ``1-2\% high precision'' cosmology \citep{Pietrzyski2019, Riess2019}. This situation could improve in the next years if both the sample of ``calibrators'' (that is the number of novae inside the Milky Way with GAIA DR2 distance and characterized by good measurement of A$_V$) and the samples of novae discovered in external galaxies with reliable measurements of the MMRD parameters, will significantly increase.  

6) The nova rate in the Milky Way is currently known within a factor of two: 20--40 novae/yr. Recent new measurements ranging between 
a bulge rate of $13.8\pm 2.6$ novae/yr \citep{Mroz2015} and a global rate 50 novae/yr \citep{Shafter2017} do not help to clarify this discrepancy. On the other hand we have shown that novae in the Milky Way and M31 exhibit similar properties such as the distributions of the rates of decline and the spectroscopic classification (most MW and M31 novae belong to Fe II type). Therefore if most Galactic novae are produced in the bulge, it is unlikely that the global nova rate of the Milky Way might be significantly larger than it was measured by \citet{Mroz2016}. So, a rate of ${\rm R}_{\rm MW}=25^{+15}_{-5}$/yr seems an appropriate educated guess.   

7) A few sporadic observations have allowed to provide an estimate of the nova rate in Globular Clusters systems. An analysis of the present data finds a rate ranging between: a few $\times 10^{-4}$ and $10^{-2}$ novae ${\rm GC}^{-1}\,{\rm yr}^{-1}$. 

8) While there is a substantial agreement on LMC, SMC and M33 rates, we find a factor two discrepancy for M31, ${\rm R}_{\rm M31}=30$--$60/{\rm yr}$ and a full factor three uncertainty for the giant elliptical galaxies in Virgo cluster ${\rm R}_{\rm Virgo}=100-360$/yr. A simple explanation is that old nova survey suffer of severe incompleteness. For example the rather sparse surveys carried out by Rosino in the 1970s on M31 might suffer of such observational bias. Despite the control time has been duly estimated and the correction for field incompleteness have been taken into account, a factor $\sim$2 may be still hidden in fragmented and not homogeneous material collected over 25 years. Much more difficult is to explain a factor two incompleteness in the \citet{Arp1956} data. From his paper we can read some impressive numbers 

\begin{quote}
``Nearly 1000 plates were taken on 290 nights between June 1953 and January 1955. Observations were as continuous as possible\dots. Consequently of 30 novae discovered for only 5 were the maxima missed by more than one day.''
\end{quote}

This summary does not leave too much room for having missing one nova out of two, unless 50\% of novae in M31 are completely extinguished by absorption. We have adopted a rate of ${\rm R}_{\rm M31}=40^{+20}_{-10}$ novae/yr.  

As far as concern nova observations in Virgo ellipticals the most straight explanation is that nova surveys from ground based telescopes suffer of severe incompleteness. However, this hypothesis seems not supported by \citet{Ferrarese2003} survey on M49 which was carried out with HST. They found  a rate of $\sim 70\div 135$ novae/yr. Another possibility discussed in this paper is that the nova rate of M87 might be enhanced by the jet. The rate of this galaxy remains uncertain, in the range 100--300 novae/yr. 

The mismatch among the nova rates represents a good rationale for undertaking new dedicated surveys on novae in Local Group galaxies and beyond. Particularly, new surveys in MCs, M33 and M31 characterized by a cadence of few hours (rather than a few days) would be capable to discover the possible sub-class of ultra-fast novae predicted on theoretical grounds (e.g., \citealt{Yaron2005}). There is no reason to look for these intrinsically faint transients in distant galaxies. The use of telescopes characterized by large fields of view will help to minimize the number of correction factors that are normally applied to the observed rates.   

9) Whether or not the nova rate depends on the Hubble type of parent galaxy is a point not yet fully clarified.  One result about which all authors seem to converge is that the specific nova rate (i.e., the normalized nova rate per unit of luminosity of the host) exhibited by  blue low luminosity class galaxies, such as M33, LMC and SMC, is higher by a factor about 3 than observed in giant Spirals and Ellipticals. This result can be explained in terms of ``selection effect'' due to the different recurrence times of the nova outbursts combined with the different density of nova progenitors belonging to parent stellar populations of different ages. 

10) High resolution spectroscopic observations of galactic and MCs novae have produced two interesting results. i) In 2008, Williams and collaborators have discovered the so called THEAs (Transient Heavy Element Absorptions), which are signatures of a toroidal component of material surrounding the primary WD that is observed at low velocities during the nova outburst.
ii) \citet{Izzo2015} have observed for the first time $^7$Li in the ejecta of V1369 Cen and in the same years, \citet{Tajitsu2015,Tajitsu2016}, \citet{Molaro2016} and \citet{Izzo2018} have identified  also the presence of $^7$Be, which decays in $^7$Li with an half-life time decay of $\sim$ 53 days, finally confirming 1970s predictions that novae were Lithium factories.  

\begin{figure}[htb]
    \centering
    \includegraphics[width=\textwidth]{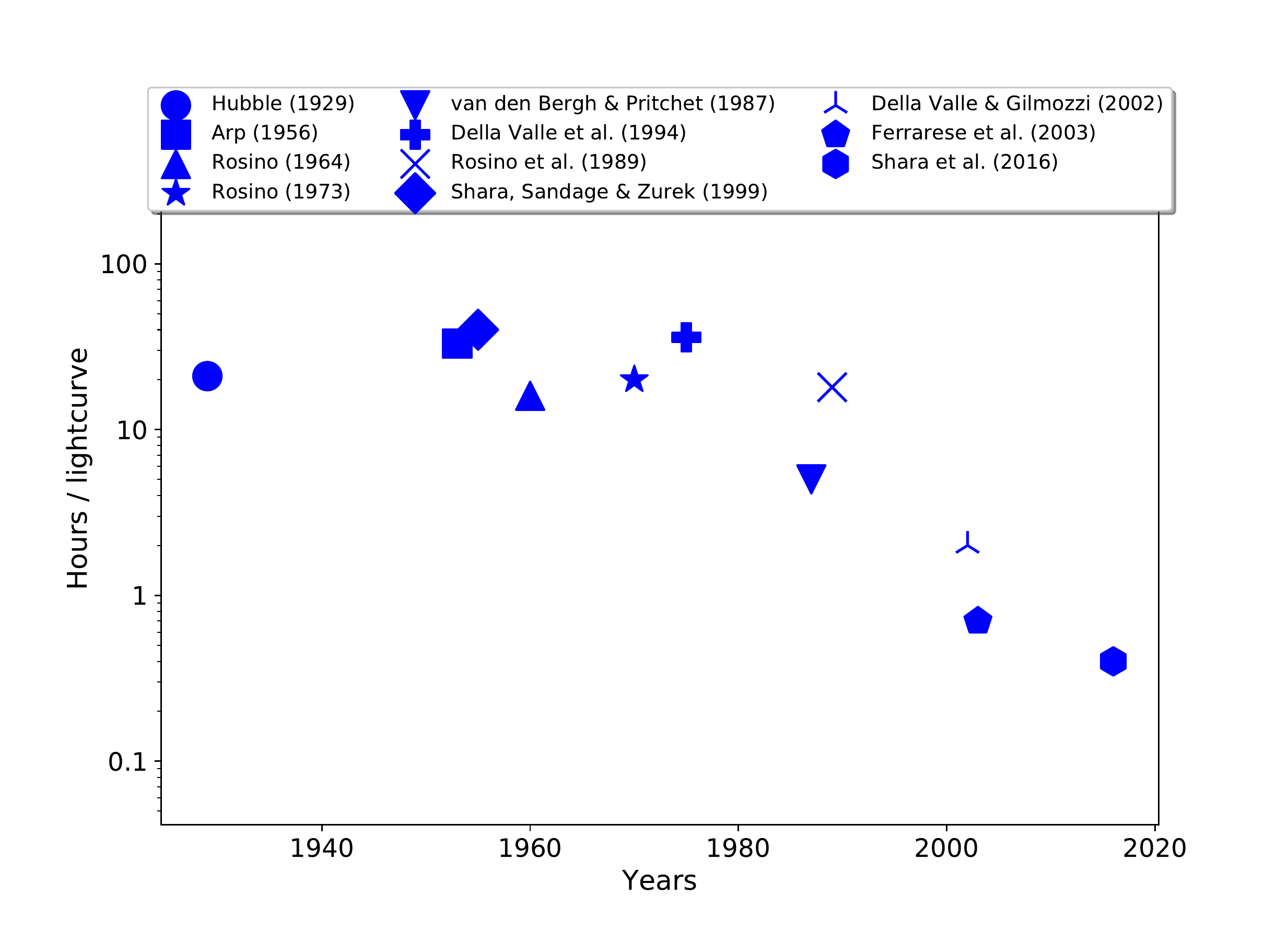}
    \caption{The nova ``time consuming'' index, that is expressed in terms of telescope time (computed in hours) needed to make the ``average'' nova lightcurve in  extragalactic systems. It is plotted vs.\ the epochs of the observations.}
    \label{fig:7}
\end{figure}

11) Novae and SNe-Ia connection. Several pieces of evidence point out that RNe and symbiotic stars should not be the major class of progenitors of Type Ia supernovae. Particularly, the radio observations of two dozen SNe-Ia with no positive detections and pre-explosion images of the SN-Ia 2011fe seem to exclude that stellar systems containing red giant donors are the main channel for SN-Ia production. Nevertheless the SD does not appear completely ruled out. Particularly, objects such as V407 Cyg and M31N 2008-12a that contain red giant donors seem to be promising candidates to explode as SNe-Ia.  We tentatively conclude that the both channels are open, although DD seems to be able to provide the main contribution.

12) Nova survey in extragalactic systems, despite their scientific interest, have not been so popular among astronomers and even less among the various time allocation committees spread over the world (although some remarkable exceptions do exist). The main reason for this is the unpredictable nature of nova events, which made nova observations considerably time consuming and difficult to schedule. However, two important facts may help to boost the interest for nova studies: i) the observing time requests at 2--4-m class telescopes have decreased in the last 20 years by about a factor 5 (survey on ESO telescopes, \citet{Patat2019} but see also \citealt{Patat2013}). Telescope of these sizes are suitable to study the nova properties in Local Group galaxies and therefore the general decrease for their quest can play in favor of proposal dedicated to novae; ii) large field of view telescopes, such as LSST that will be come operational in the next few years, have the potential to change dramatically the future of nova studies.  Figure \ref{fig:7} shows the trend of the ``time consuming'' index, that we have computed in terms of hours at the telescope needed to make the ``average'' nova lightcurve in extragalactic systems vs. the respective observing epochs.  
Obviously, this parameter depends on many factors (e.g., distance to the galaxy, size and field of view of the telescope, seeing, photographic plates or CCDs, ground-based vs. space instrumentation, intrinsic rate of events\dots), however we can recognize a general trend. From the 20--40 hours typical of the surveys carried out during the photographic era on nearby galaxies, we drop to $\sim $ 8 hours per lightcurve with 4-meter class telescope + CCD for galaxies in Virgo cluster and to 1.5h/lightcurve with 8-meter telescope + CCDs for a survey in giant ellipticals. This index has further decreased with the use of HST to less than 1h per lightcurve. In principle the study of extragalactic novae with future large size and large fields telescope such as LSST will dramatically decrease this ``index'' to a few minutes/lightcurve. It will be essentially limited by overheads (e.g., slewing telescope to target, CCD readout\dots) and the intrinsic frequency of nova outbursts of the observed galaxies.

%





\begin{acknowledgements}
The authors thank the referees for their constructive criticisms that have significantly improved the presentation and discussion of the data. We are also very grateful to Bob Williams for giving us valuable comments on the manuscript and to Roberto Gilmozzi and Tatiana Muraveva for helpful discussions about nova parallaxes.  We also  thank Mart\'in Guerrero, Patrick Schmeer and Ken Croswell for further insights. MDV is deeply grateful to Prof. Leonida Rosino for having introduced him to the realm of nova physics and to Hilmar Duerbeck for the many nights shared at the La Silla telescopes. MDV thanks ESO-Garching, where this manuscript was drafted for the creative and friendly atmosphere. LI was supported by grants from VILLUM FONDEN (project number 16599 and 25501). LI also acknowledges support from funding associated with Juan de la Cierva Incorporacion fellowship IJCI-2016-30940. 
\end{acknowledgements}

\clearpage

\appendix
\normalsize

\section{Tables}

\begin{table}[htb]
    \caption{The list of novae (Col. 1), their distance as measured from GAIA DR2 (Col. 2), the V magnitude at maximum (Col. 3) and at minimum (Col 4), the corresponding $t_2$ and $t_3$ values (Cols. 5 and 6) and the Galactic longitude $l$ and latitude $b$ (Cols. 7 and 8). Data from S18 and SG19}
    \label{tab:app1}
    \centering
    \begin{tabular}{lccccccc}
    \hline\hline
     Nova    & Distance & $V_{\max}$ & $V_{\min}$ & $t_2$ & $t_3$ & $l$ & $b$ \\
      & (kpc)   & (mag) & (mag) & (days) & (days) & (deg) & (deg)  \\
         \hline
V1494 Aql & 1.239$\pm$0.422 & 4.1 & 17.1 & 8 & 16 & 40.97 & -4.742 \\ 
V705 Cas & 2.157$\pm$0.799 & 5.7 & 16.4 & 33 & 67 & 113.7 & -4.096 \\ 
V842 Cen & 1.379$\pm$0.12 & 4.9 & 15.8 & 43 & 48 & 316.6 & 2.452 \\ 
V476 Cyg & 0.665$\pm$0.107 & 1.9 & 16.2 & 6 & 16 & 87.37 & 12.42 \\ 
V1974 Cyg & 1.631$\pm$0.261 & 4.3 & 21.0 & 19 & 43 & 89.13 & 7.819 \\ 
T Pyx & 3.185$\pm$0.607 & 6.4 & 18.5 & 32 & 62 & 257.2 & 9.707 \\ 
V732 Sgr & 1.795$\pm$0.458 & 6.4 & 16.0 & 65 & 75 & 2.53 & -1.188 \\ 
FH Ser & 1.06$\pm$0.112 & 4.5 & 16.8 & 49 & 62 & 32.91 & 5.786 \\ 
V382 Vel & 1.8$\pm$0.243 & 2.8 & 16.6 & 6 & 13 & 284.2 & 5.771 \\ 
NQ Vul & 1.08$\pm$0.169 & 6.2 & 17.2 & 21 & 50 & 55.36 & 1.29 \\ 
PW Vul & 2.42$\pm$1.337 & 6.4 & 16.9 & 44 & 116 & 61.1 & 5.197 \\ 
\hline
V603 Aql & 0.311$\pm$0.007 & -1.4 & 11.7 & 5 & 9 & 33.16 & 0.829 \\ 
T Aur & 0.857$\pm$0.038 & 4.2 & 15.2 & 80 & 100 & 177.14 & -1.7 \\ 
Q Cyg & 1.32$\pm$0.043 & 3.0 & 14.9 & 5 & 11 & 89.93 & -7.552 \\ 
HR Del & 0.932$\pm$0.031 & 3.8 & 12.1 & 160 & 230 & 63.43 & -13.97 \\ 
DN Gem & 1.318$\pm$0.146 & 3.5 & 15.5 & 17 & 37 & 184.02 & 14.71 \\ 
DQ Her & 0.494$\pm$0.006 & 1.35 & 14.3 & 67 & 94 & 73.15 & 26.44 \\ 
V446 Her & 1.308$\pm$0.13 & 3.0 & 16.9 & 6 & 15 & 45.409 & 4.708 \\ 
V533 Her & 1.165$\pm$0.044 & 3.0 & 14.5 & 25 & 44 & 69.19 & 24.273 \\ 
CP Lac & 1.129$\pm$0.054 & 2.1 & 15.5 & 5 & 10 & 102.14 & -0.837 \\ 
DI Lac & 1.569$\pm$0.051 & 4.6 & 14.7 & 20 & 40 & 103.107 & -4.855 \\ 
DK Lac & 2.296$\pm$0.391 & 5.7 & 16.8 & 40 & 60 & 105.237 & -5.352 \\ 
HR Lyr & 4.493$\pm$0.684 & 6.5 & 15.5 & 45 & 80 & 59.584 & 12.47 \\ 
GI Mon & 2.849$\pm$0.46 & 5.4 & 15.8 & 20 & 30 & 222.93 & 4.749 \\ 
V841 Oph & 0.805$\pm$0.018 & 4.2 & 13.5 & 54 & 130 & 7.621 & 17.779 \\ 
GK Per & 0.437$\pm$0.008 & 0.2 & 13.4 & 6 & 13 & 150.956 & -10.104 \\ 
RR Pic & 0.504$\pm$0.008 & 2.0 & 12.0 & 125 & 250 & 272.355 & -25.672 \\ 
CP Pup & 0.795$\pm$0.013 & 0.4 & 15.0 & 5 & 8 & 252.926 & -0.835 \\  
\hline
V1500 Cyg & 1.287$\pm$0.309 & 1.9 & 20.5 & 2 & 5 & 89.823 & -0.073 \\
       \hline
    \end{tabular}
\end{table}

\begin{table}[htb]
    \caption{The list of absortions $A_V$ used in this work. Data from Downes \& Duerbeck (2000), S18 and SG19. $^*$ The third column shows the extinction as provided by \citet{Hachisu2019}.}
    \label{tab:app2}
    \centering
    \begin{tabular}{lcc}
    \hline\hline
     Nova    & $A_V$ & $E(B-V)$$^*$\\
      &  (mag) & (mag) \\
         \hline
CI Aql & 2.6 & 1.0\\ 
V1494 Aql & 1.9 & - \\ 
V705 Cas & 1.3 & 0.45\\ 
V1330 Cyg & 2.1 & - \\ 
DN Gem & 0.5 & - \\ 
V446 Her & 1.1 & 0.40\\ 
T Pyx & 0.8 & - \\ 
V732 Sgr & 2.5 & - \\ 
V382 Vel & 0.4 & 0.25\\ 
Q Cyg & 0.81 & - \\ 
DI Lac & 0.81 & - \\ 
HR Lyr & 0.56 & - \\ 
GI Mon & 0.31 & - \\ 
V841 Oph & 1.36 & - \\ 
CP Pup & 0.79 & - \\ 
CT Ser & 0.02 & - \\ 
HR Del & 0.47 & - \\ 
NQ Vul & 2.1 & - \\ 
PW Vul & 1.73 & 0.57 \\ 
QU Vul & 1.89 & 0.55\\ 
V842 Cen & 1.73 & - \\ 
QV Vul & 1.26 & - \\ 
V1974 Cyg & 1.12 & 0.30\\ 
T Aur & 0.64 & - \\ 
GK Per & 0.95 & - \\ 
V603 Aql & 0.22 & - \\ 
V476 Cyg & 0.82 & - \\ 
RR Pic & 0.16 & - \\ 
DQ Her & 0.32 & - \\ 
CP Lac & 0.64 & - \\ 
DK Lac & 1.4 & - \\ 
V446 Her & 1.12 & 0.40\\ 
V533 Her & 0.0 & 0.038\\ 
V1229 Aql & 1.58 & -\\ 
FH Ser & 2.02 & - \\
       \hline
    \end{tabular}
\end{table}

\begin{table}[htb]
    \caption{The average expansion velocity (Cols. 3 \& 7) for the novae shown in Fig.~\ref{fig:MDV1998}. In the table are also reported the corresponding $t_2$ value (Col. 2 \& 6) and the spectral type (Col. 4 \& 8) for each nova considered. Data from Ozdonmez et al. (2018) and Della Valle \& Livio (1998).}
    \label{tab:app3}
    \centering
    \begin{tabular}{lccc|lccc}
    \hline\hline
     Nova    & $t_2$ & FWHM & Type & Nova    & $t_2$ & FWHM & Type \\
      & (days) & (km/s) & & & (days) & (km/s) &\\
         \hline
V356 Aql & 127 & 1140 & Fe II & V533 Her & 26 & 1600 & Fe II \\ 
V500 Aql & 20 & 2800 & He/N & V838 Her & 2 & 3300 & He/N \\  
V603 Aql & 4 & 4000 & He/N & CP Lac & 5 & 4200 & He/N \\
V1229 Aql & 18 & 2200 & Fe II & DK Lac & 19 & 2500 & Fe II \\ 
V1425 Aql & 27 & 1340 & Fe II & V382 Nor & 12 & 1850 & Fe II \\ 
V1494 Aql & 8 & 2470 & Fe II & V2264 Oph & 18 & 2300 & Fe II \\ 
V1663 Aql & 16 & 1900 & Fe II & V2615 Oph & 26.5 & 800 & Fe II \\
V1974 Cyg & 23 & 2300 & Fe II & V2672 Oph & 2.3 & 8000 & He/N \\ 
T Aur & 80 & 1200 & Fe II & GK Per & 6 & 4000 & He/N \\ 
V723 Cas & 263 & 600 & Fe II & RR Pic & 80 & 1600 & Fe II \\ 
V842 Cen & 43 & 760 & Fe II & CP Pup & 5 & 2000 & He/N \\ 
V868 Cen & 18 & 2300 & Fe II & V351 Pup & 10 & 2500 & Fe II \\
V1065 Cen & 11 & 2700 & Fe II & V745 Sco & 7 & 2000 & He/N \\ 
V407 Cyg & 5.9 & 2300 & He/N & V1187 Sco & 10 & 3000 & Fe II \\
V476 Cyg & 7 & 2500.0 &Fe II & V1280 Sco & 21.3 & 640 & Fe II \\
V1500 Cyg & 2.4 & 5000 & He/N & V1324 Sco & 25 & 2000 & Fe II \\
V1668 Cyg & 11 & 2400 & Fe II & V496 Sct & 59 & 1200 & Fe II \\ 
V1819 Cyg & 95 & 2742 & Fe II & FH Ser & 42 & 2100 & Fe II \\ 
V2467 Cyg & 8 & 950 & Fe II & V3890 Sgr & 10 & 2500 & He/N \\ 
V2468 Cyg & 9 & 2300 & Fe II-b & V5558 Sgr & 281 & 1000 & Fe II-b \\
V2491 Cyg & 4 & 4860 & He/N & V382 Vel & 6 & 2400 & He/N \\
HR Del & 150 & 1700 & Fe II & LV Vul & 21 & 2700 & Fe II \\ 
DQ Her & 67 & 1000 & Fe II & QU Vul & 22 & 1700 & Fe II \\
V446 Her & 5 & 2400 & He/N & QV Vul & 37 & 920 & Fe II \\ 
       \hline
    \end{tabular}
\end{table}

\begin{table}[htb]
    \caption{The distance measured from the nebular parallax method (Col. 2) by Downes \& Duerbeck (2000) compared with the recent GAIA DR2 measurements (Col. 3) as given in S18 and SG19, for a sample of novae (Col. 1).}
    \label{tab:app4}
    \centering
    \begin{tabular}{lcc}
    \hline\hline
     Nova    & Nebular distance & GAIA DR2 distance \\
      & $(kpc)$ & $(kpc)$ \\
         \hline
CP Pup & 1.70 & 0.81 \\ 
CT Ser & 1.40 & 2.77 \\ 
HR Del & 0.76 & 0.96 \\ 
NQ Vul & 1.16 & 1.08 \\ 
PW Vul & 1.80 & 2.42 \\ 
QU Vul & 1.75 & 1.79 \\ 
V842 Cen & 1.15 & 1.38 \\ 
QV Vul & 2.70 & 3.62 \\ 
V1974 Cyg & 2.00 & 1.63 \\ 
T Aur & 0.96 & 0.88 \\ 
GK Per & 0.45 & 0.44 \\ 
V603 Aql & 0.33 & 0.31 \\ 
V476 Cyg & 1.62 & 0.66 \\ 
RR Pic & 0.58 & 0.51 \\ 
DQ Her & 0.48 & 0.50 \\ 
CP Lac & 1.35 & 1.17 \\ 
DK Lac & 3.90 & 2.52 \\ 
V446 Her & 1.35 & 1.36 \\ 
V533 Her & 1.25 & 1.20 \\ 
V1229 Aql & 2.10 & 2.78 \\ 
FH Ser & 0.95 & 1.06 \\
       \hline
    \end{tabular}
\end{table}

\begin{table}[htb]
    \caption{The list of M31 novae used in this work. Data from Arp (1956), Rosino (1964), Rosino (1973) and Rosino et al. (1989). In this table novae are listed according to the original papers. $t_2$ values are reported in Cols. 2 \& 6. The photographic magnitudes $m_{pg}$ at peak and errors are reported in Cols. 3 \& 7 and 4 \& 8, respectively.}
    \label{tab:app5}
    \centering
    \begin{tabular}{lccc|lccc}
    \hline\hline
     Nova    & $t_2$ & $m_{pg}$ & $\sigma m_{pg}$ & Nova    & $t_2$ & $m_{pg}$ & $\sigma m_{pg}$\\
      &  (days) & (mag) & (mag) & &  (days) & (mag) & (mag)\\
         \hline
A01 & 2.0 & 15.73 & 0.2 & R24 & 10.0 & 16.1 & 0.3 \\ 
A02 & 2.0 & 15.67 & 0.2 & R27 & 28.6 & 16.9 & 0.3 \\ 
A03 & 5.3 & 15.9 & 0.3 & R28 & 8.0 & 14.85 & 0.1 \\ 
A04 & 10.0 & 18.2 & 0.1 & R29 & 6.7 & 16.0 & 0.3 \\ 
A05 & 11.8 & 15.9 & 0.2 & R30 & 11.8 & 16.2 & 0.1 \\ 
A06 & 8.7 & 16.0 & 0.1 & R32 & 4.0 & 15.75 & 0.3 \\ 
A07 & 13.3 & 15.9 & 0.1 & R33 & 8.0 & 16.2 & 0.2 \\ 
A08 & 8.7 & 16.0 & 0.2 & R34 & 16.7 & 17.0 & 0.3 \\ 
A09 & 11.5 & 16.0 & 0.3 & R36 & 80.0 & 16.9 & 0.3 \\ 
A10 & 6.9 & 16.0 & 0.2 & R37 & 18.2 & 16.8 & 0.2 \\ 
A12 & 11.1 & 16.1 & 0.1 & R38 & 18.2 & 16.9 & 0.1 \\ 
A13 & 26.0 & 17.0 & 0.1 & R41 & 100.0 & 17.9 & 0.2 \\ 
A14 & 12.3 & 16.2 & 0.2 & R42 & 28.6 & 16.85 & 0.2 \\ 
A15 & 15.9 & 16.4 & 0.2 & R43 & 9.1 & 16.3 & 0.1 \\ 
A16 & 12.8 & 16.7 & 0.2 & R46 & 16.0 & 16.6 & 0.2 \\ 
A17 & 29.0 & 17.2 & 0.2 & R48 & 11.0 & 17.8 & 0.1 \\ 
A18 & 34.5 & 17.5 & 0.2 & R49 & 22.2 & 16.35 & 0.3 \\ 
A19 & 28.6 & 17.6 & 0.1 & R52 & 16.0 & 16.35 & 0.1 \\ 
A20 & 33.3 & 17.2 & 0.1 & R53 & 50.0 & 16.95 & 0.1 \\ 
A21 & 26.7 & 17.4 & 0.1 & R55 & 29.9 & 17.0 & 0.3 \\ 
A22 & 29.9 & 17.6 & 0.2 & R57 & 9.1 & 14.95 & 0.1 \\ 
A23 & 43.5 & 17.4 & 0.2 & R58 & 16.0 & 16.5 & 0.3 \\ 
A24 & 33.9 & 17.8 & 0.1 & R59 & 20.0 & 17.3 & 0.3 \\ 
A25 & 32.8 & 17.6 & 0.1 & R60 & 15.4 & 16.4 & 0.3 \\ 
A26 & 46.5 & 18.0 & 0.1 & R66 & 1.5 & 16.4 & 0.2 \\ 
A27 & 83.3 & 17.8 & 0.3 & R67 & 15.4 & 16.2 & 0.1 \\ 
A28 & 105.3 & 17.8 & 0.2 & R76 & 14.3 & 16.1 & 0.2 \\ 
A29 & 117.6 & 18.0 & 0.1 & R77 & 9.1 & 15.9 & 0.1 \\ 
A30 & 117.6 & 18.1 & 0.2 & R78 & 25.0 & 16.9 & 0.3 \\ 
R04 & 15.4 & 16.4 & 0.2 & R80 & 58.8 & 17.4 & 0.1 \\ 
R05 & 3.3 & 15.6 & 0.2 & R85 & 16.7 & 15.7 & 0.1 \\ 
R06 & 14.3 & 16.5 & 0.1 & R86 & 33.3 & 17.3 & 0.3 \\ 
R07 & 5.0 & 16.2 & 0.2 & R89 & 20.0 & 16.8 & 0.3 \\ 
R09 & 13.3 & 16.8 & 0.3 & R95 & 6.7 & 16.2 & 0.3 \\ 
R12 & 40.0 & 17.6 & 0.1 & R98 & 9.5 & 16.0 & 0.3 \\ 
R13 & 16.7 & 16.8 & 0.3 & R100 & 10.0 & 16.0 & 0.3 \\ 
R14 & 25.0 & 17.3 & 0.3 & E104 & 50.0 & 18.4 & 0.3 \\ 
R15 & 18.2 & 16.3 & 0.3 & E105 & 15.4 & 16.9 & 0.3 \\ 
R16 & 12.5 & 16.25 & 0.3 & E106 & 44.4 & 17.6 & 0.3 \\ 
R17 & 11.8 & 16.7 & 0.3 & E109 & 10.0 & 16.3 & 0.2 \\ 
R18 & 22.2 & 17.1 & 0.1 & E112 & 20.0 & 17.3 & 0.2 \\ 
R19 & 80.0 & 17.9 & 0.2 & E113 & 5.0 & 16.0 & 0.3 \\ 
R20 & 22.2 & 16.9 & 0.1 & E117 & 100.0 & 17.9 & 0.3 \\ 
R21 & 22.2 & 16.7 & 0.3 & E118 & 28.6 & 16.9 & 0.2 \\ 
R23 & 13.3 & 16.85 & 0.3 & E120 & 40.0 & 17.4 & 0.2 \\ 
- & - & - & - & E129 & 22.2 & 16.8 & 0.2 \\
    \hline
    \end{tabular}
\end{table}

\begin{table}[htb]
    \caption{The list of LMC novae used in this work. Data from  Capaccioli et al. (1990), Shafter (2013). $t_2$ values, magnitudes at maximum and errors are reported in Col. 2, Col. 3 and 4, respectively.}
    \label{tab:app6}
    \centering
    \begin{tabular}{lccc}
    \hline\hline
     Nova    & $t_2$ & $V_{\rm obs}$ & $\sigma V_{\rm obs}$ \\
      & (days) & (mag) & (mag) \\
         \hline
LMC1926 & 117.6 & 12.0 & 0.3 \\ 
LMC1935 & 13.8 & 11.0 & 0.3 \\ 
LMC1936 & 17.7 & 10.5 & 0.3 \\ 
LMC1937 & 10.7 & 10.6 & 0.3 \\ 
LMC1948 & 58.8 & 12.82 & 0.3 \\ 
LMC1951 & 2.6 & 11.9 & 0.0 \\ 
LMC1968 & 2.0 & 10.4 & 0.4 \\ 
LMC1970B & 8.0 & 11.0 & 0.5 \\ 
LMC1971A & 15.7 & 11.77 & 0.1 \\ 
LMC1977B & 11.2 & 10.7 & 0.1 \\ 
LMC1978A & 3.5 & 9.75 & 0.6 \\ 
LMC1987 & 2.0 & 9.6 & 0.2 \\ 
LMC1988A & 22.2 & 10.85 & 0.1 \\ 
LMC1988B & 4.7 & 9.9 & 0.2 \\ 
LMC1990A & 3.4 & 9.5 & 0.3 \\ 
LMC1990B & 3.4 & 10.2 & 0.1 \\ 
LMC1991 & 5.7 & 9.0 & 0.1 \\ 
LMC1992 & 8.0 & 10.2 & 0.1 \\ 
LMC1995 & 11.8 & 10.4 & 0.3 \\ 
LMC2000 & 10.0 & 10.7 & 0.1 \\ 
LMC2002 & 13.3 & 10.1 & 0.2 \\ 
LMC2003 & 8.0 & 11.0 & 0.1 \\ 
LMC2004 & 11.8 & 10.9 & 0.1 \\ 
LMC2005 & 62.5 & 11.5 & 0.1 \\ 
LMC2009a & 5.4 & 10.6 & 0.1 \\ 
LMC2009b & 54.1 & 12.1 & 0.1 \\ 
LMC2012b & 10.0 & 11.5 & 0.1 \\        
         \hline
    \end{tabular}
\end{table}

\begin{table}[htb]
    \caption{The list of M31 novae taken from the analysis of \citet{Kasliwal2011} and \citet{Cao2012}. $t_2$ values and m$_V$ magnitudes are reported in Col. 2 and Col. 3, respectively.}
    \label{tab:app6b}
    \centering
    \begin{tabular}{lccc}
    \hline\hline
     Nova    & $t_2$ & $m_{V, \rm peak}$  \\
      & (days) & (mag)  \\
         \hline
M31-2007-10a & 8.6 & 17.5 \\
M31-2008-07b & 17.0 & 19.0 \\ 
M31-2008-8c & 26.3 & 17.1 \\ 
M31-2008-09a & 25.0 & 17.8 \\
M31-2008-09c & 16.6 & 16.6 \\ 
M31-2008-10b & 35.0 & 18.1 \\ 
M31-2008-11a & 7.5 & 17.0 \\
M31-2008-12b & 25.0 & 17.0 \\
\hline
M31-2009-08b & 45.0 & 17.86 \\ 
M31-2009-09a & 79.0 & 17.36 \\ 
M31-2009-10a & 20.0 & 17.05 \\ 
M31-2009-10b & 8.0 & 14.94 \\ 
M31-2009-11a & 22.0 & 17.33 \\ 
M31-2009-11b & 90.0 & 18.35 \\ 
M31-2009-11d & 25.0 & 17.82 \\ 
M31-2009-12a & 10.0 & 16.98 \\ 
M31-2010-06a & 40.0 & 16.98 \\ 
M31-2010-06b & 8.0 & 17.72 \\ 
M31-2010-06c & 19.0 & 17.57 \\ 
M31-2010-07a & 15.0 & 15.75 \\ 
M31-2010-09a & 40.0 & 17.52 \\ 
M31-2010-09b & 10.0 & 16.46 \\ 
M31-2010-10b & 57.0 & 17.14 \\ 
M31-2010-10a & 15.0 & 17.13 \\ 
M31-2010-10c & 60.0 & 17.54 \\
         \hline
    \end{tabular}
\end{table}

\begin{table}[htb]
    \caption{The list of M31 novae taken from the analysis of \citet{Hubble1929}. $t_2$ values and $m_{pg}$ magnitudes are reported in Col. 2 and Col. 3, respectively.}
    \label{tab:app6c}
    \centering
    \begin{tabular}{lccc}
    \hline\hline
     Nova    & $t_2$ & $m_{pg, \rm peak}$  \\
      & (days) & (mag)  \\
\hline
M31-Hubble-nova-1 & 100.0 & 16.7 \\ 
M31-Hubble-nova-2 & 22.2 & 15.8 \\ 
M31-Hubble-nova-3 & 16.7 & 15.8 \\ 
M31-Hubble-nova-4 & 80.0 & 17.1 \\ 
M31-Hubble-nova-5 & 33.3 & 16.2 \\ 
M31-Hubble-nova-6 & 66.7 & 16.9 \\ 
M31-Hubble-nova-7 & 15.4 & 16.2 \\ 
M31-Hubble-nova-8 & 100.0 & 17.5 \\ 
M31-Hubble-nova-9 & 28.6 & 16.8 \\ 
M31-Hubble-nova-10 & 20.0 & 16.3 \\ 
M31-Hubble-nova-11 & 25.0 & 16.4 \\ 
M31-Hubble-nova-12 & 11.1 & 16.3 \\ 
M31-Hubble-nova-13 & 36.4 & 17.3 \\ 
M31-Hubble-nova-14 & 28.6 & 17.2 \\ 
M31-Hubble-nova-15 & 15.4 & 15.3 \\ 
M31-Hubble-nova-16 & 33.3 & 17.4 \\ 
M31-Hubble-nova-17 & 133.3 & 17.5 \\
\hline
    \end{tabular}
\end{table}

\begin{table}[htb]
    \caption{The list of VIRGO novae used in this work. Data are from Pritchet \& van den Bergh (1987), Ferrarese et al. (2003) and Shara et al. (2016). $t_2$ values, magnitudes at maximum and errors are reported in Col. 2, Col. 3 and 4, respectively. $B_{\rm peak}$ mags have been transformed into $V_{\rm peak}$ mags by applying $\langle B-V \rangle = 0.23 \pm 0.06$ mag (van den Bergh \& Younger 1987).}
    \label{tab:app7}
    \centering
    \begin{tabular}{lccc}
    \hline\hline
     Nova    & $t_2$ & $V_{\rm peak}$ & $\sigma V_{\rm peak}$ \\
      & (days) & (mag) & (mag) \\
         \hline
NGC4365 & 18.2 & 23.97 & 0.30 \\ 
M49-1W & 20.0 & 22.51 & 0.11 \\ 
M49-2W & 12.6 & 21.01 & 0.57 \\ 
M49-3W & 54.0 & 24.13 & 0.21 \\ 
M49-5W & 10.5 & 23.81 & 0.12 \\ 
M49-2E & 33.2 & 23.11 & 0.10 \\ 
\hline
M49-4 & 52.0 & 23.33 & 0.13 \\ 
M49-5 & 7.5 & 23.10 & 0.52 \\ 
M49-6 & 19.3 & 23.33 & 0.28 \\ 
M49-7 & 11.7 & 22.72 & 0.36 \\ 
M49-9 & 37.2 & 23.74 & 0.22 \\ 
\hline
M87-1 & 15.2 & 21.84 &  -  \\ 
M87-2 & 11.2 & 22.25 &  -  \\ 
M87-3 & 2.0 & 22.64 &  -  \\ 
M87-4 & 7.7 & 22.74 &  -  \\ 
M87-5 & 17.1 & 22.99 &  -  \\ 
M87-6 & 11.2 & 23.18 &  -  \\ 
M87-7 & 9.3 & 23.21 &  -  \\ 
M87-8 & 3.7 & 23.37 &  -  \\ 
M87-10 & 22.3 & 23.51 &  -  \\ 
M87-11 & 28.9 & 23.57 &  -  \\ 
M87-12 & 33.1 & 23.58 &  -  \\ 
M87-13 & 6.7 & 23.67 &  -  \\ 
M87-14 & 31.5 & 23.74 &  -  \\ 
M87-16 & 32.6 & 23.77 &  -  \\ 
M87-17 & 30.4 & 23.81 &  -  \\ 
M87-19 & 9.0 & 23.90 &  -  \\ 
M87-20 & 36.3 & 23.94 &  -  \\ 
M87-21 & 29.9 & 24.07 &  -  \\
M87-23 & 8.0 & 24.16 &  -  \\
         \hline
    \end{tabular}
\end{table}

\clearpage

\bibliographystyle{spbasic}      

\end{document}